\documentclass[12pt, oneside, margin=1in]{book}
\usepackage{fancyhdr} 
\usepackage{blindtext}
\usepackage{lingmacros}
\usepackage{tree-dvips}
\usepackage[utf8]{inputenc}
\usepackage{arydshln}

\usepackage[numbers, sort&compress]{natbib}
\usepackage{multicol}
\usepackage{subfig}
\usepackage{amsfonts}
\usepackage{amssymb}
\usepackage{tikz}
\usepackage{amsthm}
\usepackage{amsmath}
\usepackage{bm}
\usepackage{thmtools,thm-restate}
\usepackage{enumitem}
\usepackage{pifont}
\usepackage{array}
\usepackage{float}
\usepackage{color}
\usepackage{caption}
\usepackage{multirow}
\usepackage{booktabs}
\usepackage{graphicx}
\usepackage{colortbl}
\usepackage{pythonhighlight}
\usepackage{comment}
\usepackage{lipsum}
\usepackage{bbm}
\usepackage{algorithm}
\usepackage[noend]{algpseudocode}
\usepackage{hyperref}
\hypersetup{citecolor=black}
 
\usepackage{xcolor}
\usepackage{anyfontsize}
\usepackage{makecell}
\usepackage{mathtools}
\usepackage[T1]{fontenc}
\usepackage[font=small,labelfont=bf,tableposition=top]{caption}
\usepackage[english]{babel}
\usepackage{rotating}
\usepackage{tablefootnote}
\usepackage{graphicx}
\usepackage[subtle]{savetrees}
\usepackage{adjustbox}
\usepackage{wrapfig}

\newcommand{\xingzhiswallow}[1]{{}}

\hypersetup{
colorlinks=true,
linkcolor=black,
citecolor=black,
urlcolor=black}
\usepackage{caption}

\DeclareMathOperator*{\argmin}{arg\,min}

\newtheorem{theorem}{Theorem}
\newtheorem{lemma}[theorem]{Lemma} 
\newtheorem{proposition}[theorem]{Proposition} 
\newtheorem{remark}[theorem]{Remark}

\newtheorem{definition}[theorem]{Definition}

\newcommand{\pluseq}{\mathrel{{+}{=}}}

\newcolumntype{P}[1]{>{\centering\arraybackslash}p{#1}}
\newcommand\todo[1]{\textcolor{red}{#1}}
\renewcommand{\paragraph}[1]{\textbf{\noindent{#1}}}

\newcommand*\colourcheck[1]{%
  \expandafter\newcommand\csname #1check\endcsname{\textcolor{#1}{\ding{52}}}%
}
\newcommand*\colourxmark[1]{%
  \expandafter\newcommand\csname #1xmark\endcsname{\textcolor{#1}{\ding{55}}}%
}
\colourcheck{blue}
\colourxmark{red}

\newcommand{\supp}{supp}

\newcommand{\mc}{\mathcal{}}

\newcommand{\R}{\mathbb R}

\newcolumntype{L}[1]{>{\raggedright\let\newline\\\arraybackslash\hspace{0pt}}m{#1}}
\newcolumntype{C}[1]{>{\centering\let\newline\\\arraybackslash\hspace{0pt}}m{#1}}
\newcolumntype{R}[1]{>{\raggedleft\let\newline\\\arraybackslash\hspace{0pt}}m{#1}}

\newcommand\versiontwo[1]{\textcolor{black}{#1}}
\newcommand{\remove}[1]{}

\usepackage[nottoc,numbib]{tocbibind}

\begin{document}
\pagenumbering{gobble}
\title{\bf{Analyzing the Evolution of Graphs and Texts}}
\vspace*{3\baselineskip}
\centerline{\bf{Analyzing the Evolution of Graphs and Texts}}
\vspace*{1\baselineskip}
\centerline{A Dissertation presented}
\vspace*{1\baselineskip}
\centerline{by}
\vspace*{1\baselineskip}
\centerline{\bf{Xingzhi Guo}}
\vspace*{1\baselineskip}

\centerline{to}
\vspace*{1\baselineskip}
\centerline{The Graduate School} \vspace*{1\baselineskip}
\centerline{in Partial Fulfillment of the} \vspace*{1\baselineskip} \centerline{Requirements} \vspace*{1\baselineskip}
\centerline{for the Degree of} \vspace*{1\baselineskip} \centerline{\bf{Doctor of Philosophy}} \vspace*{1\baselineskip}
\centerline{in}
\vspace*{1\baselineskip}
\centerline{\bf{Computer Science}} \vspace*{1\baselineskip} %\centerline{\bf{(Concentration - optional)}} \vspace*{2\baselineskip}
\centerline{Stony Brook University} \vspace*{2\baselineskip} \centerline{\bf{May 2023}}

\xingzhiswallow{
\newpage
\pagenumbering{gobble}
\vspace*{32\baselineskip}
% \centerline{\it{(include this copyright page only if you are selecting copyright through ProQuest, which is optional)}} \vspace*{1\baselineskip}
\centerline{Copyright by}
\centerline{Xingzhi Guo}
\centerline{2023}

}

\newpage
\pagenumbering{roman}
% \pagenumbering{arabic}

\setcounter{page}{2}
\centerline{\bf{Stony Brook University}} \vspace*{1\baselineskip}
\centerline{The Graduate School} \vspace*{2\baselineskip} \centerline{\textbf{Xingzhi Guo}} \vspace*{2\baselineskip}
\centerline{We, the dissertation committee for the above candidate for
the}
\vspace*{1\baselineskip}
\centerline{Doctor of Philosophy degree, hereby recommend}
\vspace*{1\baselineskip}
\centerline{acceptance of this dissertation}

\vspace*{1\baselineskip}
\centerline{\bf{Steven S. Skiena}}
\centerline{\bf{Distinguished Teaching Professor, Department of Computer Science, Stony Brook University}}
\vspace*{1\baselineskip}
\centerline{\bf{Nick Nikiforakis}} 
\centerline{\bf{Associate Professor, Department of Computer Science, Stony Brook University}}
\vspace*{1\baselineskip}
\centerline{\bf{Paul Fodor}} 
\centerline{\bf{Associate Professor of Practice, Department of Computer Science, Stony Brook University}}
\vspace*{1\baselineskip}
\centerline{\bf{Jason J. Jones}} 
\centerline{\bf{Associate Professor, Department of  Sociology, Stony Brook University}}
\vspace*{2\baselineskip}
% \centerline{\bf{Type the outside member's name last. Include discipline and affiliation.}}
% \vspace*{2\baselineskip}
\centerline{This dissertation is accepted by the Graduate School} \vspace*{2\baselineskip}
\centerline{Celia Marshik}
% \vspace*{2\baselineskip}
\centerline{Dean of the Graduate School}

\newpage
\centerline{Abstract of the Dissertation} \vspace*{1\baselineskip} \centerline{\bf{Analyzing the Evolution of Graphs and Texts}} \vspace*{1\baselineskip}
\centerline{by}
\vspace*{1\baselineskip}
\centerline{\bf{Xingzhi Guo}}
\vspace*{1\baselineskip}
\centerline{\bf{Doctor of Philosophy}}
\vspace*{1\baselineskip}
\centerline{in}
\vspace*{1\baselineskip}
\centerline{\bf{Computer Science}}
\vspace*{1\baselineskip}
% \centerline{\bf{(Concentration - optional)}}
% \vspace*{1\baselineskip}
\centerline{Stony Brook University}
\vspace*{1\baselineskip}
\centerline{\bf{2023}}
\vspace*{1\baselineskip}
% Our world keeps changing and being captured by abundant of data, especially in text and graphs. 
With the recent advance of representation learning algorithms on graphs (e.g., DeepWalk/GraphSage) and natural languages (e.g., Word2Vec/BERT) , the state-of-the art models can even achieve human-level performance over many downstream tasks, particularly for the task of node and sentence classification.
 However, most algorithms focus on large-scale models for static graphs and text corpus without considering the inherent dynamic characteristics or discovering the reasons behind the changes.
 
This dissertation aims to efficiently model the dynamics in graphs (such as social networks and citation graphs) and  understand the changes in texts (specifically news titles and personal biographies).
To achieve this goal, we utilize the renowned Personalized PageRank algorithm to create effective dynamic network embeddings for evolving graphs. Our proposed approaches significantly improve the running time and accuracy for both detecting network abnormal intruders and discovering entity meaning shifts over large-scale dynamic graphs. 
For text changes, we analyze the post-publication changes in news titles to understand   the intents behind the edits and discuss the potential impact of titles changes from information integrity perspective.
Moreover, we investigate self-presented occupational identities in Twitter users' biographies over five years, investigating job prestige and demographics effects in how people disclose jobs, quantifying over-represented jobs and their transitions  over time. 
% For future work, we envision faster algorithms for dynamic graph embeddings, a more effective job title categorization pipeline, a more comprehensive taxonomy of text edits and more powerful edits representation models to extract deeper insights.

% \newpage
% \centerline{\bf{Dedication Page}}
% \vspace*{4\baselineskip}
% This page is optional.
% \newpage
% \centerline{\bf{Frontispiece}}
% \vspace*{4\baselineskip}
% The frontispiece is generally an illustration, and is an optional page.

\newpage
%\centerline{\bf{Table of Contents}}
\hypersetup{linkcolor=black}
\tableofcontents 
\newpage

\pagenumbering{arabic}
\chapter{Introduction}

% \section{Motivation}
% A general motivation... 

% \section{Outline}
% We first introduce the dynamic graph embedding algorithms and its extension for anomaly detection, then present the studies about changing text in online news headlines and personal biographies over social network.

\section[Word and Graph Embeddings ]{Representation Learning for Languages and Graphs}

% Modern machine learning techniques are built upon  numerical dense vectors. From Support Vector Machine to deep neural networks, most predictive and clustering algorithms perform on dense low dimension vectors, typically ranging from 60 to 300 dimensions.  
% Starting from image classification models where each image is naturally represented by a dense matrix of pixels, numerous endeavors have been proposed to represent discrete symbols (English words and graph vertices). However, it's nontrivial because different from images which can be directly digitized by camera sensors, languages and graphs are constructed from discrete symbol, therefore challenging to be directly captured by dense low-dimensional vectors. 
Modern machine learning techniques rely heavily on dense numerical vectors. Predictive and clustering algorithms, including Support Vector Machines and deep neural networks, operate on low-dimensional vectors that typically range from 60 to 512 dimensions. While image classification models naturally use dense matrices of pixels to represent images, discrete symbols, such as English words and graph vertices, present a more challenging problem. Unlike images that can be directly digitized by camera sensors, languages and graphs are constructed from discrete symbols, which makes it difficult to directly capture using low-dimensional vectors.

In recent years, there has been significant progress in the field of natural language processing (NLP) and graph representation learning, which involves learning meaningful and low-dimensional dense vector representations of words and vertices. 
These vector representations are useful for a variety of NLP tasks as well as graph tasks, such as sentiment analysis, text summarization, node classification and link prediction.  We briefly introduce the two seminal works in word and node embeddings as follows.

% A common approach is to use neural networks, specifically deep learning models, to learn these vector representations by training them on large text corpora. 
% These models, such as word2vec and GloVe, have proven to be highly effective in generating vector representations of words that capture semantic and syntactic relationships between them.

\subsection{Word Embeddings: Represent words by vectors}
% Since the debut of word2vec, language embeddings or word embeddings have been the most visible advance in artificial intelligence. The key idea of word embeddings is to assign a discrete symbolic unit (e.g., a English word) into a low dimensional space (typically ranging from 60 to 300 dimensions) so that the close vectors 
% \footnote{The distance function could be either $\ell_p$ or cosine distance)
% would share similar semantic meanings. 
The advent of Word2Vec\cite{mikolov2013distributed}  has brought about significant advancements in artificial intelligence, specifically in the area of language embeddings or word embeddings. Word embeddings involve assigning a discrete symbolic unit (such as an English word) to a low-dimensional space that typically ranges from 60 to 300 dimensions. The aim is to ensure that vectors that are closely located in this space share similar semantic meanings.

\begin{figure}[ht]
    \centering
    \includegraphics[width=0.95\textwidth]{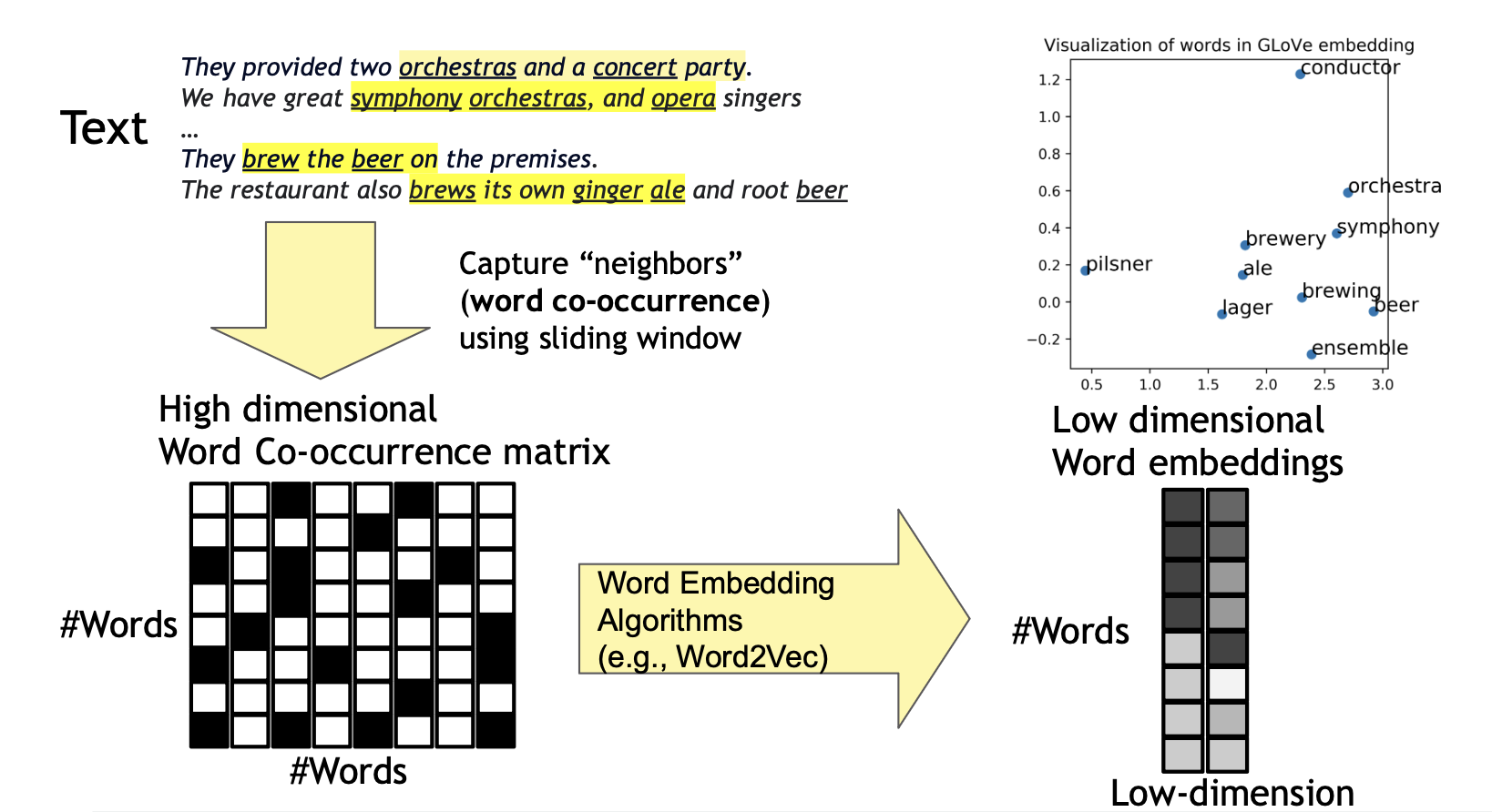}
    \caption{The illustration of word embedding algorithms. }
    \label{fig:word-emb}
\end{figure}

% By preserving the human-sense of lexical meaning within the low-dimensional vector, many downstream NLP tasks (e.g., sentiment classification, textual entailment prediction) can be tackled with human-level performance, especially after pre-training large language models (e.g., BERT or GPT). 
Word embeddings preserve the human-sense of lexical meaning within the low-dimensional vector. This property makes them useful for many downstream NLP tasks, such as sentiment classification and textual entailment prediction, which can be tackled with human-level performance. This is particularly true after pre-training large language models such as BERT or GPT.

% Technically, the word embeddings can be formulated as two stages as illustrated in Figure \ref{fig:word-emb}. For example, Word2Vec uses sliding windows to count which words are near or co-occurred with the center word within a fix context length (e.g. 5 words from left and right). 
Word embeddings can be formulated in two stages, as illustrated in Figure \ref{fig:word-emb}. The first stage, used in Word2Vec, involves counting which words are near or co-occurred with the center word within a fixed context length (e.g., 5 words from its left and right). The resulting co-occurrence matrix is large, with a size square of the vocabulary, and sparse. 
In the second stage, several self-supervised tasks are created to pre-train the neural network, leading to the low-dimensional word vectors for predicting context words given the center word (Skip-gram) or vice versa (Continuous bag-of-words).
In practice, these word embeddings demonstrate superior performance for a wide range of NLP tasks, from word similarity measurement to sentence sentiment classification.

% The resulted co-occurrence matrix is large (with the square of vocabulary size) and sparse.  In order to reduce the dimension and densify the vectors, several self-supervised tasks are created to pre-train the neural network, leading to the low-dimensional word vectors for predicting context words given center word (Skip-gram) or vice versa (Continuous bag-of-word). In practice, those word embeddings demonstrate supreme performance for wide range of NLP tasks, from word similarity measurement to sentence sentiment classification. 

\subsection{Graph Embeddings: Represent vertices by vectors}
% Network or graph embeddings is conceptually similar to word embedding, whereas it represents discrete vertices using low-dimensional vectors. 
% The seminal network embedding algorithm, DeepWalk, proposed a wise problem formulation, converting a graph into node sequences path by random walk and leading graph learning community into deep learning era. 
Network or graph embeddings are a concept that is similar to word embeddings. It involves representing discrete vertices using low-dimensional vectors. The seminal network embedding algorithm, DeepWalk \cite{perozzi2014deepwalk}, proposed a wise problem formulation that converted a graph into node sequences path by random walk, thereby leading the graph learning community into the deep learning era.

% As illustrated in Figure \ref{fig:network-emb}, DeepWalk use random walk starting from each vertex and each path ends with a fixed walk length (e.g., 80 vertex), simulating paths or \textit{"sentence"} consisting of vertices.  Afterward, the exactly same processes are applied as same as Word2Vec with context length and self-supervised pretaining. The resulted vertex embeddings are practically useful for many graph tasks, for example, node classification and missing link prediction.

\begin{figure}[ht]
    \centering
    \includegraphics[width=0.95\textwidth]{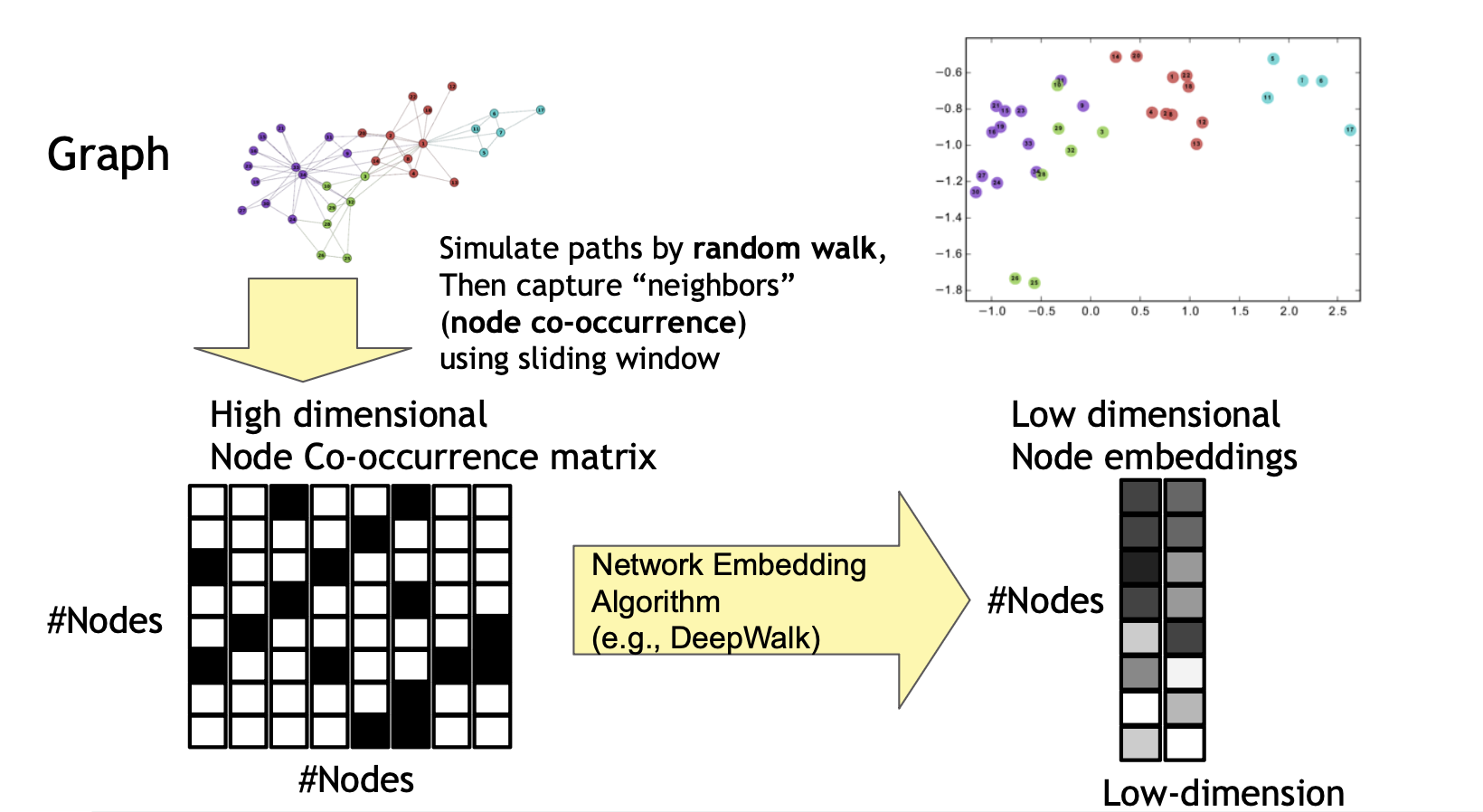}
    \caption{The illustration of network embeddings.}
    \label{fig:network-emb}
\end{figure}

As illustrated in Figure \ref{fig:network-emb}, DeepWalk utilizes random walk starting from each vertex to sample paths, where each path ends with a fixed walk length (e.g., 80 vertices) and simulates \textit{"sentences"} consisting of vertices. Afterwards, the same processes are applied as in Word2Vec with context length and self-supervised pre-training. The resulting node embeddings are practically useful for many graph tasks, such as node classification and missing link prediction.

\subsection{Our Motivations}

% These aforementioned graph algorithms only focus on static graphs while the real-world graphs are highly dynamic with structure changes over time. 
% For example, Wikipedia volunteers tirelessly update the Wiki articles as well as hyper-links to reflect the latest events happened every day. However, most network embedding algorithms were design for static graphs and can not efficiently adapt to the dynamic graph. 
The aforementioned graph algorithms only focus on static graphs, while real-world graphs are highly dynamic, with structural changes occurring over time. For instance, Wikipedia volunteers tirelessly update Wiki articles, as well as hyperlinks, to reflect the latest events that happen every day. However, most network embedding algorithms were designed for static graphs and cannot efficiently adapt to the dynamic graph's changing structure. 

Moreover, real-world graphs are usually massive, with millions of nodes, while real-world applications often focus on a few thousand nodes for clustering or prediction. Most embedding algorithms, such as DeepWalk, have to calculate whole embeddings but only use a small subset of them, which is computationally inefficient or even prohibited given the graph's large scale. Based on these two observations, we are motivated to propose three efficient graph learning frameworks, specifically for dynamic subset node embeddings.

Meanwhile, we are interested in understanding the expected and unexpected changes in texts. For example, we have monitored news headlines on major U.S. news outlets and discovered post-publication headline changes. Moreover, we have analyzed the changes in Twitter user biographies over six years to study the evolution of self-identity over time. Unlike conventional NLP tasks that predict labels given benchmarks, our aim is to utilize the latest NLP techniques to gain societal insights by examining text changes.

In the following sections, we introduce our proposed dynamic graph embedding algorithms, their extensions for graph anomaly detection applications, and an efficient graph neural network framework. Additionally, we present our studies on changing text in online news headlines and personal biographies.

% Additionally, real-world graphs are usually massive with millions of nodes while real-world applications often focus on a few thousand nodes. Most of the embedding algorithms (e.g., DeepWalk) have to calculate whole embeddings but only use a small subset of them, which is computationally inefficient or even prohibit given the large graph scale.  From these two observations, we are motivated to propose three efficient graph learning frameworks , specifically for dynamic subset node embeddings.

% % Meanwhile, we are interested in understanding the text changes that we do or do not  expect to see. 
% For example, we have monitored the news headlines on the major U.S news outlets and discovered post-publication headline changes. Meanwhile, we have looked into the biographies changes of Twitter users over 6 years, from which we study the self-identity evolution over time. Different from conventional NLP tasks to predict labels given benchmarks, here we aim to use the latest NLP techniques to gain societal insights by looking into the text changes. 

% In the rest of the section, we begin by introducing dynamic graph embedding algorithms, graph neural network models and their extensions for anomaly detection. 
% Following this, we present our studies on changing text in online news headlines and personal biographies.

\section[Subset Node Embeddings over Dynamic Graphs]{Subset Node Embeddings over Dynamic Graph}
 Dynamic graph representation learning is a task to learn node embeddings over dynamic networks, and has many important applications, including knowledge graphs, citation networks to social networks. Graphs of this type are usually large-scale but only a small subset of vertices are related in downstream tasks. Current methods are too expensive to this setting as the complexity is at best linear-dependent on both the number of nodes and edges.
 
 In Chapter \ref{chapter:dynppe} , we propose a new method, namely Dynamic Personalized PageRank Embedding (\textsc{DynamicPPE}) for learning a target subset of node representations over large-scale dynamic networks. Based on recent advances in local node embedding and a novel computation of dynamic personalized PageRank vector (PPV), \textsc{DynamicPPE} has two key ingredients: 1) the per-PPV complexity is $\mathcal{O}(m \bar{d} / \epsilon)$ where $m,\bar{d}$, and $\epsilon$ are the number of edges received, average degree, global precision error respectively. Thus, the per-edge event update of a single node is only dependent on $\bar{d}$ in average; and 2) by using these high quality PPVs and hash kernels, the learned embeddings have properties of both locality and global consistency. These two make it possible to capture the evolution of graph structure effectively.
 Experimental results demonstrate both the effectiveness and efficiency of the proposed method over large-scale dynamic networks. We apply \textsc{DynamicPPE} to capture the embedding change of Chinese cities in the Wikipedia graph during this ongoing COVID-19 pandemic \footnote{\textcolor{blue}{\url{https://en.wikipedia.org/wiki/COVID-19_pandemic}}}. Our results show that these representations successfully encode the dynamics of the Wikipedia graph.

\section[Local Anomaly Detection over
 Dynamic Graphs]{Subset Node Anomaly Detection over Dynamic Graph}
Tracking a targeted subset of nodes in an evolving graph is important for many real-world applications. Existing methods typically focus on identifying anomalous edges or finding anomaly graph snapshots in a stream way. However, edge-oriented methods cannot quantify how individual nodes change over time while others need to maintain representations of the whole graph all time, thus computationally inefficient.

Chapter \ref{chapter:DynAnom} proposes \textsc{DynAnom}, an efficient framework to quantify the changes and localize per-node anomalies over large dynamic weighted-graphs. Thanks to recent advances in dynamic representation learning based on Personalized PageRank, \textsc{DynAnom} is 1) \textit{efficient}: the time complexity is linear to the number of edge events and independent on node size of the input graph; 2) \textit{effective}: \textsc{DynAnom} can successfully track topological changes reflecting real-world anomaly; 3) \textit{flexible}: different type of anomaly score functions can be defined for various applications. Experiments demonstrate these properties on both benchmark graph datasets and a new large real-world dynamic graph. Specifically, an instantiation method based on \textsc{DynAnom} achieves the accuracy of 0.5425 compared with 0.2790, the best baseline, on the task of node-level anomaly localization while running 2.3 times faster than the baseline. We present a real-world case study and further demonstrate the usability of \textsc{DynAnom} for anomaly discovery over large-scale graphs.

\section[Efficient PPR-based GNN over Dynamic Graphs]{Efficient Contextual Node Representation Learning
over Dynamic Graphs}

Real-world graphs grow rapidly with edge and vertex insertions over time, motivating the problem of efficiently maintaining robust node representation over evolving graphs. 
Recent efficient GNNs are designed to decouple recursive message passing from the learning process, and favor Personalized PageRank (PPR) as the underlying feature propagation mechanism.  
However, most PPR-based GNNs are designed for static graphs, and efficient PPR maintenance remains as an open problem. 
Further, there is surprisingly little theoretical justification for the choice of PPR, despite its impressive empirical performance. 

In Chapter \ref{chapter:GoPPE}, we are inspired by the recent PPR formulation as an explicit $\ell_1$-regularized optimization problem and propose a unified dynamic graph learning framework based on sparse node-wise attention.  
We also present a set of desired properties to justify the choice of PPR in STOA GNNs, and serves as the guideline for future node attention designs.  
Meanwhile, we take advantage of the PPR-equivalent optimization formulation and employ the proximal gradient method (ISTA) to improve the efficiency of PPR-based GNNs by 2-5 times.  
Finally, we instantiate a simple-yet-effective model (\textsc{GoPPE}) with robust positional encodings by maximizing PPR previously used as attention.
The model performs comparably to or better than the STOA baselines and greatly outperforms when the initial node attributes are noisy during graph evolution, demonstrating the effectiveness and robustness of \textsc{GoPPE}.
% due to the great sparsity effectiveness of \textsc{GoPPE}.
%For the future, we envision a faster PPR solver under the optimization formulation and investigate better GNNs designs for dynamic graphs. 

\section[Analysis of News Title Changes]{Post-publication Modification in News Headlines}
    Digital media (including websites and online social networks) facilitate the broadcasting of news via flexible and personalized channels. Unlike conventional newspapers which become ``read-only'' upon publication, online news sources are free to arbitrarily modify news headlines \emph{after} their initial release. The motivation, frequency, and effect of post-publication headline changes are largely unknown, with no offline equivalent from where researchers can draw parallels.
    
    In Chapter \ref{chapter:title-change}, we collect and analyze over 41K pairs of altered news headlines by tracking $\sim$411K articles from major US news agencies over a six month period (March to September 2021), identifying that 7.5\% articles have at least one post-publication headline edit with a wide range of types, from minor updates, to complete rewrites.  
    We characterize the frequency with which headlines are modified and whether certain outlets are more likely to be engaging in post-publication headline changes than others. 
    We discover that 49.7\% of changes go beyond minor spelling or grammar corrections, with 23.13\% of those resulting in drastically disparate information conveyed to readers.
    Finally, to better understand the interaction between post-publication headline edits and social media, we conduct a temporal analysis of news popularity on Twitter. We find that an effective headline post-publication edit should occur within the first ten hours after the initial release to ensure that the previous, potentially misleading, information does not fully propagate over the social network.

\section[Analysis of Biography Changes]{Information and Evolution of Personal Biographies}
Occupational identity concerns the self-image of an individual's affinities and socioeconomic class, and directs how a person should behave in certain ways. Understanding the establishment of occupational identity is important to study work-related behaviors. However,  large-scale quantitative studies of occupational identity are difficult to perform due to its indirect observable nature. 
    But profile biographies on social media contain concise yet rich descriptions about self-identity. Analysis of these self-descriptions provides powerful insights concerning how people see themselves and how they change over time. 
    
    In this paper, we present and analyze a longitudinal corpus recording the self-authored public biographies of 51.18 million Twitter users as they evolve over a six-year period from 2015-2021.
    In particular, we investigate the demographics and
    social approval (e.g., job prestige and salary) effects in how people self-disclose occupational identities, quantifying over-represented occupations as well as the occupational transitions w.r.t. job prestige over time. 
    We show that males are more likely to define themselves by their occupations than females do, and that self-reported jobs and job transitions are biased toward more prestigious occupations.
    We also present an intriguing case study about how self-reported jobs changed amid COVID-19 and the subsequent \textit{"Great Resignation"} trend with the latest full year data in 2022.
    These results demonstrate that social media biographies are a rich source of data for quantitative social science studies, allowing unobtrusive observation of the intersections and transitions obtained in online self-presentation.
\section{List of Publications and Submissions}

\paragraph{Published}

\begin{enumerate}
    \item Giorgian Borca-Tasciuc, \textbf{Xingzhi Guo}, Stanley Bak, and Steven Skiena. "Provable Fairness for Neural Network Models  using Formal Verification", \textit{European Workshop on Algorithmic Fairness (EWAF 2023)}, Extended abstract \cite{borca2022provable}. 

    \item Zhuoyi Lin, Lei Feng, \textbf{Xingzhi Guo}, Rui Yin, Chee Keong Kwoh, and Chi Xu. COMET: Convolutional Dimension Interaction for Deep Matrix Factorization.  In, \textit{ACM Transactions on Intelligent Systems and Technology (TIST), 2023} \cite{lin2023comet}

    \item \textbf{Xingzhi Guo}, Baojian Zhou, and Steven Skiena. Subset Node Anomaly Tracking over Large Dynamic Graphs. In \textit{Proceedings of the 28th ACM SIGKDD Conference on Knowledge Discovery \& Data Mining (KDD '22)}., New York, NY, USA.\cite{guo2022subset} %https://doi.org/10.1145/3534678.3539389

    \item \textbf{Xingzhi Guo}, Brian Kondracki, Nick Nikiforakis, and Steven Skiena. 2022. Verba Volant, Scripta Volant: Understanding Post-publication Title Changes in News Outlets. In \textit{ Proceedings of the ACM Web Conference 2022 (WWW '22)}. ACM, New York, NY, USA. \cite{guo2022verba}%https://doi.org/10.1145/3485447.3512219
    
    \item Syed Fahad Sultan, \textbf{Xingzhi Guo}, and Steven Skiena. 2022. Low-dimensional genotype embeddings for predictive models. In \textit{Proceedings of the 13th ACM International Conference on Bioinformatics, Computational Biology and Health Informatics (BCB '22)}, ACM, New York, NY, USA, Article 52. \cite{sultan2022low} %https://doi.org/10.1145/3535508.3545507

    \item \textbf{Xingzhi Guo}, Baojian Zhou, and Steven Skiena. 2021. Subset Node Representation Learning over Large Dynamic Graphs. In \textit{Proceedings of the 27th ACM SIGKDD Conference on Knowledge Discovery \& Data Mining (KDD '21)}. ACM, New York, NY, USA.  \cite{chen2023accelerating}
    %https://doi.org/10.1145/3447548.3467393

    \item K. Gillespie, I. C. Konstantakopoulos, \textbf{X. Guo}, V. T. Vasudevan and A. Sethy, "Improving Device Directedness Classification of Utterances With Semantic Lexical Features," \textit{ICASSP 2020 - 2020 IEEE International Conference on Acoustics, Speech and Signal Processing (ICASSP)}, 2020. \cite{gillespie2020improving}

\end{enumerate}

\paragraph{In Submission}

\begin{enumerate}
    \item \textbf{Xingzhi Guo}, Baojian Zhou, and Steven Skiena. Efficient Contextual Node Representation Learning over Dynamic Graphs.

    \item Zhen Chen, \textbf{Xingzhi Guo}, Baojian Zhou, Deqing Yang, and Steven Skiena, Accelerating Personalized PageRank Vector Computation.  \cite{chen2023accelerating}

    \item \textbf{Xingzhi Guo}, Dakota Handzlik, Jason J. Jones, and Steven Skiena. Biographies and Their Evolution: Occupational Identities over Twitter.\cite{guo2024evolution}

    \item \textbf{Xingzhi Guo} and Steven Skiena. Hierarchies over Vector Space: Orienting Word and Graph Embeddings. \cite{guo2022hierarchies}

    \item \textbf{Xingzhi Guo}, Baojian Zhou, Haochen Chen, Sergiy Verstyuk, and Steven Skiena. Why Do Embedding Spaces Look as They do?

\end{enumerate}

\chapter{Personalized PageRank}

% \section[Preliminaries for PageRank]{Personalized PageRank}
\label{chapter-prelim}

Personalized PageRank (PPR) \cite{page1999pagerank} is a renowned algorithm that forms the foundation of the proposed algorithms in this thesis. In this section, we present the standard notations and preliminaries to provide an overview of the recent advancements and the current state-of-the-art PPR research.

\begin{figure}[ht]
    \centering
    \includegraphics[width=0.99\textwidth]{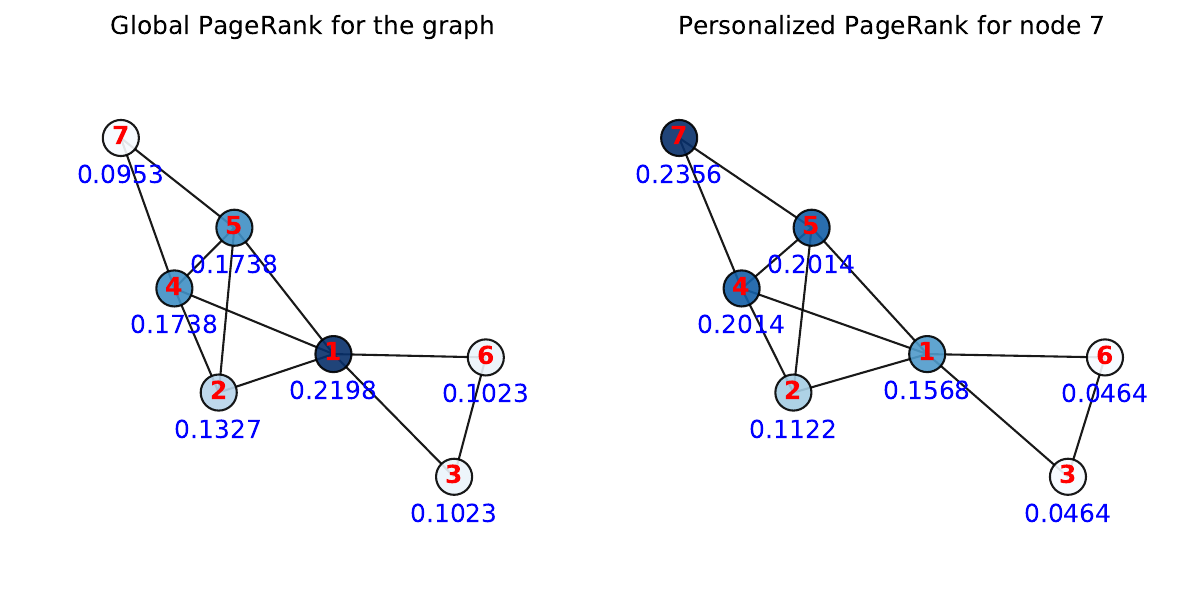}
    \caption{The example of global PageRank (Left) and Personalized PageRank for node 7 (Right). Their PageRank values are below the nodes and colored blue. Personalized PageRank measures the neighbor importance given one starting node, while Global PageRank measures the overall importance of nodes.}
    \label{fig:ppr-intro}
\end{figure}

PageRank is an algorithm used to measure the centrality of nodes in a graph. This algorithm was developed by Google to measure the importance of web pages and to improve their search engine results. The PageRank of a node is calculated by measuring the probability of landing on that node during a random walk on the graph. Figure \ref{fig:ppr-intro} illustrates an example of PageRank in the left subfigure where node 1 is the most important nodes of high PageRank value.

% Although PageRank is an effective way to measure centrality in a graph, there are other methods to compute this using linear algebra. One such method is the use of eigenvectors to calculate the importance of nodes in a graph.

Personalized PageRank is an extension of the PageRank algorithm that is used to measure the importance of nodes in a graph relative to a particular starting node. In other words, it measures the centrality of nodes in an "ego net" around a starting node.
An ego net is a sub-graph that consists of a focal node (the starting node) and its neighbors. 
To calculate PPR, one starts with a random walk on the graph, beginning at the focal node (or starting node). However, to prevent the walk from getting away from the local area, the algorithm also includes a probability of randomly jumping back to the starting node. This ensures that the algorithm explores the entire graph while still focusing on the ego net around the starting node. 
Likewise,  the PPR value of a node is calculated by measuring the probability of landing on that node during this modified random walk.

In Figure \ref{fig:ppr-intro} (right),  we illustrate the PPR values given the starting node is node 7. Compare to PageRank, although the graph structure is the same,  The PPR values are significantly different from PageRank as it favors the nodes around the starting node 7. 
%PPR is used to measure the importance of everyone in the universe of the focal node.

% Figure \ref{fig:ppr-intro} illlustrates two examples for Global PageRank and its variant -- Personalized PageRank. 
% If we start a random walk uniformly from a node, the convergence of final probability is the global PageRank. 

Typically, PPR can be solved analytically using linear algebra and power iteration method. However, these algorithms can be computationally expensive on large graphs. Therefore, various approximation techniques have been developed to compute the PPR more efficiently. These techniques include using a algorithmic local forward push operations to approximate PPR and utilizing numerical optimization techniques to solve a PPR-equivalent optimization problem.  
In the subsequent sections, we will present a formal discussion of the PPR definition and its approximation methods.

\section{Notations}
We use bold capitalized letter (e.g. $\bm A$ and $\bm X$)  to represent matrics and bold lower letters for vectors (e.g., $\bm a$ and $\bm x$). 
Let $\mathcal{G}(\mathcal{V},\mathcal{E},\mathcal{W}, \bm X )$ be a directed weighted-graph where $\mathcal{V}=\{1,2,\ldots,n\}$ is the node set, $\mathcal{E}\subseteq \mathcal{V}\times \mathcal{V}$ is the set of edges, and $\mathcal{W}$ is corresponding edge weights of $\mathcal{E}$. 
If presented, $\bm X \in \mathbb{R} ^ { d \times |\mathcal{V}| }$ is the nodes' $d$-dim raw attributes or features. 
For each node $v$, $\operatorname{Nei}_{out}(v)$ stands for out-neighbors of $v$. For all $v\in \mathcal{V}$, the generalized out-degree vector of $\mathcal{V}$ is $\bm d$ where $v$-th entry $d(v) \triangleq \sum_{u \in \operatorname{Nei}(v)}w_{(v,u)}$ and $w_{(v,u)}$ could be the weight of edge $(v,u)$ in graph $\mathcal{G}$. 
The generalized degree matrix is then denoted as $\bm D := \operatorname{diag}(\bm d)$ and the adjacency matrix is written as $\bm A$ with $\bm A(u,v) = w_{(u,v)}$.
To simplify our notation, we use $d(i)$ or $d_i$ to denote the degree of node $i$. Let $\bm P = \bm D^{-1} \bm A$ be the random walk matrix and then $\bm P^\top = \bm A^\top \bm D^{-1}$ is the column stochastic matrix. 

\section{Personalized PageRank Formulation}
\label{sec:ppr}
% \vspace{-1mm}
As a measure of the relative importance among nodes, Personalized PageRank (PPR), a generalized version of original PageRank \cite{page1999pagerank}, plays an important role in many graph mining tasks. In this thesis, most proposed algorithms are built on PPR. Specifically, the Personalized PageRank vector (PPV) of a node $s$ in $\mathcal{G}$ is defined as the following:

\begin{definition}[Personalized PageRank Vector
(PPV)] Given a graph $\mathcal{G}(\mathcal{V}, \mathcal{E},\mathcal{W})$ with $|\mathcal{V}| = n $ and $ |\mathcal{E}| = m$. Define the lazy random walk transition matrix $\bm P \triangleq (1-\beta) \bm D^{-1} \bm A + \beta  \bm I_n$, $\beta \in [0,1)$ where $\bm D$ is the generalized degree matrix and $\bm A$ is the adjacency matrix of $\mathcal{G}$.\footnote{Our PPV is defined based on the lazy random walk transition matrix. This definition is equivalent to the one using $\bm D^{-1}\bm A$ but with a different parameter $\alpha$. Without loss of generality, we use $\beta=0$ throughout the paper.} 
Given teleport probability $\alpha \in [0,1)$ and the source node $s$, the Personalized PageRank vector of $s$ is defined as:
\begin{equation}
\bm \pi_{\alpha, s} = (1-\alpha) \bm P^\top \bm \pi_{\alpha, s}  + \alpha \bm 1_s, \label{equ:static-ppr-static-graph}
\end{equation}
where the teleport probability $\alpha$ is a small constant (e.g. $\alpha=.15$), and $\bm 1_s$ is an indicator vector of node $s$, that is, $s$-th entry is 1, 0 otherwise. We simply denote PPV of $s$ as $\bm \pi_s$.
\label{def:static-ppr-static-graph}
\end{definition}

Clearly, $\bm \pi_s$ can be calculated in a closed form, i.e., $\bm \pi_{s} = \alpha  (\bm I_n - (1-\alpha ) \bm P^\top)^{-1} \bm 1_s$ but with time complexity $\mathcal{O}(n^3)$. The most commonly used method is \textit{Power Iteration} \cite{page1999pagerank}, which approximates $\bm \pi_{s}$ iteratively: $\bm \pi_{s}^{(t)} = (1-\alpha)  \bm P^\top \bm \pi_{s}^{(t-1)} + \alpha \bm 1_s$. After $t =\lceil \log_{1-\alpha}\epsilon\rceil$ iterations, one can achieve $\| \bm \pi_s - \bm \pi_s^{(t)} \|_1 \leq \epsilon$. Hence, the overall time complexity of power iteration is $\mathcal{O}(m \log_{1-\alpha} \epsilon)$ with $\mathcal{O}(m)$ memory requirement. However, the per-iteration of the power iteration needs to access the whole graph which is time-consuming. Even worse, it is unclear how one can efficiently use power-iteration to obtain an updated $\bm \pi_s$ from $\mathcal{G}_t$ to $\mathcal{G}_{t+1}$. Other types of PPR can be found in \citep{wang2017fora,wei2018topppr,wu2021unifying} and references therein.

\section{Approximate PPR using Forward Push Algorithm}

\textit{forward push algorithm} \cite{andersen2006local}, a.k.a. the bookmark-coloring algorithm \cite{berkhin2006bookmark}, approximates $\pi_{s}(v)$ locally via an approximate $p_s(v)$. The key idea is to maintain solution vector $\bm p_s$ and a residual vector $\bm r_s$ (at the initial $\bm r_s = \bm 1_s, \bm p_s = \bm 0$).

    % There are several works on computing PPVs for static graph \cite{andersen2006local,berkhin2006bookmark,andersen2007local}. The idea is based a local push operation proposed in \cite{andersen2006local}. 
    %We use a modified version of it as presented in Algorithm \ref{algo:forward-local-push}.
    
    \begin{algorithm}[H]
    \caption{$\textsc{ForwardLocalPush}$ \cite{zhang2016approximate}}
    \begin{algorithmic}[1]
    \State \textbf{Input: }$\bm p_s, \bm r_s, \mathcal{G}, \epsilon, \beta = 0$
    \While{$\exists u, r_s(u) > \epsilon d(u)$}
    \State $\textsc{Push}(u)$
    \EndWhile
    \While{$\exists u, r_s(u) < -\epsilon d(u)$} 
    \State $\textsc{Push}(u)$
    \EndWhile
    \State \Return $(\bm p_s, \bm r_s)$
    \Procedure{Push}{$u$}
    \State $p_s(u) \pluseq \alpha r_s(u)$
    \For{$v \in \operatorname{Nei}(u)$}
    \State $r_s(v) \pluseq (1-\alpha) r_s(u) (1-\beta) / d(u)$
    \EndFor
    \State $r_s(u) = (1-\alpha) r_s(u) \beta $
    \EndProcedure
    \end{algorithmic}
    \label{algo:forward-local-push}
    \end{algorithm}
    
    Algorithm \ref{algo:forward-local-push} is a generalization from \citet{andersen2006local} and there are several variants of forward push \cite{andersen2006local,lofgren2015efficient,berkhin2006bookmark}, which are dependent on how $\beta$ is chosen (we assume $\beta=0$). The essential idea of forward push is that, at each \textsc{Push} step, the frontier node $u$ transforms her $\alpha$ residual probability $r_s(u)$ into estimation probability $p_s(u)$ and then pushes the rest residual to its neighbors. The algorithm repeats this push operation until all residuals are small enough \footnote{There are two implementation of forward push depends on how frontier is selected. One is to use a first-in-first-out (FIFO) queue \cite{gleich2015pagerank} to maintain the frontiers while the other one maintains nodes using a priority queue is used \cite{berkhin2006bookmark} so that the operation cost is $\mathcal{O}(1/\epsilon \alpha)$ instead of $\mathcal{O}(\log n/ \epsilon \alpha)$.}.  Methods based on local push operations have the following invariant property.
    
    \begin{restatable}[Invariant property \cite{hong2016discriminating}]{lemma}{primelemma}
    \label{lemma:1}
     \textsc{ForwardLocalPush} has the following invariant property
    \begin{equation}
    \pi_s(u) = p_s(u) + \sum_{v \in V} r_s(v) \pi_v(u), \forall u \in \mathcal{V}.
    \end{equation}
    \end{restatable}
    
    The local push algorithm can guarantee that the each entry of the estimation vector $p_s(v)$ is very close to the true value $\pi_s(v)$. We state this property as in the following
    
    \begin{lemma}[Approximation error and time complexity \cite{andersen2006local,zhang2016approximate}]
    \label{lemma:2}
    Given any graph $\mathcal{G}(\mathcal{V},\mathcal{E})$ with $\bm p_s = \bm 0, \bm r_s = \bm 1_s$ and a constant $\epsilon$, the run time for \textsc{ForwardLocalPush} is at most $\frac{1-\|\bm r_s\|_1}{\alpha \epsilon}$ and the estimation error of $\pi_s(v)$ for each node $v$ is at most $\epsilon$, i.e. $|p_s(v) - \pi_s(v)|/d(v)| \leq \epsilon$  .
    \end{lemma}

\section{Approximate PPR as \texorpdfstring{$\ell_1$}-Regularized Optimization}
Interestingly, recent research \cite{fountoulakis2019variational, fountoulakis2022open} discovered that Algorithm \ref{algo:forward-local-push} is equivalent to the Coordinate Descent algorithm, which solves an explicit optimization problem. 
Furthermore, with the specific termination condition, a $\ell_1$-regularized quadratic objective function has been proposed as the alternative PPR formulation (Lemma \ref{lemma:ppr-quad}).

\begin{lemma}[Variational Formulation of Personalized PageRank] \label{lemma:ppr-quad}  Solving PPR in Equ.\ref{equ:static-ppr-static-graph} is equivalent to solving the following $\ell_1$-regularized quadratic objective function where the PPR vector $\bm \pi := \bm D^{1/2} \bm x^{*} $: 

\begin{align}
     \bm x^{*} &= \argmin_{\bm x } f(\bm x) := \underbrace{\frac{1}{2} \bm x^{\top} \bm W \bm x + \bm x^{\top} \bm b}_{g(\bm x)}  + \underbrace{\epsilon\| \bm D^{1/2} \bm x \|_1}_{h(\bm x)}, \  where \label{eq:ppr-quad-form} \\
    \bm W &= \bm D^{-1/2}(\bm D - (1-\alpha) \bm A ) \bm D^{-1/2} \nonumber , \ \bm b = - \alpha \bm D^{-1/2} \bm e_s   \nonumber
\end{align}
We present the proof details in Appendix \ref{app:algos-proofs}.

\end{lemma}

Lemma \ref{lemma:ppr-quad} provides a unique perspective to understand PPR and tackle it with intensively studied optimization methods for the $\ell_1$-regularization problem, specifically proximal gradient methods such as ISTA and FISTA. 

\paragraph{ISTA Solver for PPR}: Equ.\ref{eq:ppr-quad-form} has the composition of a smooth function  $g(\bm x)$
%$g(\bm x) = \frac{1}{2} \bm x^{\top} \bm W \bm x + \bm x^{\top} \bm b $ 
and a non-smooth function $h(\bm x)$, 
%$h(\bm x) = \epsilon' \| \bm D^{-1/2} \bm x\|_{1}$. 
Note that 
\begin{align}
    \nabla^{2}g(\bm x) = \bm W &= \alpha \bm I_n+ (1-\alpha) \underbrace{(\bm I_n - \bm D^{-1/2} \bm A \bm D^{-1/2})}_{\text{Normalized Laplacian of $\lambda_{max} \leq 2$ }} ,
 \nonumber
\end{align}
where $\lambda_{max}$ is the largest eigenvalue. With $\| \nabla^2 g(\bm x) \|_2 \leq 2-\alpha$, let $\eta = \frac{1}{2-\alpha}$, one can convert the original optimization problem (Equ.\ref{eq:ppr-quad-form}) into proximal form using the second-order approximation:
\begin{align}
    \bm x^{(k+1)} &= \textsc{prox}_{h(\cdot)}( \bm x^{(k)} - \eta  \nabla_i f(\bm x^{(k)})), where  \nonumber \\
    \textsc{prox}_{h(\cdot)}(\bm z) &= \argmin_x \frac{1}{2 \eta} \|\bm x - \bm z\|_2^2 + h(\bm x) \nonumber
\end{align}
To this end, we can apply the Proximal Gradient Descent algorithm, specifically \textsc{ISTA} (Algorithm \ref{algo:ppr-ista}) to solve this well-defined PPR-equivalent $\ell_1$-regularized optimization problem. 

% $\argmin_{\bm \eta } f(\bm x) = \argmin_{\bm x } f() $

% $\operatorname{prox}_{(L, )}(\bm x) = \operatorname{argmin}_{\bar{\bm x}} \frac{1}{2}\|  \bm x - \bar{\bm x} \|_2^2 + \epsilon' \|\bm D^{1/2} \bm x \|_1 $

% $\operatorname{prox}_(\bm x) = \operatorname{argmin}_{\bar{\bm x}} \frac{1}{2}\|  \bm x - \bar{\bm x} \|_2^2 + \epsilon' \|\bm D^{1/2} \bm x \|_1 $

\begin{algorithm}[ht]
\caption{$\textsc{ISTA Solver for PPR-equivalent Optimization}$ }
\begin{algorithmic}[1]
\State \textbf{Input: }$ k = 0, \bm x^{(0)} = \bm 0, \mathcal{G}, \epsilon, \alpha$
% \State $\bm r^{(0)} = \left(\bm I_n - (1-\alpha) \bm P^{\top} \right) \bm D^{-1/2} \bm x ^{(0)}  - \alpha \bm e_s  = - \alpha \bm e_s$
\State $\nabla f(\bm x^{(0)}) = - \alpha \bm D^{-1/2} \bm e_s$
\State $\eta = \frac{1}{2-\alpha}$

% \While{ exists $i$ such that $\|\nabla f(\bm x)(i)\|_1 > \epsilon' d(i)^{1/2}$}
\While{$\bm x^{(k)}$ has not converged yet }
\State $ \bm z^{(k)}(i) := \bm x^{(k)}(i) - \eta  \nabla_i f(\bm x^{(k)}) $
\State $\bm x^{(k+1)(i)} = \textsc{prox}_{h(\cdot)}\left( \bm z^{(k)}(i)\right)$  can be solved analytically:
%_{\epsilon'\bm d(i)^{1/2}}
\[
    %\frac{1}{\eta} 
    \bm x^{(k+1)}(i)= 
\begin{dcases}
    \bm z^{k}(i) -  \epsilon'  d(i)^{1/2} 
    , & \text{ if }  \bm z^{k}(i) >  \epsilon  d(i)^{1/2} \\
    0 , & \text{ if }   \| \bm z^{k}(i) \|_1 \leq  \epsilon'  d(i)^{1/2} \\
    \bm z^{k}(i) +  \epsilon'  d(i)^{1/2} , 
    & \text{ if }  \bm z^{k}(i) < - \epsilon d(i)^{1/2} \\
    % -\epsilon' \bm d(i)^{1/2},  & \text{if } \bm x^{*}(i) > 0 \\
    % \epsilon' \bm d(i)^{1/2}, & \text{if } \bm x^{*}(i) < 0 \\
    % \in [-\epsilon' \bm d(i)^{1/2}, \epsilon' \bm d(i)^{1/2}] & \text{if } \bm x^{*}(i) = 0 \\
\end{dcases}
\]
\EndWhile
\State \Return $\left( \bm x^{(k)}, \nabla f(\bm x^{(k)})\right)$
\State \textcolor{black}{//Note that PPR vector $ \bm \pi = \bm D^{1/2} \bm x^{(k)}, \bm r =  \bm D^{1/2}\nabla f(\bm x^{(k)})$}
% \State \Return $\left(\bm \pi = \bm D^{1/2} \bm x^{(k)}, \bm r =  \bm D^{1/2}\nabla f(\bm x^{(k)}) \right)$

\end{algorithmic}
\label{algo:ppr-ista}
\end{algorithm}

% \paragraph{Complexity Analysis:}
% -- Meanwhile Network embedding algorithms (e.g., Deepwalk) purely depend on the network structure and capture the node proximity in low-dimensional vectors via different dimension reduction techniques (e.g, matrix factorization, random projection, hashing, ... ). 
% For example, Single Source Personalized PageRank (\textit{SSPPR}) captures the node-wise proximity. Denote SSPPR as $\bm \pi_i \in \mathbb{R}^{|V|}$ where $V$ is the vertex set of the graph, one can further employ dimension reduction function \footnote{ e.g., random projection or hash-based reduction} $f: \mathbb{R}^{|V|} \rightarrow \mathbb{R}^{d}$, such that $\|  f(x_i) - f(x_j) \|_2^2 \leq (1+\epsilon)\| x_i -x_j \|_2^2 $ 

% -- Because of the extra information encoded in node features, more trainable and usually supervised training, 

\section{Incrementally Maintain PPR}
%There are several works on computing PPVs for static graph \cite{andersen2006local,berkhin2006bookmark,andersen2007local}. The idea is based a local push operation proposed in \cite{andersen2006local}. 
% The main challenge to directly use forward push algorithm to obtain high quality PPVs in dynamic setting is that: the quality $\bm p_s$ return by forward push algorithm  will critically depend on the precision parameter $\epsilon$ which unfortunately is unknown under the dynamic problem setting. 
 For a dynamic graph, the graph snapshot at time $t$ is denoted as $\mathcal{G}^t\left(\mathcal{V}^t,\mathcal{E}^t, \mathcal{W}^t, \bm X^{(t)} \right)$.  The notation ( subscript for time $t$) are used for other graph attributes. For instance, if there is no ambiguity, the degree of a node $u$ can be denoted as $d^{(t)}_u$ or $d^{t}_(u)$. 
 
To maintain the PPVs (PPR Vectors) for a dynamic graph, the naive approach is to recalculate the PPVs from scratch. However, this approach comes at the cost of a significant computational budget.
Interestingly, \citet{zhang2016approximate}  proposed a novel updating strategy for calculating dynamic PPVs.  
Given a new edge $e_{u,v}$ with $ d^{t+1}(u) = d^{t}(u)+1$ at time $t+1$, the PPR adjustments are :
\begin{align}
\footnotesize
    \bm \pi^{t+1}(u) = \bm \pi^{t}(u) * \frac{ d^{t}(u)+1}{ d^{t}(u)} ,\quad 
    \bm r^{t+1}(u) = \bm r^{t}(u) - \frac{\bm \pi^{t}(u)}{ \alpha d(u)}, \quad
    \bm r^{t+1}(v) = \bm r^{t}(v) + \frac{1-\alpha}{\alpha} \frac{\bm \pi^{t}(u)}{ d^{t}(u)} \nonumber
\end{align}

However, the original update of per-edge event strategy proposed in \cite{zhang2016approximate} is not suitable for batch update between graph snapshots. \citet{guo2017parallel} propose to use a batch strategy, which is more practical in real-world scenario where there is a sequence of edge events between two consecutive snapshots. This motivates us to develop a new algorithms for dynamic PPVs and then use these PPVs to obtain high quality dynamic embedding vectors. In Chapter \ref{chapter:dynppe}, \ref{chapter:DynAnom}, \ref{chapter:GoPPE} we present our efficient algorithms that build upon the key preliminaries discussed above.

%We use a modified version of it as presented in Algorithm \ref{algo:forward-local-push}.However, the original update of per-edge event strategy proposed in \cite{zhang2016approximate} is not suitable for batch update between graph snapshots. \citet{guo2017parallel} propose to use a batch strategy, which is more practical in real-world scenario where there is a sequence of edge events between two consecutive snapshots. This motivates us to develop a new algorithm for dynamic PPVs and then use these PPVs to obtain high quality dynamic embedding vectors. 

\chapter[Subset Node Embeddings over Dynamic Graph]{Subset Node Embeddings over Dynamic Graph
\protect\footnote{Xingzhi Guo, Baojian Zhou, and Steven Skiena. 2021. Subset Node Representation Learning over Large Dynamic Graphs. In \textit{Proceedings of the 27th ACM SIGKDD Conference on Knowledge Discovery \& Data Mining (KDD '21)}. Association for Computing Machinery, New York, NY, USA, 516–526. https://doi.org/10.1145/3447548.3467393 \cite{xingzhi2021subset}}}
\label{chapter:dynppe}
\graphicspath{{chapter1-DynPPE/}}

\begin{figure}[ht]
\centering
\subfloat[Dynamic graph model]{\includegraphics[width=4.3cm]{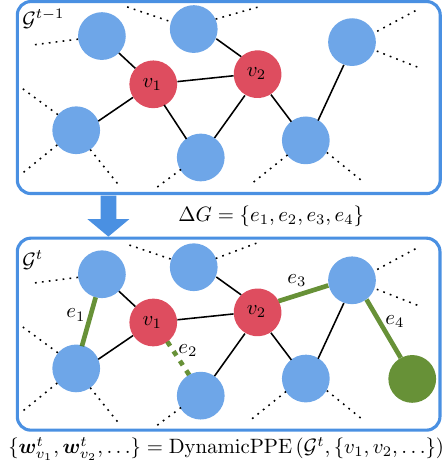}}\quad
\subfloat[An application of \textsc{DynamicPPE}]{\includegraphics[width=5.7cm]{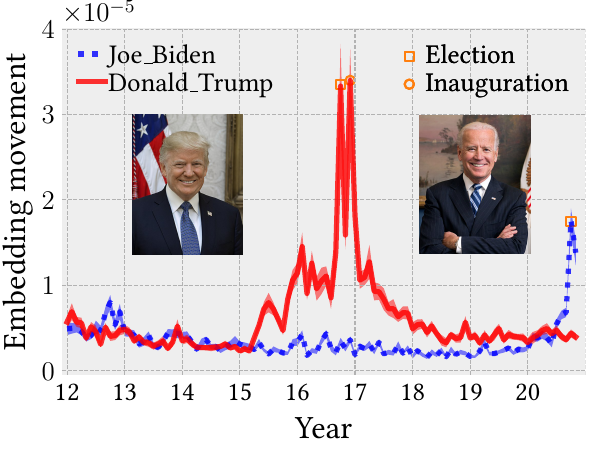}}
\caption{ (a) The model of dynamic network in two consecutive snapshots. (b) An application of \textsc{DynamicPPE} to keep track embedding movements of interesting Wikipedia articles (vertices). We learn embeddings of two presidents of the United States on the whole English Wikipedia graph from 2012 monthly, which cumulatively involves 6.2M articles (nodes) and 170M internal links (edges). The \textit{embedding movement} between two time points is defined as $1-\cos(\bm w_v^t, \bm w_v^{t+1})$ where $\cos(\cdot,\cdot)$ is the cosine similarity. The significant embedding movements may reflect big social status changes of Donald\_Trump  and Joe\_Biden \protect\footnotemark in this dynamic Wikipedia graph.\vspace{-5mm}}
\label{fig:example-model}
\end{figure}

\section{Introduction}

Graph node representation learning aims to represent nodes from graph structure data into lower dimensional vectors and has received much attention in recent years \cite{perozzi2014deepwalk,grover2016node2vec,rossi2013modeling,trivedi2018dyrep,hamilton2017inductive,kipf2017semi,hamilton2017representation}. Effective methods have been successfully applied to many real-world applications where graphs are large-scale and static \cite{ying2018graph}. However, networks such as social networks \cite{berger2006framework}, knowledge graphs \cite{ji2015knowledge, wang2022knowledge, wang2023knowledge, wang2023knowledge}, and citation networks \cite{clement2019use} are usually time-evolving where edges and nodes are inserted or deleted over time. Computing representations of all vertices over time is prohibitively expensive because only a small subset of nodes may be interesting in a particular application. Therefore, it is important and technical challenging to efficiently learn dynamic embeddings for these large-scale dynamic networks under this typical use case.

\footnotetext{Two English Wikipedia articles are accessible at \textcolor{blue}{\url{https://en.wikipedia.org/wiki/Donald_Trump}} and \textcolor{blue}{\url{https://en.wikipedia.org/wiki/Joe_Biden}}.}

Specifically, we study the following dynamic embedding problem: We are given a subset $S=\{v_1,v_2,\ldots, v_k\}$ and an initial graph $\mathcal{G}^t$ with $t=0$. Between time $t$ and $t+1$, there are edge events of insertions and/or deletions. The task is to design an algorithm to learn embeddings for $k$ nodes with time complexity independent on the number of nodes $n$ per time $t$ where $k \ll n$. This problem setting is both technically challenging and practically important. For example, in the English Wikipedia graph, one needs focus only on embedding movements of articles related to political leaders, a tiny portion of the whole Wikipedia. Current dynamic embedding methods \cite{du2018dynamic,zhou2018dynamic,zhang2018billion,nguyen2018continuous,zhu2016scalable} are not applicable to this large-scale problem setting due to the lack of efficiency. More specifically, current methods have the \textit{dependence issue} where one must learn all embedding vectors. This dependence issue leads to per-embedding update is linear-dependent on $n$, which is inefficient when graphs are large-scale. This obstacle motivates us to develop a new method. 

In this paper, we propose a dynamic personalized PageRank embedding (\textsc{DynamicPPE}) method for learning a subset of node representations over large-sale dynamic networks. \textsc{DynamicPPE} is based on an effective approach to compute dynamic PPVs \cite{zhang2016approximate}. There are two challenges of using \citet{zhang2016approximate} directly: 1) the quality of dynamic PPVs depends critically on precision parameter $\epsilon$, which unfortunately is unknown under the dynamic setting; and 2) The update of per-edge event strategy is not suitable for batch update between graph snapshots. To resolve these two difficulties, first, we adaptively update $\epsilon$ so that the \textit{estimation error} is independent of $n,m$, thus obtaining high-quality PPVs. Yet previous work does not give an estimation error guarantee. We prove that the time complexity is only dependent on $\bar{d}$. Second, we incorporate a batch update strategy inspired from \cite{guo2017parallel} to avoid frequent per-edge update. Therefore, the total run time to keep track of $k$ nodes for given snapshots is $\mathcal{O}(k \bar{d} m)$. Since real-world graphs have the sparsity property $\bar{d} \ll n $, it significantly improves the efficiency compared with previous methods. Inspired by \textit{InstantEmbedding} \cite{postuavaru2020instantembedding} for static graph, we use hash kernels to project dynamic PPVs into embedding space. Figure \ref{fig:example-model} shows an example of successfully applying \textsc{DynamicPPE} to study the dynamics of social status in the English Wikipedia graph. To summarize, our contributions are:

\begin{enumerate}[leftmargin=*]
\item We propose a new algorithm \textsc{DynamicPPE}, which is based on the recent advances of local network embedding on static graph and a novel computation of dynamic PPVs. \textsc{DynamicPPE} effectively learns PPVs and then projects them into embedding space using hash kernels.
\item \textsc{DynamicPPE} adaptively updates the precision parameter $\epsilon$ so that PPVs always have a provable estimation error guarantee. In our subset problem setting, we prove that the time and space complexity are all linear to the number of edges $m$ but independent on the number of nodes $n$, thus significantly improving the efficiency.
\item Node classification results demonstrate the effectiveness and efficiency of the proposed. We compile three large-scale datasets to validate our method. As an application, we showcase that learned embeddings can be used to detect the changes of Chinese cities during this ongoing COVID-19 pandemic articles on a large-scale English Wikipedia.
\end{enumerate}

The rest of this paper is organized as follows: In Section \ref{sec:related-work}, we give the overview of current dynamic embedding methods. The problem definition and preliminaries are in Section \ref{sec:defintions}. We present our proposed method in Section \ref{sec:proposed}. Experimental results are reported in Section \ref{sec:experiments}. The discussion and conclusion will be presented in Section \ref{sec:conclusion}. Our code and created datasets are accessible at \textcolor{blue}{\url{https://github.com/zjlxgxz/DynamicPPE}}.

\section{Related Work}
\label{sec:related-work}

There are two main categories of works for learning embeddings from the dynamic graph structure data. The first type is focusing on capturing the evolution of dynamics of graph structure \cite{zhou2018dynamic}. The second type is focusing on both dynamics of graph structure and features that lie in these graph data \cite{trivedi2019dyrep}. 
In this paper, we focus on the first type and give an overview of related works. Due to the large amount of work in this area, some related works may not be included, one can find more related works in a survey \cite{kazemi2020representation} and references therein.

\noindent\textbf{Dynamic latent space models}\quad The dynamic embedding models had been initially explored by using latent space model \cite{hoff2002latent}.  The dynamic latent space model of a network makes an assumption that each node is associated with an $d$-dimensional vector and distance between two vectors should be small if there is an edge between these two nodes \cite{sarkar2005dynamic-kdd,sarkar2005dynamic-nips}. Works of these assume that the distance between two consecutive embeddings should be small. The proposed dynamic models were applied to different applications \cite{sarkar2007latent,hoff2007modeling}.  Their methods are not scalable from the fact that the time complexity of initial position estimation is at least $\mathcal{O}(n^2)$ even if the per-node update is $\log(n)$.

\noindent\textbf{Incremental SVD and random walk based methods} \quad \citet{zhang2018timers} proposed \textsc{TIMERS} that is an incremental SVD-based method. To prevent the error accumulation, \textsc{TIMERS} properly set the restart time so that the accumulated error can be reduced. \citet{nguyen2018continuous} proposed continuous-time dynamic network embeddings, namely CTDNE. The key idea of CTDNE is that instead of using general random walks as DeepWalk \cite{perozzi2014deepwalk}, it uses temporal random walks containing a sequence of edges in order. Similarly, the work of \citet{du2018dynamic} was also based on the idea of DeepWalk. These methods have time complexity dependent on $n$ for per-snapshot update. \citet{zhou2018dynamic} proposed to learn dynamic embeddings by modeling the triadic closure to capture the dynamics. 

\noindent\textbf{Graph neural network methods} \citet{trivedi2019dyrep} designed a dynamic node representation model, namely \textsc{DyRep}, as modeling a latent mediation process where it leverages the changes of node between the node's social interactions and its neighborhoods. More specifically, \textsc{DyRep} contains a temporal attention layer to capture the interactions of neighbors. \citet{zang2020neural} proposed a neural network model to learn embeddings by solving a differential equation with ReLU as its activation function. \citep{kumar2019predicting} presents a dynamic embedding, a recurrent neural network method, to learn the interactions between users and items. However, these methods either need to have features as input or cannot be applied to large-scale dynamic graph. \citet{kumar2019predicting} proposed an algorithm to learn the trajectory of the dynamic embedding for temporal interaction networks. Since the learning task is different from ours, one can find more details in their paper.

\section{Preliminaries}
\label{sec:defintions}

\textbf{Notations for dynamic graphs}\quad We use $[n]$ to denote a ground set $[n]:=\{ 0,1, \ldots, n-1\}$. The graph snapshot at time $t$ is denoted as $\mathcal{G}^t\left(\mathcal{V}^t,\mathcal{E}^t\right)$. 
%The degree of a node $v$ is $d(v)^t$. In the rest of this paper, the average degree at time $t$ is $\bar{d}^t$ and the subset of target nodes is $S\subseteq \mathcal{V}^t$. Bold capitals, e.g. $\bm A, \bm W$ are matrices and bold lower letters are vectors $\bm w,\bm x$. 
More specifically, the embedding vector for node $v$ at time $t$ denoted as $\bm w_v^t \in \mathbb{R}^d$ and $d$ is the embedding dimension. 
The $i$-th entry of $\bm w_v^t$ is  $w_v^t(i) \in \mathbb{R}$. The embedding of node $v$ for all $T$ snapshots is written as $\bm W_v = [\bm w_v^1,\bm w_v^2,\ldots, \bm w_v^T]^\top$. 
We use $n_t$ and $m_t$ as the number of nodes and edges in $\mathcal{G}^t$ which we simply use $n$ and $m$ if time $t$ is clear in the context.

Given the graph snapshot $\mathcal{G}^t$ and a specific node $v$, the personalized PageRank vector (PPV) is an $n$-dimensional vector $\bm \pi_v^t \in \mathbb{R}^{n}$ and the corresponding $i$-th entry is $\pi_v^t(i)$.  We use $\bm p_v^t \in \mathbb{R}^{n}$ to stand for a calculated PPV obtained from a specific algorithm. Similarly, the corresponding $i$-th entry is $p_v^t(i)$. The teleport probability of the PageRank is denoted as $\alpha$. The \textit{estimation error} of an embedding vector is the difference between true embedding $\bm w_v^t$ and the estimated embedding $\hat{\bm w}_v^t$ is measure by $\| \cdot \|_1 := \sum_{i=1}^n \left| w_v^t(i) - \hat{w}_v^t(i)\right|$.

\subsection{Dynamic graph model and its embedding}

Given any initial graph (could be an empty graph), the corresponding dynamic graph model describes how the graph structure evolves over time. We first define the dynamic graph model, which is based on \citet{kazemi2020representation}.

\begin{definition}[Simple dynamic graph model \cite{kazemi2020representation}]
\label{def:dynamic-graph-model}
A simple dynamic graph model is defined as an ordered of snapshots $\mathcal{G}^0, \mathcal{G}^1$, $\mathcal{G}^2,  \ldots , \mathcal{G}^T$ where $\mathcal{G}^0$ is the initial graph. The \textit{difference} of graph $\mathcal{G}^t$ at time $t=1,2,\ldots,T$ is $\Delta G^t(\Delta \mathcal{V}^t,\Delta\mathcal{E}^t) := G^t \backslash G^{t-1}$ with $\Delta \mathcal{V}^t := \mathcal{V}^t \backslash \mathcal{V}^{t-1}$ and $\Delta \mathcal{E}^t :=\mathcal{E}^t \backslash \mathcal{E}^{t-1}$. Equivalently, $\Delta \mathcal{G}^t$ corresponds to a sequence of edge events as the following
\begin{equation}
\Delta \mathcal{G}^t = \left\{ e_1^t,e_2^t,\ldots, e_{m^\prime}^t \right\},
\end{equation}
where each edge event $e_i^t$ has two types: \textit{insertion} or \textit{deletion}, i.e, $ e_i^t = \left(u,v, \operatorname{event} \right)$ where $\operatorname{event} \in \{ \operatorname{Insert}, \operatorname{Delete}\}$ \footnote{The node insertion can be treated as inserting a new edge and then delete it and node deletion is a sequence of deleting its edges.}.
\end{definition}

The above model captures the evolution of a real-world graph naturally where the structure evolution can be treated as a sequence of edge events occurred in this graph. To simplify our analysis, we assume that the graph is undirected. Based on this, we define the subset dynamic representation problem as the following.

\begin{definition}[Subset dynamic network embedding problem]
Given a dynamic network model $\left\{\mathcal{G}^0, \mathcal{G}^1, \mathcal{G}^2, \ldots, \mathcal{G}^T\right\}$ define in Definition \ref{def:dynamic-graph-model} and a subset of target nodes $S=\left\{ v_1,v_2,\ldots,v_k \right\}$, the subset dynamic network embedding problem is to learn dynamic embeddings of $T$ snapshots for all $k$ nodes $S$ where $k \ll n$. That is, given any node $v\in S$, the goal is to learn embedding matrix for each node $v \in S$, i.e. 
\begin{equation}
\bm W_v := [\bm w_v^1,\bm w_v^2,\ldots, \bm w_v^T]^\top \text{ where }\bm w_v^t \in \mathbb{R}^d \text{ and } v \in S.
\end{equation}
\end{definition}

\xingzhiswallow{
    \subsection{Personalized PageRank}
    
    Given any node $v$ at time $t$, the personalized PageRank vector for graph $\mathcal{G}^t$ is defined as the following
    
    \begin{definition}[Personalized PageRank (PPR)]
    \label{def:ppr}
    Given normalized adjacency matrix $\bm W_t = \bm D_t^{-1} \bm A_t$ where $\bm D_t$ is a diagonal matrix with $D_t(i,i) = d(i)^t$ and $\bm A_t$ is the adjacency matrix, the PageRank vector $\bm \pi_s^t$ with respect to a source node $s$ is the solution of the following equation
    \begin{equation}
    \bm \pi_s^t = \alpha * \bm 1_s +  (1-\alpha)\bm \pi_s^t \bm W^t, \label{equ:ppr}
    \end{equation}
    where $\bm 1_s$ is the unit vector with $1_s(v)=1$ when $v =s$, 0 otherwise.
    \end{definition}
    
    There are several works on computing PPVs for static graph \cite{andersen2006local,berkhin2006bookmark,andersen2007local}. The idea is based a local push operation proposed in \cite{andersen2006local}. Interestingly, \citet{zhang2016approximate} extends this idea and proposes a novel updating strategy for calculating dynamic PPVs. We use a modified version of it as presented in Algorithm \ref{algo:forward-local-push}.
    
    \begin{algorithm}[ht]
    \caption{$\textsc{ForwardPush}$ \cite{zhang2016approximate}}
    \begin{algorithmic}[1]
    \State \textbf{Input: }$\bm p_s, \bm r_s, \mathcal{G}, \epsilon, \beta = 0$
    \While{$\exists u, r_s(u) > \epsilon d(u)$}
    \State $\textsc{Push}(u)$
    \EndWhile
    \While{$\exists u, r_s(u) < -\epsilon d(u)$} 
    \State $\textsc{Push}(u)$
    \EndWhile
    \State \Return $(\bm p_s, \bm r_s)$
    \Procedure{Push}{$u$}
    \State $p_s(u) \pluseq \alpha r_s(u)$
    \For{$v \in \operatorname{Nei}(u)$}
    \State $r_s(v) \pluseq (1-\alpha) r_s(u) (1-\beta) / d(u)$
    \EndFor
    \State $r_s(u) = (1-\alpha) r_s(u) \beta $
    \EndProcedure
    \end{algorithmic}
    \label{algo:forward-local-push}
    \end{algorithm}
    
    Algorithm \ref{algo:forward-local-push} is a generalization from \citet{andersen2006local} and there are several variants of forward push \cite{andersen2006local,lofgren2015efficient,berkhin2006bookmark}, which are dependent on how $\beta$ is chosen (we assume $\beta=0$). The essential idea of forward push is that, at each \textsc{Push} step, the frontier node $u$ transforms her $\alpha$ residual probability $r_s(u)$ into estimation probability $p_s(u)$ and then pushes the rest residual to its neighbors. The algorithm repeats this push operation until all residuals are small enough \footnote{There are two implementation of forward push depends on how frontier is selected. One is to use a first-in-first-out (FIFO) queue \cite{gleich2015pagerank} to maintain the frontiers while the other one maintains nodes using a priority queue is used \cite{berkhin2006bookmark} so that the operation cost is $\mathcal{O}(1/\epsilon \alpha)$ instead of $\mathcal{O}(\log n/ \epsilon \alpha)$.}.  Methods based on local push operations have the following invariant property.
    
    \begin{restatable}[Invariant property \cite{hong2016discriminating}]{lemma}{primelemma}
    \label{lemma:1}
     \textsc{ForwardPush} has the following invariant property
    \begin{equation}
    \pi_s(u) = p_s(u) + \sum_{v \in V} r_s(v) \pi_v(u), \forall u \in \mathcal{V}.
    \end{equation}
    \end{restatable}
    
    The local push algorithm can guarantee that the each entry of the estimation vector $p_s(v)$ is very close to the true value $\pi_s(v)$. We state this property as in the following
    
    \begin{lemma}[\cite{andersen2006local,zhang2016approximate}]
    \label{lemma:2}
    Given any graph $\mathcal{G}(\mathcal{V},\mathcal{E})$ with $\bm p_s = \bm 0, \bm r_s = \bm 1_s$ and a constant $\epsilon$, the run time for \textsc{ForwardLocalPush} is at most $\frac{1-\|\bm r_s\|_1}{\alpha \epsilon}$ and the estimation error of $\pi_s(v)$ for each node $v$ is at most $\epsilon$, i.e. $|p_s(v) - \pi_s(v)|/d(v)| \leq \epsilon$  
    \end{lemma}
    
    The main challenge to directly use forward push algorithm to obtain high quality PPVs in our setting is that: 1) the quality $\bm p_s$ return by forward push algorithm  will critically depend on the precision parameter $\epsilon$ which unfortunately is unknown under our dynamic problem setting. Furthermore, the original update of per-edge event strategy proposed in \cite{zhang2016approximate} is not suitable for batch update between graph snapshots. \citet{guo2017parallel} propose to use a batch strategy, which is more practical in real-world scenario where there is a sequence of edge events between two consecutive snapshots. This motivates us to develop a new algorithm for dynamic PPVs and then use these PPVs to obtain high quality dynamic embedding vectors.

}

\section{Proposed method}
\label{sec:proposed}

To obtain dynamic embedding vectors, the general idea is to obtain dynamic PPVs and then project these PPVs into embedding space by using two kernel functions \cite{weinberger2009feature,postuavaru2020instantembedding}.
In this section, we present our proposed method \textsc{DynamicPPE} where it contains two main components: 1) an adaptive precision strategy to control the estimation error of PPVs. We then prove that the time complexity of this dynamic strategy is still independent on $n$. With this quality guarantee, learned PPVs will be used as proximity vectors and be "projected" into lower dimensional space based on ideas of \textit{Verse} \cite{tsitsulin2018verse} and \textit{InstantEmbedding} \cite{postuavaru2020instantembedding}. We first show how can we get high quality PPVs and then present how use PPVs to obtain dynamic embeddings. Finally, we give the complexity analysis.

\subsection{Dynamic graph embedding for single batch}

For each batch update $\Delta G^t$, the key idea is to dynamically maintain PPVs where the algorithm updates the estimate from $\bm p_v^{t-1}$ to $\bm p_v^t$ and its residuals from $\bm r_v^{t-1}$ to $\bm r_v^t$. Our method is inspired from \citet{guo2017parallel} where they proposed to update a personalized contribution vector by using the local reverse push \footnote{One should notice that, for undirected graph, PPVs can be calculated by using the invariant property from the contribution vectors. However, the invariant property does not hold for directed graph. It means that one cannot use reverse local push to get a personalized PageRank vector directly. In this sense, using forward push algorithm is more general for our problem setting.}. The proposed dynamic single node embedding, \textsc{DynamicSNE} is illustrated in Algorithm \ref{algo:dynamic-sne}. It takes an update batch $\Delta \mathcal{G}^t$ (a sequence of edge events), a target node $s$ with a precision $\epsilon^t$, estimation vector of $s$ and residual vector as inputs. It then obtains an updated embedding vector of $s$ by the following three steps: 1) It first updates the estimate vector $\bm p_s^t$ and $\bm r_s^t$ from Line 2 to Line 9; 2) It then calls the forward local push method to obtain the updated estimations, $\bm p_s^t$; 3) We then use the hash kernel projection step to get an updated embedding. This projection step is from \textit{InstantEmbedding} where two universal hash functions are defined as $h_d: \mathbb{N} \rightarrow [d]$ and $h_{\operatorname{sgn}} : \mathbb{N} \rightarrow \{\pm 1\}$ \footnote{For example, in our implementation, we use MurmurHash \textcolor{blue}{\url{https://github.com/aappleby/smhasher}}}. Then the hash kernel based on these two hash functions is defined as $H_{h_{\operatorname{sgn}},h_d}(\bm x): \mathbb{R}^n \rightarrow \mathbb{R}^d$ where each entity $i$ is $\sum_{j \in h_d^{-1}(i)} x_j h_{\operatorname{sgn}}(j)$. Different from random projection used in RandNE \cite{zhang2018billion} and FastRP \cite{chen2019fast},  hash functions has $\mathcal{O}(1)$ memory cost while random projection based method has $\mathcal{O}(d n)$ if the Gaussian matrix is used. Furthermore, hash kernel keeps unbiased estimator for the inner product \cite{weinberger2009feature}.

In the rest of this section, we show that the time complexity is $\mathcal{O}(m \bar{d} / \epsilon)$ in average and the estimation error of learned PPVs measure by $\| \cdot \|_1$ can also be bounded. Our proof is based on the following lemma which follows from \citet{zhang2016approximate,guo2017parallel}.

\begin{lemma}
\label{lemma:6}
Given current graph $\mathcal{G}^t$ and an update batch $\Delta \mathcal{G}^t$, the total run time of the dynamic single node embedding, \textsc{DynamicSNE} for obtaining embedding $\bm w_s^t$ is bounded by the following
\begin{equation}
T_t \leq \frac{\|\bm r_s^{t-1}\|_1 - \|\bm r_s^t\|_1}{\alpha \epsilon^t} + \sum_{u\in \Delta \mathcal{G}^t}\frac{2-\alpha}{\alpha} \frac{p_s^{t-1}(u)}{d(u)}
\end{equation}
\end{lemma}
\begin{proof}
We first make an assumption that there is only one edge update $(u,v,\operatorname{event})$ in $\Delta \mathcal{G}^t$, then based Lemma 14 of \cite{zhang2016approximate-extended}, the run time of per-edge update is at most:
\begin{equation}
\frac{\|\bm r_s^{t-1}\|_1 - \|\bm r_s^{t-1}\|_1}{\alpha \epsilon^t} + \frac{\Delta_t(s)}{\alpha \epsilon^t},
\end{equation}
where $\Delta_t(s) = \frac{2-\alpha}{\alpha} \frac{p_s^{t-1}(u)}{d(u)}$. Suppose there are $k$ edge events in $\Delta \mathcal{G}^t$. We still obtain a similar bound, by the fact that forward push algorithm has monotonicity property: the entries of estimates $ p_s^t(v)$ only increase when it pushes positive residuals (Line 2 and 3 of Algorithm \ref{algo:forward-local-push}). Similarly, estimates $p_s^t(v)$ only decrease when it pushes negative residuals (Line 4 and 5 of Algorithm \ref{algo:forward-local-push}). In other words, the amount of work for per-edge update is not less than the amount of work for per-batch update. Similar idea is also used in \cite{guo2017parallel}.
\end{proof}

\newcommand{\eqplus}{=\mathrel{+}}
\begin{algorithm}[H]
\caption{$\textsc{DynamicSNE}(\mathcal{G}^t, \Delta \mathcal{G}^t,  s,\bm p_s^{t-1},\bm r_s^{t-1}, \epsilon^t,\alpha)$}
\begin{algorithmic}[1]
\State \textbf{Input: } graph $\mathcal{G}^t, \Delta \mathcal{G}^t$, target node $s$, precision $\epsilon$, teleport $\alpha$
\For{$(u,v,\operatorname{op}) \in \Delta G^{t}$}
\If{$\operatorname{op}==\textsc{Insert}(u,v)$}
\State $\Delta_p = p_s^{t-1}(u) / (d(u)^t -1)$
\EndIf
\If{$\operatorname{op} == \textsc{Delete}(u,v)$}
\State $\Delta_p = -p_s^{t-1}(u) / (d(u)^t + 1)$
\EndIf
\State $p_s^{t-1}(u) \leftarrow p_s^{t-1}(u) + \Delta_p$
\State $r_s^{t-1}(u) \leftarrow r_s^{t-1}(u) - \Delta_p / \alpha$
\State $r_s^{t-1}(v) \leftarrow r_s^{t-1}(v) + \Delta_p / \alpha - \Delta_p$
\EndFor
\State $\bm p_s^t = \textsc{ForwardPush}(\bm p_s^{t-1},\bm r_s^{t-1},\mathcal{G}^t,\epsilon^t,\alpha)$
\State $\bm w_s^t = \bm 0$
\For{ $i \in \{ v : p_s^t(v) \ne 0, v \in \mathcal{V}^t\}$} 
\State $w_s^{t}(h_d(i)) \pluseq h_{\operatorname{sgn}}(i) \max\left(\log \left(p_s^t(i) n^t\right), 0\right)$
\EndFor
\end{algorithmic}
\label{algo:dynamic-sne}
\end{algorithm}

\begin{theorem}
Given any graph snapshot $\mathcal{G}^t$ and an update batch $\Delta G^t$ where there are $m_t$ edge events and suppose the precision parameter is $\epsilon^t$ and teleport probability is $\alpha$, \textsc{DynamicSNE} runs in $\mathcal{O}(m_t/\alpha^2 + m_t \bar{d}^t / (\epsilon\alpha^2) + m_t / (\epsilon\alpha))$ with $\epsilon^t = \epsilon / m_t$
\label{thm:run-time}
\end{theorem}
\begin{proof}
Based lemma \ref{lemma:6}, the proof directly follows from Theorem 12 of \cite{zhang2016approximate-extended}.
\end{proof}

\begin{figure}[H]
\centering
\includegraphics[width=0.8\textwidth]{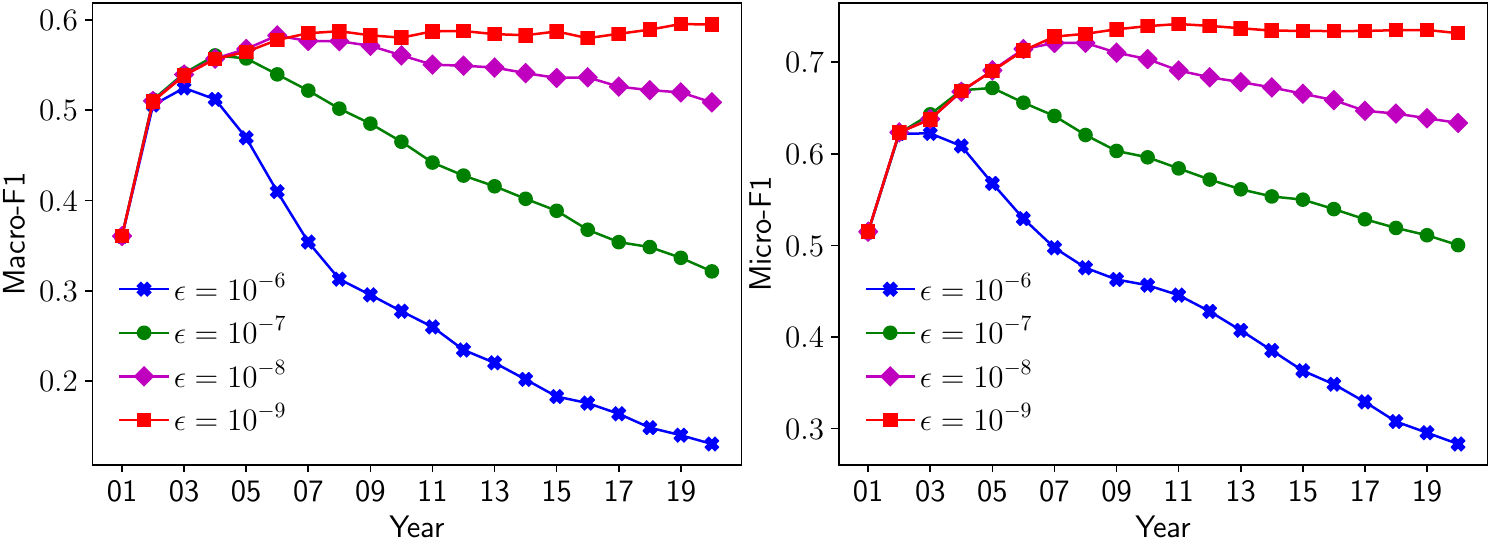}
\caption{$\epsilon$ as a function of year for the task of node classification on the English Wikipedia graph. Each line corresponds to a fixed precision strategy of \textsc{DynamicSNE}. Clearly, when the precision parameter $\epsilon$ decreases, the performance of node classification improves.}
\label{fig:epsilon-fix}
\end{figure}

The above theorem has an important difference from the previous one \cite{zhang2016approximate}. We require that the precision parameter will be small enough so that $\| \bm p_s^t - \bm \pi_s^t\|_1$ can be bound ( will be discussed later). As we did the experiments in Figure \ref{fig:epsilon-fix}, the fixed epsilon will make the embedding vector bad. We propose to use a dynamic precision parameter, where $\epsilon^t \sim \mathcal{O}(\epsilon^\prime /m_t)$, so that the $\ell_1$-norm can be properly bounded. Interestingly, this high precision parameter strategy will not cause too high time complexity cost from $\mathcal{O}(m/\alpha^2)$ to $\mathcal{O}(m \bar{d} / (\epsilon\alpha^2))$. The bounded estimation error is presented in the following theorem.

\begin{theorem}[Estimation error]
Given any node $s$, define the estimation error of PPVs learned from \textsc{DynamicSNE} at time $t$ as $\| \bm p_s^t - \bm \pi_s^t \|_1$, if we are given the precision parameter $\epsilon^t = \epsilon / m_t$, the estimation error can be bounded by the following
\begin{equation}
\| \bm p_s^t - \bm \pi_s^t \|_1 \leq \epsilon,
\end{equation}
where we require $\epsilon \leq 2$ \footnote{By noticing that $\|\bm p_s^t - \bm \pi_s^t\|_1 \leq \|\bm p_s^t\|_1 + \| \bm \pi_s^t\|_1 \leq 2$, any precision parameter larger than 2 will be meaningless. } and $\epsilon$ is a global precision parameter of \textsc{DynamicPPE} independent on $m_t$ and $n_t$.
\end{theorem}
\begin{proof}
Notice that for any node $u$, by Lemma \ref{lemma:2}, we have the following inequality
\begin{equation}
| p_s^t(u) - \pi_s^t(u) | \leq \epsilon d^t(u). \nonumber
\end{equation}
Summing all these inequalities over all nodes $u$, we have
\begin{align*}
\| \bm p_s^t - \bm \pi_s^t \|_1 &= \sum_{u\in \mathcal{V}^t} \left| p_s^t(u) - \pi_s^t(u)\right| \\
&\leq \sum_{u \in \mathcal{V}^t} \epsilon^t d^t(u) = \epsilon^t m_t = \frac{\epsilon}{m_t} m_t = \epsilon.
\end{align*}
\end{proof}

The above theorem gives estimation error guarantee of $\bm p_s^t$, which is critically important for learning high quality embeddings. First of all, the dynamic precision strategy is inevitable because the precision is unknown parameter for dynamic graph where the number of nodes and edges could increase dramatically over time. To demonstrate this issue, we conduct an experiments on the English Wikipedia graph where we learn embeddings over years and validate these embeddings by using node classification task. As shown in Figure \ref{fig:epsilon-fix}, when we use the fixed parameter, the performance of node classification is getting worse when the graph is getting bigger. This is mainly due to the lower quality of PPVs. Fortunately, the adaptive precision parameter $\epsilon / m_t$ does not make the run time increase dramatically. It only dependents on the average degree ${\bar d}^t$. In practice, we found $\epsilon =0.1$ are sufficient for learning effective embeddings.

\subsection{\textsc{DynamicPPE}}

Our finally algorithm \textsc{DynamicPPE} is presented in Algorithm \ref{algo:dynamic-ppe}. At every beginning, estimators $\bm p_s^t$ are set to be zero vectors and residual vectors $\bm r_s^t$ are set to be unit vectors with mass all on one entry (Line 4). The algorithm then call the procedure \textsc{DynamicSNE} with an empty batch as input to get initial PPVs for all target nodes \footnote{For the situation that some nodes of $S$ has not appeared in $\mathcal{G}^t$ yet, it checks every batch update until all nodes are initialized. } (Line 5). From Line 6 to Line 9, at each snapshot $t$, it gets an update batch $\Delta \mathcal{G}^t$ at Line 7 and then calls \textsc{DynamicSNE} to obtain the updated embeddings for every node $v$.

\begin{algorithm}[H]
\caption{$\textsc{DynamicPPE}(\mathcal{G}_0, S,\epsilon,\alpha)$}
\begin{algorithmic}[1]
\State \textbf{Input: } initial graph $\mathcal{G}^0$, target set $S$, global precision $\epsilon$, teleport probability $\alpha$
\State $t=0$
\For{$s \in S := \{v_1,v_2,\ldots, v_k\}$}
\State $\bm p_s^t = \bm 0,\quad \bm r_s^t = \bm 1_s$
\State \textsc{DynamicSNE}$(\mathcal{G}^0, \emptyset, s, \bm p_s^t, \bm r_s^t, 1/ m_0, \alpha)$
\EndFor
\For{$t \in \{1,2,\ldots, T\}$}
\State \text{ read a sequence of edge events $\Delta \mathcal{G}^t := \mathcal{G}^t \backslash \mathcal{G}^{t-1}$}
\For{$s \in S := \{v_1,v_2,\ldots, v_k\}$}
\State $ \bm w_s^t = \textsc{DynamicSNE}(\mathcal{G}^t, \Delta \mathcal{G}^t, s, \bm p_s^{t-1}, \bm r_s^{t-1}, \epsilon/m_t,\alpha)$
\EndFor
\EndFor
\State \textbf{return} $\bm W_s^t = [\bm w_s^1,\bm w_s^2,\ldots, \bm w_s^T], \forall s\in S$. 
\end{algorithmic}
\label{algo:dynamic-ppe}
\end{algorithm}

Based on our analysis, \textsc{DynamicPPE} is an dynamic version of \textit{InstantEmbedding}. Therefore, \textsc{DynamicPPE} has two key properties observed in \cite{postuavaru2020instantembedding}: \textit{locality} and \textit{global consistency}. The embedding quality is guaranteed from the fact that \textit{InstantEmbedding} implicitly factorizes the proximity matrix based on PPVs \cite{tsitsulin2018verse}.

\subsection{Complexity analysis}

\textbf{Time complexity}\quad The overall time complexity of \textsc{DynamicPPE} is the $k$ times of the run time of \textsc{DynamicSNE}. We summarize the time complexity of \textsc{DynamicPPE} as in the following theorem
\begin{theorem} 
The time complexity of \textsc{DynamicPPE} for learning a subset of $k$ nodes is $\mathcal{O}(k \frac{m_t}{\alpha^2} + k \frac{m_t \bar{d}^t}{\alpha^2} + \frac{m_t}{\epsilon} + k T \min\{n, \frac{m}{\epsilon \alpha}\})$
\end{theorem}

\begin{proof}
We follow Theorem \ref{thm:run-time} and summarize all run time together to get the final time complexity.
\end{proof}

\noindent\textbf{Space complexity}\quad The overall space complexity has two parts: 1) $\mathcal{O}(m)$ to store the graph structure information; and 2) the storage of keeping nonzeros of $\bm p_s^t$ and $\bm r_s^t$. From the fact that local push operation \cite{andersen2006local}, the number of nonzeros in $\bm p_s^t$ is at most $\frac{1}{\epsilon \alpha}$. Thus, the total storage for saving these vectors are $\mathcal{O}(k \min\{n,\frac{m }{\epsilon \alpha}\})$. Therefore, the total space complexity is $\mathcal{O}(m + k \min(n,\frac{m}{\epsilon\alpha}))$.

\noindent\textbf{Implementation}\quad Since learning the dynamic node embedding for any node $v$ is independent with each other, \textsc{DynamicPPE} is  are easy to parallel. Our current implementation can take advantage of multi-cores and compute the embeddings of $S$ in parallel.

\section{Experiments}
\label{sec:experiments}

To demonstrate the effectiveness and efficiency of \textsc{DynamicPPE}, in this section, we first conduct experiments on several small and large scale real-world dynamic graphs on the task of node classification, followed by a case study about changes of Chinese cities in Wikipedia graph during the COVID-19 pandemic.

\subsection{Datasets}

We compile the following three real-world dynamic graphs, more details can be found in Appendix \ref{appendix::data}. 

\noindent \textbf{Enwiki20} {English Wikipedia Network\quad} We collect the internal Wikipedia Links (WikiLinks) of English Wikipedia from the beginning of Wikipedia, January 11th, 2001, to December 31, 2020 \footnote{We collect the data from the dump \url{https://dumps.wikimedia.org/enwiki/20210101/}}. The internal links are extracted using a regular expression proposed in \cite{consonni2019wikilinkgraphs}. During the entire period, we collection 6,151,779 valid articles\footnote{A valid Wikipedia article must be in the 0 namespace}. We generated the WikiLink graphs only containing edge insertion events. We keep all the edges existing before Dec. 31 2020, and sort the edge insertion order by the creation time. There are  6,216,199 nodes and 177,862,656 edges during the entire period. Each node either has one label (\textit{Food, Person, Place,...}) or no label.
%2) \textit{Edge insertion-deletion model}. The edges in the Wikipedia network may be deleted or inserted. It is possible that an edge deleted or inserted multiple times. In this case, we only keep track of the first time insertion and the last event (either insertion or deletion). There are 255,811,491 edge insertion events where 80,893,315 edges are finally deleted during the entire period. 

\noindent \textbf{Patent} ({US Patent Graph}) \quad The citation network of US patent\cite{hall2001nber} contains 2,738,011 nodes with 13,960,811 citations range from year 1963 to 1999. For any two patents $u$ and $v$, there is an edge $(u,v)$ if the patent $u$ cites patent $v$. Each patent belongs to six different types of patent categories. We extract a small weakly-connected component of 46,753 nodes and 425,732 edges with timestamp, called \textit{Patent-small}.

\noindent \textbf{Coauthor} \quad We extracted the co-authorship network from the Microsoft Academic Graph \cite{sinha2015overview} dumped on September 21, 2019. We collect the papers with less than six coauthors, keeping those who has more than 10 publications, then build undirected edges between each coauthor with a timestamp of the publication date. In addition, we gather temporal label (e.g.: \textit{Computer Science, Art, ... }) of authors based on their publication's field of study. Specifically we assign the label of an author to be the field of study where s/he published the most up to that date. The original graph contains 4,894,639 authors and 26,894,397 edges ranging from year 1800 to 2019. We also sampled a small connected component containing 49,767 nodes and 755,446 edges with timestamp, called \textit{Coauthor-small}.

\noindent \textbf{Academic} The co-authorship network is from the academic network \cite{zhou2018dynamic,tang2008arnetminer} where it contains 51,060 nodes and 794,552 edges. The nodes are generated from 1980 to 2015. According to the node classification setting in \citet{zhou2018dynamic}, each node has either one binary label or no label.

\subsection{Node Classification Task}
\noindent\textbf{Experimental settings\quad}
We evaluate embedding quality on binary classification for Academic graph (as same as in \cite{zhou2018dynamic}), while using multi-class classification for other tasks. We use balanced logistic regression with $\ell_2$ penalty in on-vs-rest setting, and report the Macro-F1 and Macro-AUC (ROC) from 5 repeated trials with 10\% training ratio on the labeled nodes in each snapshot, excluding dangling nodes. Between each snapshot, we insert new edges and keep the previous edges.

We conduct the experiments on the aforementioned small and large scale graph. 
In small scale graphs, we calculate the embeddings of all nodes and compare our proposed method (\textbf{DynPPE.}) against other state-of-the-art models from three categories\footnote{Appendix \ref{appendix::parameter} shows the hyper-parameter settings of baseline methods}.
1) Random walk based static graph embeddings (\textbf{Deepwalk}\footnote{\url{https://pypi.org/project/deepwalk/}} \cite{perozzi2014deepwalk}, \textbf{Node2Vec}\footnote{\url{https://github.com/aditya-grover/node2vec}} \cite{grover2016node2vec}); 
2) Random Projection-based embedding method which supports online update: \textbf{RandNE}\footnote{\url{https://github.com/ZW-ZHANG/RandNE/tree/master/Python}} \cite{zhang2018billion}; 
3) Dynamic graph embedding method: \textbf{DynamicTriad} (\textbf{DynTri.}) \footnote{\url{https://github.com/luckiezhou/DynamicTriad}} \cite{zhou2018dynamic}  which models the dynamics of the graph and optimized on link prediction.

\begin{table}[!h]
\centering
\caption{Node classification task on the \textit{Academic, Patent Small, Coauthor Small} graph on the final snapshot}
\footnotesize
\begin{tabular}{p{0.10\textwidth}|p{0.10\textwidth}|p{0.08\textwidth}|p{0.08\textwidth}|p{0.08\textwidth}|p{0.08\textwidth}|p{0.08\textwidth}|p{0.08\textwidth} }
\toprule
  \multicolumn{2}{l|}{}   & \multicolumn{2}{l|}{Academic} & \multicolumn{2}{l|}{Patent Small} & \multicolumn{2}{l}{Coauthor Small}\\
%\cline{1-8}
  \multicolumn{2}{l|}{} &  F1 &  AUC  &  F1 &  AUC  &  F1 &  AUC \\
 \cline{1-8}
\multicolumn{8}{c}{$d=128$} \\
\cline{1-8}
  \multirow{2}{0mm}{Static method} & Node2Vec &0.833 &\textbf{0.975} &0.648 &0.917 &0.477 &\textbf{0.955}\\
\cline{2-8}
  &Deepwalk &\textbf{0.834} &\textbf{0.975} &\textbf{0.650} &\textbf{0.919} &\textbf{0.476} &0.950\\
\cline{1-8}
  \multirow{4}{0mm}{Dynamic method}   &DynTri. &0.817 &0.969 &0.560 &0.866 &0.435 &0.943\\
\cline{2-8}
%    &TNE &0.503 &0.791 &0.283 &0.570 &0.136 &0.630\\
%\cline{2-8}
    &RandNE &0.587 &0.867 &0.428 &0.756 &0.337 &0.830\\
\cline{2-8}
    &DynPPE. &0.808 &0.962 &0.630 &0.911 &0.448 &0.951\\
\hline
    \multicolumn{8}{c}{$d=512$} \\
\hline
  \multirow{2}{0mm}{Static method} & Node2Vec &0.841 &\textbf{0.975} &0.677 &0.931 &0.486 &0.955 \\
\cline{2-8}
  &Deepwalk &\textbf{0.842} &\textbf{0.975} &0.680 &0.931 &0.495 &0.955 \\
\cline{1-8}
  \multirow{4}{0mm}{Dynamic method}   &DynTri. &0.811 &0.965 &0.659 &0.915 &0.492 &0.952 \\
\cline{2-8}
%    &TNE &0.x &0.x &0.281 &0.565 &0.151 &0.654 \\
%\cline{2-8}
    &RandNE &0.722 &0.942 &0.560 &0.858 &0.493 &0.895 \\
\cline{2-8}
    &DynPPE. &\textbf{0.842} &0.973 &\textbf{0.682} & \textbf{0.934} &\textbf{0.509} &\textbf{0.958}\\
% tne 281
 \bottomrule
\end{tabular}
\label{tab:nc-small-table}
\end{table}

\begin{table}[!h]
\caption{Total CPU time for small graphs (in second). RandNE-I is with orthogonal projection, the performance is slightly better, but the running time is significantly increased. RandNE-II is without orthogonal projection.}

\centering
\begin{tabular}{p{0.2\textwidth}|R{0.17\textwidth}|R{0.17\textwidth}|R{0.17\textwidth}}
\toprule
  & Academic & Patent Small & Coauthor Small \\
\hline 
Deepwalk\footnotemark & 498,211.75& 181,865.56 & 211,684.86 \\
\hline 
Node2vec & 4,584,618.79& 2,031,090.75 & 1,660,984.42 \\
\hline 
 DynTri. & 247,237.55 & 117,993.36 & 108,279.4 \\
\hline 
 RandNE-I & 12,732.64 & 9,637.15 & 8,436.79 \\
\hline 
 RandNE-II &  1,583.08 & 9,208.03 & 177.89 \\
\hline 
 DynPPE. & 18,419.10 & 3,651.59 & 21,323.74 \\
\bottomrule
\end{tabular}
\label{tab:runtime-small}
\end{table}
\footnotetext{ We ran static graph embedding methods over a set of sampled snapshots and estimate the total CPU time for all snapshots.}

\begin{figure*}[h!]
	\centering
	\subfloat{\includegraphics[width=.31\textwidth]{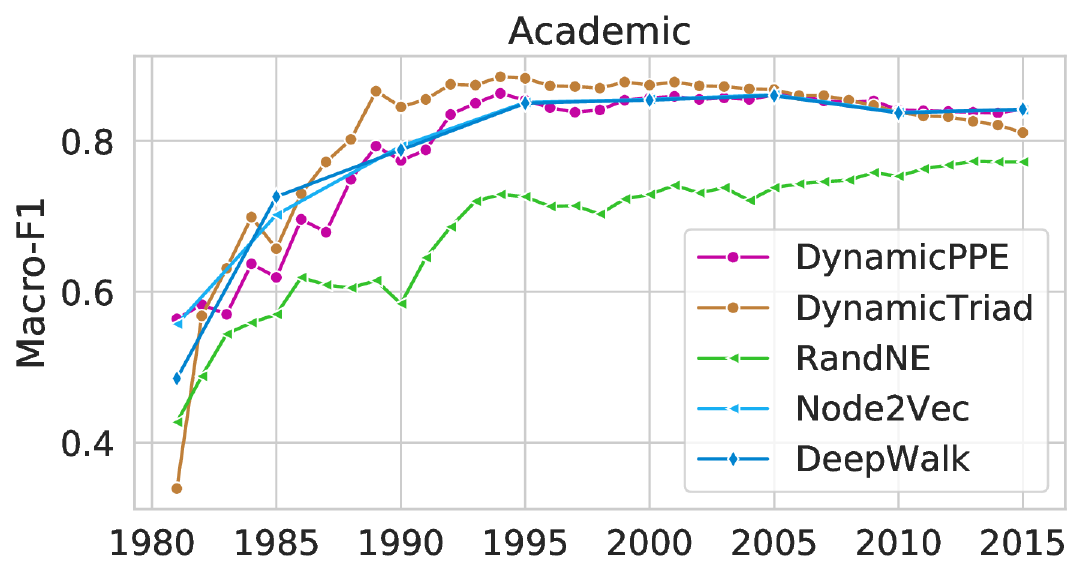}}\quad
	\subfloat{\includegraphics[width=.31\textwidth]{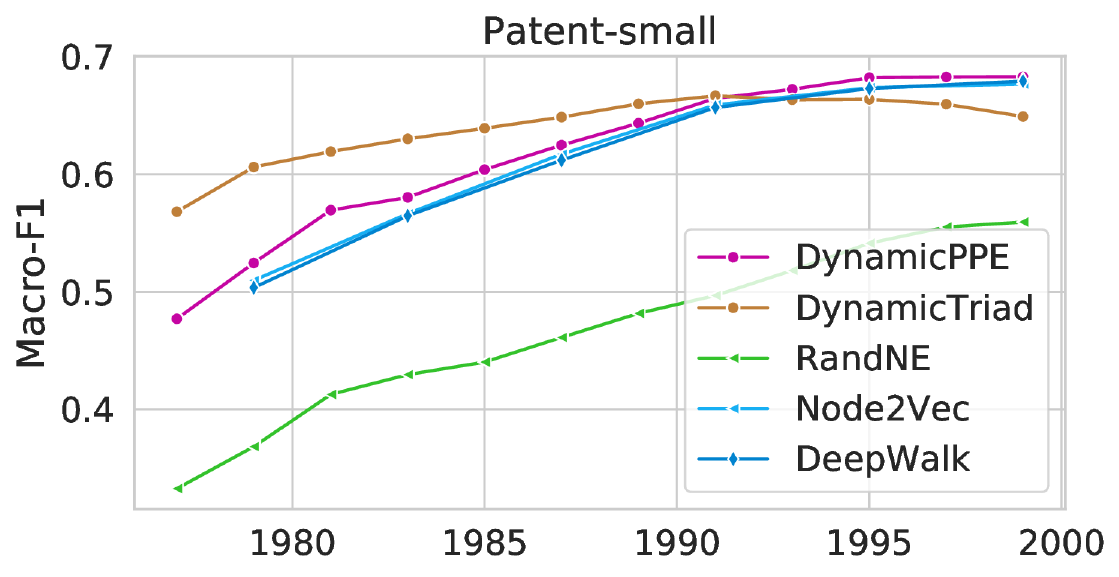}}\quad
	\subfloat{\includegraphics[width=.31\textwidth]{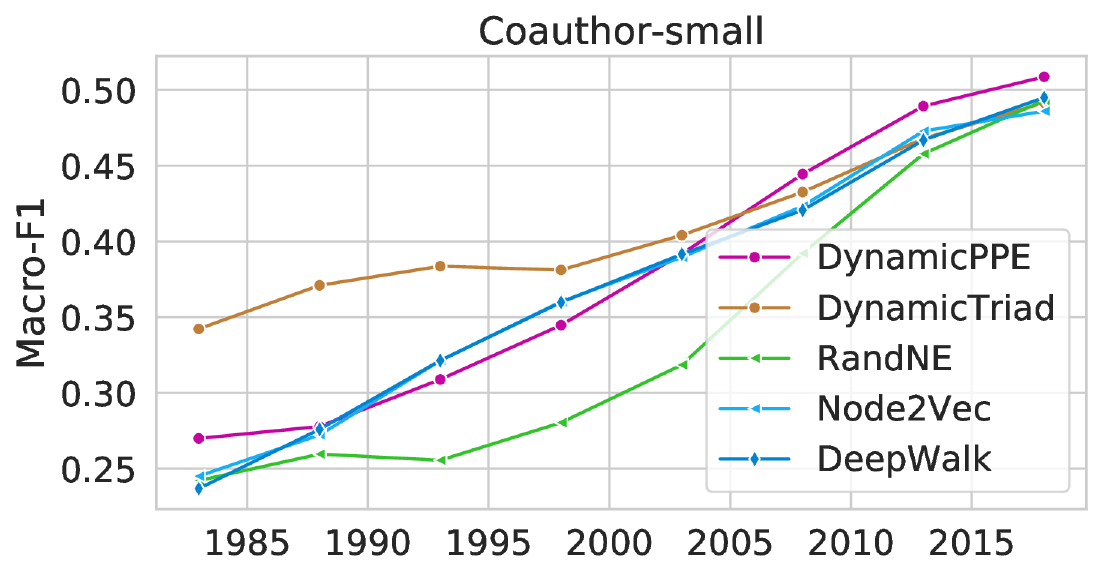}}
	
	\subfloat{\includegraphics[width=.31\textwidth]{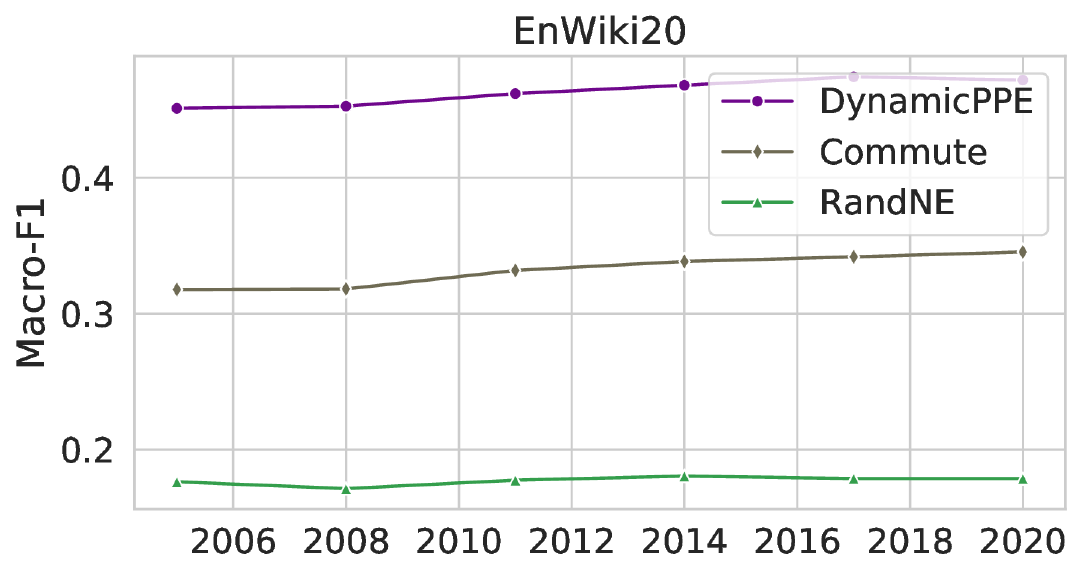}}\quad
	\subfloat{\includegraphics[width=.31\textwidth]{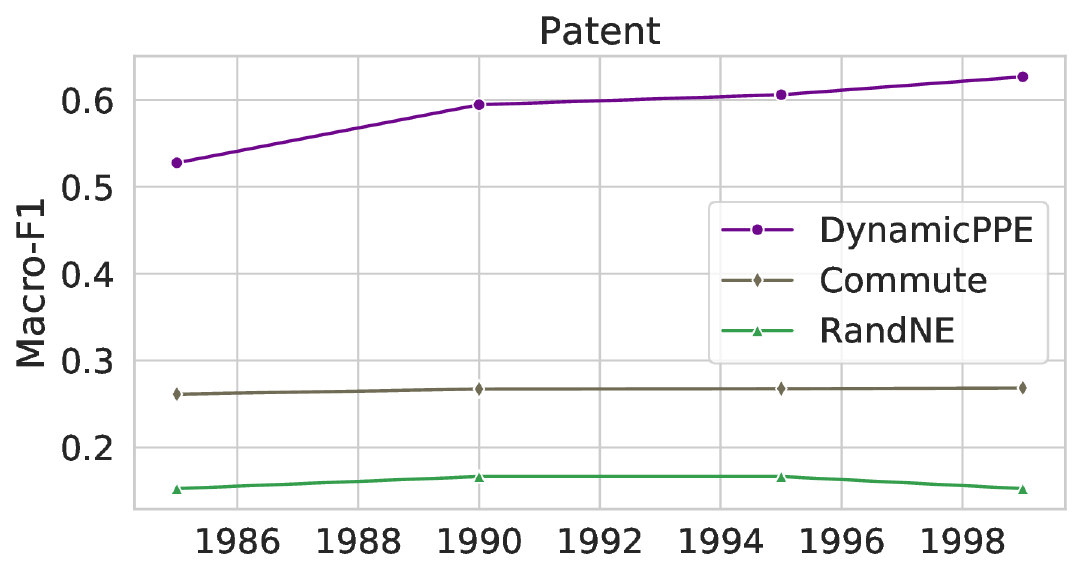}}\quad
	\subfloat{\includegraphics[width=.31\textwidth]{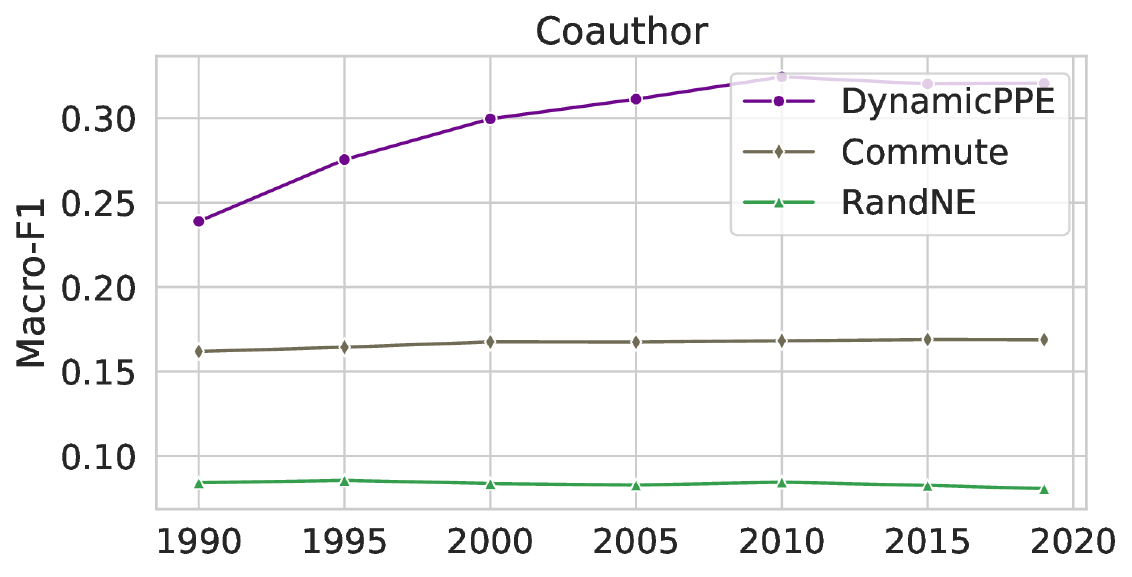}}

    \caption{Macro-F1 scores of node classification as a function of time. The results of the small and large graphs are on the first and second row respectively (dim=512). Our proposed methods achieves the best performance on the last snapshot when all edges arrived and performance curve matches to the static methods as the graph becomes more and more complete.}
    \label{fig:nc-full-macro-f1-over-time}
\end{figure*}

Table \ref{tab:nc-small-table} shows the classification results on the final snapshot. When we restrict the dimension to be 128, we observe that static methods outperform all dynamic methods. However, the static methods independently model each snapshot and the running time grows with number of snapshots, regardless of the number of edge changes.
In addition, our proposed method (DynPPE.) outperforms other dynamic baselines, except that in the \textit{academic} graph, where DynTri. is slightly better. However, their CPU time is 13 times more than ours as shown in Table \ref{tab:runtime-small}. 
According to the Johnson-Lindenstrauss lemma\cite{johnson1984extensions,dasgupta1999elementary}, we suspect that the poor result of RandNE is partially caused by the dimension size. As we increase the dimension to 512, we see a great performance improvement of RandNE. Also, our proposed method takes the same advantage, and outperform all others baselines in F1-score. Specifically, the increase of dimension makes hash projection step (Line 12-13 in Algorithm \ref{algo:dynamic-sne}) retain more information and yield better embeddings. We attach the visualization of embeddings in Appendix.\ref{tsne-embedding}.

The first row in the Figure \ref{fig:nc-full-macro-f1-over-time} shows the F1-scores in each snapshot when the dimension is 512. We observe that in the earlier snapshots where many edges are not arrived, the performance of DynamicTriad \cite{zhou2018dynamic} is better. One possible reason is that it models the dynamics of the graph, thus the node feature is more robust to the "missing" of the future links. While other methods, including ours, focus on an online feature updates incurred by the edge changes without explicitly modeling the temporal dynamics of a graph. Meanwhile, the performance curve of our method matches to the static methods, demonstrating the embedding quality of the intermediate snapshots is comparable to the state-of-the-art.

Table \ref{tab:runtime-small} shows the CPU time of each method (Dim=512). As we expected, static methods is very expensive as they calculate embeddings for each snapshot individually. Although our method is not blazingly fast compared to RandNE, it has a good trade-off between running time and embedding quality, especially without much hyper-parameter tuning. Most importantly, it can be easily parallelized by distributing the calculation of each node to clusters.

We also conduct experiment on large scale graphs (\textit{EnWiki20, Patent, Coauthor}). We keep track of the vertices in a subset containing $|S|=3,000$ nodes randomly selected from the first snapshot in each dataset, and similarly evaluate the quality of the embeddings of each snapshot on the node classification task. Due to scale of the graph, we compare our method against \textbf{RandNE} \cite{zhang2018billion} and an fast heuristic \textbf{Algorithm \ref{algo:commute}}. 
 Our method can calculate the embeddings for a subset of useful nodes only, while other methods have to calculate the embeddings of all nodes, which is not necessary under our scenario detailed in Sec. \ref{sec:case-study}. The second row in Figure \ref{fig:nc-full-macro-f1-over-time} shows that our proposed method has the best performance. 
%To our surprise, the heuristic method beats RandNE. \textbf{to discuss...}. %A possible explanation is that our initialization is better?

\begin{table}[!h]
\caption{Total CPU time for large graphs (in second) }
\centering
\begin{tabular}{p{0.15\textwidth}|R{0.18\textwidth}|R{0.18\textwidth}|R{0.18\textwidth}}\toprule 
  & Enwiki20 & Patent & Coauthor \\
\hline 
 Commute & 6,702.1 & 639.94 & 1,340.74 \\
\hline 
 RandNE-II & 47,992.81 & 6,524.04 & 20,771.19 \\
\hline 
 DynPPE. & 1,538,215.88 & 139,222.01 & 411,708.9 \\
\hline 
 DynPPE (Per-node) & 512.73 & 46.407 & 137.236 \\
 \bottomrule
\end{tabular}
\label{tab:runtime-large}
\end{table}

Table \ref{tab:runtime-large} shows the total CPU time of each method (Dim=512). Although total CPU time of our proposed method seems to be the greatest, the average CPU time for one node (as shown in row 4) is significantly smaller. This benefits a lot of downstream applications where only a subset of nodes are interesting to people in a large dynamic graph. For example, if we want to monitor the weekly embedding changes of a single node (e.g., the city of Wuhan, China) in English Wikipedia network from year 2020 to 2021, we can have the results in roughly 8.5 minutes. 
Meanwhile, other baselines have to calculate the embeddings of all nodes, and this expensive calculation may become the bottleneck in a downstream application with real-time constraints. To demonstrate the usefulness of our method, we conduct a case study in the following subsection.

\subsection{Change Detection}
\label{sec:case-study}
%https://www.who.int/news/item/27-04-2020-who-timeline---covid-19
%https://en.wikipedia.org/wiki/Extinction:_The_Facts
%https://en.wikipedia.org/wiki/The_Pandemic_Special
%https://en.wikipedia.org/wiki/Timeline_of_the_COVID-19_pandemic_in_October_2020
%https://china.usembassy-china.org.cn/embassy-consulates/chengdu/
% about hangzhou: 
%https://en.wikipedia.org/wiki/Ingrid_Wilm
%https://en.wikipedia.org/wiki/Kimberly_Ince
%https://en.wikipedia.org/wiki/Aela_Janvier

Thanks to the contributors timely maintaining Wikipedia, we believe that the evolution of the Wikipedia graph reflects how the real world is going on. We assume that when the structure of a node greatly changes, there must be underlying interesting events or anomalies. Since the structural changes can be well-captured by graph embedding, we use our proposed method to investigate whether anomalies happened to the Chinese cities during the COVID-19 pandemic (from Jan. 2020 to Dec. 2020).

\begin{table*}[!ht]
\caption{The top cities ranked by the z-score along time. The corresponding news titles are from the news in each time period. $d(v)$ is node degree, $\Delta d(v)$ is the degree changes from the previous timestamp.}
\centering
\footnotesize
\begin{tabular}{p{0.08\textwidth}|p{0.085\textwidth}|p{0.05\textwidth}|p{0.07\textwidth}|p{0.055\textwidth}|p{0.41\textwidth}}
\toprule 
 Date & City & $d(v)$ & $\Delta d(v)$ & Z-score & Top News Title \\
\hline 
 1/22/20 & Wuhan & 2890 & 54 & 2.210 & NBC: "New virus prompts U.S. to screen passengers from Wuhan, China" \\
\hline 
 2/2/20 & Wuhan & 2937 & 47 & 1.928 & WSJ: "U.S. Sets Evacuation Plan From Coronavirus-Infected Wuhan" \\
\hline 
 2/13/20 & Hohhot & 631 & 20 & 1.370 & Poeple.cn: "26 people in Hohhot were notified of dereliction of duty for prevention and control, and the director of the Health Commission was removed" (Translated from Chinese). \\
 %(Original text: "呼和浩特26人因防控失职被通报 卫健委主任被免职") \\
\hline 
 2/24/20 & Wuhan & 3012 & 38 & 2.063 & USA Today: "Coronavirus 20 times more lethal than the flu? Death toll passes 2,000" \\
\hline 
 3/6/20 & Wuhan & 3095 & 83 &  1.723 & Reuters: "Infelctions may drop to zero by end-March in Wuhan: Chinese government expert" \\
\hline 
 3/17/20 & Wuhan & 3173 & 78 & 1.690 & NYT: "Politicians Use of 'Wuhan Virus' Starts a Debate Health Experets Wanted to Avoid" \\
\hline 
 3/28/20 & Zhang Jia Kou & 517  & 15 & 1.217 & "Logo revealed for Freestyle Ski and Snowboard World Championships in Zhangjiakou" \\
\hline 
 4/8/20 & Wuhan & 3314  & 47               & 2.118 & CNN:"China lifts 76-day lockdown on Wuhan as city reemerges from conronavirus crisis" \\
\hline 
 4/19/20 & Luohe          & 106 & 15 & 2.640 & Forbes: "The Chinese Billionaire Whose Company Owns Troubled Pork Processor Smithfield Foods"  \\
\hline 
 4/30/20 & Zunhua & 52 & 17 & 2.449 & XINHUA: "Export companies resume production in Zunhua, Hebei" \\
\hline 
 5/11/20  & Shulan & 88 & 46      & 2.449 & CGTN: "NE China's Shulan City to reimpose community lockdown in 'wartime' battle against COVID-19" \\
 \bottomrule
\end{tabular}
\label{tab:z-score-city-changes}
\end{table*}

\noindent \textbf{Changes of major Chinese Cities\quad} We target nine Chinese major cities (\textit{Shanghai, Guangzhou, Nanjing, Beijing, Tianjin, \textbf{Wuhan}, Shenzhen, Chengdu, Chongqing)} and keep track of the embeddings in a 10-day time window.  From our prior knowledge, we expect that \textit{Wuhan} should greatly change since it is the first reported place of the COVID-19 outbreak in early Jan. 2020.

\begin{figure}[!h]
	\centering
	\subfloat{\includegraphics[width=.45\textwidth]{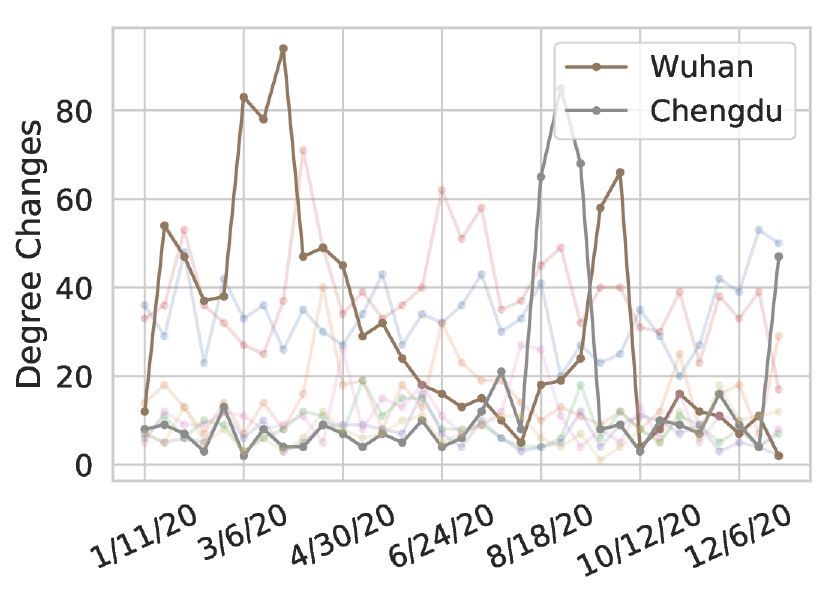}}\quad%
	\subfloat{\includegraphics[width=.50\textwidth]{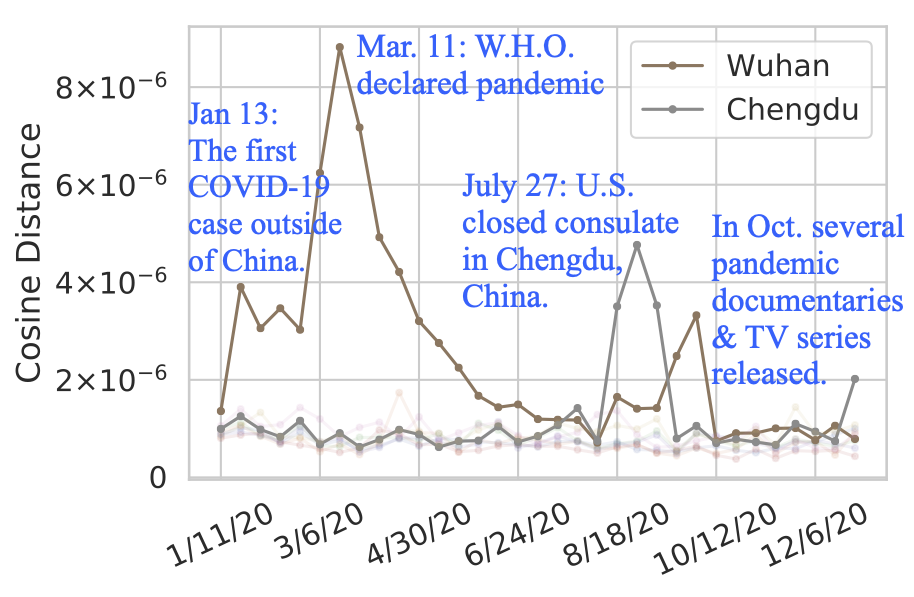}}

% 	\begin{subfigure}{.45\textwidth}
% 		\includegraphics[width=6.3cm]{figs/Case-Study/cs-Wuhan-Chengdu-deg-change.pdf}%
		%\caption{}
% 	\end{subfigure}
% 	\begin{subfigure}{.45\textwidth}
% 		\includegraphics[width=6.8cm]{figs/Case-Study/cs-Wuhan-Chengdu-cos.png}%
% 		%\caption{}
% 	\end{subfigure}
	%\begin{subfigure}{.3\textwidth}
	%	\includegraphics[width=5cm]{figs/Case-Study/cs-Wuhan-Chengdu-z-score-date.pdf}%
	%	\caption{Changes in Z-score}
	%\end{subfigure}
    \caption{The changes of major Chinese cities in 2020. \textit{Left}:Changes in Node Degree. \textit{Right}: Changes in Cosine distance }
    \label{fig:cs-changes}
\end{figure}

Figure.\ref{fig:cs-changes}(a) shows the node degree changes of each city every 10 days. The curve is quite noisy, but we can still see several major peaks from \textit{Wuhan} around 3/6/20 and \textit{Chengdu} around 8/18/20. When using embedding changes \footnote{Again, the embedding movement $\operatorname{Dist}(\cdot,\cdot)$ is defined as $1-\cos(\cdot,\cdot)$}  as the measurement,  Figure.\ref{fig:cs-changes} (b) provides a more clear view of the changed cities. We observe strong signals from the curve of \textit{Wuhan}, correlating to the initial COVID-19 outbreak and the declaration of pandemic \footnote{COIVD-19: \url{https://www.who.int/news/item/27-04-2020-who-timeline---covid-19}}. 
In addition, we observed an peak from the curve of \textit{Chengdu} around 8/18/20 when U.S. closed the consulate in Chengdu, China, reflecting the U.S.-China diplomatic tension\footnote{US Consulate:  \url{https://china.usembassy-china.org.cn/embassy-consulates/chengdu/}}.

\noindent \textbf{Top changed city along time\quad} We keep track of the embedding movement of 533 Chinese cities from year 2020 to 2021 in a 10-day time window, and filter out the inactive records by setting the threshold of degree changes (e.g. greater than 10). The final results are 51 Chinese cities from the major ones listed above and less famous cities like \textit{Zhangjiakou, Hohhot, Shulan, ...}.

Furthermore, we define the z-score of a target node $u$ as $Z_t(u)$ based on the embedding changes within a specific period of time from $t-1$ to $t$.
%{\tiny
\begin{align*}
 Z_t(u|u\in S) &= \frac{Dist(w_u^{t}, w_u^{t-1}) - \mu  }{\sigma} \\
    \mu = \frac{1}{|S|} \sum_{u' \in S} Dist(w_{u'}^{t}, w_{u'}^{t-1})&,\  \ 
    \sigma = \sqrt{\frac{1}{|S|} \sum_{u' \in S} (Dist(w_{u'}^{t} w_{u'}^{t-1}) - \mu)^2 } \\
\end{align*}
%}

In Table \ref{tab:z-score-city-changes}, we list the highest ranked cities by the z-score from Jan.22 to May 11, 2020. In addition, we also attach the top news titles corresponding to the city within each specific time period. We found that Wuhan generally appears more frequently as the situation of the pandemic kept changing. Meanwhile, we found the appearance of \textit{Hohhot} and \textit{Shulan} reflects the time when COVID-19 outbreak happened in those cities. We also discovered cities unrelated to the pandemic. For example, \textit{Luohe}, on 4/19/20, turns out to be the city where the headquarter of the organization which acquired Smithfield Foods (as mentioned in the news). In addition, \textit{Zhangjiakou}, on 3/28/20, is the city, which will host World Snowboard competition, released the Logo of that competition.

\section{Discussion and Conclusion}
\label{sec:conclusion}

In this paper, we propose a new method to learn dynamic node embeddings over large-scale dynamic networks for a subset of interesting nodes. Our proposed method has time complexity that is linear-dependent on the number of edges $m$ but independent on the number of nodes $n$. This makes our method applicable to applications of subset representations on very large-scale dynamic graphs. For the future work, as shown in \citet{trivedi2019dyrep}, there are two dynamics on dynamic graph data, \textit{structure evolution} and dynamics of \textit{node interaction}.  It would be interesting to study how one can incorporate dynamic of node interaction into our model.  It is also worth to study how different version of local push operation affect the performance of our method.

\paragraph{Acknowledges}
This work was partially supported by NSF grants IIS-1926781, IIS-1927227, IIS-1546113 and OAC-1919752.

% \bibliographystyle{ACM-Reference-Format}
% \bibliography{references.bib}

% \clearpage
% \appendix

\section{Appendix}
\subsection{Proof of lemmas}

To prove Lemma \ref{lemma:1}, we first introduce the property of uniqueness of PPR $\bm \pi_s$ for any $s$.

\begin{proposition}[Uniqueness of PPR \citep{andersen2007using}]
\label{prop:1}
For any starting vector $\bm 1_s$, and any constant $\alpha \in (0,1]$, there is a unique vector $\bm \pi_s$ satisfying (\ref{equ:static-ppr-static-graph}).
\end{proposition}
\begin{proof}
Recall the PPR equation
\begin{equation}
\bm \pi_s = \alpha \bm 1_s + (1-\alpha) \bm D^{-1}\bm A \bm \pi_s. \nonumber
\end{equation}
We can rewrite it as $(\bm I - (1-\alpha)\bm D^{-1} \bm A) \bm \pi_s = \alpha  \bm 1_s$. Notice the fact that matrix $\bm I - (1-\alpha)\bm D^{-1} \bm A$ is  strictly diagonally dominant matrix. To see this, for each $i \in \mathcal{V}$, we have $1 - (1-\alpha)\sum _{j\neq i}|1/d(i)| = \alpha > 0$. By \cite{lei2003schur}, strictly diagonally dominant matrix is always invertible.
\end{proof}

\begin{proposition}[Symmetry property \citep{lofgren2015bidirectional}]
Given any undirected graph $\mathcal{G}$, for any $\alpha \in (0,1)$ and for any node pair $(u,v)$, we have
\begin{equation}
    d(u) \pi_u(v) = d(v) \pi_v (u).
\end{equation}
\label{proposition:symmetry-property}
\end{proposition}

\begin{proof}[Proof of Lemma 7]
Assume there are $T$ iterations. For each forward push operation $t=1,2,\ldots T$, we assume the frontier node is $u_t$, the run time of one push operation is then $d(u_t)$. For total $T$ push operations, the total run time is $\sum_{i=1}^T d(u_i)$. Notice that during each push operation, the amount of $\|\bm r_s^{t-1}\|_1$ is reduced at least $\epsilon \alpha d(u_t)$, then we always have the following inequality
\begin{equation}
\epsilon \alpha d(u_t) < \| \bm r_s^{t-1} \|_1 - \| \bm r_s^t\|_1 \nonumber
\end{equation}
Apply the above inequality from $t=1,2,$ to $T$, we will have
\begin{equation}
\epsilon\alpha \sum_{t=1}^T d(u_t) \leq \|\bm r_s^0\| - \|\bm r_s^T\|_1 = 1 - \|\bm r_s \|_1,
\end{equation}
where $\bm r_s$ is the final residual vector. The total time is then $\mathcal{O}(1/\epsilon\alpha)$. To show the estimation error, we follow the idea of \cite{lofgren2015efficient}. Notice that, the forward local push algorithm always has the invariant property by Lemma \ref{lemma:1}, that is 
\begin{equation}
\pi_s(u) = p_s(u) + \sum_{v \in V} r_s(v) \pi_v(u), \forall u \in \mathcal{V}.
\end{equation}
By proposition \ref{proposition:symmetry-property}, we have
\begin{align*}
\pi_s(u) &= p_s(u) + \sum_{v \in V} r_s(v) \pi_v(u), \forall u \in \mathcal{V} \\
&= p_s(u) + \sum_{v \in V} r_s(v) \frac{d(u)}{d(v)} \pi_u(v), \forall u \in \mathcal{V} \\
&\leq p_s(u) + \sum_{v \in V} \epsilon d(u) \pi_u(v), \forall u \in \mathcal{V} = p_s(u) + \epsilon d(u),
\end{align*}
where the first inequality by the fact that $r_s(v) \leq \epsilon d(v)$ and the last equality is due to $\| \pi_u \|_1 = 1$.
\end{proof}

\begin{proposition}[\cite{zhang2016approximate}]
Let $\mathcal{G}=(V,E)$ be undirected and let $t$ be a vertex of $V$, then $\sum_{x\in V} \frac{\pi_x(t)}{d(t)} \leq 1$.
\end{proposition}
\begin{proof}
By using Proposition \ref{proposition:symmetry-property}, we have
\begin{align*}
\sum_{x\in V} \frac{\pi_x(t)}{d(t)} = \sum_{x\in V} \frac{\pi_t(x)}{d(x)} \leq \sum_{x\in V} \pi_t(x) = 1. 
\end{align*}
\end{proof}
\subsection{Heuristic method: Commute}

We update the embeddings by their pairwise relationship (resistance distance). The commute distance (i.e. resistance distance) $C_{u v} = H_{u v} + H_{v u}$, where  rescaled hitting time $H_{u v}$ converges to $1/d(v)$. As proved in~\cite{von2014hitting}, when the number of nodes in the graph is large enough, we can show that the commute distance tends to $1/d_v + 1/d_u$.

\begin{minipage}{0.9\textwidth}
\begin{algorithm}[H]
\caption{\textsc{Commute}}
\label{algo:commute}
\begin{algorithmic}[1]
\State \textbf{Input:} An graph $\mathcal{G}^0(\mathcal{V}^0,\mathcal{E}^0)$ and embedding $\bm W^0$, dimension $d$.
\State \textbf{Output:} $\bm W^T$
\For {$e^t(u,v,t) \in \left\{e^1(u^1,v^1,t_1), \ldots, e^T(u^T,v^T,t_T)\right\}$}
\State Add $e^t$ to $\mathcal{G}^t$
\State \textbf{If } $u\notin V^{t-1}$ \textbf{ then } 
\State \qquad generate $\bm w_u^t = \mathcal{N}(\bm 0,0.1\cdot\bm I_d)$ or $ \mathcal{U}(-0.5,0.5)/d$
\State \textbf{If } $v\notin V^{t-1}$ \textbf{ then } 
\State \qquad generate $\bm w_v^t = \mathcal{N}(\bm 0,0.1\cdot\bm I_d)$ or $\mathcal{U}(-0.5,0.5)/d$
\State $\bm w_u^t = \frac{d(u)}{d(u)+1} \bm w_u^{t-1} + \frac{1}{d(u)} \bm w_v^t$
\State $\bm w_v^t = \frac{d(v)}{d(v)+1} \bm w_v^{t-1} + \frac{1}{d(v)} \bm w_u^t$
\EndFor
\State \textbf{Return} $\bm W_T$
\end{algorithmic}
\end{algorithm}
\end{minipage}

One can treat the Commute method, i.e. Algorithm \ref{algo:commute}, as the first-order approximation of RandNE \cite{zhang2018billion}. The embedding generated by RandNE is given as the following
\begin{equation}
    \bm U = \left( \alpha_0 \bm I + \alpha_1 \bm A + \alpha_2 \bm A^2 + \ldots + \alpha_q \bm A^q \right) \bm R,
\end{equation}
where $\bm A$ is the normalized adjacency matrix and $\bm I$ is the identity matrix. At any time $t$, the dynamic embedding of node $i$ of Commute is given by
\begin{align*}
\bm w_i^t &= \frac{d(u)}{d(u)+1} \bm w_i^{t-1} + \frac{1}{d(u)} \bm w_v^t    \\
&= \frac{1}{d(u)+1} \bm w_i^{0} + \frac{1}{|\mathcal{N}(i)|} \sum_{j\in \mathcal{N}(i)} \frac{1}{d(u)} \bm w_v^t    \\
\end{align*} 
\subsection{Details of data preprocessing}
\label{appendix::data}
In this section, we describe the preprocessing steps of three datasets. \textbf{Enwiki20}: In Enwiki20 graph, the edge stream is divided into 6 snapshots, containing edges before 2005, 2005-2008, ..., 2017-2020. The sampled nodes in the first snapshot fall into 5 categories. \textbf{Patent}: In full patent graph, we divide edge stream into 4 snapshots, containing patents citation links before 1985, 1985-1990,..., 1995-1999. In node classification tasks, we sampled 3,000 nodes in the first snapshot, which fall into 6 categories. 
In patent small graph, we divide into 13 snapshots with a 2-year window. All the nodes in each snapshot fall into 6 categories.%
%patent full: [1,2,3,4,5,6]
%academic: [4]
%mag_coauthor_small: 14 classes[0,3,4,5,7,9,10,15,17,6,11,12,14,16]
%mag_coauthor large: 9 classes [1,2,3,8,11,14,16,17,18]
%enwiki: [0,1,2,4,6]
%patent: [1,2,3,4,5,6]
\textbf{Coauthor graph}: In full Coauthor graph, we divide edge stream into 7 snapshots (before 1990, 1990-1995, ..., 2015-2019). The sampled nodes in the first snapshot fall into 9 categories. 
In Coauthor small graph, the edge stream is divided into 9 snapshots (before 1978, 1978-1983,..., 2013-2017). All the nodes in each snapshot have 14 labels in common.
\subsection{Details of parameter settings}
\label{appendix::parameter}

\textbf{Deepwalk}: number-walks=40, walk-length=40, window-size=5\\
\textbf{Node2Vec}: Same as Deepwalk, p = 0.5, q = 0.5\\
\textbf{DynamicTriad}: iteration=10, beta-smooth=10, beta-triad=10. Each input snapshot contains the previous existing edges and newly arrived edges. \\
\textbf{RandNE}: q=3, default weight for node classification $[1,1e2,1e4,1e5]$, input is the transition matrix, the output feature is normalized (l-2 norm) row-wise. \\
\textbf{DynamicPPE}: $\alpha=0.15$,  $\epsilon=0.1$, projection method=hash. our method is relatively insensitive to the hyper-parameters. \\
\textbf{Infrastructure}: 64 CPUs with 16 cores on each (Intel(R) Xeon(R) Silver 4216 CPU @ 2.10GHz) with 376 GB memory.
\subsection{Visualizations of Embeddings }
\label{tsne-embedding}
\begin{figure}[htbp]
	\centering
	\subfloat{\includegraphics[width =.3\textwidth]{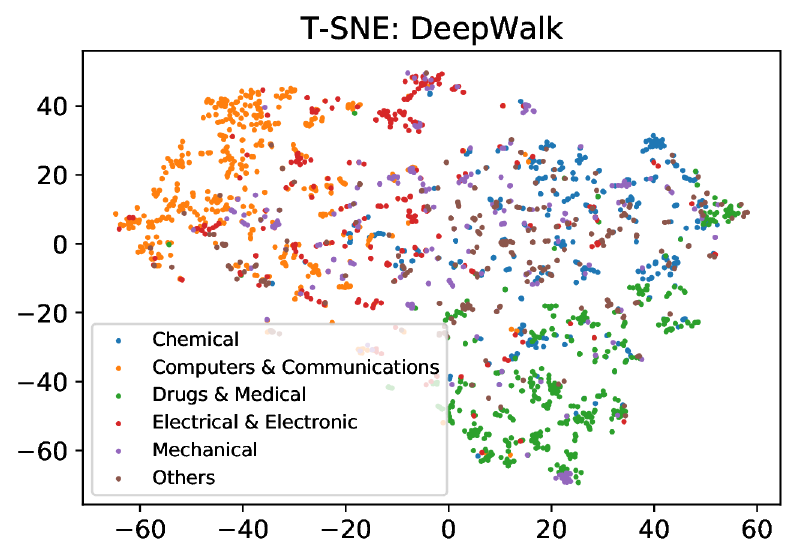}}\quad
	\subfloat{\includegraphics[width =.3\textwidth]{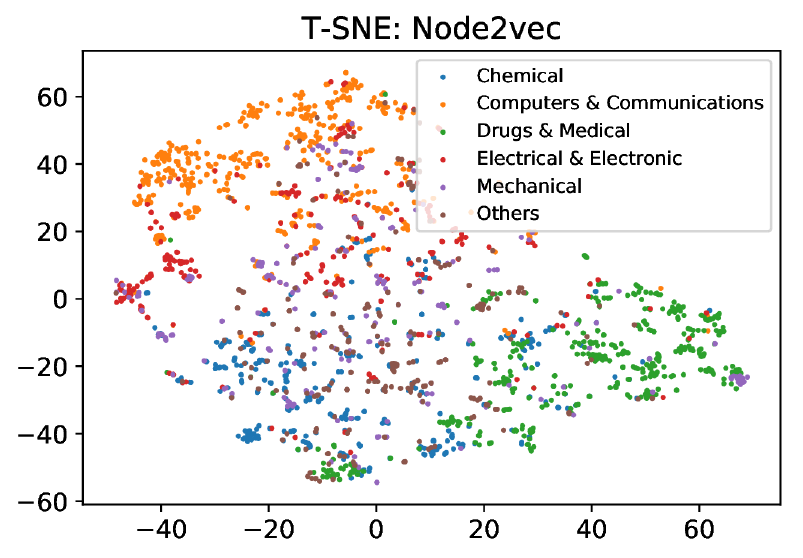}}\quad
	\subfloat{\includegraphics[width =.3\textwidth]{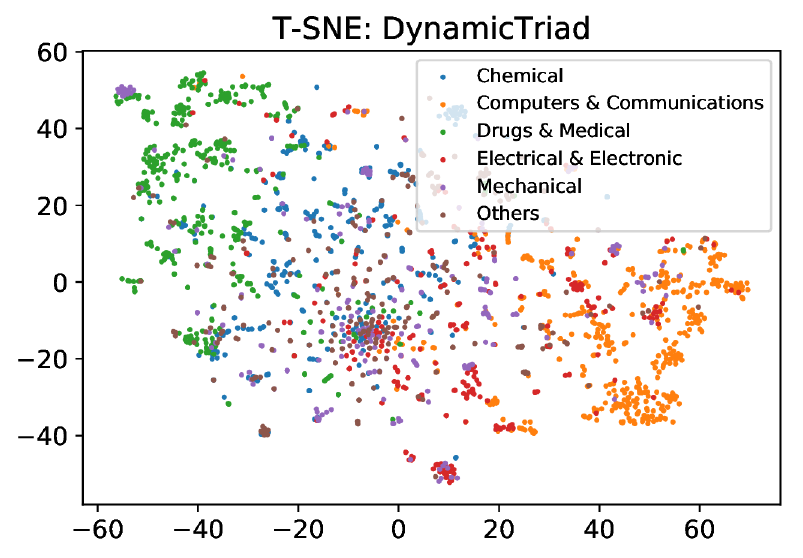}}\quad
	\subfloat{\includegraphics[width =.3\textwidth]{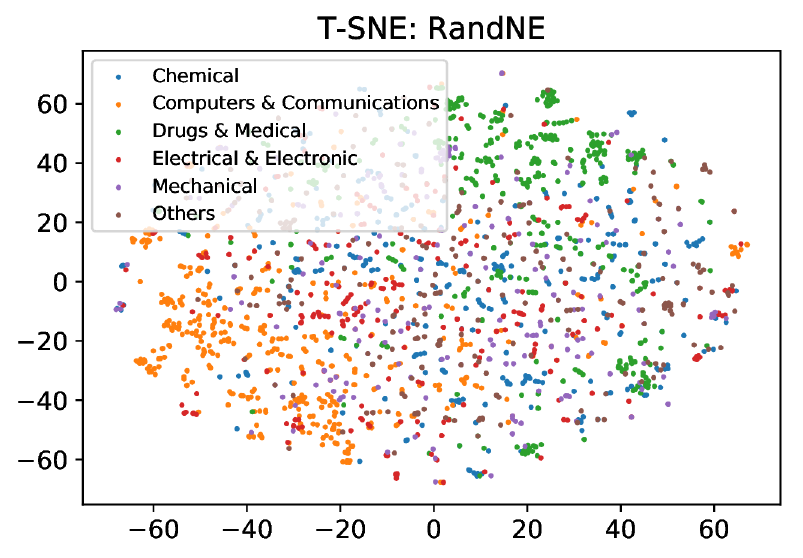}}\quad
	\subfloat{\includegraphics[width =.3\textwidth]{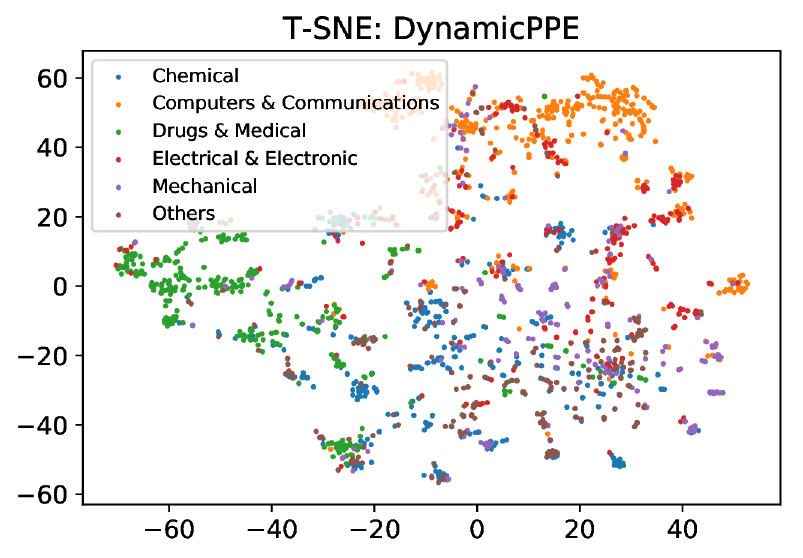}}

    \caption{We randomly select 2,000 nodes from patent-small and visualize their embeddings using T-SNE\cite{van2008visualizing}}.
    \label{fig:nc-tsne-patent-small}
\end{figure}

\begin{figure}[htbp]
	\centering
	\subfloat{\includegraphics[width =.3\textwidth]{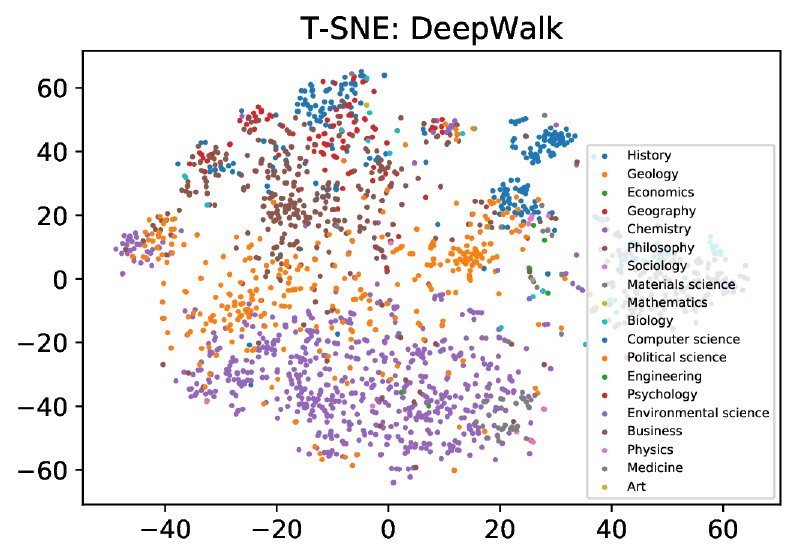}}\quad
	\subfloat{\includegraphics[width =.3\textwidth]{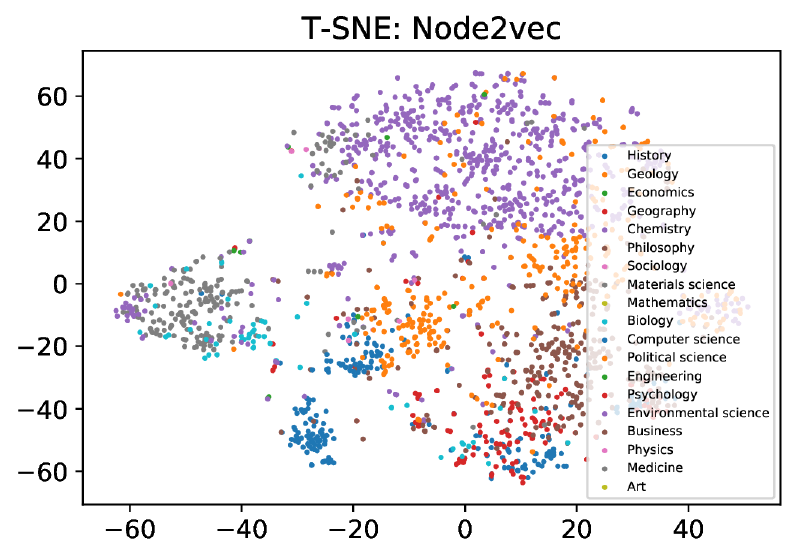}}\quad
	\subfloat{\includegraphics[width =.3\textwidth]{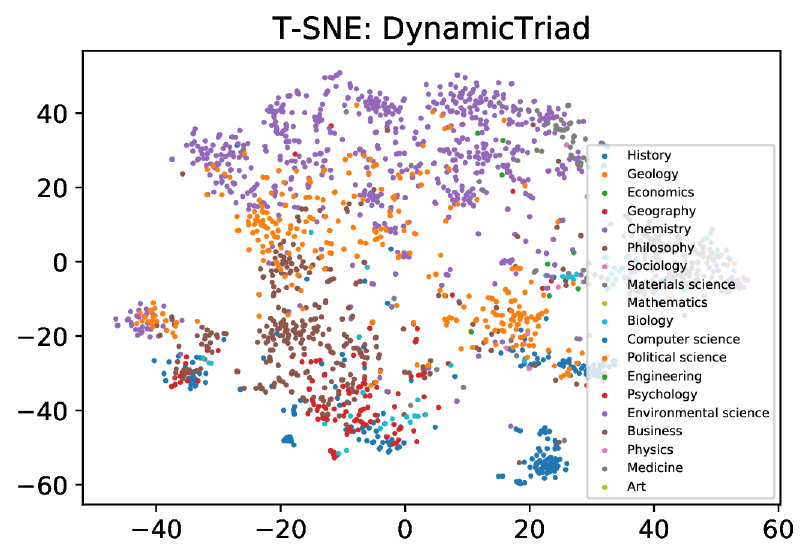}}\quad
	\subfloat{\includegraphics[width =.3\textwidth]{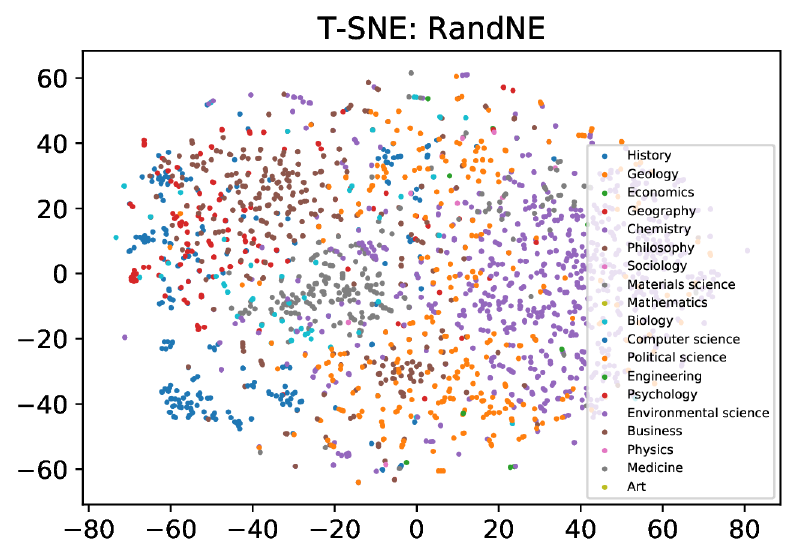}}\quad
	\subfloat{\includegraphics[width =.3\textwidth]{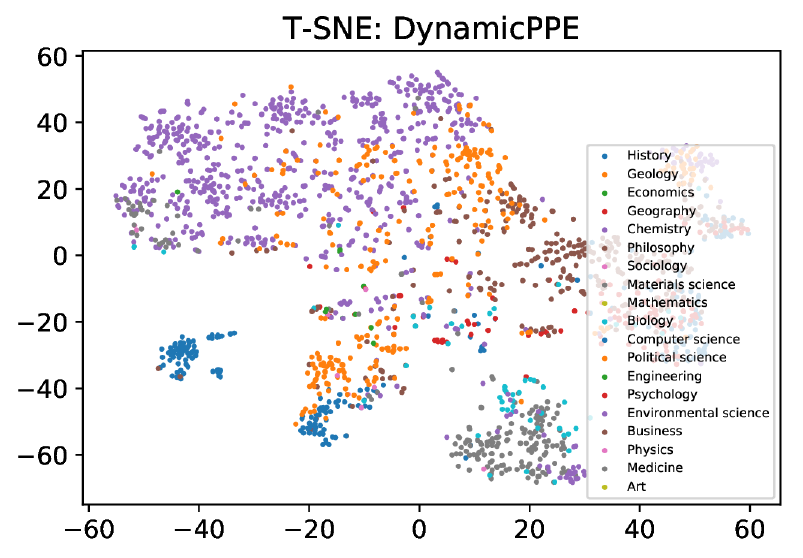}}

    \caption{We randomly select 2,000 nodes from Coauthor-small graph and visualize their embeddings using T-SNE\cite{van2008visualizing}}.
    \label{fig:nc-tsne-coauthor-small}
\end{figure}

\begin{figure}[htbp]
% 	\centering
% 	\begin{subfigure}{.23\textwidth}
% 		\includegraphics[width=3cm]{figs/TSNE-Plot/TSNE-academic-DeepWalk.pdf}
% 		%\caption{Changes in Node Degree}
% 	\end{subfigure}
% 	\begin{subfigure}{.23\textwidth}
% 		\includegraphics[width=3cm]{figs/TSNE-Plot/TSNE-academic-Node2vec.pdf}
% 		%\caption{Changes in Cosine distance}
% 	\end{subfigure}
% 	\begin{subfigure}{.23\textwidth}
% 		\includegraphics[width=3cm]{figs/TSNE-Plot/TSNE-academic-DynamicTriad.pdf}
% 		%\caption{Changes in Cosine distance}
% 	\end{subfigure}
% 	\begin{subfigure}{.23\textwidth}
% 		\includegraphics[width=3cm]{figs/TSNE-Plot/TSNE-academic-RandNE.pdf}
% 		%\caption{Changes in Cosine distance}
% 	\end{subfigure}
% 	\begin{subfigure}{.23\textwidth}
% 		\includegraphics[width=3cm]{figs/TSNE-Plot/TSNE-academic-DynamicPPE.pdf}
% 		%\caption{Changes in Cosine distance}
% 	\end{subfigure}
	
	\centering
	\subfloat{\includegraphics[width =.3\textwidth]{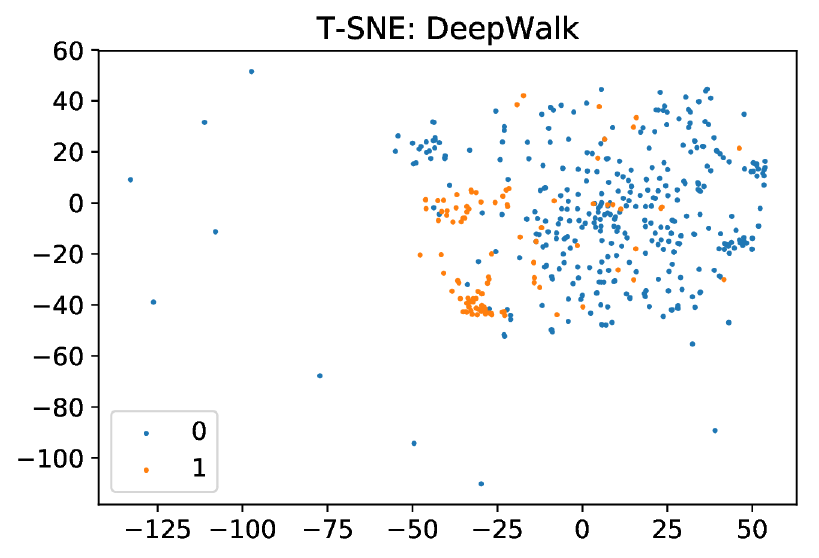}}\quad
	\subfloat{\includegraphics[width =.3\textwidth]{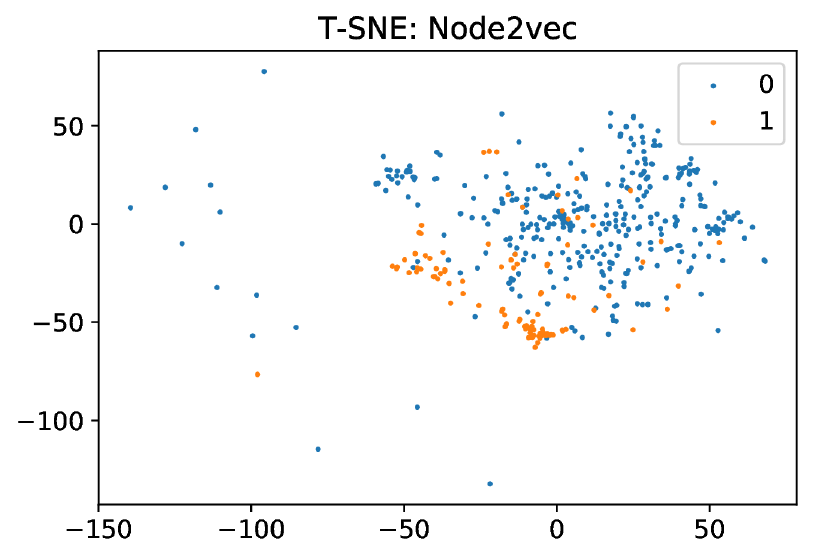}}\quad
	\subfloat{\includegraphics[width =.3\textwidth]{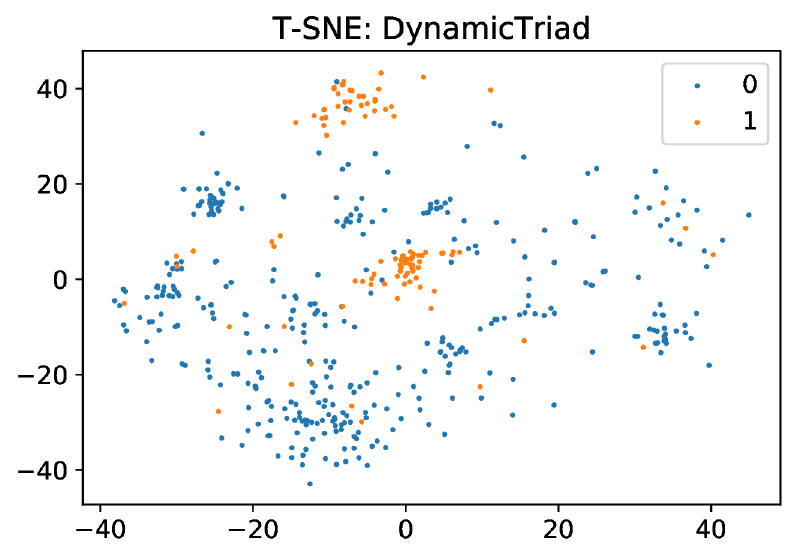}}\quad
	\subfloat{\includegraphics[width =.3\textwidth]{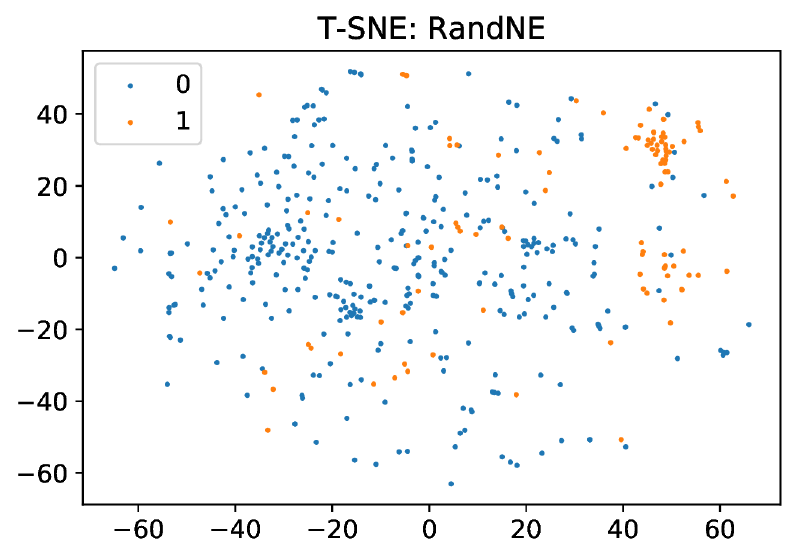}}\quad
	\subfloat{\includegraphics[width =.3\textwidth]{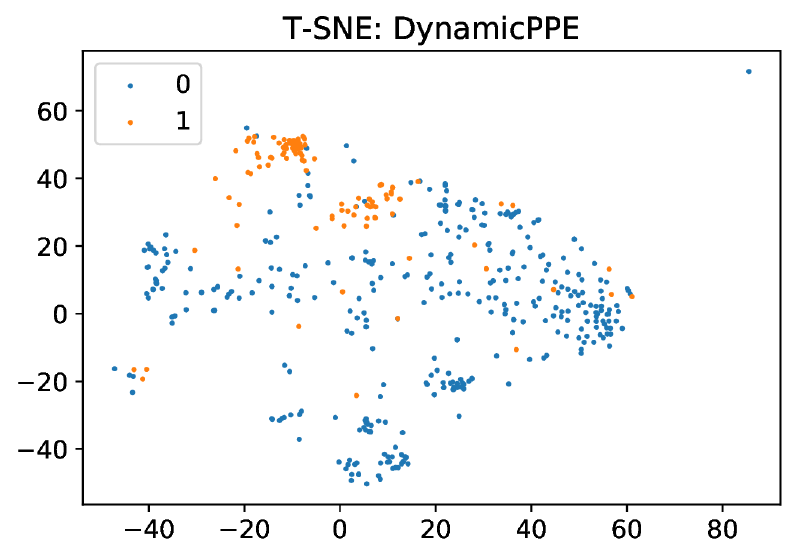}}

    \caption{We select all labeled nodes from Academic graph and visualize their embeddings using T-SNE\cite{van2008visualizing}}.
    \label{fig:nc-tsne-academic-small}
\end{figure}

We visualize the embeddings of small scale graphs using T-SNE\cite{van2008visualizing} in Fig.\ref{fig:nc-tsne-patent-small},\ref{fig:nc-tsne-coauthor-small},\ref{fig:nc-tsne-academic-small}.

% \end{document}
% \endinput

\chapter[Subset Node Anomaly Detection over Dynamic Graph]{Subset Node Anomaly Detection over Dynamic Graph\protect\footnote{Xingzhi Guo, Baojian Zhou, and Steven Skiena. Subset Node Anomaly Tracking over Large Dynamic Graphs.In \textit{Proceedings of the 28th ACM SIGKDD Conference on Knowledge Discovery \& Data Mining (KDD '22)}. Association for Computing Machinery, New York, NY, USA, 475–485. https://doi.org/10.1145/3534678.3539389 \cite{DynAnom}}}
\label{chapter:DynAnom}
\graphicspath{{chapter2-DynAnom/}}

\section{Introduction}

\begin{figure}[t]
\centering
    \subfloat[Dynamic Network] {
        \begin{tikzpicture}[x=0.75pt,y=0.75pt,yscale=-0.95,xscale=0.80]
        %uncomment if require: \path (0,270); %set diagram left start at 0, and has height of 270
        
        %Rounded Rect [id:dp40437996572522916] 
        \draw   (15,27.2) .. controls (15,15.49) and (24.49,6) .. (36.2,6) -- (167.8,6) .. controls (179.51,6) and (189,15.49) .. (189,27.2) -- (189,90.8) .. controls (189,102.51) and (179.51,112) .. (167.8,112) -- (36.2,112) .. controls (24.49,112) and (15,102.51) .. (15,90.8) -- cycle ;
        %Shape: Ellipse [id:dp2646329502205623] 
        \draw   (49.41,32.59) .. controls (49.41,27.3) and (54.49,23.02) .. (60.74,23.02) .. controls (67,23.02) and (72.07,27.3) .. (72.07,32.59) .. controls (72.07,37.88) and (67,42.16) .. (60.74,42.16) .. controls (54.49,42.16) and (49.41,37.88) .. (49.41,32.59) -- cycle ;
        %Shape: Ellipse [id:dp7428451069893581] 
        \draw  [fill={rgb, 255:red, 0; green, 0; blue, 255 }  ,fill opacity=1 ] (84.66,57.41) .. controls (84.66,52.12) and (89.74,47.84) .. (96,47.84) .. controls (102.25,47.84) and (107.33,52.12) .. (107.33,57.41) .. controls (107.33,62.7) and (102.25,66.98) .. (96,66.98) .. controls (89.74,66.98) and (84.66,62.7) .. (84.66,57.41) -- cycle ;
        %Shape: Ellipse [id:dp18836392620264886] 
        \draw   (136,33.82) .. controls (136,28.53) and (141.08,24.25) .. (147.33,24.25) .. controls (153.59,24.25) and (158.67,28.53) .. (158.67,33.82) .. controls (158.67,39.11) and (153.59,43.39) .. (147.33,43.39) .. controls (141.08,43.39) and (136,39.11) .. (136,33.82) -- cycle ;
        %Shape: Ellipse [id:dp44787195692399373] 
        \draw   (128.75,96.84) .. controls (128.75,91.55) and (133.82,87.26) .. (140.08,87.26) .. controls (146.34,87.26) and (151.41,91.55) .. (151.41,96.84) .. controls (151.41,102.12) and (146.34,106.41) .. (140.08,106.41) .. controls (133.82,106.41) and (128.75,102.12) .. (128.75,96.84) -- cycle ;
        %Shape: Ellipse [id:dp8172237305157681] 
        \draw   (43.96,93.22) .. controls (43.96,87.93) and (49.03,83.65) .. (55.29,83.65) .. controls (61.55,83.65) and (66.62,87.93) .. (66.62,93.22) .. controls (66.62,98.51) and (61.55,102.79) .. (55.29,102.79) .. controls (49.03,102.79) and (43.96,98.51) .. (43.96,93.22) -- cycle ;
        %Straight Lines [id:da7774860552266111] 
        \draw    (60.74,42.16) -- (55.29,83.65) ;
        %Straight Lines [id:da029859917521621204] 
        \draw    (63.53,87) -- (88.39,64) ;
        %Straight Lines [id:da7533427358418158] 
        \draw    (107.33,57.41) -- (135.5,36.39) ;
        %Straight Lines [id:da06407721853912296] 
        \draw    (72.07,32.59) -- (136,33.82) ;
        %Straight Lines [id:da7273513737814138] 
        \draw [fill={rgb, 255:red, 100; green, 100; blue, 100 }  ,fill opacity=1 ] [dash pattern={on 0.84pt off 2.51pt}]  (50.51,27) -- (24.47,10) ;
        %Straight Lines [id:da25395336086823206] 
        \draw [fill={rgb, 255:red, 100; green, 100; blue, 100 }  ,fill opacity=1 ] [dash pattern={on 0.84pt off 2.51pt}]  (51.69,38) -- (16.18,49) ;
        %Straight Lines [id:da8462491471677348] 
        \draw [fill={rgb, 255:red, 100; green, 100; blue, 100 }  ,fill opacity=1 ] [dash pattern={on 0.84pt off 2.51pt}]  (43.96,93.22) -- (15,90.8) ;
        %Straight Lines [id:da49624875504666466] 
        \draw [fill={rgb, 255:red, 100; green, 100; blue, 100 }  ,fill opacity=1 ] [dash pattern={on 0.84pt off 2.51pt}]  (56.43,112) -- (55.29,100.79) ;
        %Straight Lines [id:da2068543421394059] 
        \draw [fill={rgb, 255:red, 100; green, 100; blue, 100 }  ,fill opacity=1 ] [dash pattern={on 0.84pt off 2.51pt}]  (174.8,109) -- (151.41,96.84) ;
        %Straight Lines [id:da8213255990163468] 
        \draw [fill={rgb, 255:red, 100; green, 100; blue, 100 }  ,fill opacity=1 ] [dash pattern={on 0.84pt off 2.51pt}]  (153.49,7) -- (147.33,24.25) ;
        %Straight Lines [id:da9555794669364805] 
        \draw [fill={rgb, 255:red, 100; green, 100; blue, 100 }  ,fill opacity=1 ] [dash pattern={on 0.84pt off 2.51pt}]  (185.45,47) -- (157.04,40) ;
        %Straight Lines [id:da7408053432401054] 
        \draw    (147.33,43.39) -- (140.08,87.26) ;
        %Rounded Rect [id:dp1751852590819647] 
        \draw   (15,167.2) .. controls (15,155.49) and (24.49,146) .. (36.2,146) -- (167.8,146) .. controls (179.51,146) and (189,155.49) .. (189,167.2) -- (189,230.8) .. controls (189,242.51) and (179.51,252) .. (167.8,252) -- (36.2,252) .. controls (24.49,252) and (15,242.51) .. (15,230.8) -- cycle ;
        %Shape: Ellipse [id:dp3042691730331203] 
        \draw   (50.13,172.59) .. controls (50.13,167.3) and (55.31,163.02) .. (61.7,163.02) .. controls (68.08,163.02) and (73.26,167.3) .. (73.26,172.59) .. controls (73.26,177.88) and (68.08,182.16) .. (61.7,182.16) .. controls (55.31,182.16) and (50.13,177.88) .. (50.13,172.59) -- cycle ;
        %Shape: Ellipse [id:dp7177157173574817] 
        \draw  [fill={rgb, 255:red, 0; green, 0; blue, 255 }  ,fill opacity=1 ] (86.12,197.41) .. controls (86.12,192.12) and (91.29,187.84) .. (97.68,187.84) .. controls (104.07,187.84) and (109.25,192.12) .. (109.25,197.41) .. controls (109.25,202.7) and (104.07,206.98) .. (97.68,206.98) .. controls (91.29,206.98) and (86.12,202.7) .. (86.12,197.41) -- cycle ;
        %Shape: Ellipse [id:dp6617481785890633] 
        \draw   (138.52,173.82) .. controls (138.52,168.53) and (143.7,164.25) .. (150.09,164.25) .. controls (156.48,164.25) and (161.66,168.53) .. (161.66,173.82) .. controls (161.66,179.11) and (156.48,183.39) .. (150.09,183.39) .. controls (143.7,183.39) and (138.52,179.11) .. (138.52,173.82) -- cycle ;
        %Shape: Ellipse [id:dp8768372644245978] 
        \draw   (131.12,236.84) .. controls (131.12,231.55) and (136.3,227.26) .. (142.69,227.26) .. controls (149.08,227.26) and (154.25,231.55) .. (154.25,236.84) .. controls (154.25,242.12) and (149.08,246.41) .. (142.69,246.41) .. controls (136.3,246.41) and (131.12,242.12) .. (131.12,236.84) -- cycle ;
        %Shape: Ellipse [id:dp9011893255045064] 
        \draw   (44.56,233.22) .. controls (44.56,227.93) and (49.74,223.65) .. (56.13,223.65) .. controls (62.52,223.65) and (67.69,227.93) .. (67.69,233.22) .. controls (67.69,238.51) and (62.52,242.79) .. (56.13,242.79) .. controls (49.74,242.79) and (44.56,238.51) .. (44.56,233.22) -- cycle ;
        %Straight Lines [id:da6196040367703768] 
        \draw    (61.7,182.16) -- (56.13,223.65) ;
        %Straight Lines [id:da33831908137175126] 
        \draw    (64.54,227) -- (89.92,204) ;
        %Straight Lines [id:da761365624150331] 
        \draw [color={rgb, 255:red, 144; green, 19; blue, 254 }  ,draw opacity=1 ][fill={rgb, 255:red, 144; green, 19; blue, 254 }  ,fill opacity=1 ][line width=3]    (109.25,197.41) -- (140.67,180) ;
        %Straight Lines [id:da6546773853993254] 
        \draw    (73.26,172.59) -- (138.52,173.82) ;
        %Straight Lines [id:da6618011229363271] 
        \draw  [dash pattern={on 0.84pt off 2.51pt}]  (51.25,167) -- (24.67,150) ;
        %Straight Lines [id:da08305733605818688] 
        \draw  [dash pattern={on 0.84pt off 2.51pt}]  (52.46,178) -- (16.21,189) ;
        %Straight Lines [id:da7212189005480917] 
        \draw  [dash pattern={on 0.84pt off 2.51pt}]  (44.56,233.22) -- (15,230.8) ;
        %Straight Lines [id:da008440198894443363] 
        \draw  [dash pattern={on 0.84pt off 2.51pt}]  (57.29,252) -- (56.13,240.79) ;
        %Straight Lines [id:da16770559556698839] 
        \draw  [dash pattern={on 0.84pt off 2.51pt}]  (178.13,249) -- (154.25,236.84) ;
        %Straight Lines [id:da3828076021315745] 
        \draw  [dash pattern={on 0.84pt off 2.51pt}]  (156.38,147) -- (150.09,164.25) ;
        %Straight Lines [id:da5223613448627548] 
        \draw  [dash pattern={on 0.84pt off 2.51pt}]  (189,187) -- (160,180) ;
        %Straight Lines [id:da5632994136521178] 
        \draw    (150.09,183.39) -- (142.69,227.26) ;
        %Straight Lines [id:da9923953282093889] 
        \draw [color={rgb, 255:red, 144; green, 19; blue, 254 }  ,draw opacity=1 ][fill={rgb, 255:red, 0; green, 0; blue, 0 }  ,fill opacity=1 ]   (105.63,205) -- (132.21,231) ;
        %Straight Lines [id:da6862349118069943] 
        \draw [line width=1.5]    (34.32,112) -- (34.32,143) ;
        % \draw [shift={(34.32,146)}, rotate = 270] [color={rgb, 255:red, 0; green, 0; blue, 0 }  ][line width=1.0]    (14.21,-4.28) .. controls (9.04,-1.82) and (4.3,-0.39) .. (0,0) .. controls (4.3,0.39) and (9.04,1.82) .. (14.21,4.28)   ;
        \draw [shift={(34.32,146)}, rotate = 270] [fill={rgb, 255:red, 0; green, 0; blue, 0 }  ][line width=0.5]  [draw opacity=0] (8.93,-4.29) -- (0,0) -- (8.93,4.29) -- cycle    ;

        % Text Node
        \draw (88.58,48) node [anchor=north west][inner sep=0.75pt]  [color={rgb, 255:red, 255; green, 255; blue, 255 }  ,opacity=1 ] [align=left] {u};
        % Text Node
        \draw (140.66,25) node [anchor=north west][inner sep=0.75pt]   [align=left] {v};
        % Text Node
        \draw (90.27,188) node [anchor=north west][inner sep=0.75pt]  [color={rgb, 255:red, 255; green, 255; blue, 255 }  ,opacity=1 ] [align=left] {u};
        % Text Node
        \draw (143.44,165) node [anchor=north west][inner sep=0.75pt]   [align=left] {v};
        % Text Node
        \draw (92.59,5.4) node [anchor=north west][inner sep=0.75pt]    {$\mathcal{G}_{t}$};
        % Text Node
        \draw (94.67,144.4) node [anchor=north west][inner sep=0.75pt]    {$\mathcal{G}_{t+1}$};
        % Text Node
        \draw (133.50,88) node [anchor=north west][inner sep=0.75pt]   [align=left] {k};
        % Text Node
        \draw (135.00,226) node [anchor=north west][inner sep=0.75pt]   [align=left] {k};
        % Text Node
        \draw (45,120.4) node [anchor=north west][inner sep=0.75pt]    {\tiny $ \Delta E =\{( u,v,1) ,( u,k,1)\}$};
        \end{tikzpicture}
    \label{fig:dyn-network-example}
    }
    \quad
    \subfloat[Application: Quantify changes of Biden] {\includegraphics[width=0.52\linewidth]{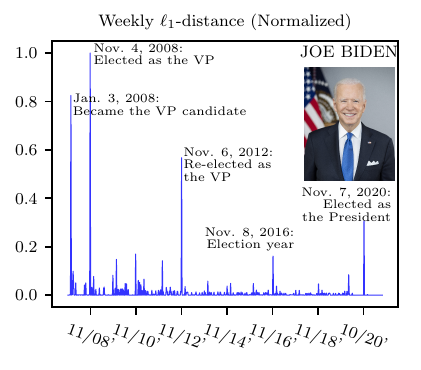}
     \label{fig:biden-life-example}
     }
\caption{ (a) Two consecutive snapshots $\mathcal{G}_t$ and $\mathcal{G}_{t+1}$ from a dynamic weighted-graph. The inserted edge $(u, v, 1)$ further strengths an existing relation, while edge $(u, k, 1)$ builds a new link.
(b) Anomaly tracking of Joe Biden over the \textit{Person Knowledge-Graph} using our proposed \textsc{DynAnom}.
We track Joe Biden on a weekly basis from 2007 to 2020, and calculate the $\ell_1$-distance between the representations on consecutive weeks. The detected peaks correlate well with significant changes in Biden's life.}
\label{fig:biden-changes}

\end{figure}
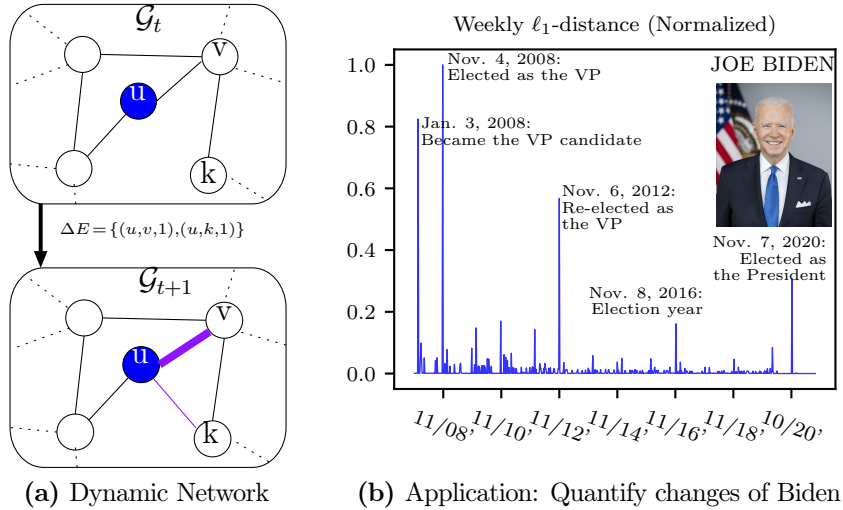

Analyzing the evolution of events in a large-scale dynamic network is important in many real-world applications. We are motivated by the following specific scenario: 
\begin{quote}
Given a person-event interaction graph containing millions of nodes with several different types (e.g., person, event, location) with a stream of weekly updates over many years, how can we identify when specific individuals significantly changed their context?
\end{quote}
The example in Fig. \ref{fig:biden-changes} shows our analysis as Joe Biden shifted career from senator to vice president and finally to president. In each transition he relates with various intensity to other nodes across time. We seek to identify such transitions through analysis of these interaction frequencies and its inherent network structure. This task becomes challenging as the graph size and time horizon scale up. For example, the raw graph data used in Fig. \ref{fig:biden-changes} contains roughly 3.2 million nodes and 1196 weekly snapshots (where each snapshot averages 4 million edges) of Person Graph from 2000 to 2022.  

Despite the extensive literature \cite{akoglu2010oddball,yu2018netwalk,bhatia2021mstream, zhang2023subanom} on graph anomaly detection, previous work focuses on different problem definitions of \textit{anomaly}. On the other hand, works \cite{yoon2019fast} on graph-level anomaly detection cannot identify individual node changes but only uncover the global changes in the overall graph structure. Most representative methods leverage tensor factorization \cite{chang2021f} and graph sketching \cite{bhatia2021sketch}, detecting the sudden dis/appearance of dense subgraphs. However, we argue that since most anomalies are locally incurred by a few anomalous nodes/edges, the global graph-level anomaly signal may overlook subtle local changes. Similarly, edge-level anomaly detection \cite{eswaran2018sedanspot} cannot identify node evolution when there is no direct edge event associated with that node. When Donald Trump became the president, his wife (\textit{Melania Trump}) changed status to become First Lady, but no explicit edge changes connected her to other politicians except Trump. According to the edge-level anomaly detection, there is no evidence for Melania's status change. 

Node representation learning-based methods such as \cite{yu2018netwalk,tsitsulin2021frede} could be helpful to identify anomaly change locally. However, it is impractical to directly apply these methods on large-scale dynamic settings due to 1) the low efficiency: re/training all node embeddings for snapshots is prohibitively expensive; 2) the missing alignment: node representations of each snapshot in the embedding space may not be inherently aligned, making the anomaly calculation difficult.

Inspired from the recent advances on local node representation methods \cite{postuavaru2020instantembedding,xingzhi2021subset} for both static and dynamic graphs, we could efficiently calculate the node representations based on approximated Personalized PageRank vectors. These dynamic node representations are keys for capturing the evolution of dynamic graph. One important observation is that time complexity of calculating approximate PPV is $\mathcal{O}(\frac{1}{\alpha \epsilon})$  for per queried node, where $\alpha$ is PageRank \cite{page1999pagerank} teleport probability and $\epsilon$ is the approximation error tolerance. As the graph evolves, the per-node update complexity is $\mathcal{O}(\frac{|\Delta E|}{\epsilon})$, where $|\Delta E|$ is the total edge events. The other key observation is that the representation space is inherently consistent and aligned with each other, thus there is a meaningful metric space defined directly over the Euclidean space of every node representation across different times. Current local node representation method \cite{xingzhi2021subset} only captures the node-level structural changes over unweighted graphs, which ignores the important interaction frequency information over the course of graph evolution. For example, in a communication graph, the inter-node interaction frequency is a strong signal, but the unweighted setting ignores such crucial feature. This undesirable characteristic makes the node representation becomes less expressive in weighted-graph. 

Based on the above observations, in this paper, we generalize the local node representation method so that it is more expressive and supports dynamic weighted-graph, and resolve the practical subset node anomaly tracking problem. We summarize our contributions as follows:
% Efficient subset node tracking algorithm over dynamic weighted graph with theoretical analysis: 
\begin{itemize}
\item We generalize dynamic forward push algorithm \cite{zhang2016approximate} for calculating Personalized PageRank, making it suitable for weighted-graph anomaly tracking. The algorithm updates node representations per-edge event (i.e., weighted edge addition/deletion), and the per-node time complexity linear to edge events ($|\Delta E|$) and error parameter($\epsilon$) which is independent on the node size.
\item We propose an efficient anomaly framework, namely \textsc{DynAnom} that can support both node and graph-level anomaly tracking, effectively localizing the period when a specified node significantly changed its context. Experiments demonstrate its superior performance over other baselines with the accuracy of 0.5425 compared with 0.2790, the baseline method meanwhile 2.3 times faster.
\item A new large-scale real-world graph, as new resources for anomaly tracking: PERSON graph is constructed as a more interpretable resource, bringing more research opportunities for both algorithm benchmarking and real-world knowledge discovery. Furthermore, we conduct a case study over it, successfully capturing the career shifts of three public figures in the U.S., showing the usability of \textsc{DynAnom}.
\end{itemize}

The rest paper is organized as follows: 
Section \ref{sec:related-work} reviews previous graph anomaly tracking methods. Section \ref{sec:notation-prelim} describes the notation and preliminaries. Section \ref{sec:problem-formulation} gives the problem formulation. We present our proposed framework -- \textsc{DynAnom} in Section \ref{sec:methods}. Section \ref{sec:experiments} gives the experiment results over standard benchmark graphs and a case study on the real-world graph. Finally, we conclude and discuss future directions in Section \ref{sec:conclusion}. Source code and datasets will be released upon publication, and included in supplementary file (compiled datasets are in \textcolor{black}{\url{https://bit.ly/3rAshBn}}).

\vspace{-2mm}
\section{Related Work}
\label{sec:related-work}
\vspace{-1mm}
We review the three most common graph anomaly tasks over dynamic graphs, namely graph, edge and node-level anomaly.

\paragraph{Graph-level and edge-level anomaly.} Graph anomaly refers to sudden changes in the graph structure during its evolution, measuring the difference between consecutive graph snapshots \cite{aggarwal2011outlier, beutel2013copycatch,eswaran2018sedanspot, eswaran2018spotlight}. \citet{aggarwal2011outlier} use a structural connectivity model to identify anomalous graph snapshots. \citet{shin2017densealert} apply incremental tensor factorization to spot dense connection formed in a short time.  \citet{yoon2019fast} use the first and second derivatives of the global PageRank Vectors as the anomaly signal, assuming that a normal graph stream will evolve smoothly. On the other hand, edge anomaly identifies unexpected edges as the graph evolves where anomalous edges adversarially link nodes in two sparsely connected components or abnormal edge weight changes \cite{ranshous2016scalable,eswaran2018sedanspot,bhatia2020midas}. Specifically, \citet{eswaran2018sedanspot} propose to use the approximated node-pair Personalized PageRank (PPR) score before and after the new edges are inserted. Most recently, \citet{chang2021f} estimate the interaction frequency between nodes, and incorporate the network structure into the parameterization of the frequency function. However, these methods cannot be able to reveal the node local anomalous changes, and cannot identify the individual node changes for those without direct edge modification as we illustrated in introduction.

\paragraph{Node-level local anomaly.} Node anomaly measures the sudden changes in individual node's connectivity, activity frequency or community shifts \cite{wang2015localizing, yu2018netwalk}. \citet{wang2015localizing} uses hand-crafted features (e.g., node degree, centrality) which involve manual feature engineering processes. Recently, the dynamic node representation learning methods \cite{yu2018netwalk,goyal2018dyngem, zhou2018dynamic,kumar2019predicting, lu2019temporal,sankar2020dysat} were proposed. For example, a general strategy of them \cite{yu2018netwalk,kumar2019predicting} for adopting such dynamic embedding methods for anomaly detection is to incrementally update node representations and apply anomaly detection algorithms over the latent space. To measure the node changes over time, a comparable or aligned node representation is critical. \citet{yu2018netwalk} uses auto-encoder and incremental random walk to update the node representations, then apply clustering algorithm to detect the node outliers in each snapshot. However, its disadvantage is that the embedding space is not aligned, making the algorithm only detects anomalies in each static snapshot. Even worse, it is inapplicable to large-scale graph because the fixed neural models is hard to expand without retraining. Similar approaches \cite{kumar2019predicting} with more complicated deep learning structure were also studied. These existing methods are not suitable for the subset node anomaly tracking. Instead, our framework is inspired from recent advances in efficient local node representation algorithms \cite{zhang2016approximate,guo2017parallel,postuavaru2020instantembedding,xingzhi2021subset}, which is successful in handling subset node representations over large-scale graph. 

\section{Forward Local Push Algorithm over Weighted Graphs}
    \label{sec:notation-prelim}

    As we introduced in Section \ref{chapter-prelim}, \textit{forward push algorithm} \cite{andersen2006local}, a.k.a. the bookmark-coloring algorithm \cite{berkhin2006bookmark}, approximates $\pi_{s}(v)$ locally via an approximate $p_s(v)$. The key idea is to maintain solution vector $\bm p_s$ and a residual vector $\bm r_s$ (at the initial $\bm r_s = \bm 1_s, \bm p_s = \bm 0$). When the graph updates from $\mathcal{G}_t$ to $\mathcal{G}_{t+1}$, a variant forward push algorithm \cite{zhang2016approximate} dynamically maintains $\bm r_s$ and $\bm p_s$. We generalize it to a weighted version of \textit{dynamic forward push} to support dynamic updates on dynamic weighted-graph as presented in Algo. \ref{algo:forward-local-push-weighted}. At each local push iteration, it pushes large residuals to neighbors whenever these residuals are significant ($ |r_s(u)| > \epsilon d(u)$). Compare with power-iteration, this operation avoids the access of whole graph hence speedup the calculations. Based on \cite{zhang2016approximate}, we consider a more general setting where the graph could be weighted-graph. Fortunately, this weighted version of \textsc{ForwardLocalPush} still has \textit{invariant property} as presented in the following lemma.
    
    \begin{algorithm}[ht]
\caption{$\textsc{ForwardLocalPush (Weighted Version)}$ \cite{zhang2016approximate}}
\begin{algorithmic}[1]
\State \textbf{Input: }$\bm p_s, \bm r_s, \mathcal{G}_t, \epsilon, \alpha$
\While{ exists $u$ such that $| r_s(u)| > \epsilon d(u)$}
\State $\textsc{Push}(u)$
\EndWhile
\State \Return $(\bm p_s, \bm r_s)$
\Procedure{Push}{$u$}
\State $p_s(u) \pluseq \alpha r_s(u)$
\For{$v \in \operatorname{Nei}(u)$}
\State $r_s(v) \pluseq \frac{(1-\alpha)  r_s(u)w_{(u, v)}}{ \sum_{v \in \operatorname{Nei}(u)} w_{(u, v)}} \textcolor{blue}{}$
\EndFor
\State $r_s(u) = 0 $
\EndProcedure
\end{algorithmic}
\label{algo:forward-local-push-weighted}
\end{algorithm}
    
    \begin{lemma} [PPR Invariant Property in weighted graphs \cite{zhang2016approximate}] Suppose $\bm \pi_s$ is the PPV of node $s$ on graph $\mathcal{G}_t$. Let $\bm p_s$ and $\bm \pi_s$ be returned by the weighted version of \textsc{DynamicForwardPush} presented in Algo. \ref{algo:forward-local-push-weighted}. Then, we have the following invariant property.
    \begin{align}
    \pi_s(u) &= p_s(u) + \sum_{x \in \mathcal{V}}r_s(x) \pi_s(x),  \text{ for all } u \in \mathcal{V},  \allowdisplaybreaks\\
    p_s(u) + \alpha r_s(u) &= (1-\alpha) \sum_{x \in Nei(u)} \frac{w_{(u,x)}p_s(x)}{d(x)} + \alpha 1_{u=s},\allowdisplaybreaks
    \end{align}
    where $1_{u=s} = 1$ if $u = s$, 0 otherwise.
    \label{lemma:ppr-invar}
    \end{lemma}
    
    \paragraph{From PPVs to node representations.} Obtained PPVs are not directly applicable to our anomaly tracking problem as operations on these $n$-dimensional vectors is time-consuming. To avoid this, we treat PPVs as the intriguing similarity matrix and project them into low-dimensional spaces using transformations such as SVD \cite{qiu2019netsmf}, random projection \cite{zhang2018billion, chen2019fast}, matrix sketch\cite{tsitsulin2021frede}, or hashing \cite{postuavaru2020instantembedding} so that the original node similarity is well preserved by the low dimension node representations. We detail this in our main section.

\section{Problem formulation}
\label{sec:problem-formulation}
Before we present the subset node anomaly tracking problem. We first define the edge events in the dynamic graph model as the following: A set of edge events $\Delta E_t$ from $t$ to $t+1$ is a set of operations on edges, i.e., edge insertion, deletion or weight adjustment while the graph is updating from $\mathcal{G}_t$ to $\mathcal{G}_{t+1}$. Mathematically, $\Delta E_t \triangleq \{ (u_0, v_0, \Delta w_{(u_0, v_0)}, (u_1, v_1, \Delta w_{(u_1, v_1)})$, $\ldots, (u_i, v_i, \Delta w_{(u_i, v_i)})\}$ where each $\Delta w_{(u_i, v_i)}$ represents insertion (or weight increment) if it is positive, or deletion (decrement) if it is negative. Therefore, our dynamic graph model can be defined as a sequence of edge events. We state our formal problem definition as the following.

\begin{definition}[Subset node Anomaly Tracking Problem]
Given a dynamic weighted-graph $\mathcal{G}_t = \langle \mathcal{V}_t,\mathcal{E}_t, \mathcal{W}_t \rangle, \forall t \in [0,T]$, consisting of initial snapshot $\mathcal{G}_0$ and the following T snapshots that have edge events $\Delta E_t, |\Delta E_t| \geq 0$ (e.g., edge addition, deletion, weight adjustment from $t-1$ to $t$). We want to quantify the status change of a node $v$ in the targeted node subset $v \in \mathcal{S} \triangleq \{ v_0, v_1, ..., v_k \} $ from $\mathcal{G}_{t-1}$ to $\mathcal{G}_{t}$ with the anomaly measure function $f(\cdot, \cdot)$ so that the measurement is consistent to human judgement, or reflects the ground-truth of node status change if available. To illustrate this edge event process, Table \ref{tab:node-track-def} presents the process for better understanding.
\label{def:node-track-def}
\end{definition}

\begin{table}[!ht]
\caption{Illustration of subset node anomaly tracking problem. At each time $t$, node representations $\bm x_i^t$ are provided, and anomaly of node changes is quantified by $f(\bm x_i^{t-1}, \bm x_i^t$), which is expected to correlated with the anomaly label if available.}
\centering
\begin{tabular}{p{3cm}||c|c|c|c|c}
\toprule
 Timestamp & 0 & 1 & 2 & 3 & ...  \\
\hline 
 Edge events & - & $\Delta E_1$ & $\Delta E_2$ & $\Delta  E_3$ & ...  \\
\hline 
 Snapshots & $G_0$ & $G_1$ & $G_2$ & $G_3$ & ... \\
\hline 
 Score for $v_i \in \mathcal{S}$ & - & \small{$f(x_i^{0},x_i^{1})$} &  \small{$f(x_i^{1},x_i^{2})$}&  \small{$f(x_i^{2},x_i^{3})$} & ... \\
\hline 
 Label for $v_i \in \mathcal{S}$ & - & Normal &  Normal & Abnormal & ... \\
\bottomrule
\end{tabular}
\label{tab:node-track-def}
\end{table}

\section{The Proposed Framework}
\label{sec:methods}

In order to localize the node anomaly across time, the basic idea is to incrementally obtain the PPVs as the node representation at each time, then design anomaly score function $f(\cdot, \cdot)$ to quantify the PPVs changes from time to time. This section presents our proposed framework $\textsc{DynAnom}$ which has three components: 1) A provable dynamic PPV update strategy extending \cite{zhang2016approximate} to weighted edge insertion/deletion; 2) Node level anomaly localization based on incremental PPVs; and 3) Graph level anomaly detection based on approximation of global PageRank changes. We first present how local computation of PPVs can be generalized to dynamic weighted graphs, then present the unified framework for node/graph-level anomaly tracking, finally we analyze the complexity of an instantiation algorithm.
 
\subsection{Maintenance of Dynamic PPVs}

Multi-edges record the interaction frequency among nodes, and reflect the evolution of the network not only in topological structure, but also in the strength of communities. Previous works \cite{zhang2016approximate,xingzhi2021subset} focus only on the structural changes over unweighted graph, ignoring the multi-edge cases. In a communication graph (e.g., Email graph), the disadvantage of such method is obvious: as more interactions/edges are observed, the graph becomes denser, or even turn out to be fully-connected in an extreme case. Afterwards, all the upcoming interactions will not change the node status since the ignorance of interaction frequency. This aforementioned scenario motivates us to generalize dynamic PPV maintenance to weighted graph, expanding its usability for more generic graphs and diverse applications. In order to incorporate edge weights into the dynamic forward push algorithm to obtain PPVs, we modify the original algorithm \cite{andersen2006local,zhang2016approximate} by adding a weighted push mechanism as presented in Algo. \ref{algo:forward-local-push-weighted}. Specifically, for a specific node $v$, at each iteration of the push operation, we update its neighbor residual $r_s(v)$ as the following
\begin{equation}
r_s(v) \pluseq \frac{(1-\alpha) r_s(u) w_{(u,v)}}{\sum_{v\in \operatorname{Nei}(u)} w_{(u,v)}}. \label{equ:update-rule}
\end{equation}
The modified update of $r_s(v)$ in Equ. \eqref{equ:update-rule} efficiently approximate the PPR over static weighted graph with same time complexity as the original one.  At time snapshot $\mathcal{G}_t$, the weight of each edge event $\Delta w_{(u, v)}$ could be either positive or negative, representing edge insertion or deletion) at time $t$ so that the target graph $\mathcal{G}_t$ updates to $\mathcal{G}_{t+1}$. However, this set of edge events will break up the invariant property shown in Lemma \ref{lemma:ppr-invar}. To maintenance the invariant property on weighted graphs, we present the following Thm. \ref{theorem:dyn-adjust-weighted-graph}, a generalized version from \cite{zhang2016approximate}. The key idea is to update $\bm p_s, \bm r_s$ so that updated weights satisfy the invariant property. We present the theorem as following.

\begin{figure}[t]
\centering
\includegraphics[width=1\textwidth]{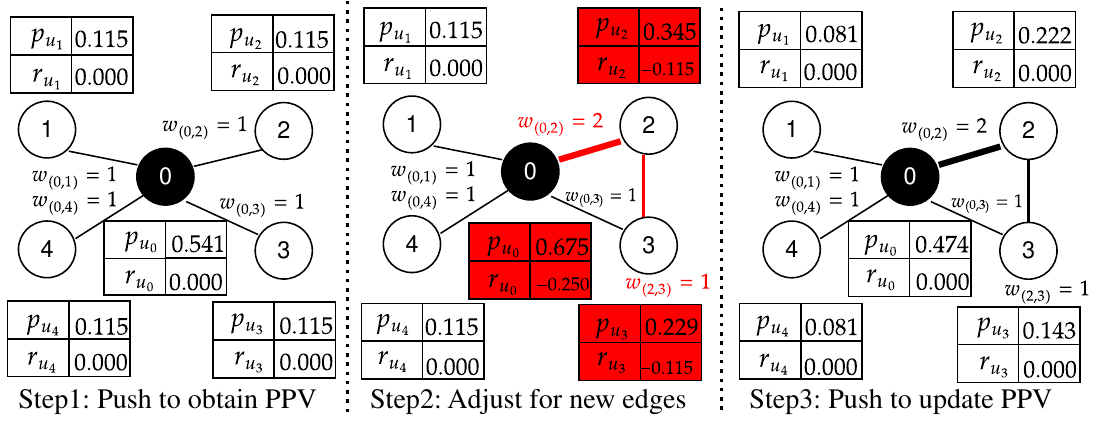}
\caption{Illustration of the PPR adjustment over a  weighted graph of five nodes. \textit{Step1}: Apply Algo.\ref{algo:forward-local-push-weighted} to calculate the initial $p_s,r_s$ where $s=v_0, \alpha = 0.15$ over the initial graph. \textit{Step2}: After inserting new edges  $(v_0,v_2,1), ((v_2,v_3,1))$, the strength between $v_0$ and $v_2$ increases, and $v_2$ builds a new connection to $v_3$, which both break the invariant in Lemma \ref{lemma:ppr-invar}. We adjust to recover the invariant by applying Theorem \ref{theorem:dyn-adjust-weighted-graph} in the red-colored blocks. \textit{Step3:} We re-apply Algo.\ref{algo:forward-local-push-weighted} and update $p_s$ for better approximation error. %The updated $p_s$ is same to the PPV calculated over the updated graph snapshot. 
}
\label{fig:new-edge-example}
\end{figure}

\paragraph{Generalized PPR update rules.} Figure \ref{fig:new-edge-example} illustrates the dynamic maintenance of PPV over a weighted graph where two new edges are added, incurring edge weight increment between node 0 and 2, and new connection between node 2 and 3. Note that one only needs to have $\mathcal{O}(1)$ time to update $p_s(u), r_s(u)$, and $r_s(v)$ for each edge event. As proved, the total cost of $m$ edge events will be linearly dependent on, i.e., $\mathcal{O}(m)$. This efficient update procedure is much better than naive updates where one needs to calculate all quantities from scratch.

\begin{theorem} [Generalized PPR update rules]
Given an edge event $(u,v, \Delta w_{(u,v)})$. We use the the following update rules
\begin{align}
p'_s(u) &= p_s(u) \frac{\sum_{v \in \operatorname{Nei}(u)} w_{(u,v)}+\Delta w_{(u,v)}}{\sum_{v \in \operatorname{Nei}(u)} w_{(u,v)}}, \label{eq:p-update-weighted-body} \\
r_s'(u) &=  r_s(u) -\frac{\Delta w_{(u,v)} p_s(u)}{\alpha \sum_{v \in \operatorname{Nei}(u)} w_{(u,v)}}, \label{eq:r-update-weighted-body} \\
r_s'(v) &= r_s(v) + \frac{(1-\alpha)}{\alpha} \frac{\Delta w_{(u,v)}p_s(u)}{\sum_{v \in \operatorname{Nei}(u)} w_{(u,v)}}. \label{eq:r-v-update-weighted-body}
\end{align}
After applying a sequence of updating rules above and using \textsc{DynamicForwardPush} algorithm, then $\bm p_s$ and $\bm r_s$ satisfy invariant property in Lemma \ref{lemma:ppr-invar} and accurately approximate PPV. We detail the proof in Appendix \ref{appendix-proof}.
\label{theorem:dyn-adjust-weighted-graph}
\end{theorem}

\paragraph{From dynamic PPVs to node representations.} The above theorem provides efficient rules to incrementally update PPVs for dynamic weighted graphs. To obtain dynamic node representation, two universal hash functions $h_{\operatorname{d}}$ and $h_{\operatorname{sgn}}$ are used (see details in \cite{postuavaru2020instantembedding}).  In the following section, we describe how to leverage PPVs as node representation for tracking node/graph-level anomaly.
 
\subsection{Anomaly Tracking Framework: \textsc{DynAnom}}

The key idea of our proposed framework is to measure the individual node anomaly by the embedding space based on Personalized PageRank \citep{postuavaru2020instantembedding,xingzhi2021subset}, and aggregate the individual anomaly scores to represent graph-level anomaly. By the fact that the PPV-based embedding space is aligned, we could directly measure the node anomaly by existing distance functions.\footnote{In this paper, we explored $\ell_1$-distance function where given $\bm x,\bm y$, $f(\bm x, \bm y) = \| \bm x - \bm y\|_1$.} We describe our design in following subsections.

\subsubsection{Node-level Anomaly Tracking}
Since the inherently local characteristic of our proposed framework, we can efficiently measure the status of arbitrary nodes in two consecutive snapshots in an incrementally manner. Therefore, we could efficiently solve the problem defined in Def. \ref{def:node-track-def} and localize where the anomaly actually happens by ranking the anomalous scores. To do this, the key ingredient is to measure node difference from $t-1$ to $t$ as $\delta_s^t = f(\bm x_s^{t-1} , \bm x_s^{t})$ where $\bm x_s^{t}$ is the node representation of node $s$ at time $t$. Based on the above motivation, we first present the node-level anomaly detection algorithm \textsc{DynAnom} in Algo. \ref{algo:dynanom-node} where the score function $f(\cdot,\cdot)$ is realized by $\ell_1$ norm in our experiments. There are three key steps: \textit{Step 1.} calculating dynamic PPVs for  all nodes in $S$ over $T$ snapshots (Line 3); \textit{Step 2.} obtaining node representations of $S$ by local hash functions. Two universal hash functions\footnote{We use \textit{sklearn.utils.murmurhash3\_32} in our implementation.} are used in \textsc{DynNodeRep} (Line 5); \textit{Step 3.} calculating node anomaly scores based on node representation $\bm x_s$ for all $s \in S$.

Noted that our design directly use the first-order difference ($\nabla x_s$) between two consecutive time. Although there are more complex design of $f(\cdot, \cdot)$ under this framework, such as, second-order measurement $\nabla^2 x_s$ similar as in \citet{yoon2019fast}, clustering based anomaly detection \cite{yu2018netwalk}, or supervised anomaly learning. We restrict our attention on the simplest form by setting $p=1 \text{ or } 2$ with the first-order measure. In our framework, the definition of anomaly score can be highly flexible for various applications.

\setlength{\columnsep}{1.5cm}
\hspace{0cm}\scalebox{1.0}{
    \begin{minipage}{1.0\linewidth}
    \begin{algorithm}[H]
    \scriptsize
    \caption{$\textsc{DynAnom}(\mathcal{G}_0, \Delta E_{1, \ldots, T} , \mathcal{S}, \epsilon, \alpha, p)$ }
    \begin{multicols}{2}
    \begin{algorithmic}[1]
    \State \textbf{Input: }  Initial graph $\mathcal{G}_0$, Edge events $ \Delta E_{1, \ldots, T}$ , Subset target nodes $\mathcal{S}$,  PPV quality $\epsilon$,  teleport factor $\alpha $.
    \State \textcolor{black}{//Step 1: Incrementally calculate PPVs}
    \State $\bm p_{s\in \mathcal{S}} = \textsc{IncrementPush}(\mathcal{G}_0, \Delta E_{1, \ldots, T} , S, \epsilon, \alpha)$
    \State \textcolor{black}{//Step 2: Obtain node representations in target set $\mathcal{S}$ }
    \State $\bm x_{s\in \mathcal{S}} = \textsc{DynNodeRep}(p_s^0, ... ,P_s^T)$
    \State \textcolor{black}{//Step 3: Calculate node anomaly for all nodes in $\mathcal{S}$}
    \For{$s \in S, t \in [1,T] $}
        \State $\delta_s^t = f( x_s^{t-1}, x_s^{t}) = \left( \sum_{i} \left| x_s^{t}(i) - x_s^{t-1}(i) \right|^p \right)^ {1/p}$
    \EndFor
    \State \textbf{return} $\bm \delta_s = [\delta_s^1,\delta_s^2,\ldots, \delta_s^T], \forall s \in S$
    
    \noindent\hrulefill
    \Procedure{IncrementPush}{$\mathcal{G}_0, \Delta E_{1, \ldots, T} , \mathcal{S}, \epsilon, \alpha$}
    \State $t=0$
    \For{$s \in \mathcal{S} := \{v_1,v_2,\ldots, v_k\}$}
    \State $p_s^t = \bm 0, \quad r_s^t = 1_s $
    \State $p_s^t, r_s^t=$\textsc{DynamicForwardPush}$(\mathcal{G}^0, s, p_s^t, r_s^t, \epsilon, \alpha)$
    \EndFor
    
    \For{$t \in [1,T]$}
        \For{$s \in \mathcal{S}$}
            \For{$(u,v,\Delta w_{(u,v)}) \in \Delta E_t$}
                \State update $p_s^t(u), r_s^t(u), r_s^t(v)$ uses rules in Thm. \ref{theorem:dyn-adjust-weighted-graph}.
            \EndFor
        \State $p_s^t, r_s^t=$\textsc{DynamicForwardPush}$(\mathcal{G}^0, s, p_s^t, r_s^t, \epsilon, \alpha)$
        \EndFor
        \State $\mathcal{G}_t += (u,v,\Delta w_{(u,v)})$
    \EndFor
    \State \textbf{return} $ \bm p_s =[ p_s^0, ... , p_s^T ], \forall s \in S$ 
    \EndProcedure
    
    \noindent\hrulefill
    \Procedure{DynNodeRep}{$\bm p_s^0, \ldots ,\bm p_s^T$}
    \State $\epsilon_{c}= \textsc{min}(\frac{1}{|\mathcal{V}|}, \text{1e-5}), dim = 1024$
    \For{$t \in [1,T)$}
        \For{$i \in \cup_{t'\in\{t, t-1\}} \textsc{supp}(p_s^{t'})$}:
        \State $p_s^{t-1}(i) = 0$ if $p_s^{t-1}(i) \leq \epsilon_c $
        \State $p_s^t(i) = 0$ if $p_s^t(i) \leq \epsilon_c $
        \EndFor
        \State $x_s^{t} = \textsc{ReduceDim}(\frac{p_s^t}{\|p_s^t\|_1}, dim)$
        \State $x_s^{t-1} = \textsc{ReduceDim}( \frac{p_s^{t-1}}{\|p_s^{t-1}\|_1}, dim)$
        \State \textbf{return} $ \bm x_s =[ x_s^0, ... , x_s^T ], \forall s \in S$ 
    \EndFor
    \EndProcedure
    \noindent\hrulefill
    \Procedure{ReduceDim}{$x,dim$}
    \State // Hash function $h_{dim}(i): \mathbb{N}\to [dim]$, ash function $h_{\operatorname{sgn}}(i): \mathbb{N}\to \{ \pm 1\}$
    \If{$\textsc{dim}(x) \leq dim  $}
        \State \textbf{return} $x$
    \Else
        \State $\bar x = \bm 0 \in \mathcal{R}^{dim}$
        \For{$i \in \textsc{Supp}(x)$}
            \State $\bar x(h_{dim}(i)) \pluseq h_{\operatorname{sgn}}(i) \log \left(x(i) \right)$
        \EndFor
        \State \textbf{return} $ \frac{\bar x}{\| \bar x \|_1}$
    \EndIf
    \EndProcedure
    \end{algorithmic}
    \end{multicols}
    \label{algo:dynanom-node}
    \end{algorithm}
    \end{minipage}
}

\subsubsection{Graph-level Anomaly Tracking}
Based on the node-level anomaly score proposed in Algo. \ref{algo:dynanom-node}, we propose the graph-level anomaly. The key property we used for graph-level anomaly tracking is the linear relation property of PPV. More specifically, let $\bm \pi$ be the PageRank vector. We note that the PPR vector is linear to personalized vector $\bm \pi_s$. Therefore, global PageRank could be calculated by the weighted average of single source PPVs.
\begin{align}
\bm \pi = \alpha \frac{\bm \vec 1}{|\mathcal{V}|} \sum_{i = 0}^{\infty}(1-\alpha)^{i} \bm P^{i}  = \sum_{ s \in \mathcal{V} } \frac{1_s}{|\mathcal{V}|} \bm \pi_{s} = \frac{1}{|\mathcal{V}|} \sum_{ s \in \mathcal{V} } \bm \pi_{s}, \allowdisplaybreaks
\end{align}
From the above linear equality, we formulate the global PPV, denoted as $\pi_{, g}$,as the linear combination of single-source PPVs as following:
\begin{align}
\bm \pi_{g} = \sum_{ s \in \mathcal{V} } \gamma_s \pi_{s} , \gamma_s = \frac{d(s)}{m}, m= \sum_{i \in \mathcal{V}} d(i) \allowdisplaybreaks
\end{align}

In order to capture the graph-level anomaly, we use the similar heuristic weights \cite{yoon2019fast} $\gamma_s = \frac{d(s)}{m}$, which implies that $\bm \pi_{\alpha}$ is dominated by high degree nodes, thus the anomaly score is also dominated by the greatest status changes from high degree nodes as shown below:
% \vspace{-1.5mm}
\begin{align}
    \| \bm \pi_{g}^t - \bm \pi_{g}^{t-1} \|_{1} & = 
    \| \sum_{ s \in \mathcal{V} } \gamma_s^{t-1} \bm \pi^{t-1}_{ s}  - \sum_{ s \in \mathcal{V} } \gamma_s^t \bm \pi^t_{ s} \|_1  \nonumber \allowdisplaybreaks\\
    &= \| \sum_{ s \in \mathcal{V} } \frac{d_s^{t-1}}{m^{t-1}} \bm \pi^{t-1}_{ s}  - \sum_{ s \in \mathcal{V} } \frac{d_s^{t}}{m^{t}} \bm \pi^t_{ s} \|_1 \nonumber \allowdisplaybreaks\\
    &\approx   \sum_{ s \in \mathcal{V}^t_{high}}  \frac{d_s^{t}}{m^{t}} \| \bm \pi^{t-1}_{ s}  - \bm \pi^t_{ s} \|_1 \nonumber \allowdisplaybreaks\\
    &\leq \sum_{ s \in \mathcal{V}^t_{high}}  \| \bm \pi^{t-1}_{ s}  - \bm \pi^t_{ s} \|_1,
\end{align}
where $\mathcal{V}_{high}^t$ denotes the set of high-degree nodes. Note that the above approximated upper bound of $\| \bm \pi_g^t - \bm \pi_g^{t-1}\|_1$ can lower bounded by the following
\begin{align}
\sum_{ s \in \mathcal{V}^t_{high}}  \| \bm \pi^{t-1}_{ s}  - \bm \pi^t_{ s} \|_1 \geq   \textsc{max}(\{ \left\| \bm \pi^{t-1}_{ s} - \bm \pi^t_{ s} \right\|_1 ,\forall s \in \mathcal{V}^t_{high} \} ).
    \label{eq:graph-anomaly-score}
\end{align}
By the approximation of global PPV difference in Equ. \eqref{eq:graph-anomaly-score}, we discover that the changes of high degree nodes will greatly affect $\ell_1$-distance, and similarly for $\ell_2$-distance. The above analysis assumes $d_s^{t-1} / m^{t-1} \approx d_s^{t} / m^{t}$.  We could also hypothesize that the changes in high-degree node will flag the graph-level anomaly signal. Therefore, we propose to use high node tracking strategy for graph-level anomaly detection:
\begin{align}
\textsc{DynAnom}_{graph}^t &:= \textsc{max}(\{ \| \bm \pi^{t-1}_{ s} - \bm \pi^t_{ s} \|_1 ,\forall s \in \mathcal{V}_{high} \} ). \allowdisplaybreaks
\end{align}
In practice, we incrementally add high degree nodes (e.g., top-50) into the tracking list from each snapshots and extract the maximum PPV changes among them as the graph-level anomaly score. This proposed score could capture the anomalous scenario where one of the top nodes (e.g., popular hubs) encounter great structural changes, similar to the DDoS attacks on the internet.

\subsubsection{Comparison between \textsc{DynAnom} and other methods.}
The significance of our proposed framework compared with other anomaly tracking methods is illustrated in Table \ref{tab:algo-comparison}.  \textsc{DynAnom} is capable of detecting both node and graph-level anomaly, supporting all types of edge modifications, efficiently updating node representations (independent of $|\mathcal{V}|$), and working with flexible anomaly score functions customized for various downstream applications.

\newcolumntype{D}[1]{>{\centering\let\newline\\\arraybackslash\hspace{0pt}}m{#1}}

\begin{table}[ht]
        \centering
        \small
\caption{The comparison of supported features between DynAnom and other methods.}
% \vspace{-2mm}
\begin{tabular}{@{\extracolsep{4pt}}p{0.16\linewidth}p{0.04\linewidth}p{0.04\linewidth}p{0.06\linewidth}p{0.06\linewidth}p{0.09\linewidth}p{0.05\linewidth}p{0.06\linewidth}p{0.06\linewidth}@{}}
\toprule 
  \multirow{2}{*}{\diaghead(3,-2){\hskip 18mm} {Method}{Feature}}  & \multicolumn{2}{D{0.14\linewidth}}{Anomaly} & \multicolumn{3}{D{0.25\linewidth}}{Edge Event Types} & \multicolumn{3}{D{0.24\linewidth}}{ Algorithm  } \\
\cmidrule{2-3} \cmidrule{4-6}\cmidrule{7-9}

& Node level & Graph level & Edge Stream & Add/ Delete & Weight Adjust & $|\mathcal{V}|$ Ind. & Align Repr.  & Flex. Score  \\
\midrule 
 SedanSpot & \redxmark  & \redxmark & \bluecheck & \redxmark & \redxmark & \bluecheck & \redxmark & \redxmark \\
 AnomRank & \redxmark & \bluecheck & \redxmark & \bluecheck & \bluecheck & \redxmark & \redxmark & \redxmark \\
 NetWalk & \bluecheck & \redxmark & \redxmark & \bluecheck & \redxmark & \redxmark & \redxmark & \redxmark \\
 DynPPE & \bluecheck & \redxmark & \bluecheck & \bluecheck & \redxmark & \bluecheck & \bluecheck & \redxmark \\
 \hline
 \textbf{DynAnom} & \bluecheck & \bluecheck & \bluecheck & \bluecheck & \bluecheck & \bluecheck & \bluecheck & \bluecheck \\
 \bottomrule
\end{tabular}
\label{tab:algo-comparison}
\end{table}

\paragraph{Time Complexity Analysis.}
The overall complexity of tracking subset of $k$ nodes across $T$ snapshots depends on run time of three main steps as shown in Algo. \ref{algo:dynanom-node}: 1. the run time of \textsc{IncrementPush} for nodes $\mathcal{S}$; 2. the calculation of dynamic node representations, which be finished in $\mathcal{O}(k T |\operatorname{supp}(\bm p_{s^\prime})|)$ where $|\operatorname{supp}(\bm p_{s^\prime})|$ is the maximal support of all $k$ sparse vectors, i.e. $|\operatorname{supp}(\bm p_{s^\prime})| = \max_{s\in \mathcal{S}} |\operatorname{supp}(\bm p_{s})|$. The overall time complexity of our proposed framework is stated as in the following theorem.
\begin{theorem}[Time Complexity of \textsc{DynAnom}]
Given the set of edge events where $\mathbb{E} = \{ e_1, e_2,\ldots, e_m\}$ and $T$ snapshots and subset of target nodes $\mathcal{S}$ with $|\mathcal{S}| = k$, the instantiation of our proposed \textsc{DynAnom} detects whether there exist events in these $T$ snapshots in $\mathcal{O}( k m/\alpha^2 + k \bar{d^t} / (\epsilon \alpha^2) +  k/(\epsilon \alpha) + k T s)$ where $\bar{d^t}$ is the average node degree of current graph $\mathcal{G}_t$ and $s\triangleq \max_{s\in\mathcal{S}} |\operatorname{supp}(\bm p_s)|$.
\label{thm:time-complexity}
\end{theorem}

Note that the above time complexity is dominated by the time complexity of \textsc{IncrementPush} following from \cite{xingzhi2021subset}. Hence, the overall time complexity is linear to the size of edge events. Notice that in practice, we usually have $s \ll n$ due to the sparsity we controlled in \textsc{DynNodeRep}.

\section{Experiments}
\label{sec:experiments}

In this section, we demonstrate the effectiveness and efficiency of \textsc{DynAnom} for both node-level and graph-level anomaly tasks over three state-of-the-art baselines, followed by one anomaly visualization of ENRON benchmark, and another case study of a large-scale real-world graph about changes in person's life.

\subsection{Datasets}

\paragraph{DARPA}: The network traffic graph \cite{lippmann20001999}. Each node is an IP address, each edge represents network traffic. Anomalous edges are manually annotated as network attack (e.g., DDoS attack). 
\paragraph{ENRON}\cite{nr-aaai15, cohen2005enron}: The email communication graph. Each node is an employee in Enron Inc. Each edge represents the email sent from one to others. There is no human-annotated anomaly edge or node. Following the common practice used in \cite{yoon2019fast, chang2021f}, we use the real-world events to align the discovered highly anomalous time period.
\paragraph{EU-CORE}: A small email communication graph for benchmarking. Since it has no annotated anomaly, we use two strategies (Type-s and Type-W), similar to \cite{eswaran2018sedanspot},  to inject anomalous edge into the graph. \textbf{EU-CORE-S} includes anomalous edges, creating star-like structure changes in 20 random snapshots. Likewise, \textbf{EU-CORE-L:} is injected with multiple edges among randomly selected pair of nodes, simulating  node-level local changes.  We describe details in appendix. \ref{sec:appendix-data}. 
\paragraph{Person Knowledge Graph (PERSON)}: We construct PERSON graph from EventKG \cite{gottschalk2019eventkg, gottschalk2018eventkg}. Each node represents a person or an event. The edges represent a person's participation history in events with timestamps. The graph has no explicit anomaly annotation. Similarly, we align the detected highly anomalous periods of a person with the real-world events, which reflects the person's significant change as his/her life events update. We summarize the statistics of dataset in the following table:

\xingzhiswallow{
\begin{table}[ht]
        \centering
        \small
\begin{tabular}{p{0.05\textwidth}|p{0.05\textwidth}p{0.07\textwidth}p{0.05\textwidth} p{0.06\textwidth} p{0.06\textwidth}}
\toprule 
%   &  \multicolumn{2}{l}{Quantitatively studied} & \multicolumn{3}{l}{Qualitatively studied }  \\
%  \hline 
 Dataset & DARPA & EU-CORE & ENRON & PERSON & COUNTRY \\
\midrule 
 $\displaystyle |V|$ & 25,525 & 986 & 151 & 609,930 & 206 \\
\hline 
 $\displaystyle |E|$ & 4,554,344 & 333,734 & 50,572 & 8,782,630 & 2,013,931 \\
\hline 
%  Duration & 87,726 seconds & 69,459,255 s & \textasciitilde 3 years & 43 years from 1980 & 43 years from 1979\\
% Duration & 87,726 seconds & 69,459,255 s & 98,284,400 s (\textasciitilde 3 years) & 43 years from 1980 & 43 years from 1979\\
% \hline 
 $|\mathcal{G}_{0,...,T}|$ & 1463 & 115 & 1138 & 43 & 43 \\
\hline 
 Init. $t$ & 256 & 25 & 256 & 20  & 11 \\
\bottomrule 
%  Initial T for $\mathcal{G}_0$ & 256 & 25 & 256 & 20 ( year 2000) & 11 (year 1990) \\
% \hline 
%  Tasks & Node and Graph Anomaly  & Node Anomaly  & Case Study &  Case Study &  Case Study \\
% \hline
\end{tabular}
\caption{Dataset Statistics. We convert them into undirected graph by repeating edges with reversed source and destination node.}
\end{table}

}

\begin{table}[ht]
\centering
\caption{Dataset Statistics. Graphs are converted into undirected by repeating edges with reversed source and destination node. We detect anomalies after the initial snapshot (Init. $t$).}
% \vspace{-2mm}
\begin{tabular}{p{0.10\textwidth}|R{0.15\textwidth}R{0.15\textwidth}R{0.15\textwidth} R{0.15\textwidth} }
\toprule 
%   &  \multicolumn{2}{l}{Quantitatively studied} & \multicolumn{3}{l}{Qualitatively studied }  \\
%  \hline 
 Dataset & DARPA & EU-CORE & ENRON & PERSON  \\
\midrule 
 $\displaystyle |V|$ & 25,525 & 986 & 151 & 609,930  \\
\hline 
 $\displaystyle |E|$ & 4,554,344 & 333,734 & 50,572 & 8,782,630 \\
\hline 
%  Duration & 87,726 seconds & 69,459,255 s & \textasciitilde 3 years & 43 years from 1980 & 43 years from 1979\\
% Duration & 87,726 seconds & 69,459,255 s & 98,284,400 s (\textasciitilde 3 years) & 43 years from 1980 & 43 years from 1979\\
% \hline 
 $|\mathcal{G}_{0,...,T}|$ & 1463 & 115 & 1138 & 43  \\
\hline 
 Init. $t$ & 256 & 25 & 256 & 20   \\
\bottomrule 
%  Initial T for $\mathcal{G}_0$ & 256 & 25 & 256 & 20 ( year 2000) & 11 (year 1990) \\
% \hline 
%  Tasks & Node and Graph Anomaly  & Node Anomaly  & Case Study &  Case Study &  Case Study \\
% \hline
\end{tabular}

\end{table}
% \vspace{-8mm}

%dataset_gdelt-country-level-m-conflict-edge-197901-202112-num-art-5-gs-scale-7.0-initSS_1-timeStep_31536000
\subsection{Baseline Methods}
We compare our proposed method to four representative state-of-the-art methods over dynamic graphs:
\paragraph{Graph-level method: AnomRank}\cite{yoon2019fast} \footnote{We discovered potential bugs in the official code, which make the results different from the original paper. We have contacted authors for further clarification.} uses the global Personalized PageRank and use 1-st and 2-nd derivative to quantify the anomaly score of graph-level.  We use the node-wise AnomRank score for the node-level anomaly,
\paragraph{Edge-level method: SadenSpot} \cite{eswaran2018sedanspot} approximates node-wise PageRank changes before and after inserting edges as the edge-level anomaly score. 
\paragraph{Node-level method: NetWalk} \cite{yu2018netwalk} incrementally updates the node embeddings by extending random walks. However, since the \textit{anomaly} defined in the original paper is the outlier in each static snapshot, we re-alignment \cite{gower1975generalized, wang2008manifold} NetWalk embeddings and apply $\ell_2$-distance for node-level anomaly. \footnote{We use Procrustes analysis to find optimal scale, translation and rotation to align}

\paragraph{Node-level DynPPE} \cite{xingzhi2021subset} used the local node embeddings based on PPVs for unweighted graph. We extend its core PPV calculation algorithm for weighted graphs. The detailed hyper-parameter settings are listed in Table \ref{tab:param-config} in appendix. In the following experiment, we demonstrate that our proposed algorithm significantly outperform over the strong baselines.

\subsection{Exp1: Node-level Anomaly Localization}
\label{sec:exp1}

\paragraph{Experiment Settings:} As detailed in Def.\ref{def:node-track-def}, given a sequence of graph snapshots $  \mathcal{\bm G} = \{ \mathcal{G}_0, \mathcal{G}_1, ..., \mathcal{G}_T \}$ and a tracked node subset $\bm S = \{ u_0, u_1, ..., u_i \}$. 
We denote the set of anomalous snapshots of node u as ground-truth, such that $\bm Y_u$, containing $k_u$ timestamps where there is at least one new anomalous edge connected to node $u$ in that time.
We calculate the node anomaly score for each node. For example for node u: $\bm \delta_u = \{ \hat \delta_u^1, \hat \delta_u^2, \ldots, \hat \delta_u^T \}$, and rank its anomaly scores across time. 
Finally, we use the set of the top-$k_u$ highest anomalous snapshots $ \hat{\bm Y_u} $ as the predicted anomalous snapshot \footnote{In experiment, we assume that the number of anomalies is known for each node to calculate the prediction precision.} and calculate the averaged prediction precision over all nodes as the final performance measure as shown below:
%we identify at which snapshots did the node anomalies happen. We annotate node u's anomaly of when anomalous edges occurs in the node at timestamp t 
%It is a proxy for node-level anomaly, and easier for quantitative experiments.
%For example, in a graph of 5 snapshots, for the node $u_k$, it has anomalous edges in snapshot $\{2,4\}$ and non-anomalous edges at $0,1,3$, we rank the node-level anomaly score of $u_k$ for every snapshot, and take the top-2 snapshot with highest snapshots as the anomaly localization results.  
% We rank each node by its anomaly scores across time, and take the top-k snapshots $\bm \hat T_u^{k}$ as the predicted anomalous snapshot \footnote{In experiment, we assume that the number of anomalies is known for each node to calculate the prediction accuracy.}, then calculated the averaged prediction accuracy over all nodes as the final performance measure. 
% \vspace{-1mm}
\begin{align}
    Precision_{avg} = \frac{1}{|\bm S|} \sum_{u \in S}\frac{ \hat{\bm Y_u} \cap \bm  Y_u  }{ |\hat{\bm Y_u}| } \nonumber
\end{align}
% \vspace{-1mm}
% In order to leverage the annotated edge anomaly as the ground-truth,  we define the time (snapshot) of node-level anomaly to be the time when the anomalous edges get connected to the end-points. Note that, though this ground-truth is related to edge anomaly, we independently measure the node-level anomaly score of each node (not the edge). 
For DARPA dataset, we track 151 nodes\footnote{We track totally 200 nodes, but 49 nodes have all anomalous edges before the initial snapshot, so we exclude them in precision calculation.} which have anomalous edges after initial snapshot. Similarly we track 190,170 anomalous nodes over EU-CORE-S and EU-Core-L.
% Since EU-CORE dataset does not have annotated anomalous edges,  we randomly select 20 snapshots to inject artificial anomalous edges using two popular injection methods,  which are similar to the approach used in \cite{yoon2019fast}. For each selected snapshot in \textit{EU-CORE-S}, we uniformly select one node,$u_{high}$,  from the top 1\% high degree nodes, and injected 70 multi-edges connecting the selected node to other 10 random nodes which were not connected to $u_{high}$ before, which simulates the structural changes.  In each selected snapshot in \textit{EU-CORE-L}, we uniformly select 5 pairs of nodes, and injected edge connect each pair with totally 70 multi-edges as anomaly, which simulates the sudden peer-to-peer communication.  
We exclude edge-level method SedanSpot for node-level task because it can not calculate node-level anomaly score.% of node, but we include it for global anomaly detection. 
 We present the precision and running time in Table \ref{tab:node-anomaly-loc-accuracy} and Table \ref{tab:time-node-anomaly}.

\begin{table}[ht]
    \centering
    \caption{The average precision of node-level anomaly localization. Our proposed \textsc{DynAnom} outperforms other baselines.}
    % \vspace{-3mm}
    % \small
    \begin{tabular}{lrrr}
    \toprule 
      & DARPA & EU-CORE-S  & EU-CORE-L  \\
    \midrule 
     Random  & 0.0075 & 0.0142 & 0.0088 \\
    \hline 
     AnomRank & 0.2790 & 0.2173 & 0.4019 \\
    \hline 
     AnomRankW & 0.2652 & 0.2213 & 0.4078 \\
    \hline 
     NetWalk  & OOM & 0.0421 & 0.0416 \\
    \hline 
     DynPPE &  0.1701 & 0.2478 & 0.2372 \\
    \hline 
     \textbf{DynAnom} & \textbf{0.5425} & \textbf{0.4242} & \textbf{0.5215} \\
     \bottomrule
    \end{tabular}
\label{tab:node-anomaly-loc-accuracy}
\end{table}
% \vspace{-2mm}

\paragraph{Effectiveness.} In Table \ref{tab:node-anomaly-loc-accuracy}, the low scores of \textit{Random}\footnote{For Random baseline, we randomly assign anomaly score for each node across snapshots} demonstrate the task itself is very challenging. We note that our proposed \textsc{DynAnom} outperforms all other baselines, demonstrating its effectiveness for node-level anomaly tracking.

\paragraph{Scalability.} NetWalk hits Out-Of-Memory (OOM) on DARPA dataset due to the Auto-encoder (AE) structure where the parameter size is related to $|\mathcal{V}|$. Specifically, the length of input/output one-hot vector is $|\mathcal{V}|$, making it computationally prohibitive when dealing with graphs of even moderately large node set.  Moreover, since the AE structure is fixed once being trained, there is no easy way to incrementally expand neural network to handle the new nodes in dynamic graphs. While \textsc{DynAnom} can easily track with any new nodes by adding extra dimensions in $\bm p_s, \bm r_s$ and $\bm d$ and apply the same procedure in an online fashion.

\paragraph{Power of Node-level Representation.} The core advantage of \textsc{DynAnom} over AnomRank(W) is the power of node-level representation. AnomRank(W) essentially represents individual node's status by one element in the PageRank vector. While \textsc{DynAnom} uses node representation derived from PPVs, thus having far more expressivity and better capturing  node changes over time.

\paragraph{Advantage of Aligned Space.} Despite that NetWalk hits OOM on DARPA dataset, we hypothesize that the reason of its  poor performance on EU-CORE\{S,L\} is the embedding misalignment. The incremental training makes the embedding space not directly comparable from time to time even after extra alignment. Although it's  useful for classification or clustering within one snapshot, the misaligned embeddings may cause troublesome overhead for downstream tasks, such as training separate classifiers for every snapshot.

\paragraph{Importance of Edge Weights.} Over all datasets, our proposed \textsc{DynAnom} consistently outperforms DynPPE, which is considered as the unweighted counterpart of \textsc{DynAnom}. It demonstrates the validity and versatility of our proposed framework.

\begin{table}[ht]
\centering
\caption{For comparing running time (in seconds), excluding the time for data loading and graph update. }
\begin{tabular}{lrrr}
\toprule 
  & DARPA & EU-CORE-S & EU-CORE-L \\
\midrule 
 AnomRank(W) & 905.983  &  26.084& 26.865 \\
\hline 
 NetWalk & OOM &  3649.14 &  3644.58\\
\hline 
DynPPE &  84.247 &  33.835 & 30.933 \\
\hline 
 DynAnom & 379.334 & 30.054 & 26.359 \\
 \bottomrule
\end{tabular}
\label{tab:time-node-anomaly}
\end{table}

\paragraph{Efficiency.} Table \ref{tab:time-node-anomaly} presents the wall-clock time \footnote{We track function-level running time by CProfile, and remove those time caused by data loading and dynamic graph structure updating}. Deep learning based NetWalk is the slowest. Although it can incrementally update random walk for the new edges, it has to  re-train or fine-tune for every snapshot. 
At the first glance, DynPPE achieves the shortest running time on DAPRA dataset, but it is an illusive victory caused by dropping numerous multi-edge updates at the expense of precision. Furthermore, the Power Iteration-based AnomRank(W) is slow on DARPA as we discussed in Section \ref{sec:notation-prelim}, but the speed difference may not be evident when the graph is small (e.g., DARPA has 25k nodes, and EU-CORE has less than 1k nodes). Moreover, our algorithm can independently track individual nodes as a result of its local characteristic,  it could be further speed up by embarrassingly parallelization on a computing cluster with massive cores.
% Moreover, on \textit{large} graph (DAPRA of 25k nodes), but the speed advantage may not be evident when the graph is small (less than 1k nodes).

% \paragraph{Observations from Table \ref{tab:time-node-anomaly}}: \textbf{I.} Although it seems that DynPPE achieves the shortest running time on DAPRA dataset, it is an illusive victory caused by dropping numerous multi-edges and sacrifice performance as presented in \ref{tab:node-anomaly-loc-accuracy}. 

% \textbf{II.} AnomRank(W) is based on Power Iteration, which could be slower as we discussed in Section \ref{sec:notation-prelim}. On the other hand, our local method could be faster on \textit{large} graph (DAPRA of 25k nodes), but the speed advantage may not be evident when the graph is small (less than 1k nodes).

\subsection{Exp2: Graph-level Anomaly Detection }
\label{sec:exp2}
\paragraph{Experiment Setting}:
Given a sequence of graph snapshots $ \mathcal{ \bm G} = \{ \mathcal{G}_0, \mathcal{G}_1, ..., \mathcal{G}_T \}$, we calculate the snapshot anomaly score and consider the top ones as anomalous snapshots, adopting the same experiment settings in \cite{yoon2019fast}. 
We use both $\ell_1$ and $\ell_2$ as distance function to test the practicality of \textsc{DynAnom}.
We modify edge-level SedanSpot for Graph-level anomaly detection by aggregating edge anomaly score with  $\textsc{Mean}()$ \footnote{ For a fair comparison, we tried $\{ Mean(), Median(), Min(), Max() \}$ operators for SedanSpot, and found that $Mean()$ yields the best results.} in each snapshot. We describe detailed hyper-parameter settings in Appendix. \ref{sec:appendix-param-config}.
For DAPRA dataset, as suggested by \citet{yoon2019fast}, we take the precision at top-$250$ \footnote{ $k=250$ reflects its practicality since it is the closest $k$ without exceeding total 288 graph-level anomalies} as the performance measure in Table \ref{tab:top-k-precision}, and present Figure \ref{fig:darpa-rank-topk-precision} of recall-precision curves as we vary the parameter $k \in \{ 50,100,\ldots,800 \}$  for top-$k$ snapshots considered to be anomalous. 
% Enron emails records notorious Enron scandal has been intensively studied in \red{cite!}. 
% For the ENRON graph without anomaly ground-truth, the common evaluation is to visualize the anomalous scores against time, then interpret or associate the peaks with the real-world events, as listed in Table \ref{tab:enron-events} in Appendix.  

\paragraph{Practicality for Graph-level Anomaly}: 
The precision-recall curves show that \textsc{DynAnom} have a good trade-off, achieving the competitive second/third highest precision by predicting the highest top-$150$ as anomalous snapshots, and the precision decreases roughly at a linear rate when we includes more top-$k$  snapshots as anomalies. 
Table \ref{tab:top-k-precision} presents the precision at top-$250$, and we observe that \textsc{DynAnom} could outperform all other strong baselines, including its unweighted counterpart -- \textsc{DynPPE}. It further demonstrates the high practicality of the proposed flexible framework even with the simplest $\ell_1 \text{and }\ell_2$ distance function. 

\begin{table}[ht]
	\begin{minipage}{0.6\linewidth}
		\centering
		\includegraphics[width=1\textwidth]{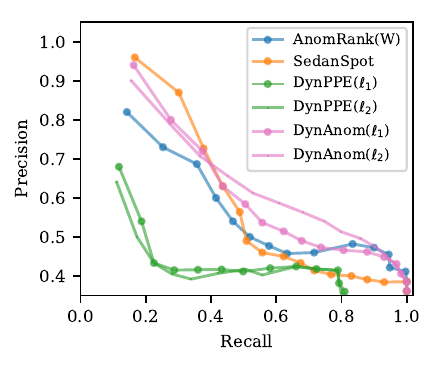}
		\vspace{-10mm}
		\captionof{figure}{Precision-Recall Curve}
		\label{fig:darpa-rank-topk-precision}
	\end{minipage}%
	\hfill
	\begin{minipage}{0.39\linewidth}
	\small
		\caption{The precision of top-250 anomalies}
		\vspace{-4mm}
		\label{tab:top-k-precision}
		\centering
        \begin{tabular}{lr}
            \toprule
                Algorithms &  Precision \\
            \midrule
                NetWalk &     OOM  \\
                AnomRankW &     0.5400  \\
                SedanSpot &     0.5640  \\
                DynPPE($\ell_1$) &     0.4160 \\
                DynPPE($\ell_2$) &     0.3920 \\
                DynAnom($\ell_1$) &     0.5840 \\
                \textbf{DynAnom($\ell_2$)} &     \textbf{0.6120} \\
            \bottomrule
        \end{tabular}
	\end{minipage}

\end{table}

% \xingzhiswallow{
% \begin{minipage}{0.5\textwidth}
%     \hspace{-7mm}
%     \begin{minipage}[b]{0.60\textwidth}
%         \includegraphics[width=1\textwidth]{figs/fig-res-dataset_darpa-initSS_256-timeStep_60-topk-recall-precision-v2.pdf}
%     \vspace{-9mm}
%     \captionof{figure}{Precision-Recall Curve}
%     \label{fig:darpa-rank-topk-precision}
%     \end{minipage}
%     %
%     \begin{minipage}[b]{0.39\textwidth}
%     \small
%     %\vspace{-8mm}
%         \captionof{table}{The precision of top-250 anomalies.}
%         \vspace{-2mm}
%         \begin{tabular}{lr}
%             \toprule
%                 Algorithms &  Precision \\
%             \midrule
%                 NetWalk &     OOM  \\
%                 AnomRankW &     0.5400  \\
%                 SedanSpot &     0.5640  \\
%                 DynPPE($\ell_1$) &     0.4160 \\
%                 DynPPE($\ell_2$) &     0.3920 \\
%                 DynAnom($\ell_1$) &     0.5840 \\
%                 \textbf{DynAnom($\ell_2$)} &     \textbf{0.6120} \\
%             \bottomrule
%         \end{tabular}
%         \label{tab:top-k-precision}
%     \end{minipage}
% \end{minipage}

% }

% Enron emails records notorious Enron scandal has been intensively studied in \red{cite!}. 
% For the ENRON graph without anomaly ground-truth, the common evaluation is to visualize the anomalous scores against time, then interpret or associate the peaks with the real-world events, as listed in Table \ref{tab:enron-events} in Appendix.  

\begin{figure}[htbp]
    \centering
    \includegraphics[width=0.8\textwidth]{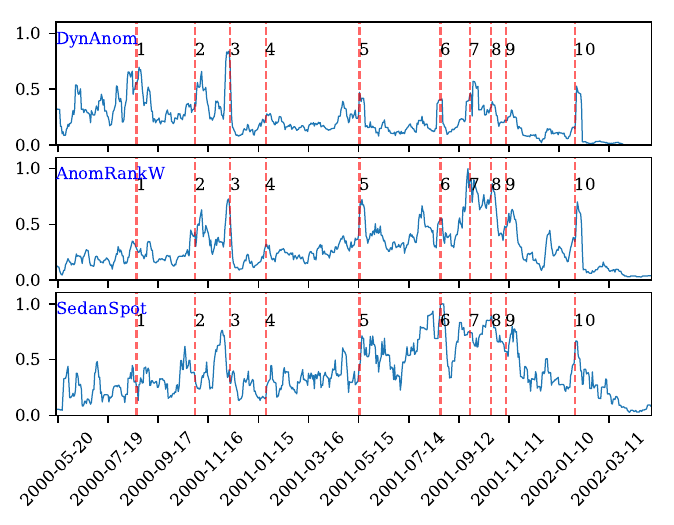}
    \vspace{-3mm}
    \caption{ The anomaly scores of ENRON graph. \textsc{DynAnom} well correlates with other baselines, and detect novel peaks at beginning.}
    \label{fig:enron-curve}
\end{figure}

ENRON graph records the emails involved in the notorious Enron company scandal, and it has been intensively studied in \cite{eswaran2018sedanspot,chang2021f, yoon2019fast}.  Although ENRON graph has no explicit anomaly annotation, a common practice \cite{chang2021f,yoon2018fast} is to visualize the detected anomaly score against time and discover the associated real-world events. Likewise, Figure \ref{fig:enron-curve} plots the detected peaks side-by-side with other strong baselines \footnote{All values are post-processed by dividing the maximum value. For SedanSpot, we use $Mean()$ to aggregate edge anomaly score for the best visualization. Other aggregator (e.g., $Max()$) may produce flat curve, causing an unfair presentation. }. We find that \textsc{DynAnom} shares great overlaps with meaningful peaks and detects more prominent peak at Marker.1 when Enron got investigated for the first time,  demonstrating its effectiveness for sensible change discovery . 
%\footnote{All values are post-processed by dividing the maximum value. }
% In Figure \ref{fig:enron-curve}, We compare our method $DynAnom(\ell_2)$ to other strong baselines, and find that they share great overlaps with meaningful peaks, demonstrating the effusiveness of our method. 
% and annotate with the events for interpretation. 
For better interpretation, we annotate the peaks associated to the events collected from the Enron Timeline \footnote{Full events are included in Table \ref{tab:enron-events} in Appendix. Enron timeline: \url{https://www.agsm.edu.au/bobm/teaching/BE/Enron/timeline.html}}, and present some milestones as followings:
\begin{itemize}
\small
\item[1]:(2000/08/23) Investigated Enron  when stock was high.
% \item[5]:(2001/05/17) Enron Chairman had a "Secret"  meeting with state governor. 
\item[6]:(2001/08/22) CEO was warned of accounting irregularities.
\item[10]:(2002/01/30) CEO was replaced after bankruptcy.
\end{itemize}
\vspace{-5mm}
% \vspace{-4mm}

% \paragraph{Observations}: We observe 

% We compare AnomRank, SadenSpot and our proposed score for global graph-level anomaly detection.

% \red{add different epsilon? or mean ?  discussion about findings }

% \textcolor{red}{Figure \ref{fig:darpa-score-timeline} and Figure \ref{fig:enron-score-timeline} } show the anomaly scores from baselines on DARPA and ENRON dataset.

% NetWalk is inefficient if we have finer-grained snapshots, it cannot finish in 10 hours, given ~1K snapshots over DARPA dataset.

% darpa precision-recall
\xingzhiswallow{

\begin{figure}[ht]
    \centering
    \includegraphics[width=0.35\textwidth]{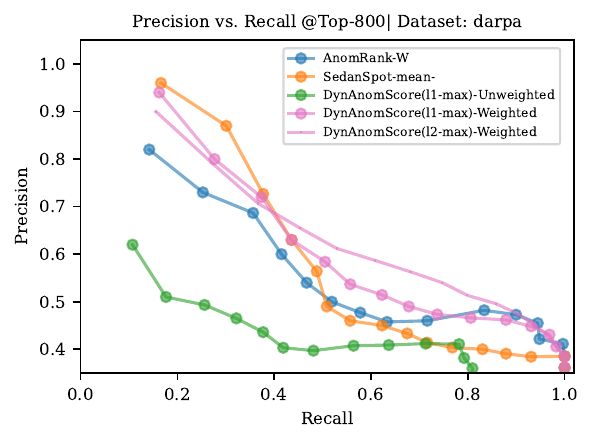}
    \caption{DARPA network: Precision-Recall for top-800}
    \label{fig:darpa-rank-topk-precision}
\end{figure}

\begin{table}[ht]
    \centering
    \small
\begin{tabular}{lr}
\toprule
          Anomaly score &  DARPA \\
\midrule
    NetWalk &     OOM  \\
    AnomRankW &     0.5400  \\
    SedanSpot &     0.5640  \\
    DynPPE &     0.4360 \\
    \textbf{DynAnom} &     \textbf{0.6120} \\

\bottomrule
\end{tabular}
    \caption{The precision of top-250 detected anomalous snapshots on DARPA dataset.}
\end{table}

}

\xingzhiswallow{

    \begin{figure*}[ht]
    \begin{subfigure}{0.5\textwidth}
       \centering
        \includegraphics[width=1.0\linewidth]{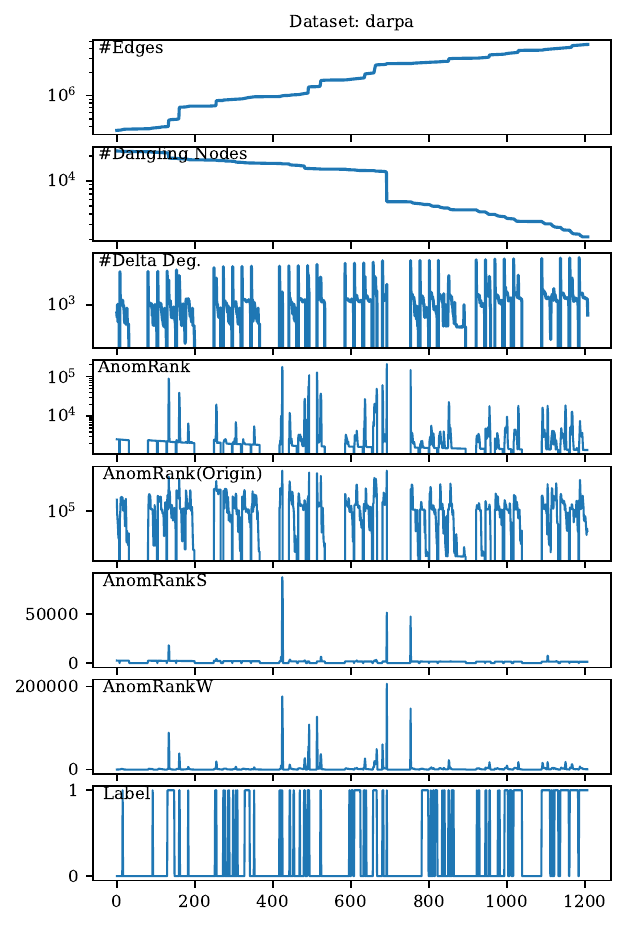}
        \caption{DARPA network: AnomRank Anomaly Scores vs. Ground-truth. }
        \label{fig:darpa-anomrank-scores}
    \end{subfigure}\hfill
    \begin{subfigure}{.5\textwidth}
        \centering
        \includegraphics[width=1.0\linewidth]{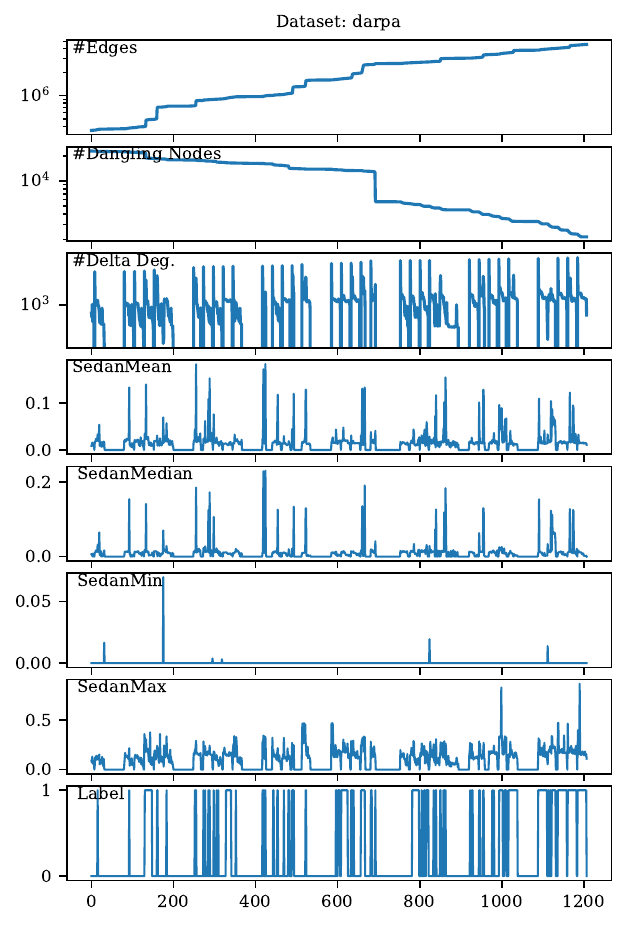}
        \caption{DARPA network: SedanSpot Anomaly Scores vs. Ground-truth}
        \label{fig:darpa-sedanSpot-scores}
    \end{subfigure}
    \caption{The baselines on DARPA dataset}
    \label{fig:darpa-score-timeline}
    \end{figure*}

\begin{figure*}[ht]
\begin{subfigure}{0.5\textwidth}
    \centering
    \includegraphics[width=0.7\linewidth]{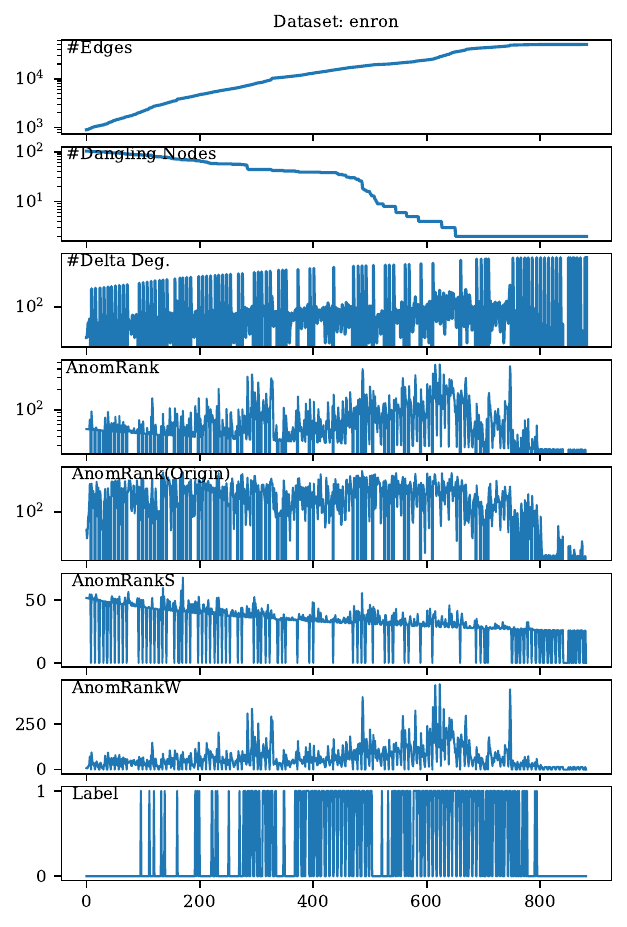}
    \caption{Enron network: AnomRank Anomaly Scores (No ground truth).}
    \label{fig:enron-anomrank-scores}
\end{subfigure}\hfill
\begin{subfigure}{.5\textwidth}
    \centering
    \includegraphics[width=0.7\linewidth]{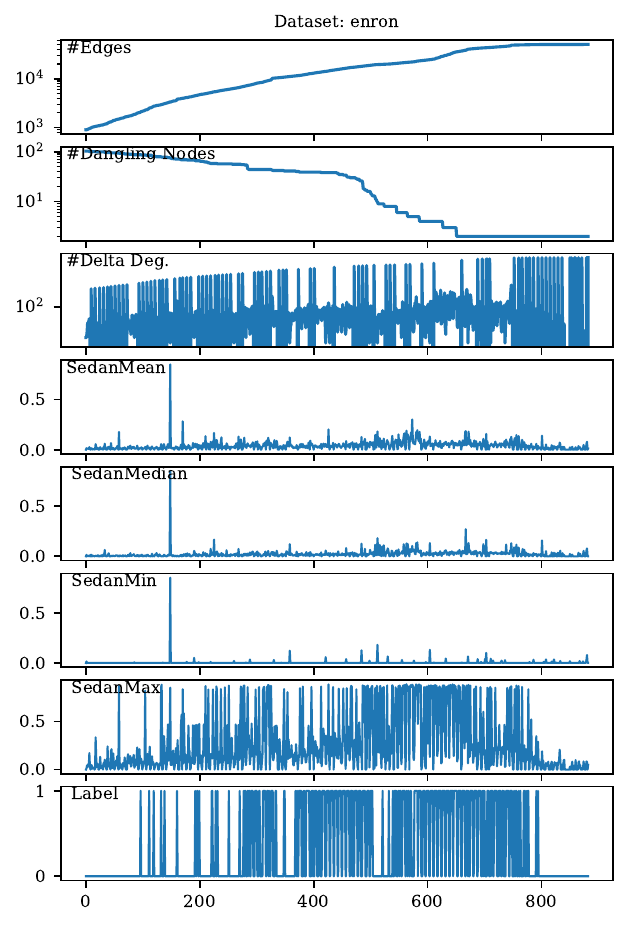}
    \caption{Enron network: SedanSpot Anomaly Scores (No ground truth).}
    \label{fig:enron-sedanSpot-scores}
\end{subfigure}
\begin{subfigure}{.5\textwidth}
    \centering
    \includegraphics[width=0.7\linewidth]{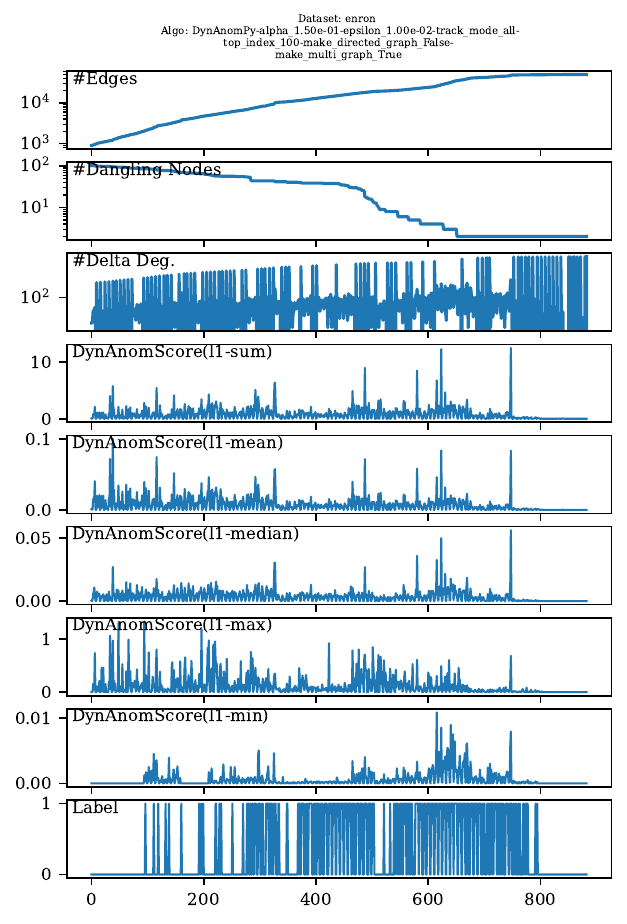}
    \caption{Enron network: \textcolor{red}{The Proposed} DynAnom Anomaly Scores (No ground truth).}
    \label{fig:enron-dynAnom-scores}
\end{subfigure}\hfill
\begin{subfigure}{.5\textwidth}
    \centering
    \includegraphics[width=1.3\linewidth]{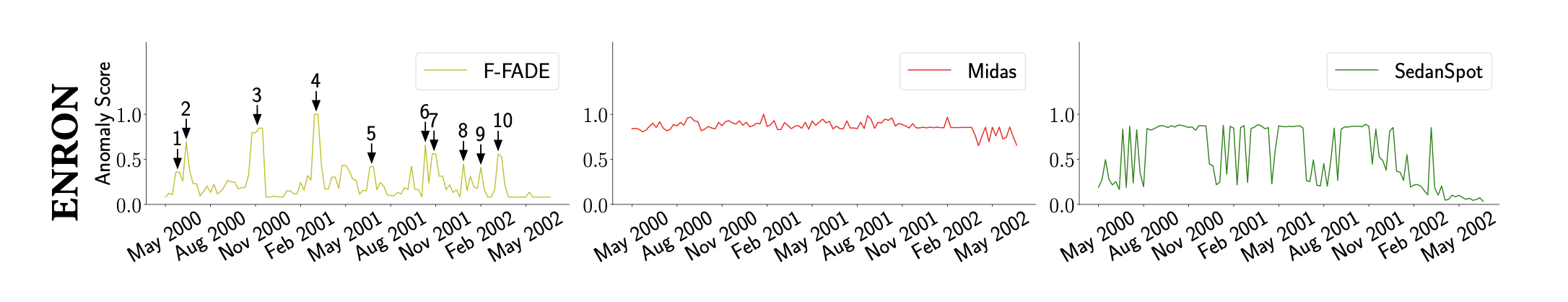}
    \caption{Enron network: From F-Fade, which is believed to be the best detection over Enron dataset.}
    \label{fig:enron-f-fade-scores}
\end{subfigure}

\caption{The baselines on Enron dataset. Our proposed method detects anomalies as peaks shown in Figure \ref{fig:enron-dynAnom-scores}}.
\label{fig:enron-score-timeline}
\end{figure*}

}

% darpa 
\xingzhiswallow{
\begin{figure}[ht]
    \centering
    \includegraphics[width=0.7\linewidth]{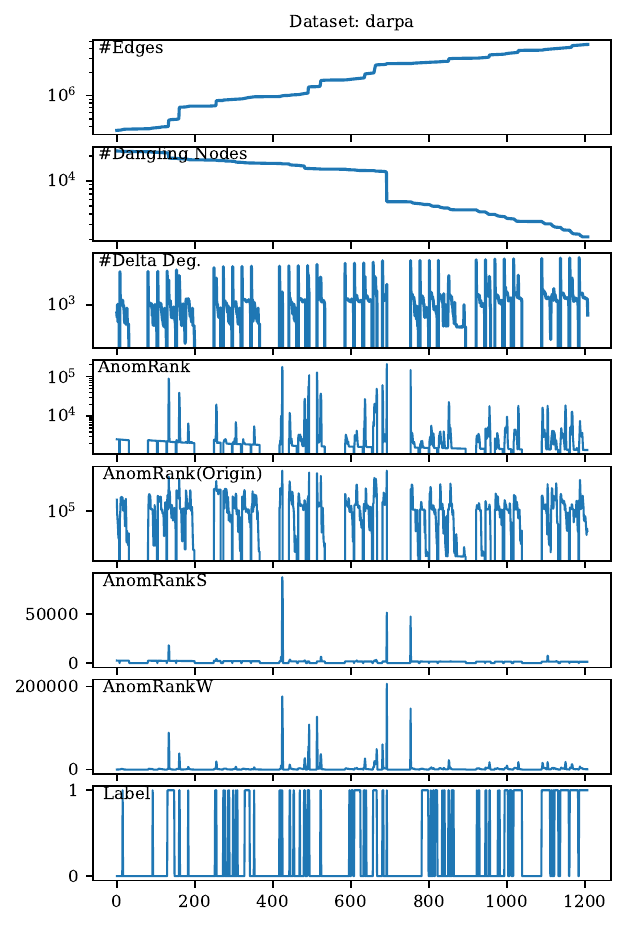}
    \caption{DARPA network: Anomaly Scores vs. Ground-truth. }
    \label{fig:darpa-anomrank-scores}
\end{figure}

\begin{figure}[ht]
    \centering
    \includegraphics[width=0.7\linewidth]{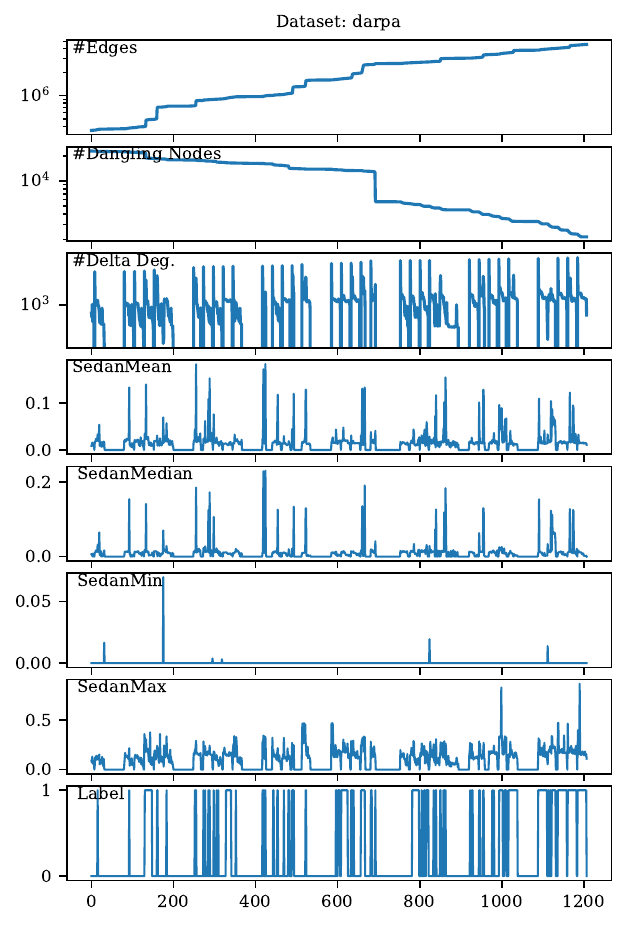}
    \caption{Enron network: SedanSpot Anomaly Scores (No ground truth).}
    \label{fig:darpa-sedanSpot-scores}
\end{figure}

}

% enron 
\xingzhiswallow{
\begin{figure}[ht]
    \centering
    \includegraphics[width=0.7\linewidth]{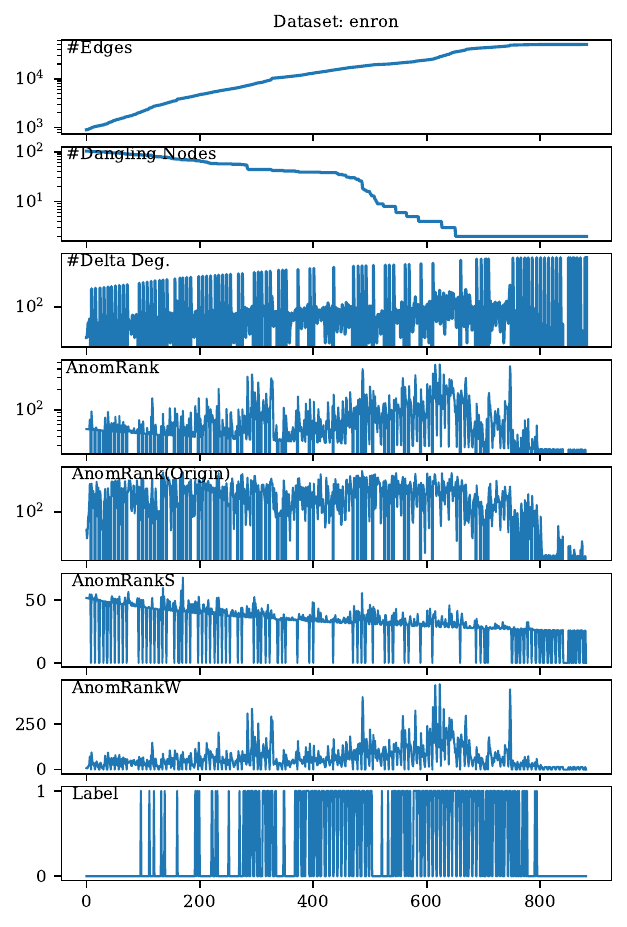}
    \caption{Enron network: Anomaly Scores (No ground truth).}
    \label{fig:enron-anomrank-scores}
\end{figure}

\begin{figure}[ht]
    \centering
    \includegraphics[width=0.7\linewidth]{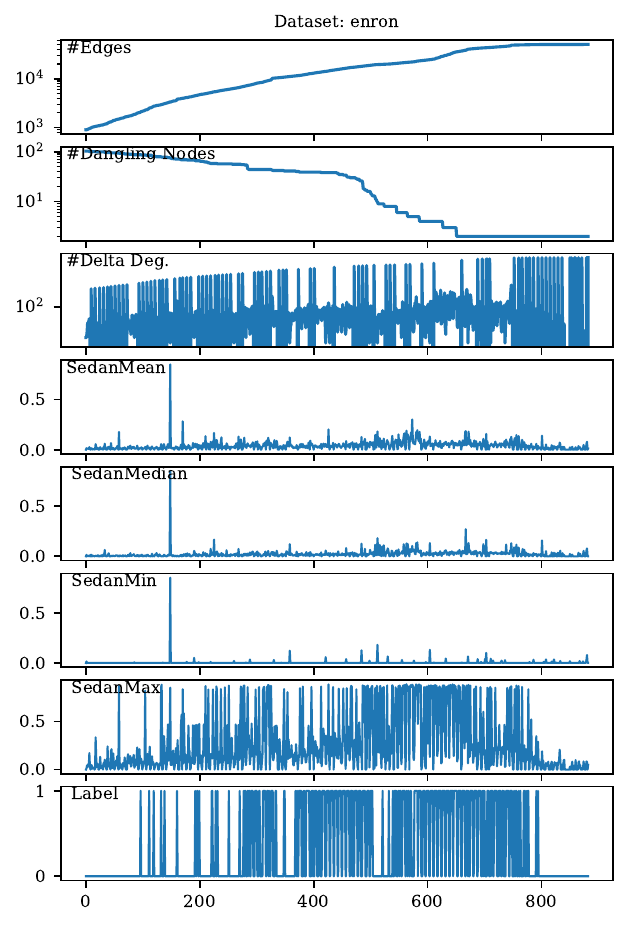}
    \caption{Enron network: SedanSpot Anomaly Scores (No ground truth).}
    \label{fig:enron-sedanSpot-scores}
\end{figure}

}

% \paragraph{ENRON Email Dataset}

% NetWalk\cite{yu2018netwalk} experiment setting:  Use first 50\% edges without any injected fake edge, then train clusters. Inject anomaly (fake) edges in the last 50\% and use embeddings to predict whether it's far from any trained cluster centers. For each testing snapshot, it use both previous fake and real edges to update the node embeddings. It is more similar to link prediction instead of anomaly detection.

\subsection{Case Study: Localize Changes in Person's Life over Real World Graph}

\begin{figure}[ht]
    \centering
    \includegraphics[width=1.0\linewidth]{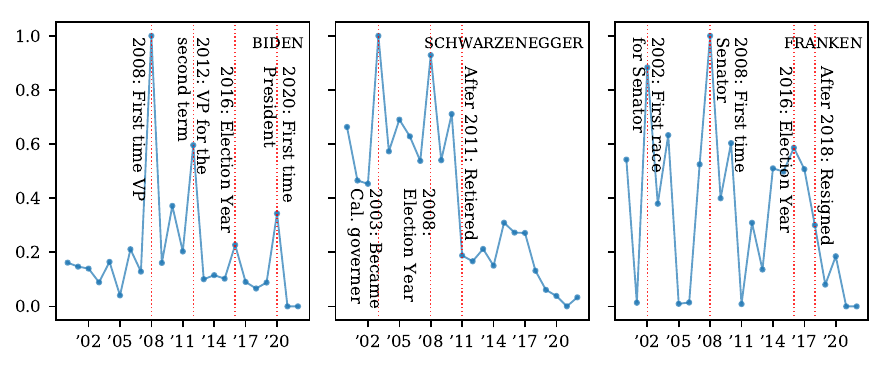}
    \vspace{-6mm}
    \caption{The yearly changes of public figures in PERSON graph from 2000-2022: The detected peaks are well correlated to the major events of the individuals as we annotated. }
    \label{fig:eventkg-person}
\end{figure}

% In our case studies, we want to detect the changes of individual entities in real-world graphs. Specifically 
To demonstrate the scalability and usability of \textsc{DynAnom},  we present an interesting case study over our constructed large-scale PERSON graph, which records the structure and intensity of person-event interaction history. 
We apply \textsc{DynAnom} to track public figures (Joe Biden, Arnold Schwarzenegger, Al Franken) of the U.S. from 2000 to 2022 on a yearly basis, and visualize their changes by $\ell_1$-distance, together with annotated peaks in Figure \ref{fig:eventkg-person}.

% Figure \ref{fig:eventkg-person} shows three individuals with major career changes in 22 years from 2000 to 2022, where we plot the yearly difference \footnote{All values are post-processed by dividing the maximum value. } and annotate the peaks for each individual.

% Specifically, we track public figures (e.g., Joe Biden, Arnold Schwarzenegger, Al Franken) of the U.S. from 2000 to 2021 in a year basis. 
% \vspace{-2mm}

As we can see, \textit{Biden} has his greatest peak in year 2008 when he became the 47-th U.S. Vice President for the very first time, which is considered to be his first major advancement. The second peak occurs in 2012 when he won the re-election. This time the peak has smaller magnitude because he was already the VP, thus the re-election caused less difference in him, compared to his transition in 2008. The third peak is in 2020 when he became the president, the  magnitude is even smaller as he had been the VP for 8 years without huge context changed. %, although having great political impact.
% In year 2020, Biden has a great peak in degree change, but the status may node change so much since he had been already Vice President since 2008, where the major peak occurred. 
Similarly, the middle sub-figure captures  \textit{Arnold Schwarzenegger}'s transition to be California Governor in 2003, and high activeness in election years. 
%
%shows \textit{Arnold Schwarzenegger} who transited from an actor to a California Governor in 2003 with the highest peak. Followed by election year 2008 when he joined campaigning as the key endorsement.  %After 2011, Arnold reached his term limit as Governor and returned to acting.
%
\textit{Al Franken} is also a famous comedian who shifted career to be a politician in 2008, the peaks well capture the time when he entered political area and the election years. Table \ref{tab:person-events} lists full events in Appendix \ref{sec:appendix-timelines}. This resources could bring more opportunities for knowledge discovery and anomaly mining studies in the graph mining community. 

Moreover, based on our current Python implementation, it took roughly 37.8 minutes on a 4-core CPU machine to track the subset of three nodes over the PERSON graph, which has more than 600k nodes and 8.7 million edges.  Both presented results demonstrate that our proposed method has great efficiency and usability for practical subset node tracking.

% where we observed peaks related to the individual when s/he had major career events.
% demonstrating that the detected peaks are closely correlated to the well-known real life events. 

% Specifically, we track two major countries (China and the U.S.) in \textit{COUNTRY} graph from 1980 to 2022, where we observed several peaks associated to the country involving in major conflicts. 
% Meanwhile, we also track public figures (e.g., Joe Biden, Arnold Schwarzenegger, Al Franken) of the U.S. in the \textit{PERSON} graph from 2000 to 2021, where we observed peaks related to the individual when s/he has major career events.

% \paragraph{PERSON Graph}
%2245.452 (edge adjust) + 10.516 (push )

% \vspace{-2mm}

%  17565258 2229.887    0.000 2245.452    0.000 DynamicPPE.py:287(dynamic_adjust)
%       43  218.194    5.074 2661.072   61.885 DynamicPPE.py:445(get_approx_ppv_with_new_edges_batch_multip_source)
%  17565258  114.060    0.000  114.060    0.000 DynamicPPE.py:416(update_graph)
% 105394548   33.739    0.000   33.739    0.000 typedlist.py:280(_numba_type_)
%         1   25.702   25.702   46.287   46.287 graph_data_utils.py:12(read_data)
%  52697651   15.566    0.000   15.566    0.000 serialize.py:29(_numba_unpickle)
%  26957907   14.991    0.000   26.550    0.000 std.py:1159(__iter__)
%         1   14.639   14.639 2737.509 2737.509 test_dynmaicPPE_exp3_case_studies.py:109(test_graph_stream)
%       43   10.516    0.245   10.516    0.245 DynamicPPE.py:151(push_numba_parallel)

% \paragraph{Wiki-Graph Dataset}

\xingzhiswallow{
\textbf{COUNTRY Graph} records the conflict between countries and entities. We investigated China and the U.S. and found interesting peaks associated to global/regional tension as presented in Figure \ref{fig:gdelt-china-us} and full event list in Table \ref{tab:country-events} in appendix. 

\vspace{-2mm}
\begin{figure}[ht]
    \centering
    \includegraphics[width=1.0\linewidth]{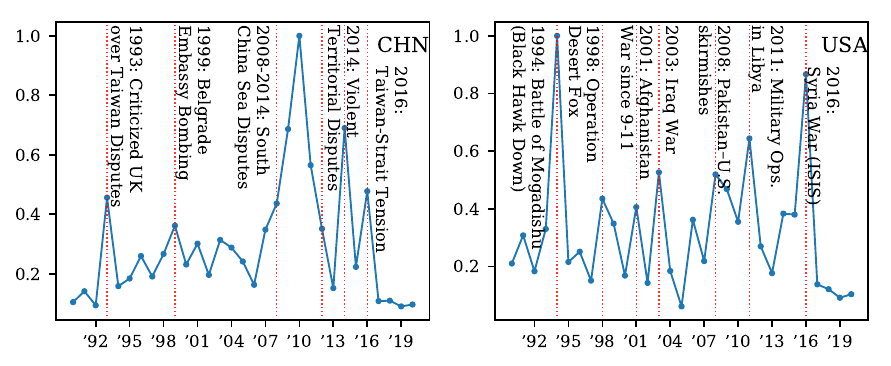}
    \vspace{-8mm}
    \caption{ The yearly changes of two countries in  \textit{COUNTRY} graph from 1990-2020 : The detected peaks reflect the major conflicts as the countries shift their diplomatic/militant behaviors.  }
    \label{fig:gdelt-china-us}
\end{figure}
}

% \begin{figure}[ht]
%     \centering
%     \includegraphics[width=1.0\linewidth]{figs/fig-anold-temporal-gamma-1.pdf}
%     \caption{Arnold Schwarzenegger: 1979-2021}
%     \label{fig:eventkg-arnold}
% \end{figure}

\section{Discussion and Conclusion}
\label{sec:conclusion}

In this paper, we propose an unified framework \textsc{DynAnom} for subset node anomaly tracking over large dynamic graphs. This framework can be easily applied to different graph anomaly detection tasks from local to global with a flexible score function customized for various application. 
Experiments show that our proposed framework outperforms current state-of-the-art methods by a large margin, and has a significant speed advantage (2.3 times faster) over large graphs.  
We also present a real-world PERSON graph with an interesting case study about personal life changes, providing a rich resource for both knowledge discovery and algorithm benchmarking.
For the future work, it remains interesting to explore different type of score functions, automatically identify the interesting subset of nodes as better strategies for tracking global-level anomaly, and further investigate temporal-aware PageRank as better node representations.

%\textcolor{red}{future work, score function, S for graph, temporal pagerank.}

% \bibliographystyle{ACM-Reference-Format}
% \bibliography{references}

% \clearpage
% \appendix
\section{Appendix}
 
\subsection{Proof of Weighted Dynamic Push}
%\ref{theorem:dyn-adjust-weighted-graph}
\label{appendix-proof}
\begin{proof}
By the invariant property in Lemma \ref{lemma:ppr-invar}, we have
\begin{align}
p_s(i) + \alpha r_s(i) = (1-\alpha)\sum_{x \in N^{in}(i)}  \frac{ w_{(x,i)} p_s(x)}{d(x)} + \alpha \times 1_{i=s}, \forall i \in \mathcal{V},  \label{eq:u-invaraint-appendix} 
\end{align}

Initially, this invariant holds and keeps $ \frac{r_s(i)}{d(i)} \leq \epsilon$ after applying Algo.\ref{algo:forward-local-push-weighted}. 
When the new edge $(u,v,\Delta w_{(u,v)})$ arrives, it changes the out-degree of $u$, breaking up the balance for all invariant involving $d(u)$. Our goal is to recover such invariant by adjusting small amount of $p_s, r_s$. While such adjustments may compromise the quality of PPVs, we could incrementally invoke Algo.\ref{algo:forward-local-push-weighted} to update $p_s$ and $r_s$ for better accuracy afterwards. We denote the initial vectors as $p_s, r_s, d$ and post-change vectors as $p'_s, r'_s, d'$. 

As $d'(u)$ is involved in $p_s(u) / d'(u)$, one need to have $p'_s(u)$ such that $ p_s(u) / d(u) = p'_s(u) / d'(u)$, i.e. the invariant maintenance of Equ. \eqref{eq:u-invaraint-appendix}. The updated amount of weight is $\delta w_(u,v)$, which indicates $p'_s(u) =( d(u)+\Delta w_{(u,v)} ) p_s(u) / d(u)$. So, we have
\begin{align}
p'_s(u) = p_s(u) \frac{\sum_{v \in \operatorname{Nei}(u)} w_{(u,v)} + \Delta w_{(u,v)}}{\sum_{v \in \operatorname{Nei}(u)} w_{(u,v)}}. \label{eq:p-update-weighted-appendix}
\end{align}
Hence, we reach
Equ. \ref{eq:p-update-weighted-appendix} implies that we should scale up $p_s(u)$ to recover balance. This strategy is the general case of \citet{zhang2016approximate} for unweighted graphs where $\Delta w_{(u,v)} = 1$.

However, once we assign $p'_s(u)$, it breaks the invariant for $u$ since $p_s(u)+\alpha r_s(u) \neq p'_s(u)+\alpha r_s(u)$. Likewise, we maintain this equality by introducing $r'_s(u)$:
\begin{align}
p_s(u) + \alpha r_s(u) & = p_s'(u) + \alpha r_s'(u) \nonumber\\
&\text{Substitute $p'_s(u)$ from eq.\ref{eq:p-update-weighted-body}} \nonumber \allowdisplaybreaks\\
& = p_s(u) + \frac{\Delta w_{(u,v)} p_s(u)}{d(u)} + \alpha r_s'(u) , \nonumber \allowdisplaybreaks \\
\Longrightarrow \alpha r_s'(u) - \alpha r_s(u) &= -\frac{\Delta w_{(u,v)} p_s(u)}{d(u)},\nonumber \allowdisplaybreaks\\
r_s'(u) &=  r_s(u) -\frac{\Delta w_{(u,v)} p_s(u)}{\alpha d(u)} 
\label{eq:r-update-weighted-appendix}
\end{align}

Equ. \eqref{eq:r-update-weighted-appendix} implies that we should decrease $r_s(u)$. From a heuristic perspective, it's consistent to the behavior of BCA-algorithm which keeps taking mass from $r_s$ to $p_s$. In this updating case, we artificially create mass for $p'_s$ at the expense of $r_s$'s decrements. When $\Delta w_{(u,v)} = 1$ in case of unweighted graphs, this update rule is also equivalent to \citet{zhang2016approximate}.

So far, we have the updated $p'_s(u)$ and $r'_s(u)$ so that all nodes (without direct edge update, except for $v$) keep the invariant hold. However, due to the introduction of $\Delta w_{(u,v)}$, the shares of mass pushed to $v$ is changes, breaking the invariant for $v$ as shown below:
\begin{align}
    p_s(v) + \alpha r_s(v) & \neq (1-\alpha) \Bigl( \frac{ (w_{(u,v)} + \Delta w_{(u,v)})  p'_s(u)}{d'(u)} \Bigr. \nonumber\\
    & \Bigl. + \sum_{x \in N^{in}(v) \setminus \{ u \} }  \frac{w_{(u,x)} p_s(x)}{d(x)} \Bigr) +\alpha \times 1_{t=s} \nonumber
\end{align}

In order to recover the balance with minimal effort, we should update $r_s(v)$ instead of $p_s(v)$. The main reason is that any change in $p_s(v)$ will break the balance for $v$'s neighbors, similar to the breaks incurred by the change of $p_s(u)$. We present the updated $r'_s(v) = r_s(v)+\Delta $ as following:
\begin{align}
p_s(v) + \alpha \Bigl(r_s(v)+\Delta \Bigr) = (1-\alpha) \Bigl( \frac{\Delta w_{(u,v)} p'_s(u)}{d'(u)} + \Bigr. \nonumber\\
\Bigl. \sum_{x \in N^{in}(v) }  \frac{w_{(u,x)} p_s(x)}{d(x)} \Bigr) +\alpha 1_{t=s}  \label{eq:invar-v-body}.    
\end{align}
Note that $p'_s(u) / d'(u) = p_s(u) / d(u)$, and
\[
p_s(v) + \alpha r_s(v) = (1-\alpha)\sum_{x \in N^{in}(v)}  \frac{ w_{(x,v)} p_s(x)}{d(x)}.
\]
Reorganize Equ. \eqref{eq:invar-v-body} and cancel out $p_s(v), r_s(v)$, we have
\begin{align}
\Delta &= \frac{(1-\alpha)}{\alpha} \frac{ \Delta_{w_{(u,v)}} p_s(u)}{d(u)} \nonumber \allowdisplaybreaks\\
r_s'(v)  &= r_s(v) + \frac{(1-\alpha)}{\alpha} \frac{\Delta_{w_{(u,v)}}p_s(u)}{d(u)} \label{eq:r-v-update-weighted-body-appendix}
\end{align}
Combining Equ.  \eqref{eq:p-update-weighted-appendix},\eqref{eq:r-update-weighted-appendix}, and \eqref{eq:r-v-update-weighted-body-appendix}, we prove the theorem.
\end{proof}
\begin{remark}
The above proof mainly follows from \cite{zhang2016approximate} where unweighted graph is considered while we consider weighted graph in our problem setting.
\end{remark}

\subsection{Proof of Time Complexity}
%. \ref{thm:time-complexity}
Before we proof the theorem, we present the known time complexity of \textsc{IncrementPush} as in the following lemma.
\begin{lemma}[Time complexity of \textsc{IncrementPush} \cite{xingzhi2021subset}]
Suppose the teleport parameter of obtaining PPR is $\alpha$ and the precision parameter is $\epsilon$. Given current weighted graph snapshot $\mathcal{G}_t$ and a set of edge events $\Delta E_t$ with $|\Delta E_t| = m$ , the time complexity of \textsc{IncrementPush} is $\mathcal{O}(m/\alpha^2 + \bar{d}^t / (\epsilon \alpha^2) +  1/(\epsilon \alpha))$ where $\bar{d}^t$ is the average node degree of current graph $\mathcal{G}_t$.
\label{lemma:run-time-increment-push}
\end{lemma}
\begin{proof}
The total time complexity of our instantiation \textsc{DynAnom} has components, which correspond to three steps in Algo. \ref{algo:dynanom-node}. The run time of step 1, \textsc{IncrementPush} is $\mathcal{O}(m/\alpha^2 + \bar{d}^t / (\epsilon \alpha^2) + 1/(\epsilon \alpha))$ by Lemma \ref{lemma:run-time-increment-push}. The run time of step 2, \textsc{DynNodeRep} is bounded by $\mathcal{O}(n)$ as the component of \textsc{DynNodeRep} is the procedure of the dimension reduction by using two hash functions. The calculation of two has functions given $\bm p_s$ is linear on $|\operatorname{supp}(\bm p_s)|$, which is $|\operatorname{supp}(\bm p_s)| \leq n$. Finally, the instantiation of our score function is also linear on $n$. Therefore, the whole time complexity is dominated by $\mathcal{O}(k m/\alpha^2 + k \bar{d^t} / (\epsilon \alpha^2) +  k /(\epsilon \alpha) + k T s)$ where $s$ is the maximal allowed sparsity defined in the theorem. We proof the theorem.
\end{proof}
 
\subsection{Timelines of real world graphs}
\label{sec:appendix-timelines}
We list the real-world events for the ENRON and PERSON graph in table \ref{tab:enron-events} and \ref{tab:person-events}.

\begin{table}[h!]
    \centering
    \caption{The events in Enron scandal timeline}
    \small
\begin{tabular}{p{0.05\linewidth} p{0.15\linewidth} p{0.65\linewidth} }
\toprule
 Index &       Date &  Event Description \\
\midrule
     %1 & 2000/08/23 &           Stock hits all-time high of \$90.56. Market valuation of \$70 billion. FERC (the Federal Energy Regulatory Commission) orders an investigation into strategies designed to drive electricity prices up in California. \\
    1 & 2000/08/23 &           Stock hits all-time high of \$90.56. the Federal Energy Regulatory Commission orders an investigation. \\

     2 & 2000/11/01 &                                                                                                                                                         FERC investigation exonerates Enron for any wrongdoing in California. \\
     3 & 2000/12/13 &                                                               Enron announces that president and chief operating officer Jeffrey Skilling will take over as chief executive in February. Kenneth Lay will remain as chairman. \\
     4 & 2001/01/25 &                                                                   Analyst Conference in Houston, Texas. Skilling bullish on the company. Analysts are all convinced. \\
     5 & 2001/05/17 &                                                                                                                                                     "Secret" meeting at Peninsula Hotel in LA -- Schwarzenegger, Lay, Milken. \\
     6 & 2001/08/22 &                                                                                                                   Ms Watkins meets with Lay and gives him a letter in which she says that Enron might be an "elaborate hoax." \\
     7 & 2001/09/26 &                                                                                                              Employee Meeting. Lay tells employees: Enron stock is an "incredible bargain." "Third quarter is looking great." \\
     8 & 2001/10/22 &                                                                   Enron acknowledges Securities and Exchange Commission inquiry into a possible conflict of interest related to the company's dealings with the partnerships. \\
     9 & 2001/11/08 & Enron files documents with SEC revising its financial statements to account for \$586 million in losses. The company starts negotiations to sell itself to head off bankrutcy. \\
    10 & 2002/01/30 &                                                                                                                                                                 Stephen Cooper takes over as Enron CEO. \\
\bottomrule
\end{tabular}
    \label{tab:enron-events}
\end{table}

\begin{table}[h!]
    \centering
    \small
    \caption{The person events of \textit{Schwarzenegger} and \textit{Franken}}
\begin{tabular}{p{0.1\linewidth} p{0.77\linewidth}}
\toprule
 Year & Major Event \\
\midrule 
 \multicolumn{2}{c}{Arnold Schwarzenegger} \\
\midrule 
 2003 & Arnold Schwarzenegger won the California gubernatorial election as his first politician role\\
 2008 & Arnold began campaigning with McCain as the key endorsement for McCain's presidential campaign\\
 2010 & mid-term election\\
 2011 & Arnold reached his term limit as Governor and returned to acting\\
\midrule 
 \multicolumn{2}{c}{Al Franken} \\
\midrule 
 2002 & Al Franken consider his first race for office due to the tragedy of Minnesota Sen. Paul Wellstone. \\
  2004 & The Al Franken Show aired \\
 2007 & Al Franken announced for candidacy for Senate. \\
 2008 & Al Franken won election. \\
 2012 & Al Franken won re-election. \\
 2018 & Al Franken resigned \\
 \bottomrule
\end{tabular}
    \label{tab:person-events}
\end{table}

\subsection{Hyper-parameter Settings}
we (re)implement the algorithms in Python to measure comparable running time.
\label{sec:appendix-param-config}
We list the hyper-parameter in Table \ref{tab:param-config}.
\begin{table}[h]
    \footnotesize
    \centering
\caption{The hyper-parameter configurations of algorithms in experiments. }
\begin{tabular}{p{0.20\linewidth}p{0.65\linewidth}}
\toprule 
 Algorithm & Hyper-parameter Settings \\
\midrule 
 AnomRank & alpha = 0.5 (default ) , \ epsilon = 1e-2 (default ) or 1e-6 (for better accuracy)  \\
\hline 
 SedanSpot & sample-size = 500 (default settings) \\
\hline 
 NetWalk & epoch\_per\_update=10, dim=128, default settings for the rest parameters \\
\hline 
 DynAnom and DynPPE & For exp1 and exp2: alpha = 0.15, minimum epsilon = 1e-12 , dim=1024; For exp2: we keep track of the top-100 high degree nodes for graph-level anomaly.  For case studies: alpha = 0.7, the rest are the same. \\
 \bottomrule
\end{tabular}
\label{tab:param-config}
\end{table}

\subsection{Experiment Details}
\label{sec:appendix-data}

\paragraph{Infrastructure}: We conduct our experiment on machine with 4-core Intel i5-6500 3.20GHz CPU, 32 GB memory, and GeForce GTX 1070 GPU (8 GB memory) on Ubuntu 18.04.6 LTS.

\paragraph{Datasets:}
For DARPA dataset: we totally track 200 anomalous nodes, and 151 of them have at least one anomalous edge after initial snapshot, the ground-truth of node-level anomaly is derived from the annotated edge as aforementioned. 
For EU-CORE dataset: Since EU-CORE dataset does not have annotated anomalous edges,  we randomly select 20 snapshots to inject artificial anomalous edges using two popular injection methods,  which are similar to the approach used in \cite{yoon2019fast}. For each selected snapshot in \textit{EU-CORE-S}, we uniformly select one node,$u_{high}$,  from the top 1\% high degree nodes, and injected 70 multi-edges connecting the selected node to other 10 random nodes which were not connected to $u_{high}$ before, which simulates the structural changes.  In each selected snapshot in \textit{EU-CORE-L}, we uniformly select 5 pairs of nodes, and injected edge connect each pair with totally 70 multi-edges as anomaly, which simulates the sudden peer-to-peer communication.  We include all datasets in our supplementary materials.
\let\cleardoublepage\clearpage
%\blindtext{4}

\chapter[Efficient Contextual Node Representation Learning
over Dynamic Graphs]{Efficient Contextual Node Representation Learning
over Dynamic Graphs \protect\footnote{Xingzhi Guo, Baojian Zhou, and Steven Skiena. Efficient Contextual Node Representation Learning
over Dynamic Graphs. In Submission.}}
\label{chapter:GoPPE}
\graphicspath{{chapter2-Z-GoPPE}}

\section{Introduction}
\label{sec:intro}

Many real-world graphs evolve with structural changes\cite{wen2022disentangled, hajiramezanali2019variational} and node attributes refinement over time. 
Considering Wikipedia as an evolving graph where Wiki pages (nodes) are interconnected by hyperlinks, each node is associated with the article's descriptions as attributes. 
For example, the node \textit{"ChatGPT"} initially
\footnote{The initial ChatGPT Wikipedia page: \url{https://en.wikipedia.org/w/index.php?title=ChatGPT&oldid=1125621134}} 
contained only had a few links redirecting to the nodes \textit{OpenAI} and \textit{Chatbot}, before being gradually enriched \footnote{The refined ChatGPT Wikipedia page shortly after the creation: \url{https://en.wikipedia.org/w/index.php?title=ChatGPT&oldid=1125717452}
}
with more hyperlinks and detailed descriptions.  Such examples motivate the following problem:
\begin{quote}
    How can we design an \textbf{efficient} and \textbf{robust} node representation learning model that can be updated quickly to reflect graph changes, even when node attributes are limited or noisy? 
\end{quote}

The proposed problem consists of two parts-- \textbf{1):} how to design a high-quality and robust node representation algorithm and \textbf{2):}  how to efficiently update the node representation along with the graph changes.  
%
%
%
% Although Graph Neural Networks (GNNs) \todo{add-ref} have been intensively studied, there is little research on  node representations learning over dynamic graphs. Specifically, the current research lacks of especially toward the direction of model efficiency and  robustness. 
%

One possible approach is through Graph Neural Networks (GNNs). Pioneered by SGC\cite{wu2019simplifying}, APPNP \cite{gasteiger2018predict}, PPRGo\cite{bojchevski2020scaling}, and AGP \cite{wang2021approximate},
they replace recursive local message passing with an approximate feature propagation matrix. For example, SGC formulates the node classification problem into $\bar{\bm y} = \operatorname{Softmax}(\bm S^{k} \bm X \bm \Theta)$ where $\bm S^k$ is the $k^{th}$ power of the normalized adjacency matrix,  $\bm X$ is the node attribute matrix and $\bm \Theta$ is the learnable classifier parameter.  This formulation decouples propagation from the learning process because $\bm S^k$ can be explicitly calculated independent of the neural model.

More importantly, this formulation provides a novel interpretation of GNN as a feature propagation process, opening doors to new propagation mechanism designs. 
In particular, Personalized PageRank (PPR) is a popular choice, and PPR-based models significantly improve GNN's efficiency and scalability. 
However, our interested problem is in dynamic settings. Although one can naively train multiple GNNs on every graph snapshot, the efficiency is compromised because they are designed for static graphs and have to re-calculate the propagation from scratch as the graph changes over time.

Recently, several works \cite{zheng2022instant,guo2021subset, guo2022subset, fu2021sdg} have adopted PPR in dynamic graph learning, and use incremental PPR maintenance algorithm \cite{zhang2016approximate} to update PPR as the graph evolves. 
Despite the fact that PPR is a very successful vertex proximity measure and provides great empirical performance, there is no clear formal framework to justify the choice of using PPR, and little theoretical guidance on the propagation mechanism design. 
In addition, the PPR-based GNN's efficiency largely depends on the PPR solver's speed. To our best knowledge,  all the previous works favored \textit{ForwardPush} as the underlying PPR method, which still has a large room  for improvement, especially under dynamic settings.

On the other hand, recent research \cite{fountoulakis2019variational,fountoulakis2022open} proposed a variational formulation of PPR as an explicit $\ell_1$-regularized quadratic optimization problem.  
Different from the traditional PPR interpretation as the random surfer model that can be solved algorithmically by power iteration\cite{page1999pagerank} and \textit{ForwardPush} \cite{andersen2006local}, the PPR-equivalent $\ell_1$-regularized optimization problem bridges two seemingly disjoint subjects (graph learning and optimization). 
One benefit is that this well-defined PPR-equivalent optimization problem can be solved efficiently by standard proximal gradient methods such ISTA and FISTA\cite{daubechies2004iterative, beck2009fast}, which have great potential for more efficient GNNs.
Most importantly, the $\ell_1$-regularization reveals PPR's inherent sparse property, which is similar to the well-known \textit{LASSO} regression \cite{tibshirani1996regression} and online learning \cite{zhou2019dual,zhang2018dual,zhang2016symmetrical,zhang2017fast, zhang2017fast2} for robustness and interpretability.
While \textit{LASSO} keeps the most important subset of feature dimensions, the $\ell_1$-regularizer in PPR  enforces to only keep a subset of active or important nodes, especially in large graphs. Therefore, the resulting models are robust by only aggregating the most important \textit{local} features.
% Therefore, our designed model has great robustness to noise.

 \vspace{-3mm}
\begin{figure}[htbp]
    \centering
    \includegraphics[width=1.0\linewidth]{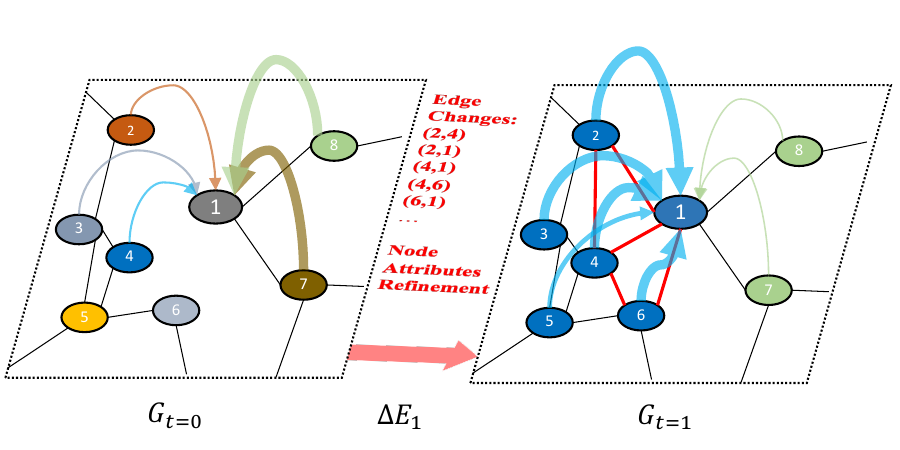}
    \vspace{-9mm}
    \caption{A cartoon illustrating contextualized node representation over a dynamic graph. 
    In the beginning, the graph has few edges  (black line) and relatively noisy node attributes (various colors). For example, the node $u_1$ only has two edges and receives features (arrows over nodes) mainly from the right-hand side. 
    As the graph evolves, the edges and node attributed become more complete, represented by red new edges and more consistent colored attributes.
    Accordingly, the node-wise attention for $u_1$ changes, favoring the nodes on its left side. We adopt $u_1$'s Personalized PageRank or $\bm \pi_{1}$ as the attention design which is sparse, informative, and can be justified by our proposed desired properties in our framework. 
    }
    \label{fig:goppe-overview}
\end{figure}

These findings inspire us to formulate PPR-based GNNs as a unified contextual node representation learning framework (shown in Figure \ref{fig:goppe-overview}) together with a new design of node positional encodings. %with a sparse node selection mechanism
Within this framework, we propose an elegant and fast solution to maintain PPR as the graph evolves and discuss several desired properties to justify the PPR choice with the theoretical ground. 
Finally, we instantiate a simple-yet-effective model (\textsc{GoPPE}) which has PPR-based node positional encodings, making the representation robust and distinguishable even when the node attributes are very noisy. 
The experiments demonstrate the potential of this framework for tackling the dynamic graph problem we are interested in.

% We propose a unified node embedding framework for attributed dynamic graphs, together with an instantiated simple-yet-effective model called $\operatorname{GoPPE}$. 
% The proposed framework consists of a Dynamic Personalized PageRank (PPR) based propagation step, followed by a learnable feature aggregation step. 
%between them the most obvious difference is whether to use node's raw feature.
%

We summarize our main contributions as follows: 
\begin{itemize}
    \item \textit{A novel contextual node representation framework for dynamic graphs} -- We propose a unified framework based on global node attention, which can explain the designs in the STOA propagation-based GNNs. 
    We identify the desired properties (e.g., attention sparsity, bounded $\ell_p$-norm, update efficiency) for the attention design, providing a guideline to justify the use of PPR and for designing the future attention-based GNN.  
    % Within this framework, we also proposed the desired properties of attention design, which justifies PPR as a well-design choice.
    %Specifically, we draw the connection between generic PPR-based network embeddings and neural network-based feature aggregations.
    \item \textit{Improved efficiency of PPR-based dynamic GNNs through an optimization lens} -- Inspired by the new PPR optimization formulation, we improve the efficiency of PPR-based GNNs using the proximal gradient methods, specifically \textit{Iterative Shrinkage-Thresholding Algorithm} or \textit{ISTA}, demonstrating roughly three times speedup in dynamic settings. 
    \item \textit{Instantiating \textsc{GoPPE}: a simple, effective and robust method} -- Within the proposed framework, we instantiate \textsc{GoPPE}, a simple-yet-effective method which is comparable in quality to and faster than the STOA baselines. Specifically, our novel PPR-based node positional encodings and the attention-based framework significantly outperform on noisy graphs, demonstrating its robustness against noise attributes during graph evolution.

\end{itemize}

The rest of this paper is organized as follows: Section \ref{sec:related-work}  clarifies different settings for dynamic graphs. 
Section \ref{sec:prelim}-\ref{sec:problem} discuss the preliminaries and problem formulation. 
We present our proposed node contextual learning framework in Section \ref{sec:method}, and discuss experimental results in Section \ref{sec:exp} .
Finally, we conclude and discuss future directions in Section \ref{sec:conclusion}.
\textit{We have made our code accessible at \textcolor{blue}{\url{http://bit.ly/3JQCUcX}} for review and will make it publicly available upon publication.   }

\section{Related Work}
\label{sec:related-work}

Learning over dynamic graphs is an emerging topic, and has been studied from three perspectives with different problem settings: 

\paragraph{Co-evolution Models over Dynamic Graphs}: The problem of dynamic co-evolution focuses on the relations between node attributes and the growing network topology. 
For example, DynRep \cite{trivedi2019dyrep} is the pioneering model.
A typical task is to predict whether two vertices will be connected or interact in the future given the current graph topology and the node attributes.  
The key is to learn a node-wise relation model to represent the interaction probability among vertices, rather than updating the node embedding w.r.t. the graph changes. 
Due to the nature of $\mathcal{O}(|\mathcal{V}|^2)$ complexity, these models are not scalable when the node size is large.

% \paragraph{Evovlving graph \textit{v.s.} Temporal model}: Typically as the graph evolves, $\mathcal{E}_t$ becomes more complete and $\bm X_t$ gets updated with richer information. 

\paragraph{Node Time Series Models over Dynamic Graphs}: 
Several temporal GNNs \cite{sankar2020dysat, rossi2020temporal,xu2020inductive} have been proposed for modeling the node embeddings w.r.t. the graph changes in \textit{discrete or continuous time}, then predict the anomaly using time series analysis \cite{zhang2017fast,zhang2017fast2,zhang2016symmetrical}. 
A typical example is \textsc{JODIE}\cite{kumar2019predicting}, which models the \textit{trajectory} of node embeddings across time and predicts its \textit{position} at a future timestamp. Similar to the co-evolution models, the essence is to obtain a compact embedding from the node history, and use it to extrapolate for future prediction. However, these models cannot update node embeddings quickly and usually involve data re-sampling and model fine-tuning while the graph changes.

\paragraph{Maintainable Node Embeddings Models over Dynamic Graphs}: Maintainable or incremental node embedding models focus on efficient embedding updates as the graph changes. 
Starting from the dynamic network embeddings which do not consider the node's raw attributes, previous works focus on online matrix factorization methods (e.g., incremental SVD \cite{brand2002incremental}), and incremental fine-tuning algorithms to adapt the static node embedding algorithms into dynamic (e.g., NetWalk\cite{yu2018netwalk}).  
However, they are not scalable because of the embedding dependency issue. For example, using the re-sample and fine-tune strategy,  one must update all node embeddings \textit{globally} to have one node embedding updated. 

Fortunately, \textit{localized} node embedding algorithms \cite{postuavaru2020instantembedding, guo2021subset, guo2022subset} adopt PPR which can be calculated for a subset of nodes individually with the complexity independent of node size. 
Meanwhile, one node's PPR is incrementally updatable and avoids calculating from scratch when the graph updates. 
Due to PPR's nature of local computation and maintainability, there have been follow-up works on the applications of efficient node anomaly tracking and efficient feature propagation GNNs over dynamic graphs. 
Specifically, DynPPE \cite{guo2021subset} employs incremental \textsc{ForwardPush} algorithm to maintain PPR and apply local hashing method to obtain high-quality PPR-based embeddings, showing promising results in both node classification and anomaly tracking tasks.   
Another recent development is \textsc{InstantGNN} \cite{zheng2022instant}, which applies the same incremental PPR solver, but uses PPR as the  feature aggregation weights for GNNs design. 
Like its counterparts for static graphs (\textsc{PPRGo} \cite{bojchevski2020scaling} and \textsc{AGP} \cite{wang2021approximate}), the model obtains high-quality node representation while achieving high efficiency with the increment PPR algorithm. 

\paragraph{Our approach:}
Following this research line, we are curious about how to justify PPR as a good design choice for dynamic GNNs. Despite the fact that it can be interpreted as a specific GNN design with infinite layers, what are the desired properties (e.g., sparsity, locality) for one propagation mechanism to be suitable and efficient?  We seek to propose such properties within a contextual graph learning framework, justify PPR as a good choice and provide guidelines for future GNN designs.

In addition, as PPR has been widely adopted with \textsc{ForwardPush} as its solver, we are curious about its efficiency under various dynamic settings (e.g., with large/small graph changes), and how to further improve GNN's efficiency. 
Finally, as the motivating scenario suggests, since the node attributes may be noisy during graph evolution, we aim to build a robust node representation against noisy attributes.  Specifically, we can re-use PPR (previously as the node attention) and convert it as node positional encodings with constant computation cost.

\begin{table}[ht]
\centering
\caption{Different STOA models can be realized within the proposed framework of various design choices. a : \textsc{PPRGo} only selects the top-$k$ PPR as attention weights.
b:  \textsc{InstantGNN} uses the generalized transitional matrix $\bm D^{\alpha} \bm A \bm D^{1-\alpha}$.
c: \textit{"S"} stands for Static PPR calculated from scratch in each snapshot. d: {\textit{"D"} means maintain PPR dynamically or incrementally.}
}
\vspace{-2mm}
\label{tab:design-compare}
\small
\begin{tabular}{
p{0.23\textwidth} |p{0.12\textwidth}p{0.12\textwidth} p{0.12\textwidth}p{0.08\textwidth}p{0.14\textwidth}}
\toprule
 Method & Dynamic &  Attribute & Attention & PE & Solver \\
\midrule
 \textsc{InstantEmb.} & \redxmark & \redxmark & \redxmark & PPR & Push-S$^\text{c}$ \\
 \textsc{PPRGo} & \redxmark & \bluecheck & PPR$^\text{a}$ & \redxmark & Push-S \\
 \textsc{DynamicPPE} & \bluecheck & \redxmark & \redxmark & PPR & Push-D$^\text{d}$ \\
 \textsc{InstantGNN} & \bluecheck & \bluecheck & PPR$^{\text{b}}$ & \redxmark & Push-D  \\
\midrule
 \textbf{\textsc{GoPPE-\{S,D\}}} & \bluecheck & \bluecheck & PPR & PPR & ISTA-\{S,D\} \\
\bottomrule
\end{tabular}
\end{table}

\section{Preliminaries}
\label{sec:prelim}

% \paragraph{Notations}: 
% Given a undirected unweighted graph $\mathcal{G} = (\mathcal{V}, \mathcal{E}, \bm X)$ where $\mathcal{V}=\{1,2,\ldots,n\}$ is the node set,  $\mathcal{E} \subseteq \mathcal{V} \times \mathcal{V}$ denotes the edge set and $\bm X \in \mathbb{R} ^ { d \times |\mathcal{V}| }$ is the nodes' $d$-dim attributes.  
% For each node $u$, we define $\operatorname{Nei}(u)$ to be the set of $v$'s neighbors; 
% $\bm d$ denotes node degree vector where each element $d(i) = |\operatorname{Nei}(i)|$. 
% The degree matrix is $\bm D := \operatorname{diag}(\bm d)$ and $\bm A$ is the adjacent matrix where $A_{i,j} \in \{0, 1\}$ and without self-loop. 
% To simplify our notation, we use $d_i$ to denote the degree of node $i$. Let $\bm P = \bm D^{-1} \bm A$ be the random walk matrix and then $\bm P^\top = \bm A^\top \bm D^{-1}$ is the column stochastic matrix. 

% \todo[inline]{Describe SGC, APPNP, PPRGo, message passing}. The goal is to mix the information from context in proximity.

\subsection{Message Passing as Approx. Propagation}
%One direct approach is to calculate  

In many successful and popular GNNs designs \cite{kipf2016semi, hamilton2017inductive}, the node attributes $\bm x_i $  serve as the message or graph signal, propagating to its local neighbors $\operatorname{Nei{}}(v_i)$ defined explicitly by the network structure $\mathcal{G}$. Meanwhile, $v_i$ locally \textit{convolutes} the incoming neighbors' messages as its hidden representation, denoted as $\bm h_i$. 
Typically, a set of trainable convolutional kernels $\bm W^{(l)}  \in \mathbb{R}^{h \times d} $ is used in a permutation-invariant function recursively ($l$ is the recursion level). For example, 
$\bm h_i ^{(l+1)} = \frac{1}{d_i} \sum_{j \in {Nei{}}(v_i)} \bm W^{(l)} \bm h^{(l)}_{j}$,
where $\bm h^{l=0}_i := \bm x_i$. The GNN models are usually trained in a supervised setting, learning local message aggregation and label prediction end-to-end. 

Interestingly, SGC \cite{wu2019simplifying} has simplified GNN design and decoupled the feature propagation from label prediction learning: 
\vspace{-1mm}
\begin{align}
    %\bm H ^{(L)}(i) &= \bm  \sum_{j \in {Nei{}}(v_i)}  W^{(L-1)} \bm h^{(L-1)}_{j} \nonumber\\
    \bm H ^{(L)} &= \bm  W^{(L-1)} \bm H^{(L-1)} \bm P^{\top}, \bm P = \bm D^{-1} \bm A\nonumber\\
    \bm H ^{(L-1)} &= \bm  W^{(L-2)} \bm H^{(L-2)} \bm P^{\top} \nonumber\\
    \ldots \nonumber \\
    % ...\red{BJ: use vdots}\nonumber\\
    \bm H^{(1)} &= \bm W^{(0)} \bm H^{(0)} \bm P^{\top} \nonumber\\
    %= \bm W^{(0)} \bm X \bm P^{\top} \\
    \bm H^{(0)} &:= \bm X \in \mathbb{R}^{d \times |\mathcal{V}|}, \text{ by definition.}\nonumber\\ 
    \bm H ^{(L)} &= \bm  W^{(L-1)} \bm  W^{(L-2)} \ldots \bm W^{(0)} \bm X \bm P^{\top} \ldots \bm P^{\top} \bm P^{\top} \nonumber\\
    % & = \Pi_{l=0}^{L-1} \bm W^{(l)} \bm X (\bm P^{L})^{\top}, \nonumber \\
    & = \hat{\bm  W} \bm X \hat{\bm P}^{\top}, \text{ note that  } \hat{\bm  W} := \prod_{l=0}^{L-1} \bm W^{(l)},  \hat{\bm P} := \bm P^{L}, \label{eq:sgc}
\end{align}
%\paragraph{GNNs and Random Surfer}: 
where Equ.\eqref{eq:sgc} reveals strong connections between the GNN message-passing mechanism and well-studied vertex proximity measures in network science. 
Specifically,  $\hat{\bm P} := (\bm D^{-1} \bm A)^{L}$ is the core component in power-method \cite{golub2013matrix} for problems like 
%Markov Chain Equilibrium \todo{add-ref} and 
PageRank \cite{page1999pagerank}. 
Let $\bm p_i$ is $i^{th}$ row of $\hat{\bm P}$, where $\| \bm p_i \|_{1} = 1$ and each element $\bm p_i[j]$ represents the  probability of a random surfer stops at $v_j$ from $v_i$ by taking $L$ times random walk, represnting a sense of proximity between $v_i$ and $v_j$.
%Given the proximity matrix $\hat{\bm P}$ and treat $\bm X \hat{\bm P}^{\top}$ as a whole, 
As $L \rightarrow \infty$,  Equ.\ref{eq:sgc} essentially represents each node feature $\hat{\bm x}_i$ by \textit{Global Feature Aggregation}: $\hat{\bm x}_i = \bm X \hat{\bm P}^{\top}[:, i] = \sum_{j=0}^{|\mathcal{V}|} \bm p_i [j] \bm x_j$ where the probability vector $\bm p_i$ serves as aggregation weights.

Alternatively, by taking into account 
$\hat{\bm  W} \bm X$ as a whole, 
Equ. \ref{eq:sgc} can be interpreted as \textit{global label smoothing} and each node is \textit{pre-classified} without accessing the graph structure: 
$\bm Y^{'} = \hat{\bm  W} \bm X$ where

$ \bm Y^{'} \in \R^{|\mc{C}| \times |\mc{V}|} $, 

$\mc{C}$ is the label set, then smooth the labels according to the vertex proximity: 
$\bm Y[:,i] = \bm Y^{'} \hat{\bm P}^{\top}[:,i] = \sum_{j = 0} ^{|\mathbf{V}|} \bm p_i[j] \bm Y[:,j]$. 
Both interpretations give a \textit{contextualized} node representation for $v_i$ over $\mathcal{V}$. 
However, as $L \rightarrow \infty$, 
$(\bm D^{-1} \bm A)^{L}$ will converge w.r.t. its first eigenvector (stationary property), making the output less distinctive and causing the over-smoothing issue \cite{Oono2020Graph}.

\paragraph{PPR as a node proximity design for GNNs}:  The celebrated algorithm Personalized Pagerank (PPR) provides a successful \textit{localized} design of node proximity.
%defined as $\pi_{(\alpha, s)} = (1-\alpha) \bm P^{\top} \pi_s  + \alpha \bm e_{s} $, where personalized vector $\bm e_{s}$ controls where a random surfer teleport with probability $\alpha \in [0, 1)$.  
One intriguing property of PPR is the \textit{locality} governed by the teleport component, making PPR vector with quantities concentrated \textit{around} the source node. 
Starting from APPNP \cite{gasteiger2018predict} , PPR became a popular GNN design choice and empirically achieves outstanding performance. However, there is little theoretical framework or justification for using PPR. 
In section \ref{sec:ppr-gnn-pe}, we will propose a novel GNN framework and discuss the desired properties of proximity design, then justify the choice of PPR.

% \red{BJ: Our motivation is here.} 

\subsection{Calculate PPR as \texorpdfstring{$\ell_1$}-Regularized Optimization}
The key ingredient in PPR-based GNNs is PPR calculation. Many algorithms have been designed to approximate PPR while having better run-time complexity.  
In this section, we discuss the recent research on efficient PPR approximation, including traditional algorithmic approaches and newly emerging optimization perspectives.

\xingzhiswallow{
    \begin{definition}[Personalized PageRank] Given an undirected unweighted graph $\mathcal{G}(\mathcal{V}, \mathcal{E})$ with $|\mathcal{V}| = n $ and $ |\mathcal{E}| = m$. 
    Define the random walk transition matrix $\bm P \triangleq \bm D^{-1} \bm A $, where $\bm A$ is the adjacency matrix of $\mathcal{G}$ and $\bm A = \bm A^{\top}$.
    % \footnote{Our PPV is defined based on the lazy random walk transition matrix. This definition is equivalent to the one using $\bm D^{-1}\bm A$ but with a different parameter $\alpha$.} 
    Given teleport probability $\alpha \in [0,1)$ and the source node $s$, the Personalized PageRank vector of $s$ is defined as:
    \begin{equation}
    \bm \pi_{\alpha, s} = (1-\alpha) \bm P^\top \bm \pi_{\alpha, s}  + \alpha \bm e_s, \label{equ:ppr}
    \end{equation}
    where the teleport probability $\alpha$ is a small constant (e.g. $\alpha=0.15$), $\bm e_s$ is an indicator column vector of node $s$, that is, $s$-th entry is 1, 0 otherwise. We use
    $\bm \pi_{s, \alpha}$ \footnote{ We use column vector by default and may elide $s, \alpha$ for simplicity if there is no ambiguity in the context.} 
    to denote the PPR of $s$ with teleport value $\alpha$.
    \label{def:static-ppr-static-graph}
    \end{definition}
}

\paragraph{Algorithmic approaches}: Power iteration and Forward push \cite{andersen2006local}  are designed to compute PPR algorithmically. The key idea of the push-based methods \cite{wu2021unifying,wang2022edge} is to exploit PPR invariant\cite{andersen2006local}  and gradually reduce the residual vector to approximate the high-precision PPR.

\paragraph{Approximate PPR via optimization}: Interestingly, recent research \cite{fountoulakis2019variational, fountoulakis2022open} discovered that Algorithm \ref{algo:forward-local-push} is equivalent to the Coordinate Descent algorithm, and can be solved as an explicit optimization problem (Lemma \ref{lemma:ppr-quad}).
%
% Furthermore, with the specific termination condition, a $\ell_1$-regularized quadratic objective function has been proposed as the alternative PPR formulation (Lemma \ref{lemma:ppr-quad}).
%
To this end, we can apply the Proximal Gradient Descent algorithm, specifically \textsc{ISTA} (Algorithm \ref{algo:ppr-ista}) to solve this well-defined PPR-equivalent $\ell_1$-regularized optimization problem.

\section{Problem Definition}
\label{sec:problem}

\begin{definition}[Node Classification Problem over Dynamic Graph]\label{def:prob-def}
Given an initial graph at timestamp $t=0$ $\mathcal{G}_0 = (\mathcal{V}, \mathcal{E}_0, \bm X_0)$
\footnote{For simplicity, we assume |$\mathcal{V}| = n$ does not change over time.},
 where $\bm X_0 \in \mathbb{R}^{n \times d}$ denotes the node attribute matrix, we want to predict the labels of a subset of nodes in $\mathcal{S} = \{u_0, u_1, \ldots, u_k\}$ as the graph evolves over time.
At time $t \in \{1, \ldots, T\}$, the graph $\mathcal{G}_{t-1}$ is updated with a batch of edge events $\Delta \mathcal{E}_{t}$ and a new node attribute $\bm X_t$. 
For each snapshot $\mathcal{G}_t$, we train a new classifier with updated node representation $\bm H_{t}^{\mathcal{S}} \in  \mathbb{R}^{|S| \times \bar d}$ to predict the node labels $\bm y_t$ and evaluate with the ground truth $\bm y^{*}_t$.   
\end{definition}

% \remind{add an example figure to explain it?}

\paragraph{Key Challenges}: Since 
\footnote{In practice, given a large graph we may use a small subset of nodes for training and evaluation.} 
$|\mathcal{S}| \ll |\mathcal{V}|$  the efficiency bottleneck for PPR-based models is at calculating $\bm H_{t}^{\mathcal{S}}$, rather than training prediction models given $\bm H_{t}^{\mathcal{S}}$ . 
 The key challenges under the framework of the PPR-based GNNs boil down to two questions -- \textbf{1):} \textit{How to efficiently calculate PPR as the graph evolves?} and \textbf{2):}  \textit{How to effectively fold in PPR as part of the node representation?} 
 In the following section, we answer these key questions using a justifiable PPR design and a faster ISTA-based PPR solver. 
% \todo{but not theoretically faster?..}

% update the node representation as the graph evolves.    
% In the case of PPR-based graph learning models, the problem boils down to the question \textit{"How to efficiently calculate PPR as the graph evolves"}.

% \remind{describe node/edge events, we do for incremental node embeddings (interpolation). While we leave temporal (extrapolation) for future work, which is usually done by \textsc{RNN} over incremental embeddings at each time. }

% \subsection{todo: write ISTA and double check the update rules:}

% \newpage

\section{Proposed Framework} % VertexFormer
\label{sec:method}
To efficiently and effectively learn node representations over time, we formulate PPR-based GNNs as a contextual node representation learning framework which has two key components: 
\textbf{1)}: Efficient PPR maintenance through an optimization lens; 
\textbf{2)}: Effective PPR-based node contextualization and positional encodings. 
Meanwhile, we discuss the desired properties of GNN design and justify the usage of PPR.
Then we instantiate a simple-yet-effective model called \textit{GoPPE}, analyze the overall complexity, and discuss its relations to other successful PPR-based GNN designs.

% to approximate message passing via PPR, then incrementally update PPR over time for
% decouple message passing and 

\subsection{Maintain PPR in the lens of Optimization}
Over an evolving graph, the new interactions between vertices are captured by adding edges, causing graph structure changes and making the previously calculated PPRs stale.  
To make the PPR-based GNNs reflect the latest graph structure, we propose to use ISTA as the PPR solver through an optimization lens and further explore its efficiency for dynamic graphs in various change patterns.

\paragraph{Naive Approach: Always from Scratch}:  One can simply re-calculate PPRs from scratch in each snapshot, however, this is sub-optimal because it does not leverage any previous PPR results, turning the dynamic graph into independent static graph snapshots. However, if the PPR solver is fast enough, it may still achieve reasonable efficiency.
 
% \paragraph{PPR Invariants Property}

% The core component is PPR calculation and maintaining the internal states w.r.t. edge event for example: inserting a new edge $(u, v, 1)$. 

\paragraph{Improved Approach: Warm starting with local update rules}: 
By exploiting the PPR invariant, PPR adjustment rules \cite{zhang2016approximate} have been proposed to maintain PPR w.r.t. a sequence of edge changes (add/delete edges), which can be proved equivalently to adjust the solution and gradient in the PPR's optimization formulation.
By adjusting the previous PPR and using it as the warm-starting point, we can maintain PPR incrementally. 
It provides an elegant solution to dynamically maintain PPR while enjoying a favorable convergence rate inherent in ISTA.

% \textcolor{red}{convergence/efficiency analysis? how about blockwise coordinate descent instead of full gradient? }

\begin{theorem}[PPR Adjustment Rules for Dynamic Graphs] Given a new edge inserted denoted as $(u, v, 1)$, 
The internal state of the PPR solver can be adjusted by the following rules to reflect the latest PPR but in compromised quality.
Instead of starting from $\bm x = \bm 0$, we use the adjusted solution as a warm-start for ISTA-solver (Algorithm. \ref{algo:ppr-ista}). 
\begin{align}
    \bm x'(u) &= \bm x(u) * \frac{\bm d(u)+1}{\bm d(u)} \label{equ:update-pi}\\
    \nabla f'(\bm x')(u) &= \nabla f(\bm x)(u) - \frac{\bm x(u)}{ \alpha \bm d^{1/2}(u)} \nonumber \\
    \nabla f'(\bm x')(v) &= \nabla f(\bm x)(v)  + \frac{(1-\alpha) }{\alpha} \frac{ \bm d^{-1/2}(v)*\bm x(u)}{ \bm d^{1/2}(u)} \nonumber , \\
\text{where }\bm x'(u), & \nabla f'(\bm x')\text{ denotes the value after adjustment.}  \nonumber
\end{align}
We attach the proof in Section \ref{proof:ppv-maintain} in Appendix. 
\label{theorem:update}
\end{theorem}

\begin{remark}
As defined in Appendix \ref{sec:proof-push-opt},  $\nabla f'(x')$ can be evaluated analytically, so the above rules can be reduced to a single update using Equ.\ref{equ:update-pi}. Each update is a local operation having complexity $\mathcal{O}(1)$ per edge event for each node in $\mathcal{S}$. 
% Intuitively, if $u$ has one more out-degree 
\end{remark}

%With dynamic maintenance, one can solve the problem incrementally over dynamic graphs. 
%\textcolor{red}{For large changes, the adjustment method may not work well, but works for small changes.}

\xingzhiswallow{
    \begin{algorithm}[ht]
    \caption{$\textsc{ISTA PPR Solver for Dynamic Graphs}$ }
    \begin{algorithmic}[1]
    \State \textbf{Input: }$ k = 0, \bm x^{(0)} = \bm 0, \mathcal{G}, \epsilon, \alpha$
    % \State $\bm r^{(0)} = \left(\bm I_n - (1-\alpha) \bm P^{\top} \right) \bm D^{-1/2} \bm x ^{(0)}  - \alpha \bm e_s  = - \alpha \bm e_s$
    \State $\nabla f(\bm x^{(0)}) = - \alpha \bm D^{-1/2} \bm e_s$
    \State $\eta = \frac{1}{2-\alpha}$
    
    % \While{Termination Condition }
    
    % \While{ exists $i$ such that $\|\nabla f(\bm x)(i)\|_1 > \epsilon' d(i)^{1/2}$}
    \While{Termination Condition }
    \State $ \bm z^{(k)}(i) := \bm x^{(k)}(i) - \eta  \nabla_i f(\bm x^{(k)}) $
    \State $\bm x^{(k+1)(i)} = \textsc{prox}\left( \bm z^{(k)}(i)\right)$, which is solved analytically:
    %_{\epsilon'\bm d(i)^{1/2}}
    \[
        %\frac{1}{\eta} 
        \bm x^{(k+1)}(i)= 
    \begin{dcases}
        \bm z^{k}(i) -  \epsilon' \bm d(i)^{1/2} 
        , & \text{ if }  \bm z^{k}(i) >  \epsilon' \bm d(i)^{1/2} \\
        0 , & \text{ if }   \| \bm z^{k}(i) \|_1 \leq  \epsilon' \bm d(i)^{1/2} \\
        \bm z^{k}(i) +  \epsilon' \bm d(i)^{1/2} , 
        & \text{ if }  \bm z^{k}(i) < - \epsilon' \bm d(i)^{1/2} \\
        % -\epsilon' \bm d(i)^{1/2},  & \text{if } \bm x^{*}(i) > 0 \\
        % \epsilon' \bm d(i)^{1/2}, & \text{if } \bm x^{*}(i) < 0 \\
        % \in [-\epsilon' \bm d(i)^{1/2}, \epsilon' \bm d(i)^{1/2}] & \text{if } \bm x^{*}(i) = 0 \\
    \end{dcases}
    \]
    \EndWhile
    
    \State \Return $\left(\bm \pi = \bm D^{1/2} \bm x^{(k)}, \bm r =  \bm D^{1/2}\nabla f(\bm x^{(k)}) \right)$
    \end{algorithmic}
    \label{algo:dyn-ista}
    \end{algorithm}

}

\subsection{PPR-based Contextual Node Representation and Positional Encodings}
\label{sec:ppr-gnn-pe}
Local message passing is the key ingredient in many GNN designs. Inspired by SGC, we provide another interpretation of GNN as a global node contextualization process. 
It aggregates node features in a global manner, in contrast to the viewpoint of recursive local neighbor message passing.   
We start the analysis from the simplified local message passing design. Recall the serialized GNNs in Equ. \ref{eq:sgc}:
\begin{align}
    \bm H ^{(L)} = \hat{\bm  W} \bm X \hat{\bm P}^{\top},\ 
    \hat{\bm  W} := \Pi_{l=0}^{L-1} \bm W^{(l)}, \ 
    \hat{\bm P} := (\bm D^{-1} \bm A)^{L}, \nonumber
\end{align}
where $L$ is the hyperparameter controlling how many hops should the GNN recursively aggregates the local neighbors' features. 
We can further organize Equ.\ref{eq:sgc} into $\bm H ^{(L)} = \hat{\bm  W}  \hat{\bm H}$, where $\hat{\bm H} = \bm X \hat{\bm P}^{\top}$. 
Let $\hat{\bm h_i}, \hat{\bm p_i}$ denote the $i^{th}$ column of $\hat{\bm H}$ and $\hat{\bm P}^{\top}$. $\hat{\bm p_i}(j)$ refers to its $j^{th}$ element. 
We re-organize Equ.\ref{eq:sgc} into vector form:
\begin{align}
    \hat{\bm h_i} = \bm X \hat{\bm p_i} &= \sum_{j \in \mathcal{V} } \hat{\bm p_i}(j) \bm x_j  \nonumber\\
        &= \sum_{j \in \mathcal{V} } \frac{ \hat{\bm p_i}(j)}{ z_i} \bm x_j, \text{ where} \label{equ:global-agg}\\
        z_i = \sum_{k \in \mathcal{V}} \hat{\bm p_i}(k) &= \| \hat{\bm p_i}\|_1 = 1 \text{ and } \bm p_i(k) \geq 0 \label{equ:global-normalizer}
\end{align}

Regardless of the exact form of $\hat{\bm p_i}$, Equ.\ref{equ:global-agg} shows that $\hat{\bm h_i}$ can be decomposed into a weighted-sum formula with a proper normalizer $z_i$, perhaps thought of as \textit{"global attention over other nodes"} in deep learning terms. In this specific case, the attention weight $\hat{\bm p_i}$ is in its simplest form without any learnable parameter but it guarantees\footnote{$(\bm D^{-1} \bm A)^{k}$ guarantees to have row-wise $\ell_1$-norm equals to 1, and same for the column of its transpose. } that $\| \hat{\bm p_i}\|_1 = 1$ for any $L>0$ so that we do not need explicitly evaluate $z_i$ in a potentially expensive $|\mathcal{V}|$-length loop when $|\mathcal{V}|$ is large.
This observation provides an insightful viewpoint to interpret and design GNNs w.r.t. the node-wise attention weight $\hat{\bm p_i}(j)$. 
In the following, we summarize the desired properties of $\hat{\bm p_i}$ and justify our choice of using PPR (letting $\hat{\bm p_i} = \bm \pi_i$), then enhance the feature expressivity with PPR-based positional encodings.

\begin{definition}   
(The desired properties of node-wise attention design\label{def:att-property}) Given the formulation in Equ.\ref{equ:global-agg}, a well-designed node-wise attention vector $\hat{\bm p_i}$ should have the following properties:

\begin{itemize}
    \item Locality: Under the assumption of homophily graphs, $\hat{\bm p_i}(j)$ should capture the \textit{locality/proximity}  between node $u_i$ and $u_j$, making the feature aggregation informative. For example, the self-weight $\hat{\bm p_i}(i)$ should be emphasized.
    \item Efficiency: $\hat{\bm p_i}$ should be computed and maintained quickly, in particular, independently of node size $|\mathcal{V}|$.
    \item Sparsity: $\hat{\bm p_i}$  should be sparse $|\supp(\hat{\bm p_i})| \ll |\mathcal{V}|$ so that feature aggregation is memory bounded by $|\supp(\hat{\bm p_i})|$.
    %For example, given a large graph, iterating all node pairs of  $\hat{\bm p_i}(j)$
    \item Well-bounded normalizer $\| \hat{\bm p_i} \|_p$: The $\ell_p$-norm of $\hat{\bm p_i}$ should be bounded to avoid exhaustively calculating the normalizer $z_i$ over $\mathcal{V}$ when  $|\mathcal{V}|$ is large.
\end{itemize}

\end{definition}

\paragraph{A bad naive design}: One widely-adopted attention design is  $\hat{\bm p_i}(j):= \operatorname{CosSim}(\bm W \bm x_i, \bm W \bm x_j)$ where $\bm W$ is the learnt feature transformation. However, the resulting $\hat{\bm p_i}$ is neither sparse nor bounded and involves the node-wise calculation of complexity $\mathcal{O}(|\mathcal{V}|)$ per node, not to mention the ignorance of the graph structure.

\paragraph{PPR-based Attention Design and Justifications}:
To fulfill the desired properties, we use PPR vector as the attention design ($\hat{\bm p_i}:=\bm \pi_i$) with the following justifications:
1). PPR vector is a well-designed proximity measure over the graph with locality controlled by the teleport parameter $\alpha$, specifically, emphasizing the significance of node $u_i$ with the personalized vector $\bm e_i$.
2). PPR vector can be maintained efficiently with complexity $\mathcal{O}(\frac{1}{\epsilon \alpha})$ and independent of $|\mathcal{V}|$.
3). PPR vector is inherently sparse, explicitly revealed by the $\ell_1$-regularization in Equ.\ref{eq:ppr-quad-form} through optimization lens.
4). The $\ell_1$-norm of the PPR vector is  guaranteed to be 1, making the normalizer $z_i$ a constant and free of computation.  
Finally, we have a very simple-yet-effective global contextual node formulation :  
\begin{align}
    \hat{\bm h_i} = \bm X \hat{\bm \pi_i}
\end{align}

Although there are other designs for node-wise attention, %even employing PPR to aggregate features was not first proposed by this paper. 
To the best knowledge, we are the first to propose this framework with the desired properties and justify the choice of using PPR. 
This novel framework provides insights for future GNN design and motivates us to propose the following PPR-based positional encodings.

% \textcolor{red}{locality} is important. 
\paragraph{Robust Node Positional Encodings for Enhanced Locality Feature}:
The concept of \textit{"locality"} is crucial in many GNN designs as well as PPR formulation. 
In the original GCN design, the neighbor features are aggregated by $\bm D^{-1} (\bm I + \bm A)$ with the added self-loop to enhance its own features.  
On the other hand, PPR has the restart probability $\alpha \bm e_s$ to govern the locality.
%as defined in Equ.\ref{equ:ppr}.
%
%, where $\alpha$ controls how concentrated the quantity, is spread out over the graph. 
%For example, $\alpha=1.0$ means $\pi_s = \bm e_s$ or all quantities are on the node $u_s$ itself.
%
Given the importance of the node locality information and the global contextualization framework, we seek to design a node Positional Encodings (PE) $\bm p_i \in \mathbb{R}^{d_{pe}}$ for node $u_i$. It is analogous to the token PE \cite{vaswani2017attention}  in NLP community to distinguish the tokens in different positions whereas node PE tells vertex's \textit{position} or \textit{address} in the graph.

We present the two simplest formulations of PE \footnote{We elide $\hat{\bm h_i}$ for simplicity; $\oplus$ is the element-wise addition operator. $||$ is the vector concatenation operator.} in Equ.\ref{equ:goppe} and then discuss its desired properties. Note that the node positional encodings are fused into the node $u_i$ to enhance its own locality feature, providing an attribute-agnostic representation.

\begin{align}
    \bm h_i =   
    \begin{dcases}
        \bm X \bm \pi_i \oplus \bm W_{pe} \bm p_i , \text{Additive PE} \\
        \bm X \bm \pi_i \  || \ \bm W_{pe} \bm p_i , \text{Concatenative PE}\\
    \end{dcases} \label{equ:goppe} 
\end{align}
where $\bm W_{pe} \in \mathbb{R}^{d \times d_{pe}}$ is the learnable mapping parameter to align dimensions with attributes for fusion. 

\begin{definition}[The Desired Properties of Node Positional Encodings \label{def:pos-encode}] Given the graph $\mathcal{G}$,  the positional encodings $\bm p_i$ for node $u_i$ is well-designed if it has the following properties:
\begin{itemize}
    \item Effectiveness: $\bm p_i$ should represent the node's \textit{"position"} or \textit{"locality"} over the graph and be distinguishable among vertices even if the node attributes are indistinguishable. 
    For example, $\bm p_i \neq \bm p_j \text{ while } \bm x_i = \bm x_j, \forall i,j \in \mathcal{V}$.
    \item Efficiency:  $\bm p_i$ should be computationally cheap $\ll \mathcal{O}(|\mathcal{V}|)$ and low-dimensional with $d_{pe} \ll |\mathcal{V}|$ if $|\mathcal{V}|$ is large.
\end{itemize}
\end{definition}

%%%%% long algorithm 
% \input{chapter2-Z-GoPPE/goppe-algo.tex}

\begin{algorithm}[bt]
\caption{$\textsc{GoPPE}(\mathcal{G}_0, \Delta E_{1, \ldots, T}, \bm X_{1, \ldots, T}, \mathcal{S}, \epsilon, \alpha)$ }
    \scriptsize
    \begin{multicols}{2}
    \begin{algorithmic}[1]
    \State \textbf{Input: }  
    Initial graph $\mathcal{G}_0 = (\mathcal{V}, \mathcal{E}_0, \bm X_0)$, 
    Edge events $ \Delta E_{1, \ldots, T}$, 
    Node attributes $ \bm X_{1, \ldots, T}$ , 
    The train/test node sets, denoted as $\mathcal{\bar{S}} \text{ and }\mathcal{\hat{S}}$, respectively. And $\mathcal{S} = \mathcal{\bar{S}} \cup \mathcal{\hat{S}}$. 
    PPR teleport factor $\alpha $,
    PPR Vector $\ell_1$-error control parameter $\epsilon$.
    
    \State $\bm \pi_{u, init} = \bm 0, \forall u \in \mathcal{S}$
    \State \textcolor{black}{// Calculate Initial PPR for each node $u$ in $\mathcal{S}$ using Algorithm \ref{algo:ppr-ista}.}
    \State $\bm x_{u, t=0}, \nabla_{u, t=0} = \textsc{PPRIsta}(\mathcal{G}_0, \bm \pi_{u, init}),  \forall u \in \mathcal{S} $ 
    \State $\bm \pi_{u, t=0} =  \sqrt{\bm d(u)} \times \bm x_{u, t=0}  ,  \forall u \in \mathcal{S} $
    \State \textcolor{black}{// Get Contextualized Node Features and Positional Encodings}
    \State $\bm h_{u,t=0} = \textsc{FeatureFormer}(\bm \pi_{u, t=0},\bm X_0),  \forall u \in \mathcal{S} $ 
    \State \textcolor{black}{// Train classifier with the training samples $\bar{u} \in \mathcal{\bar{S}} \subset \mathcal{S} $}
    \State $f_{\bm \Theta,t=0}(\cdot) = \textsc{Train}(\bm h_{\bar{u}, t=0}, \bm y(\bar{u})),  \forall \bar{u} \in \mathcal{\bar{S}}$ 
    \State \textcolor{black}{// Infer for the testing samples $\bm h_{\hat{u},t=0}, \forall \hat{u} \in \mathcal{\hat{S}}$}
    \State $ \bar{y}_{\hat{u}, t=0} = f_{\bm \Theta,t=0}(\bm h_{\hat{u}}), \forall \hat{u} \in \mathcal{\hat{S}} $ 
    \State \textcolor{black}{// Update graphs, maintain PPR, then train/evaluate repeatedly.}
    \For{$t \in [1,T] $}
        \State $ \bar{\bm x}_{u, t-1} =\textsc{PPRAdjust}(\bm x_{u, t-1}, \Delta E_{t})$
        \State $\mathcal{G}_{t} = \textsc{GraphUpdate}(\mathcal{G}_{t-1}, \Delta E_t, \bm X_t)$
        \State $\bm x_{u, t}, \nabla_{u, t} = \textsc{PPRIsta}(\mathcal{G}_1, \bar{\bm x}_{u, t-1}),  \forall u \in \mathcal{S} $ 
        \State $\bm \pi_{u, t} =  \sqrt{\bm d(u)} \times \bm x_{u, t}  ,  \forall u \in \mathcal{S} $
        \State $\bm h_{u,t} = \textsc{FeatureFormer}(\bm \pi_{u, t},\bm X_t),  \forall u \in \mathcal{S} $
        \State $f_{\bm \Theta,t}(\cdot) = \textsc{Train}(\bm h_{\bar{u}, t}, \bm y(\bar{u})),  \forall \bar{u} \in \mathcal{\bar{S}}$ 
        \State $ \bar{y}_{\hat{u},t} = f_{\bm \Theta,t}(\bm h_{\hat{u}}), \forall \hat{u} \in \mathcal{\hat{S}} $ 
    \EndFor
    
    \State \textbf{return} $\bar{y}_{\hat{u},t \in \{ 0, 1, \ldots, T \}}, \forall \hat{u} \in \mathcal{\hat{S}}$
    
    \noindent\hrulefill
    \Procedure{FeatureFormer}
    {$\bm \pi_u,\bm X$}
    \State \textcolor{black}{// PPR-guided Node Contextualization}
    \State $\bm h^c_u =  \bm X \bm \pi_u $
    % \State \textcolor{red}{// There are many other aggregation designs }
    \State \textcolor{black}{// PPR-based Node Positional Encodings}
    \State $\bm p_i = \textsc{HashReduceDim}(\bm \pi_u, \bar{d}=512)$
    \State $\bm h^p_u = \bm W_{pe} \bm p_i $
    \State \textbf{return}  Apply \textit{Concatenative PE }in Equ.\ref{equ:goppe} 
    \EndProcedure
    
    \noindent\hrulefill
    \Procedure{PPRAdjust}{$ \bm x, \Delta E$}
    % \State $t=0$
    \For{$(u ,v)\in \Delta E$}
        \State $\bar{\bm x}_u:=$ update $\bm x_u$ Using Theorem \ref{theorem:update}.
        % \State \textcolor{blue}{Note that only $\bm x$}
    \EndFor
    
    \State \textbf{return} $ \bar{\bm x}$ 
    \EndProcedure
    
    \noindent\hrulefill
    \Procedure{HashReduceDim}{$x,d$ } \label{proc:HashReduce}
    \State // Hash function $h_{d}(i): \mathbb{N}\to [d]$
    \State // Hash function $h_{\operatorname{sgn}}(i): \mathbb{N}\to \{ \pm 1\}$
    \State $\bar x = \bm 0 \in \mathbb{R}^{d}$
    \For{$i \in \textsc{Supp}|x|$}
        \State $j = h_{dim}(i)$
        \State $\bar x(j) \pluseq h_{\operatorname{sgn}}(i) \log \left(x(i) \right)$
    \EndFor
    \State \textbf{return} $ \frac{\bar x}{\| \bar x \|_1}$
    \EndProcedure
    
    \end{algorithmic}
    \end{multicols}
    
    \label{algo:goppe}
\end{algorithm}

%%%%%

\paragraph{PPR-based node positional encodings and justifications}: 
Interestingly, as the by-product of the PPR-based node contextualization, we re-use the unique and sparse PPR vector $\bm \pi_i \in \mathbb{R}^{|\mathcal{V}|}, \forall i \in \mathcal{S}$ as the source to derive the node positional encoding $\bm p_i \in \mathbb{R}^{d_{pe}}$ using  simple dimension reduction operations. 
Specifically, we can use sparse random projection (Equ.\ref{equ:rand-proj} in Appendix) 
% \todo{add-ref, JL lemma? } 
or hashing-based dimension reduction (Line \ref{proc:HashReduce} in Algorithm \ref{algo:goppe}).
The reduction operation has time complexity $\mathcal{O}(|\supp(\bm \pi_i)|) \ll\mathcal{O}( |\mathcal{V}|)$ per node, and maximizes the usage of PPR vector. 
Meanwhile, the resulting $\bm p_i$ encodes a strong vertex positional signal per se. In an extreme case, where the node attributes are weak (noisy), $\bm p_i$ alone can provide good-quality embeddings from pure network structure, consistently providing a robust node representation.
%the importance of locality information it embedded.

% \remind{2. Fold network embedding with node feature}
% \paragraph{Baselines}
% Idea: Use full PPV as weights to agg features (InstantGNN -> but it does not work caused by over-smoothing, InstantGNN is not released as of Jan 4, 2023).

% GoPPE: Global PageRank Contextualization and Positional Encodings.
\paragraph{GoPPE as an instantiated model}: Putting all together, we instantiate a very simple-yet-effective GNN called \textit{GoPPE} for dynamic graphs. Although there are more sophisticated neural network designs under the proposed framework, we aim to demonstrate its usefulness with the simplest choices. We present the detailed model components in Algorithm \ref{algo:goppe}.

\paragraph{Relations to other PPR-based GNNs}:
Interestingly, many successful PPR-based GNNs 
\footnote{we assume that all use the standard PPR formulation with transitional matrix $\bm D^{-1} \bm A$} 
fit the proposed framework with various design choices. 
For example, \textsc{DynamicPPE} and \textsc{InstantEmb} can be considered as the feature-less \textsc{GoPPE} with only positional encodings, which is robust when the node attribute is noisy.  Meanwhile, \textsc{InstantGNN} is equivalent to \textsc{GoPPE} without positional encodings and chooses \textsc{ForwardPush} as the PPR solver. 
We summarize their design choices in Table \ref{tab:design-compare} within our proposed framework.

% \todo{Table \ref{tab:design-compare} belongs in the introduction.}

\paragraph{Complexity Analysis}:
Given total $T$ graph snapshots with $m$ edge events and  $|\mathcal{S}| = K$, the complexity of \textsc{GoPPE} ( Algorithm \ref{algo:goppe}) involves four parts:

\begin{itemize}
    \item 1: $\mathcal{O}(\frac{KT}{\epsilon \alpha})$ for the \textsc{ISTA} solver for PPR calculation 
    . %\todo{is it true for ista?} Yes
    \item 2: $ \mathcal{O}(KT \operatorname{max}(|\supp(\pi_i)|), \forall i \in \mathcal{S}) $ for the positional encoding calculation hashing method. 
    \item 3: $ \mathcal{O}(KTm) $ for the PPR adjustment of $m$ edge events.
    \item 4: $ \mathcal{O}(KT) $ for neural network training/inference. Since we use the simple stacked MLPs, we consider roughly $ \mathcal{O}(1) $ per sample.
\end{itemize}
To sum up, the total complexity is $ \mathcal{O}(K T m + \frac{K T}{\epsilon \alpha}) $. Note that the complexity is independent of $|\mathcal{V}|$, indicating the algorithm only involves efficient local computation and is scalable to large graphs.

% \todo{should check}
%  the neural network training cost which all methods share so that we elide it for simplicity.
% Given the total number of graph snapshots as $T$ with $m$ edge events, the overall complexity for \textsc{GoPPE} is $\mathcal{O}( \frac{k}{\alpha \epsilon} + \red{km +  } )$
% adjustment: which has $\mathcal{O}(k) $ for $k$ nodes per edge event;

% total number of nodes $k = |S|$ in train/test sets, 
% the total number of updated edge events $\Delta m$, 
% the sparsity of obtained PPR vector $\supp(\bm \pi_i)$ with precision control $\epsilon$.
% The overall complexity is 

%is $\mathcal{O}()$.  

\xingzhiswallow{
    
    The overall complexity of tracking subset of $k$ nodes across $T$ snapshots depends on run time of three main steps as shown in Algo. \ref{algo:dynanom-node}: 1. the run time of \textsc{IncrementPush} for nodes $\mathcal{S}$; 2. the calculation of dynamic node representations, which be finished in $\mathcal{O}(k T |\supp(\bm p_{s^\prime})|)$ where $|\supp(\bm p_{s^\prime})|$ is the maximal support of all $k$ sparse vectors, i.e. $|\supp(\bm p_{s^\prime})| = \max_{s\in \mathcal{S}} |\supp(\bm p_{s})|$. The overall time complexity of our proposed framework is stated as in the following theorem.
    [Time Complexity of \textsc{DynAnom}]
    Given the set of edge events where $\mathbb{E} = \{ e_1, e_2,\ldots, e_m\}$ and $T$ snapshots and subset of target nodes $\mathcal{S}$ with $|\mathcal{S}| = k$, the instantiation of our proposed \textsc{DynAnom} detects whether there exist events in these $T$ snapshots in $\mathcal{O}( k m/\alpha^2 + k \bar{d^t} / (\epsilon \alpha^2) +  k/(\epsilon \alpha) + k T s)$ where $\bar{d^t}$ is the average node degree of current graph $\mathcal{G}_t$ and $s\triangleq \max_{s\in\mathcal{S}} |\supp(\bm p_s)|$.
}

% We ignore the neural network complexity since it's simple stacked MLPs.
% and Neural network models. 
% The overall complexity is \textcolor{red}{add}
% \begin{theorem}
%     \textcolor{red}{add complexity analysis}
% \end{theorem}

% \newpage
\section{Experiments}
\label{sec:exp}
In this section, we describe the experiment configurations and present and discuss the results of running time and prediction accuracy.  
% datasets and baseline methods, then discuss the experiment results in terms of PPR solver and overall GNN performance. 
Finally, we demonstrate the effectiveness of the introduced positional encodings when the node attributes are noisy. 

\subsection{Experiment Configurations}
\label{sec:dataset}
\paragraph{Datasets:}
To evaluate GNNs in a dynamic setting, we divide the edges of static undirected graphs into snapshots \footnote{
Although such conversion breaks the temporal correlation, we focus on a more fundamental problem of node embedding updates instead of temporal relations, which is similar to \cite{zheng2022instant}.
}.   
For node classification tasks, the first graph snapshot $\mathcal{G}_0$ contains the $50\%$ of the total edges, and the other half is evenly divided into edge events in each snapshot $\Delta \mathcal{E}_{i}, i \in \{1, \ldots, T-1\}$ for the following $\mathcal{G}_{ \{1, \ldots, T-1\} }$. 
% \todo{This 50\% seems too large to me.}
We fix the node labels in each snapshot, while the node attributes may change according to the experiment settings. In the efficiency test, we report the running time in two dynamic setting detailed in Section \ref{sec:ppr-speed}.
The statistics of datasets are presented in Table \ref{tab:dataset}.

\begin{table}[htbp]
    \centering
        \caption{Datasets used in our experiments. $|\mathcal{V}|$: number of nodes. $|\mathcal{E}|$ number of edges ,  $T$: number of snapshots for node classification, $|\mathcal{S}|$: total nodes in train/test sets, $d$: dimension of node raw feature, $|\mathcal{L}|$: number of node labels. }
    \label{tab:dataset}
    % \vspace{-3mm}

    \footnotesize
    % \begin{tabular}{llrrrrrr}
    \begin{tabular}{
    p{0.1\textwidth} 
    |p{0.03\textwidth}
    R{0.15\textwidth} 
    R{0.10\textwidth}
    R{0.05\textwidth}
    R{0.08\textwidth}
    R{0.08\textwidth}
    R{0.035\textwidth}}
    \toprule
    Dataset &   Size &  $|\mathcal{V}|$ &  $|\mathcal{E}|$ &  $T$ & $|\mathcal{S}|$  &  $d$ &  $|\mathcal{L}|$ \\
    \midrule
        cora &  small &           2,708 &                    5,277 &              5 &   1,000 &         1,433 &                    7 \\
    citeseer &  small &           3,279 &                    4,551 &              5 &   1,000 &         3,703 &                    6 \\
      pubmed &  small &          19,717 &                   44,323 &              5 &   1,000 &           500 &                    3 \\ 
      flickr & medium &          89,250 &                  449,877 &              5 &   1,000 &           500 &                    7 \\
  arvix & medium &         169,343 &                1,157,798 &              5 &   1,000 &           128 &                   40 \\ 
  % product & large &         -- &                -- &              -- &   -- &           -- &                   -- \\ 
    \bottomrule
    \end{tabular}
\end{table}

\paragraph{Data splits}: To evaluate the performance over snapshots, all train/dev/test nodes (70\%/10\%/20\%) are sampled from the first snapshots without any dangling node. 
We apply the stratified sampling and add at least one sample of each class into the train/dev/test set. The total sampled nodes are presented $|\mathcal{S}|$ in Table. \ref{tab:dataset}.

% \paragraph{Homophilic graph}: OGB-Arvix/CORA/Citeseer/Flickr/

% \subsection{Baselines, Metrics and Training Details}
\paragraph{Baselines:} Besides the Random \footnote{We use random Gaussian as feature} and MLPs as the naive baselines, we compare with the State-of-the-Art PPR-based methods as listed in Table \ref{tab:design-compare}.
Within our framework, we implement the baselines which are equivalent to \textsc{DynamicPPE}, \textsc{PPRGo} and \textsc{InstantGNN}, together with our proposed \textsc{GoPPE}.  We use the same hyperparameters 
\footnote{We use $\alpha=0.15$ and $\epsilon=1e-8$ by default.We denote \textsc{DynamicPPE} as \textsc{DynPPE}, \textsc{InstantGNN} as \textsc{InsGNN}} 
for a fair comparison.  

\paragraph{Metrics:} We evaluate the model's efficiency and prediction accuracy. Since the PPR weights calculating are the most time-consuming bottleneck, we measure the total CPU time for PPR calculation along with the graph updates. 
We use the multi-class classification setting, train the classifier separately for each snapshot
\footnote{For benchmarking purposes, we train the classifier individually for each snapshot to minimize the variability caused by continual learning.}
, then report the testing accuracy. 
We repeat all experiments three times with various random seeds and report the average value.

\paragraph{Training protocol}: We use 2-layer MLPs together with \textsc{Relu} activations with dropout rate $p=0.15$. The inputs are the node representations produced by the aforementioned methods. 
We apply the same neural classifier and hyper-parameter settings to test the embedding quality across all methods for a fair comparison. More details are presented in the Appendix \ref{app:train-detail}. 

% \paragraph{Metrics}
% \begin{itemize}
%     \item Quality: Accuracy/F1 over snapshots for base predictions, overall in each cohort and for the newly unseen nodes.
%     \item Efficiency: computation time over snapshots. for both training and testing. for ppr update and nerual model inference.
% \end{itemize}

\xingzhiswallow{
    \begin{itemize}
        % \item \textcolor{red}{Below: NE-based}
        % \item baseline: Deepwalk every snapshot
        % \item baseline: Node2Vec every snapshot
        % \item baseline: FastRP/RandNE every snapshot
        % \item baseline: DynPPE every snapshot
        \item \textcolor{red}{Below: Feature-based}
        \item baseline: MLPs
        \item baseline: DynPPE
        \item baseline: \sout{GCN}
        \item baseline: \sout{GRPGNN}
        \item baseline: PPRGO (agg top-k)
        \item baseline: InstantGNN (mixrank-all: agg all, original code is not released)
        \item \textcolor{red}{Below: our proposed baselines:}
        \item baseline: MixRank-rnd (agg rand-k)
        \item baseline: GoPPE (MixRank-all + DynPPE)
        \item baseline: \sout{MixRank-percentile (agg top-10\%,20\%, ... )}
        \item baseline: \sout{ MixRank-poly (fit a polynomial to modify ppr as agg weights. )}
\end{itemize}
}

\xingzhiswallow{
\begin{tabular}{llrrrr}
\toprule
ppr-algo &  increment &   cora &  citeseer &  pubmed &   flickr \\
\midrule
    push &      False & 288.07 &    149.89 & 3254.14 & 19035.70 \\
    ista &      False & 150.53 &    148.38 & 1376.25 &  7452.47 \\
    push &       True & 430.57 &    223.41 & 4786.90 & 31889.43 \\
    ista &       True & 137.21 &    121.29 & 1212.33 &  7249.87 \\
\bottomrule
\end{tabular}

\begin{tabular}{llrrrr}
\toprule
ppr-algo &  increment &     cora &  citeseer &   pubmed &   flickr \\
\midrule
    push &      False & 3.99e-08 &  2.25e-08 & 4.57e-08 & 4.32e-08 \\
    ista &      False & 5.22e-08 &  2.78e-08 & 6.03e-08 & 2.84e-08 \\
    push &       True & 4.06e-08 &  2.30e-08 & 4.65e-08 & 4.34e-08 \\
    ista &       True & 5.14e-08 &  2.73e-08 & 5.94e-08 & 2.80e-08 \\
\bottomrule
\end{tabular}

}

% \subsection{Node Classification over Dynamic Graphs}

\subsection{Efficiency Experiments } 
\label{sec:ppr-speed}
In PPR-based GNNs for dynamic graphs, the main computational bottleneck is the PPR maintenance over time. 
In this section, we present the running time under two cases regarding the intensity of graph changes between snapshots 
-- \textbf{1: Major change case} where we start from $50\%$ of the total edges, then add $10\%$ edges between snapshots until all edges are used. 
\footnote{Same as we described in Section \ref{sec:dataset}, we also evaluate  accuracy in this case.} 
\textbf{2: Minor change case} where we start from  a almost full graph but held out a few edges, then we add a fixed number of edges (e.g., 100) between snapshots until all edges are added. We design these two cases to evaluate the efficiency of PPR maintenance under different dynamic patterns. In the following, we discuss the insights from our empirical observations.

% \todo{The arxiv PPRGo number is not well-formed in table \ref{tab:cpu-time}}

\paragraph{GoPPE is efficient and scalable in major change case}: Table \ref{tab:cpu-time} presents the total running time of PPR calculation. Our proposed \textsc{GoPPE} is consistently faster than others while maintaining a similar PPR precision. 
Table \ref{tab:ppr-acc} in Appendix shows the consistent approximate $\ell_1$-error as specified for each dataset. 
In general, \textsc{GoPPE} can achieve $2 \sim 3$ times faster than forward push-based methods. The gain is more significant over large graphs. 
The advantage mainly comes from the ISTA-based PPR solver, which enjoys both the ISTA's favorable convergence rate and the warm-starting strategy as the incremental PPR maintenance method. 
Note that even the static \textsc{GoPPE-S} is faster than the incremental Push-based method. 

\begin{table}[ht]
    \centering
    \caption{The total CPU time (in seconds) for PPR maintenance in major change case. 
    \text{GoPPE} uses the least time to obtain the approximate PPR weights with similar $\ell_1$-error. We elide \textsc{DynPPE} since it use the same PPR solve as used in \textsc{InsGNN}.
    }
    \vspace{-2mm}
    \label{tab:cpu-time}
    % \small
    \begin{tabular}{lrrrrr}
    \toprule
    Method &     cora &  citeseer &  pubmed &  arxiv &   flickr \\
    \midrule
        % \textsc{PPRGo} & 288.07 &    149.89 & 3,254.14 &    84,312.37 & 19,035.70 \\  \hline
        % \textsc{InsGNN} &   430.57 &    223.41 & 4,786.90 &   117,039.71 & 31,889.43 \\ \hline
        % %\textsc{DynPPE}
        % \textsc{GoPPE-S}  & 150.53 &    148.38 & 1,376.25 &    42,402.10 &  7,452.47\\ \hline
        % \textbf{\textsc{GoPPE-D}} &  \textbf{137.21} &    \textbf{121.29} &  \textbf{1,212.33} &    \textbf{34,246.45} &  \textbf{7,249.87} \\

        \textsc{PPRGo} & 288.07 &    149.89 & 3,254.14 &    79,133.65 & 19,035.70 \\  \hline
        \textsc{InsGNN} &   430.57 &    223.41 & 4,786.90 &  108,048.49 & 31,889.43 \\ \hline
        %\textsc{DynPPE}
        \textsc{GoPPE-S}  & 150.53 &    148.38 & 1,376.25 &    45,320.18 &  7,452.47\\ \hline
        \textbf{\textsc{GoPPE-D}} &  \textbf{137.21} &    \textbf{121.29} &  \textbf{1,212.33} &    \textbf{38,830.99} &  \textbf{7,249.87} \\

    \bottomrule
    \end{tabular}
\end{table}

\begin{remark}
    \textsc{ForwardPush} and \textsc{ISTA} has the same theoretical worst-case complexity $\mathcal{O}(\frac{1}{\alpha \epsilon})$, the observed running time difference is caused by their formulations where algorithmic \textsc{ForwardPush} is implemented by queue, while  \textsc{ISTA} adopts numerical computing on cache-friendly continuous memory. Similar observations are also in \cite{wu2021unifying}.
\end{remark}

% \vspace{-5mm}
\begin{figure}[ht]
    \centering
    \subfloat[Major changes case\label{fig:ppr-time-large-change}]{
        \includegraphics[width=0.48\linewidth]{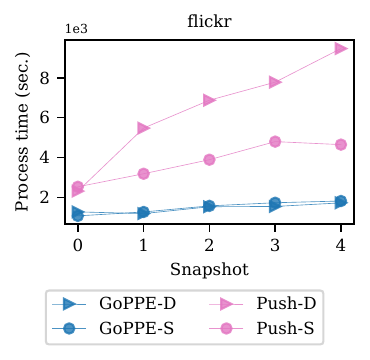}

    }
    \hfill
    \subfloat[Minor changes case \label{fig:ppr-time-small-change}]{
        \includegraphics[width=0.48\linewidth]{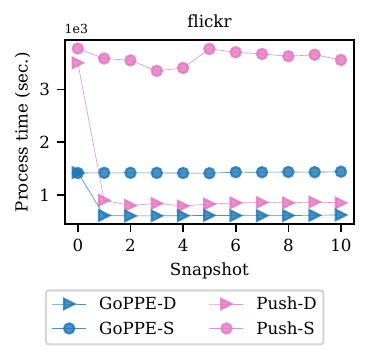}
    }
    \vspace{-3mm}
    \caption{Subfigures (a) and (b) show the CPU time  in \textit{Major/Minor} graph change experiment settings, respectively. The intensity of graph changes between snapshots (major vs minor) may greatly affect the PPR updating time. In both cases, \textsc{GoPPE} is consistently efficient in terms of total CPU time. \textsc{PPRGo} uses \textsc{Push-S} while \textsc{DynPPE} and \textsc{InsGNN} use  \textsc{Push-D}.   Note that the absolute time is not directly comparable between subfigures due to different settings.  }
    \label{fig:ppr-time-wrt-edge-changes}
\end{figure}

%Dynamic PPR update rules for 
\paragraph{The efficiency of Push-based method is limited in major change case while boosts up in minor change case}: 
One interesting observation is that the incremental push-based method (\textit{Push-D} used by \textsc{DynPPE} and \textsc{InsGNN}  ) is not faster than their static counterparts (\textit{Push-S} used by \textsc{PPRGo}), as presented in Table \ref{tab:cpu-time} and Figure \ref{fig:ppr-time-large-change}.
The root reason relates to how \textit{"different"} the graph will be after the edge events. In major change case, we cause the graph changes drastically between snapshots by adding a large number of edges. 
Consequently, the quality of the approximate PPR is too compromised after applying the PPR update rules. 
Unfortunately, the update rules create negative residuals and incur more push operations than their static version \footnote{Static push method only has a positive and monotonic decreasing residual.}. When the graph changes drastically, it eventually exceeds the total operations required from scratch to maintain the PPR precision. 

% \todo{Add a label explaining the x-axis to figures \ref{fig:ppr-time-wrt-edge-changes}. }

%%% 1e-6 for all
\begin{table}[ht]
    \centering
    \caption{The total CPU time (in seconds) used in the minor change case.  The incremental PPR maintenance methods (\textsc{GoPPE-D} and \textsc{InsGNN}) achieve significantly better efficiency than their static counterparts. In addition, \textsc{GoPPE-D} is steadily faster, and even the static \textsc{GoPPE-S} is comparable to the incremental push-based methods.  Note that the reported CPU time is not directly comparable to Table \ref{tab:cpu-time} due to different experimental settings. }
    % \vspace{-2mm}
    \label{tab:cpu-time-small-change}
    % \small
    \begin{tabular}{llrrrrr}
    \toprule
    Method & cora &  citeseer &  pubmed & arxiv&  flickr \\
    \midrule
       \textsc{PPRGo} &     548.24 &    135.96 & 5,852.38 & 138,513.10  & 39,692.24 \\ \hline
        \textsc{InsGNN} &   358.51 &    87.17 & 2,896.12 & 24,742.79  & 11,883.08 \\ \hline
        %\textsc{DynPPE}
       \textsc{GoPPE-S} &   319.60 &    142.20 & 2,847.84 & 75,117.79 & 15,632.07 \\ \hline
        \textbf{\textsc{GoPPE-D}} &  \textbf{190.16} &    \textbf{64.77} & \textbf{1,495.70} & \textbf{22,512.81} & \textbf{7,462.86} \\
    \bottomrule
    \end{tabular}
\end{table}

\paragraph{GoPPE is consistently faster regardless of the graph change intensity}: 
Table \ref{tab:cpu-time-small-change} and Figure \ref{fig:ppr-time-small-change} present PPR efficiency results in the minor change case where only a small number of edges are added between snapshots.
It shows that the update rules greatly boost PPR efficiency for both increment PPR maintenance methods (\textsc{Push-D} and \textsc{GoPPE-D}). 
Besides, \textsc{GoPPE} is a steadily fast method,  achieving impressive speed consistently in both cases.  
%

% \paragraph{ISTA-based PPR routine is fast in the first place}:\red{add text}

% \paragraph{Warm-start PPR is faster} \red{add text}

% \paragraph{Dynamic Push-based method may not gain speed up when the graph changes significantly}: small edge changes. 

% In both cases, the $\ell_1$-1 regularization formulation of PPR provides an empirically faster ISTA solver for GNN attention while well explains the desired sparse property in the GNN framework.

% \begin{figure*}[ht]
%     \centering
% \includegraphics[width=0.99\textwidth]{figs/exp-1-v7/fig-clf-exp-1-all-acc.pdf}
%     \caption{Accuracy of node classification: We use ISTA PPR to test the feature aggregation effectiveness.}
%     \label{fig:node-class-res-acc}
% \end{figure*}

% \begin{figure*}
%     \includegraphics[width=0.99\textwidth]{figs/exp-1-v7/fig-ppr-l1-err-all.pdf}
%     \caption{$\ell_1$-error of PPR. Note that }
%     \label{fig:ppr-l1-err}
% \end{figure*}

% \begin{figure*}
%     \includegraphics[width=0.99\textwidth]{figs/exp-1-v7/fig-ppr-time-all.pdf}
%     \caption{$\ell_1$-error of PPR. Note that PPRGo uses forward push, InstantGNN uses incremental version of Forward Push. Our GoPPE uses ISTA, significantly reducing the propagation time.}
%     \label{fig:ppr-runtime}
% \end{figure*}

% \begin{figure*}
%     \includegraphics[width=0.99\textwidth]{figs/exp-2-v1/fig-clf-exp-1-all-acc.pdf}
%     \caption{Robustness against noise. We gradually add Gaussian noise into node feature as the graph keeps evolving. PPE-based methods are robust against noise. }
%     \label{fig:ppr-runtime}
% \end{figure*}

\xingzhiswallow{

    \subsection{Results on OGB-Arvix (Homophily Graph)}
    \begin{figure}[ht]
        \centering
        \subfloat[][accuracy]{
         \includegraphics[width=0.5\textwidth]{figs/exp-1-v7/ogbn-arxiv-debug-7/fig-clf-res.pdf}
        } 
        \\
        \subfloat[][ppr running time]{
         \includegraphics[width=0.5\textwidth]{figs/exp-1-v7/ogbn-arxiv-debug-7/fig-ppr-time-alpha-0.15.pdf}
        }
        \caption{ogbn-arxiv-res}
        \label{fig:clf-ogbn-arxiv}
    \end{figure}
    \newpage

    \subsection{Results on Cora (Homophily Graph)}
    \begin{figure}[ht]
        \centering
        \subfloat[][accuracy]{
    \includegraphics[width=0.5\textwidth]{figs/exp-1-v7/cora-debug-7/fig-clf-res.pdf}
        }
        
        \subfloat[][ppr running time]{
            \includegraphics[width=0.5\textwidth]{figs/exp-1-v7/cora-debug-7/fig-ppr-time-alpha-0.15.pdf}
        }
        \caption{Cora-res}
        \label{fig:clf-cora}
    \end{figure}
    \newpage

    \subsection{Results on Citeseer (Homophily Graph)}
    \begin{figure}[ht]
        \centering
        \subfloat[][accuracy]{
            \includegraphics[width=0.45\textwidth]{figs/exp-1-v7/citeseer-debug-7/fig-clf-res.pdf}
        }
    
        \subfloat[][ppr running time]{
            \includegraphics[width=0.5\textwidth]{figs/exp-1-v7/citeseer-debug-7/fig-ppr-time-alpha-0.15.pdf}
        }
        \caption{citeseer-res}
        \label{fig:clf-citeseer}
    \end{figure}
    \newpage

    \subsection{Results on PubMed (Homophily Graph)}
    \begin{figure}[ht]
        \centering
        \subfloat[][accuracy]{
            \includegraphics[width=0.45\textwidth]{figs/exp-1-v7/pubmed-debug-7/fig-clf-res.pdf}
        }
    
        \subfloat[][ppr running time]{
            \includegraphics[width=0.5\textwidth]{figs/exp-1-v7/pubmed-debug-7/fig-ppr-time-alpha-0.15.pdf}
        }
        \caption{pubmed-res}
        \label{fig:clf-pubmed}
    \end{figure}
    \newpage

    \subsection{Results on Flickr (Homophily Graph)}
    \begin{figure}[ht]
        \centering
            \subfloat[][accuracy]{
        \includegraphics[width=0.5\textwidth]{figs/exp-1-v7/flickr-debug-7/fig-clf-res.pdf}
        }
    
        \subfloat[][ppr running time]{
        \includegraphics[width=0.5\textwidth]{figs/exp-1-v7/flickr-debug-7/fig-ppr-time-alpha-0.3.pdf}
        }
    
        \caption{flickr-res}
        \label{fig:clf-flickr}
    \end{figure}
    \newpage

    \subsection{Results on Film (Heterophily Graph)}
    \begin{figure}[ht]
        \centering
        \includegraphics[width=0.5\textwidth]{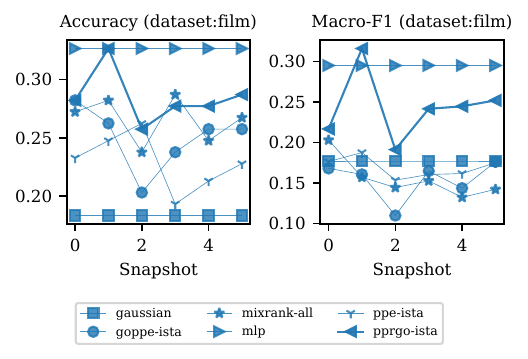}
        \caption{film-res}
        \label{fig:clf-film}
    \end{figure}
    \begin{table}[ht]
\centering
\caption{Avg. Accuracy (film)}
\label{tab:clf-res-film-avg-acc}
\begin{tabular}{lr}
\toprule
 & accuracy-mean \\
model &  \\
\midrule
mlp & 0.3267 \\
pprgo-ista & 0.2847 \\
mixrank-all & 0.2657 \\
goppe-ista & 0.2500 \\
ppe-ista & 0.2294 \\
gaussian & 0.1832 \\
\bottomrule
\end{tabular}
\end{table}

\begin{table}[ht]
\centering
\caption{Avg. Macro-F1 (film)}
\label{tab:clf-res-film-avg-macro-f1}
\begin{tabular}{lr}
\toprule
 & macro-f1-mean \\
model &  \\
\midrule
mlp & 0.2950 \\
pprgo-ista & 0.2437 \\
gaussian & 0.1772 \\
ppe-ista & 0.1692 \\
mixrank-all & 0.1554 \\
goppe-ista & 0.1539 \\
\bottomrule
\end{tabular}
\end{table}

    \newpage
    
    \subsection{Results on Chameleon (Heterophily Graph)}
    \begin{figure}[ht]
        \centering
        \includegraphics[width=0.5\textwidth]{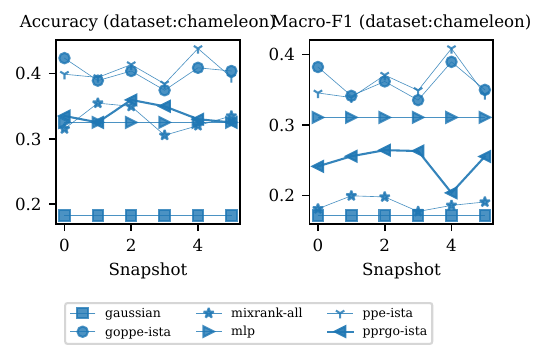}
        \caption{chameleon-res}
        \label{fig:clf-chameleon}
    \end{figure}
    \begin{table}[ht]
\centering
\caption{Avg. Accuracy (chameleon)}
\label{tab:clf-res-chameleon-avg-acc}
\begin{tabular}{lr}
\toprule
 & accuracy-mean \\
model &  \\
\midrule
ppe-ista & 0.4039 \\
goppe-ista & 0.4007 \\
pprgo-ista & 0.3374 \\
mixrank-all & 0.3300 \\
mlp & 0.3251 \\
gaussian & 0.1823 \\
\bottomrule
\end{tabular}
\end{table}

\begin{table}[ht]
\centering
\caption{Avg. Macro-F1 (chameleon)}
\label{tab:clf-res-chameleon-avg-macro-f1}
\begin{tabular}{lr}
\toprule
 & macro-f1-mean \\
model &  \\
\midrule
goppe-ista & 0.3599 \\
ppe-ista & 0.3592 \\
mlp & 0.3107 \\
pprgo-ista & 0.2472 \\
mixrank-all & 0.1888 \\
gaussian & 0.1716 \\
\bottomrule
\end{tabular}
\end{table}

    \newpage
    
    \subsection{Results on Squirrel (Heterophily Graph)}
    \begin{figure}[ht]
        \centering
        \includegraphics[width=0.5\textwidth]{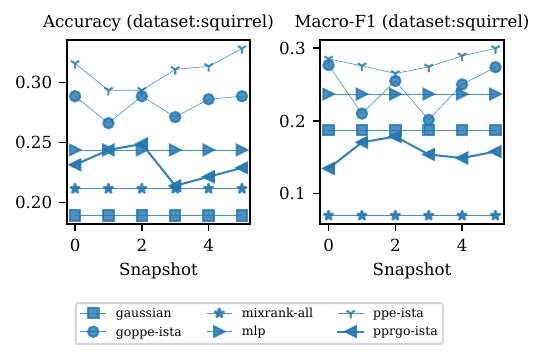}
        \caption{squirrel-res}
        \label{fig:clf-squirrel}
    \end{figure}
    \begin{table}[ht]
\centering
\caption{Avg. Accuracy (squirrel)}
\label{tab:clf-res-squirrel-avg-acc}
\begin{tabular}{lr}
\toprule
 & accuracy-mean \\
model &  \\
\midrule
ppe-ista & 0.3093 \\
goppe-ista & 0.2815 \\
mlp & 0.2438 \\
pprgo-ista & 0.2313 \\
mixrank-all & 0.2114 \\
gaussian & 0.1891 \\
\bottomrule
\end{tabular}
\end{table}

\begin{table}[ht]
\centering
\caption{Avg. Macro-F1 (squirrel)}
\label{tab:clf-res-squirrel-avg-macro-f1}
\begin{tabular}{lr}
\toprule
 & macro-f1-mean \\
model &  \\
\midrule
ppe-ista & 0.2815 \\
goppe-ista & 0.2445 \\
mlp & 0.2368 \\
gaussian & 0.1875 \\
pprgo-ista & 0.1573 \\
mixrank-all & 0.0698 \\
\bottomrule
\end{tabular}
\end{table}

    \newpage
    
}

\begin{figure*}[htbp]
    \includegraphics[width=1.0\textwidth]{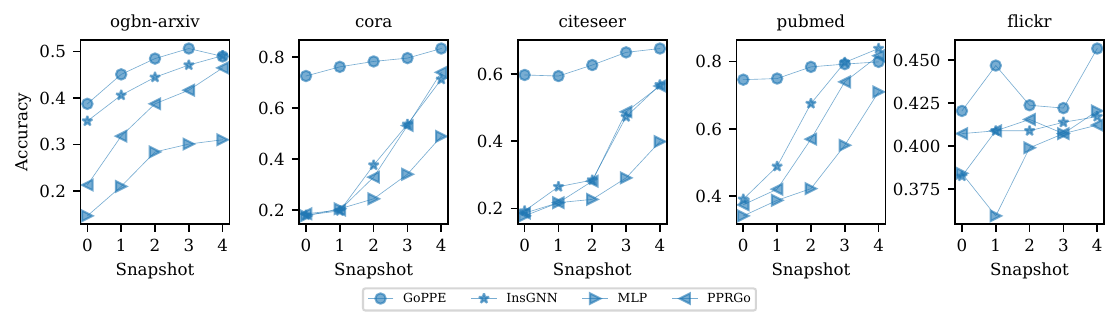}
    % \vspace{-8mm}
    \caption{Performance Robustness against noise: We gradually add Gaussian noise into the node feature as the graph keeps evolving. \textsc{GoPPE} are robust against noise and greatly outperform others at the first snapshots.
    Note that we exclude \textsc{DynPPE} because it is an attributes-free algorithm and has the same performance reported in Table. \ref{tab:clf-exp-1} and Figure \ref{fig:exp-1-clf} in Appendix. }
    \label{fig:exp-2-clf}
\end{figure*}

% \todo{Do a better job describing what people are supposed to be seeing in Figure \ref{fig:exp-2-clf}.   Presumably it is about converging faster to optimal performance or something.}

\subsection{Accuracy and Robustness Experiments}
 We have two node classification settings described as follows to evaluate the performance and robustness of \textsc{GoPPE}. In both cases, we adopt \textit{major change} as the dynamic pattern.
 \textbf{1: Intact case} where we use the intact raw node attributes for global feature aggregation; \textbf{2: Noisy case} where we add Gaussian noise to the node attributes (similar to \cite{liu2021graph}) at the beginning, then gradually reduce the noise level to simulate the node attributes getting better over time.  
The added noise is monotonically reducing defined as: 
\begin{align}
    \hat{\bm  x_t} =  \lambda_t \bm x_t + (1-\lambda_t) \mathcal{N}(\mu, \sigma^2),  \label{equ:noisy}
\end{align}
where  $\lambda_t = \frac{t}{T} + \lambda_{base} \in [0, 1]$ controls noise level,  $t$ is the snapshot stamp, $T$ is the total number of snapshots, $\lambda_{base} \in [0,1]$ controls the initial noise level, $\mu, \sigma^2 $ are the mean and variance of the elements in node attributes matrix $\bm X$, respectively. As $t$ increases from $0$ to $T$, the noisy component decays linearly. 
We use this noise model to simulate the feature refinement process, similar to the volunteers helping enrich Wikipedia articles and make the articles more clear and relevant.

% \todo{I think you need to do more in the intro and also maybe here to justify why this error model is interesting.}

\begin{table}[ht]
    \centering
    \caption{We present the average accuracy across snapshots in the intact case. $\textsc{GoPPE}$ is comparable to or better than the baselines while it's the fastest as discussed in Section \ref{sec:ppr-speed}.  We present full results in Figure \ref{fig:exp-1-clf} in Appendix. Although $\textsc{GoPPE}$ generally performs well, we will discuss its performance on \textit{pubmed} graph later. }
    \vspace{-3mm}
    \label{tab:clf-exp-1}
    \begin{tabular}{l|rrrrr}
    \toprule
    models  &   cora &  citeseer &  pubmed &  arxiv &  flickr \\
    \midrule
    % \textsc{Random}     & 0.1993 &    0.1781 &  0.3648 &      0.1343 &  0.3530 \\
    % \textsc{MLP}        & 0.4592 &    0.3758 &  0.7065 &      0.3102 &  0.4204 \\
    % \textsc{PPRGo}      & 0.6281 &    0.4964 &  \textbf{0.8156} &      0.4244 &  0.4131 \\
    % \textsc{InsGNN}     & 0.6209 &    0.4552 &  \textbf{0.8305} &      0.4772 &  0.4138 \\
    % \textsc{DynPPE}     & \textbf{0.7928} &    \textbf{0.6216} &  0.7847 &      0.4648 &  \textbf{0.4236} \\
    % \textbf{\textsc{GoPPE}}  & \textbf{0.7824} &    \textbf{0.6229} &  0.7900 &      0.4617 &  \textbf{0.4233} \\
    Random   & 0.1977 &    0.1667 &  0.3731 &      0.1343 &  0.3530 \\
    MLP      & 0.4886 &    0.3987 &  0.7098 &      0.3102 &  0.4204 \\
    PPRGo    & 0.7101 &    0.5458 &   \textbf{0.8139} &      0.4244 &  0.4128 \\
    % InsGNN   & 0.5554 &    0.4871 &  0.7438 &      0.3793 &  0.4113 \\
    InsGNN   & 0.6925 &    0.5588 &  \textbf{0.8362} &      \textbf{0.4772} &  0.4138 \\
    DynPPE   & \textbf{0.7984} &     \textbf{0.6314} &  0.7821 &       0.4648 &  \textbf{0.4220} \\ \hline
    GoPPE    &  \textbf{0.7935} &     \textbf{0.6310} &   0.7861 &       \textbf{0.4843} &   \textbf{0.4427} \\
        \bottomrule
    \end{tabular}
\end{table}

\paragraph{GoPPE achieves comparable performance with intact features }:  Table \ref{tab:clf-exp-1} present the averaged prediction accuracy across all snapshots.  
Specifically, we found the positional encoding (\textsc{DynPPE}) alone can yield strong performance, further demonstrating the importance of locality information. 
 Meanwhile, we will analyze the performance difference in \textit{pubmed} and \textit{flickr} later in Section \ref{sec:discuss}.

\begin{table}[ht]
    \centering
    \caption{GoPPE is robust to noisy node attributes. 
    We exclude $\textsc{DynPPE}$ as it is an attribute-free algorithm and has the same results as reported in Table \ref{tab:clf-exp-1}.
    }
    % \vspace{-3mm}
    \label{tab:exp-2-noisy}
    \begin{tabular}{lrrrrr}
    \toprule
    methods &   cora &  citeseer &  pubmed &  arxiv &  flickr \\
    \midrule
    % gaussian    & 0.1993 &    0.1781 &  0.3648 &      0.1343 &  0.3530 \\
    % mlp         & 0.4088 &    0.3222 &  0.6657 &      0.2583 &  0.4171 \\
    % pprgo-ista  & 0.5203 &    0.3892 &  0.8017 &      0.3904 &  0.4092 \\
    % mixrank-all & 0.5036 &    0.3879 &  0.8123 &      0.4457 &  0.4122 \\
    % ppe-ista    & 0.7928 &    0.6216 &  0.7847 &      0.4648 &  0.4236 \\
    % goppe-ista  & 0.7869 &    0.6196 &  0.7907 &      0.4552 &  0.4269 \\
    MLP      & 0.2905 &    0.2624 &  0.4816 &      0.2503 &  0.3941 \\
    PPRGo    & 0.3967 &    0.3471 &  0.5837 &      0.3599 &  0.4102 \\
    InsGNN   & 0.4000 &    0.3562 &  0.6378 &      0.4324 &  0.4062 \\ \hline
    % DynPPE   &    NaN &       NaN &     NaN &         NaN &     NaN \\
    GoPPE    & \textbf{0.7797} &    \textbf{0.6307} &  \textbf{0.7745} &      \textbf{0.4636} &  \textbf{0.4338} \\
        \bottomrule
    \end{tabular}
\end{table}

\paragraph{\textsc{GoPPE} is robust against noisy node attributes and consistently outperforms the baselines}: 
Figure \ref{fig:exp-2-clf} presents the  accuracy over snapshots where the node features are gradually refined as defined in Equ.\ref{equ:noisy}.
As expected, the models without positional encodings (MLP, PPRGo, and InsGNN) perform poorly at the beginning as they heavily depend on the node attributes.  Meanwhile, \textsc{GoPPE} demonstrates its robustness by having positional encodings and consistently outperforms others.

% \paragraph{Positional encodings provide unique strong signal regardless of attribute noise}: 

\subsection{Discussions and Future Works} 
\label{sec:discuss}
\paragraph{Robust positional encodings achieves better accuracy over graphs of \textit{weak} node attributes }: Table \ref{tab:clf-exp-1} shows that \textsc{MLPs} provides poor performance on \textit{cora} and \textit{citeseer}, implying that these graphs have weak node attributes for prediction. It makes \textsc{PPRGo} and  \textsc{InsGNN} harder to extract signals from attributes. Meanwhile, the positional encodings successfully capture the node context and achieve better performance.   
On the other hand, on the attribute-rich graphs, even \textsc{MLP} performs relatively well, for example, on \textit{pubmed} graph. It is because \textit{pubmed} has dense and informative node attributes, which makes the PPR-based feature aggregation (\textsc{PPRGo},\textsc{InsGNN}) produce a succinct and predictive signal. 

% Node attributes together with graph structure are important, especially over the homophily graphs. 

% 2. The quality of node attributes determines how GNNs work. In the worst case where node attributes are noisy, the positional encodings provide robust node interpretation.

\paragraph{How to adapt PPR propagation over dynamic heterotrophic graph remains an open problem}: 
Besides the attribute-rich property in \textit{pubmed},  it is also known for its homophily property \cite{mcpherson2001birds} which makes the nodes of the same labels well-connected.  
As we discussed in the Definition \ref{def:att-property}, one of the assumptions is graph homophily, which makes PPR-based aggregation predictive with local \textit{peer} nodes.  However, \textit{flickr} is a heterotrophic graph \cite{kim2022find}. Table \ref{tab:clf-exp-1} shows that the  accuracy of all models is worse or just slightly better than \textsc{MLP}. 
Inspired by \citet{chien2020adaptive}, we are interested in adapting PPR for dynamic heterotrophic graphs.

\paragraph{Faster PPR-GNN with efficient PPR solver}:
Given the impressive efficiency of ISTA-based PPR solver, it motivates us to explore more toward faster optimization methods for dynamic PPR settings. For example, FISTA or Blockwise Coordinate Descent methods with faster rule and active set \cite{nutini2017let}.  Such research directions have the potential for even faster GNN designs for dynamic graphs and tackle larger-scale problems.

\section{Conclusion }
\label{sec:conclusion}

In this paper, we proposed an efficient framework for node representation learning over dynamic graphs. Within the framework, we propose the desired properties in node attention design for global feature aggregation, and we justify the use of PPR under its $\ell_1$-regularized formulation. Besides, we proposed to use ISTA as the PPR solver and use warm-start as an elegant solution to maintain PPR over dynamic graphs. Finally, we introduce the PPR-based node positional encodings, maximizing the usage of PPR for robust representation against noisy node attributes.
The experiments show that our instantiated model \textsc{GoPPE} is 2-3 times faster than the current state-of-the-art models while achieving comparable or better prediction accuracy over graphs with intact node attributes. In the noisy environment, \textsc{GoPPE} significantly outperforms the others, demonstrating the effectiveness of the proposed positional encodings.

% \section*{Acknowledgement}
% This work was partially supported by several anonymous grants.

\xingzhiswallow{
In this paper, we propose a unified framework \textsc{DynAnom} for subset node anomaly tracking over large dynamic graphs. This framework can be easily applied to different graph anomaly detection tasks from local to global with a flexible score function customized for various applications. 
Experiments show that our proposed framework outperforms current state-of-the-art methods by a large margin, and has a significant speed advantage (2.3 times faster) over large graphs.  
We also present a real-world PERSON graph with an interesting case study about personal life changes, providing a rich resource for both knowledge discovery and algorithm benchmarking.
For future work, it remains interesting to explore a different type of score functions, automatically identify the interesting subset of nodes as better strategies for tracking global-level anomaly, and further investigate temporal-aware PageRank as better node representations.

}

% \newpage
% \bibliographystyle{ACM-Reference-Format}
% \balance
% \bibliography{references}
% \clearpage

\section{Appendix}

% \subsection{Algorithms and Proofs}

\subsection{Formulate Quadratic Objective from  Forward Push }
\label{sec:proof-push-opt}
\label{app:algos-proofs}

\xingzhiswallow{

\begin{lemma}[$\ell_1$-regularized Quadratic Optimization Problem of PPR ] 
\todo{add-ref}Applying the Forward Push algorithm is equivalent to applying the Coordinate Descent algorithm to optimize the following objective function in quadratic form: 

\begin{align}
    \bm x^{*} &= \argmin_{\bm x } \left( \frac{1}{2} \bm x^{\top} \bm W \bm x - \bm (\alpha \bm D^{-1/2} \bm e_s)^{\top} \bm x  \right), where \label{eq:ppr-quad-form-app}\\
    \bm W &= \bm D^{-1/2}(\bm D - (1-\alpha) \bm A ) \bm D^{-1/2}\\
    \bm \pi &= \bm D^{1/2} \bm x^{*}
    %f(\bm x) &= \frac{1}{2} \bm x^{\top} \bm W \bm x + \bm b^{\top} \bm x + \gamma , \label{eq:ppr-quad-form}\\
    % \nabla f(\bm x) &= \bm W \bm x + \bm b ,
    % \text{where }\bm b &= - \alpha \bm D^{-1/2} \bm e_s
\end{align}

% \begin{align}
    % x* = 
    %\nabla f(\bm x) = \bm D^{-\frac{1}{2}} \left(  \bm D - \frac{1-\alpha}{2} (\bm D + \bm A )\right) \bm D^{-\frac{1}{2}} \bm x - \alpha \bm D^{-\frac{1}{2}} \bm e_s,
% \end{align}

\end{lemma}
}

\paragraph{From Push to the objective function }: The proofs follows \cite{fountoulakis2019variational}.
One can rewrite Forward Push algorithm into $\textsc{CD}$:
% \input{algo-1-cd.tex}
% \begin{align}
%     \pi_{\alpha, s}=\left(\bm I_n -(1-\alpha) \bm P^{\top}\right)^{-1} \bm e_s  \label{eq:appr-linear}
% \end{align}

Given Equ.\ref{equ:static-ppr-static-graph}, we define Residual Vector $\bm r \in \mathbb{R}^{n}$ as followings:
\begin{align}
    \bm r := (\bm I_n - (1-\alpha) \bm A \bm D^{-1} ) \bm \pi - \alpha \bm e_s  \label{eq:push-update}
\end{align}

By multiplying $\bm D^{-1/2}$ on both sides of Equ.\ref{eq:push-update} and  $\bm \pi := \bm D^{1/2} \bm x$:
\begin{align}
    % \bm D^{-1/2} \bm r &= \bm D^{-1/2}(\bm I_n - (1-\alpha) \bm A \bm D^{-1}) \bm D^{-1/2} \bm x - \alpha \bm D^{-1/2} \bm e_s  \nonumber \\
    \bm D^{-1/2} \bm r &= \bm W \bm x + \bm b,   %\text{  Substituted with $\bm W$ and $\bm x$}, 
    \label{eq:grad-q}\\
    \text{where } \bm W &= \bm D^{-1/2}(\bm D - (1-\alpha) \bm A) \bm D^{-1/2}, \nonumber \\
    \bm b &= - \alpha \bm D^{-1/2} \bm e_s \nonumber
\end{align}

Equ.\ref{eq:grad-q} has a linear form w.r.t. $\bm x$ which can be interpreted as the gradient $\nabla f(\bm x) := \bm D^{-1/2} \bm r = \bm W \bm x + \bm b $ of a quadratic function $f(\cdot)$ such that:
\begin{align}
    f(\bm x) &:= \frac{1}{2} \bm x^{\top} \bm W \bm x + \bm b^{\top} \bm x  , \label{eq:ppr-quad-form-app2}
    % \nabla f(\bm x) &= \bm W \bm x + \bm b ,
\end{align}

The most intriguing finding is that by applying coordinate descent to Equ.\ref{eq:ppr-quad-form} and minimizing the objective function, the solution $\bm x^{*} = \operatorname{argmin}_{\bm x}f(\bm x)$ is the approximate PPR given the fact that $\bm \pi = \bm D^{1/2} \bm x^{*}$. Such formulation bridges two seemingly disjoint fields and provides more opportunities to solve PPR through the lens of optimization. 

\paragraph{Rewrite Coordinate Descent for PPR}

Note that $\bm r = \bm D^{1/2}\nabla f(\bm x) $, $\bm \pi = \bm D^{1/2} \bm x,$ 
\begin{algorithm}[ht]
\caption{$\textsc{CD Solver for PPR}$ }
\begin{algorithmic}[1]
\State \textbf{Input: }$ k = 0, \bm x^{(0)} = \bm 0, \mathcal{G}, \epsilon, \alpha$
% \State $\bm r^{(0)} = \left(\bm I_n - (1-\alpha) \bm P^{\top} \right) \bm D^{-1/2} \bm x ^{(0)}  - \alpha \bm e_s  = - \alpha \bm e_s$
\State $\nabla f(\bm x^{(0)}) = - \alpha \bm D^{-1/2} \bm e_s$

% \While{ exists $i$ such that $\| \bm D^{1/2}\nabla f(\bm x)(i)\|_1 > \epsilon d(i)$}
% \While{ exists $i$ such that $\|\nabla f(\bm x)(i)\|_1 > \epsilon d(i) \bm d(i)^{-1/2}$}
\While{ exists $i$ such that $\|\nabla f(\bm x)(i)\|_1 > \epsilon' d(i)^{1/2}$}

\State $\textsc{CD}(i)$
\EndWhile
\State \Return $\left(\bm \pi = \bm D^{1/2} \bm x^{(k)}, \bm r =  \bm D^{1/2}\nabla f(\bm x^{(k)}) \right)$
\Procedure{CD}{$i$}
% \State $\bm \pi^{(k+1)}(i) = \bm \pi^{(k)}(i) - \bm r^{(k)}(i) $ 
% \State $\bm D^{1/2} \bm x^{(k+1)}(i) = \bm D^{1/2} \bm x^{(k)}(i) - \bm D^{1/2}\nabla f(\bm x^{(k+1)}) $ 
\State  \textcolor{black}{/*apply coordinate descent*/} 
\State $\bm x^{(k+1)}(i) =  \bm x^{(k)}(i) -  \nabla_i f(\bm x^{(k)}) $
% \State $\bm r^{(k+1)}(i) = 0$
\State  \textcolor{black}{/*update gradient*/} 
\State $\nabla f(\bm x^{(k+1)})(i) = 0$
\For{$j \in \operatorname{Nei}_{out}(i)$}
% \State $\bm r^{(k+1)}(j) = \bm r^{(k)}(j) + (1-\alpha) \bm A_{(i,j)} \frac{\bm r^{(k)}(i)}{\bm d(i)} $
% \State $ \bm d(j)^{1/2} \bm \nabla_j f(\bm x^{(k+1)})  = \bm d(j)^{1/2} \nabla_j f(\bm x^{(k)})  + (1-\alpha) \bm A_{(i,j)} \frac{ \bm d(i)^{1/2} \nabla_i f(\bm x^{(k)})}{\bm d(i)} $
% \State $  \bm \nabla_j f(\bm x^{(k+1)})  = \nabla_j f(\bm x^{(k)})  + (1-\alpha) \bm A_{(i,j)} \frac{ \bm d(j)^{-1/2} \bm d(i)^{1/2} \nabla_i f(\bm x^{(k)})}{\bm d(i)} $
\State $  \bm \nabla_j f(\bm x^{(k+1)})  = \nabla_j f(\bm x^{(k)})  + \frac{(1-\alpha) \bm A_{(i,j)}  }{\bm d(i)^{1/2}\bm d(j)^{1/2}} \nabla_i f(\bm x^{(k)}) $
\EndFor
\State $k = k+1$
\EndProcedure
\end{algorithmic}
\label{algo:forward-local-push-coord}
\end{algorithm}

By looking closely to the termination condition: 
$\|\nabla f(\bm x)(i)\|_1 \leq \epsilon' d(i)^{1/2} 
\Rightarrow 
\nabla f(\bm x)(i) \geq - \epsilon' d(i)^{1/2}$ 
given $\nabla f_i(\bm (x) \leq 0, \forall i$, 
\xingzhiswallow{

\begin{lemma}[PPR as a $\ell_1$-regularized Quadratic Optimization]
the termination condition is equivalent to the first-order optimality condition of the following objective function with $\ell_1$-regularization:

\begin{align}
    \min_{\bm x} \psi(\bm x) := f(\bm x) + \epsilon'\| \bm D^{1/2} \bm x \|_1
\end{align}

% \[
%     \nabla_i f(\bm x^{*})= 
% \begin{dcases}
%     -\epsilon' \bm d(i)^{1/2},  & \text{if } \bm x^{*}(i) > 0 \\
%     \epsilon' \bm d(i)^{1/2}, & \text{if } \bm x^{*}(i) < 0 \\
%     \in [-\epsilon' \bm d(i)^{1/2}, \epsilon' \bm d(i)^{1/2}] & \text{if } \bm x^{*}(i) = 0 \\
% \end{dcases}
% \]
    
\end{lemma}

}
Note that the optimality conditions imply the termination condition, and $\nabla_i f(\bm x^{*}) \in [ -\epsilon' \bm d(i)^{1/2},0 ] \text{ if }  \bm x^{*}(i) = 0 $ implies its sparsity.
$\operatorname{prox}(\bm x) = \operatorname{argmin}_{\bar{\bm x}} \frac{1}{2}\|  \bm x - \bar{\bm x} \|_2^2 + \epsilon' \|\bm D^{1/2} \bm x \|_1 $

\subsection{PPV maintenance on dynamic graphs}
\label{proof:ppv-maintain}
From \cite{zhang2016approximate}: Given a new edge $e_{u,v}$, the adjustments are :
\begin{align}
    \bm \pi'(u) &= \bm \pi(u) * \frac{\bm d(u)+1}{\bm d(u)}\\
    \bm r'(u) &= \bm r(u) - \frac{\bm \pi(u)}{ \alpha \bm d(u)}\\
    \bm r'(v) &= \bm r(v) + \frac{1-\alpha}{\alpha} \frac{\bm \pi(u)}{ \bm d(u)}
\end{align}

Note that, in PPR-equivalent quadratic formulation, we define:
\begin{align}
     \bm r = \bm D^{1/2} \nabla f(\bm x) \text{ and } \bm \pi = \bm D^{-1/2} \bm x 
\end{align}
In addition, the PPR invariant property still holds for the results from the solution.
When a new edge event arrives, the update rules adjust $\bm x$ and $\nabla f(x)$ for further updating. We substitute them back to the update rules and have:

\begin{align}
    \bm d^{-1/2}(u) * \bm x'(u) &= \bm d^{-1/2}(u) * \bm x(u) * \frac{\bm d(u)+1}{\bm d(u)} \nonumber \\
    \Rightarrow \bm x'(u) &= \bm x(u) * \frac{\bm d(u)+1}{\bm d(u)}\nonumber \\
    \bm d^{1/2}(u) * \nabla f'(\bm x')(u) &= \bm d^{1/2}(u)* \nabla f(\bm x)(u) - \frac{\bm d^{-1/2}(u) * \bm x(u)}{ \alpha \bm d(u)}\nonumber \\
    \Rightarrow \nabla f'(\bm x')(u) &= \nabla f(\bm x)(u) - \frac{\bm x(u)}{ \alpha \bm d^{1/2}(u)}\nonumber \\
    \bm d^{1/2}(v) * \nabla f'(\bm x')(v) &= \bm d^{1/2}(v)*\nabla f(\bm x)(v)  + \frac{1-\alpha}{\alpha} \frac{\bm d^{1/2}(u) * \bm x(u)}{ \bm d(u)}\nonumber \\
     \Rightarrow \nabla f'(\bm x')(v) &= \nabla f(\bm x)(v)  + \frac{(1-\alpha) }{\alpha} \frac{ \bm d^{-1/2}(v)*\bm x(u)}{ \bm d^{1/2}(u)}\nonumber 
\end{align}

\subsection{Dimension Reduction using Random Projection }
The sparse Gaussian random projection can be constructed:
\begin{align}
    \bm p_i = \bm R \bm \pi_i, \text{ and }
    \bm R \in \mathbb{R}^{d_i \times |\mathcal{V}|}  := 
    \begin{dcases}
        \small
        +\sqrt{3} \text{ with probability } \frac{1}{6}\\
        \sqrt{0} \text{ with probability } \frac{2}{3}\\
        -\sqrt{3} \text{ with probability } \frac{1}{6}\\
    \end{dcases}
    \label{equ:rand-proj}
\end{align}

\subsection{Experiment Details and Reproducibility}
\label{app:train-detail}

\begin{table}[ht]
    \centering
    \caption{We maintain a similar precision level ($\epsilon \approx 1e-6$ or $1e-8$ ) across all graphs. The error was calculated against the power iteration results with a smaller $\epsilon$ setting for higher precision.}
    \label{tab:ppr-acc}
    \footnotesize
\begin{tabular}{llrrrrr}
\toprule
ppr-algo &  increment &     cora &  citeseer &   pubmed &  ogbn-arxiv &   flickr \\
\midrule
    push &      False & 3.99e-08 &  2.25e-08 & 4.57e-08 &    4.76e-06 & 4.32e-08 \\
    ista &      False & 5.22e-08 &  2.78e-08 & 6.03e-08 &    6.08e-06 & 2.84e-08 \\
    push &       True & 4.06e-08 &  2.30e-08 & 4.65e-08 &    4.85e-06 & 4.34e-08 \\
    ista &       True & 5.14e-08 &  2.73e-08 & 5.94e-08 &    6.00e-06 & 2.80e-08 \\
\bottomrule
\end{tabular}
\end{table}

\xingzhiswallow{
    \begin{table}[ht]
        \centering
        \footnotesize
        \begin{tabular}{llrrrr}
        \toprule
        ppr-algo &  increment &     cora &  citeseer &   pubmed &   flickr \\
        \midrule
            push &      False & 3.99e-08 &  2.25e-08 & 4.57e-08 & 4.32e-08 \\
            ista &      False & 5.22e-08 &  2.78e-08 & 6.03e-08 & 2.84e-08 \\
            push &       True & 4.06e-08 &  2.30e-08 & 4.65e-08 & 4.34e-08 \\
            ista &       True & 5.14e-08 &  2.73e-08 & 5.94e-08 & 2.80e-08 \\
        \bottomrule
        \end{tabular}
        \caption{The PPR precision}
        \label{tab:my_label}
    \end{table}
    
    }

\subsection{Infrastructure}
All the experiments are done on a machine equipped with 4 processors of Intel(R) Xeon(R) CPU E5-2630 v4 @ 2.20GHz, where each processor has 10 cores, resulting in 40 cores in total and  125GB main memory. Besides, we have 4 Nvidia-Titan-XP GPUs where each has 12GB GPU memory.

\subsection{PPR Implementation} We implemented all PPR routines (Power iteration, Forward Push (w/o dynamic adjustment), ISTA) in Python and ensured that all are Numba-accelerated ($\operatorname{@njit}$). Numba acceleration compiles python routines to LLVM native code, which has a comparable speed to C/C++. 
We particularly chose Python+Numba as the main programming tool because it's easier to maintain and is gradually being accepted by the scientific computing community.

\subsection{Neural Model Configurations }
We implemented the neural networks using PyTorch. Table \ref{tab:model-param}. 
We use $\textsc{Linear}\rightarrow\textsc{Relu} \rightarrow\textsc{Dropout}$ as the basic neural compoent for classification. 
% and Figure \ref{fig:ml_model} present the detailed model architecture and configurations.  
Besides. we set dropout ratio $p=0.15$ during training.
During training, we use \textsc{Adam} optimizer and force all models to train for 50/100 epochs before early exit.
More details can be found in the training scripts in our codebase included in this submission.

\begin{table}[!h]
\centering
\footnotesize
\caption{To have a fair comparison, we use the same model configurations except for the proposed feature aggregation function. Note that we denote the dimension of the node's raw feature as $|Feat|$ and the total node label as $|L|$. We use the presented hidden state sizes (128, 32, 16) in MLP-\{0,1,2\}  for small graphs, while use larger size (128, 512, 256) for larger graph. Likewise, all methods share the same architecture configs. }
\label{tab:model-param}
\begin{tabular}{|p{0.15\textwidth}|p{0.13\textwidth}|p{0.13\textwidth}|p{0.13\textwidth}|p{0.13\textwidth}|p{0.13\textwidth}|}
\hline 
  & MLP-Feat & MLP-PPE & MLP-0 & MLP-1 & MLP-2 \\ \hline 
 Gaussian & |Feat| * 128 & - & 128*32 & 32*16 & 16*|L| \\
\hline 
 PPE & - & |PPE| * 128 & 128*32 & 32*16 & 16*|L| \\
\hline 
 MLP & |Feat| * 128 & - & 128*32 & 32*16 & 16*|L| \\
\hline 
 PPRGo & |Feat| * 128 & - & 128*32 & 32*16 & 16*|L| \\
\hline 
 InstantGNN & |Feat| * 128 & - & 128*32 & 32*16 & 16*|L| \\
\hline 
 \textbf{GoPPE} & |Feat| * 128 & |PPE| * 128 & 128*32 & 32*16 & 16*|L| \\
 \hline
\end{tabular}
\end{table}

% \begin{figure}[ht]
%     \centering
%     \includegraphics[width=0.45\textwidth]{figs/ml-model.pdf}
%     \caption{The overview of neural network classifier after obtaining the contextualized node embeddings and PPR-based Positional Encodings (PPE). }
%     \label{fig:ml_model}
% \end{figure}

\subsection{Reproducibility}
We repeated the experiments with three different random seeds for all stochastic parts (e.g., python, numpy and PyTorch) during training and testing. We included the source code and the scripts needed to reproduce the results and will release them to the public upon publication.

\subsection{Edge event sampling}
% \remind{snapshots and node sample}
We create dynamic graphs from static graphs. First, we associate each edge event with a distinctive timestamp from 1 to $|\mathcal{E}|$. Then, we use the first edges $\Delta\mathcal{E}_0$ (e.g., with timestamp < 100) to construct $\mathcal{G}_0$ as the base graph. Finally, we inject a fixed number of edges  $\Delta\mathcal{E}_1$ (e.g. with the timestamp from 100 to 120) to simulate graph evolution to create the next graph snapshot $\mathcal{G}_1 = \mathcal{G}_0 + \Delta\mathcal{E}_1$. Iteratively, we create all snapshots. Note that the created dynamic graph may not have temporal characteristics and temporal modeling is not in the scope of this paper.

\xingzhiswallow{
\subsection{Node sampling}
We split data into train/dev/test sets using stratified sampling w.r.t. the node label. To avoid undefined metrics in evaluation, we ensure that each node label has at least one sample appearing in the train, dev, and test sets (a total of 3 different samples).

}

\subsection{Extra results}

\begin{figure*}[htbp]
    \includegraphics[width=1.0\textwidth]{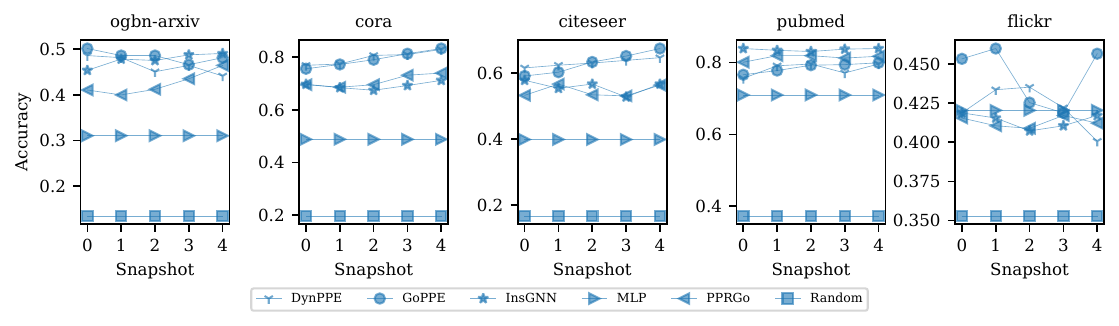}
    \caption{Prediction accuracy on intact features across snapshots. \textsc{GoPPE} has comparable or better accuracy than other baselines.}
    \label{fig:exp-1-clf}
\end{figure*}

\begin{figure*}[htbp]
    \includegraphics[width=1.0\textwidth]{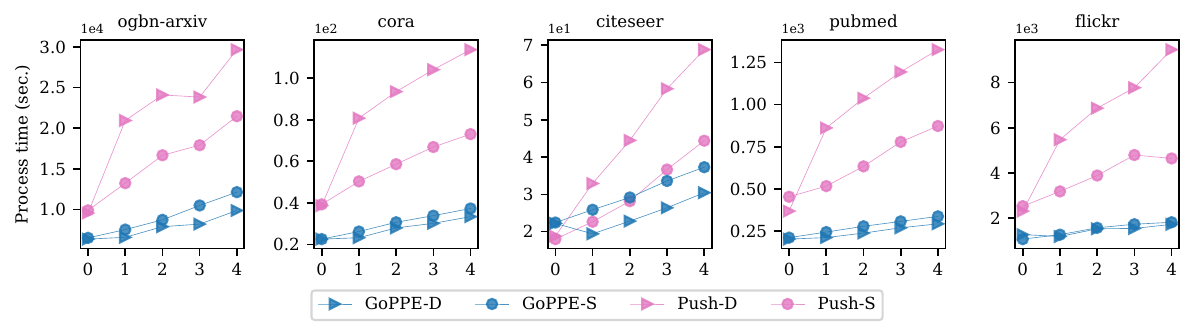}
    \caption{PPR calculation time in major change case.}
    \label{fig:exp-3-ppr-time-major}
\end{figure*}

\begin{figure*}[htbp]
    \includegraphics[width=1.0\textwidth]{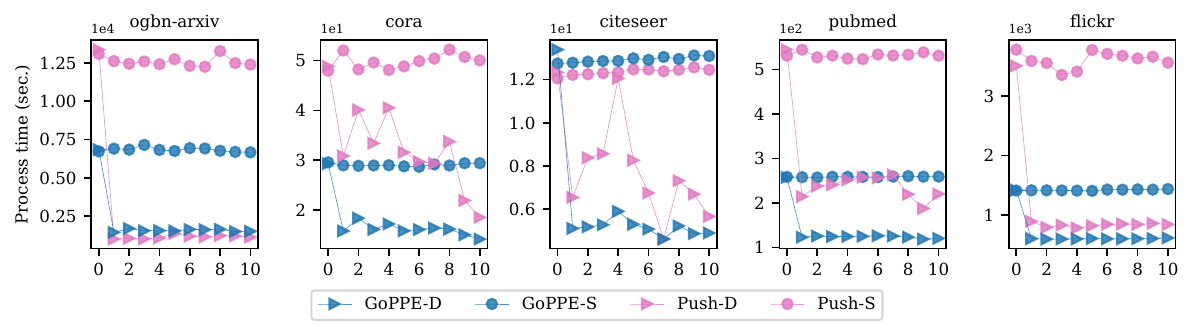}
    \caption{PPR calculation time in minor change case.}
    \label{fig:exp-3-ppr-time-minor}
\end{figure*}

% \begin{figure*}[ht]
%     \centering
% \includegraphics[width=0.99\textwidth]{figs/exp-1-v7/fig-clf-exp-1-all-macro-f1.pdf}
%     \caption{Macro-F1 of node classification}
%     \label{fig:node-class-res-f1}
% \end{figure*}

% (dropout): Dropout(p=0.15, inplace=False)

% \begin{verbatim}

% VertexFormer(
%   (mlp_mlp): Linear(in_features=1433, out_features=128, bias=True)
  
%   (mlps): ModuleList(
%     (0): Linear(in_features=128, out_features=32, bias=True)
%     (1): Linear(in_features=32, out_features=16, bias=True)
%     (2): Linear(in_features=16, out_features=7, bias=True)
%   )
% \end{verbatim}

% \end{document}

\chapter[Post-publication Modification in News Outlets]{Post-publication Modification in News Outlets\protect\footnote{Xingzhi Guo, Brian Kondracki, Nick Nikiforakis, and Steven Skiena. 2022. Verba Volant, Scripta Volant: Understanding Post-publication Title Changes in News Outlets. In \textit{ Proceedings of the ACM Web Conference 2022 (WWW '22)}. Association for Computing Machinery, New York, NY, USA, 588–598. https://doi.org/10.1145/3485447.3512219}}
\label{chapter:title-change}
\graphicspath{{chapter3-TitleModification/}}
\newcommand\numArticlesWithTitleChange{30,930\xspace}
\newcommand\datasetSize{41,906\xspace}

\newcommand\dataCollectionStart{March 1st, 2021\xspace}
\newcommand\dataCollectionEnd{August 31st, 2021\xspace}

\section{Introduction}
% For hundreds of years, people have counted on news agencies to learn of
% current events and information affecting their lives. As technology has
% advanced, the speed in which these agencies can distribute information
% has drastically increased. What began in a printed form, distributed 
% physically, has evolved with each new communication media. Today, 86\% 
% of people consume news on digital devices over the Internet~\cite{newsSources}.

The growth of the Internet has shifted the distribution of news articles from traditional channels such as print and radio to online 
websites and social media platforms. Today, 86\% of people consume news on digital devices via the Internet~\cite{newsSources}. This new medium has 
greatly reduced the time between the occurrence of newsworthy events and the 
publishing of articles reporting on them.
%, and the consumption of said articles. 
One of the strategies that news agencies appear to be regularly using is to publish articles with incomplete information, taking advantage of their ability to modify online articles \emph{after} their publication \cite{guo2022verba}.
 Consequently, readers who view an article or headline immediately after publication may be exposed to radically different information and therefore arrive at different conclusions, compared to readers consuming the same article at a later time. For example, \emph{The Atlantic} published the following article on February 11th, 2021 with the title:

%\todo{These headlines reflect something different (headline emphasis) rather than the article actually changing.    If the gist of the story changes (vs. the headline) describe this change in the text.}

\begin{quote}
    Original title\cite{trumptitlea}: \textit{"I Miss the Thrill of Trump"}  %\footnote{\url{https://web.archive.org/web/20210211110734if_/https://www.theatlantic.com/ideas/archive/2021/02/i-miss-thrill-trump/617993/}} 
    %https://web.archive.org/web/20210211110734if_/https://www.theatlantic.com/ideas/archive/2021/02/i-miss-thrill-trump/617993/
\end{quote}

\noindent However, five hours later the title of the article changed to:

\begin{quote}
    
    Altered title\cite{trumptitleb} : \textit{"I Was an Enemy of the People"} %\footnote{\url{https://web.archive.org/web/20210223004426if_/https://www.theatlantic.com/ideas/archive/2021/02/i-miss-thrill-trump/617993/}} 
    %https://web.archive.org/web/20210223004426if_/https://www.theatlantic.com/ideas/archive/2021/02/i-miss-thrill-trump/617993/
\end{quote}

Given the documented tendency of users to skim through titles rather than read the content of every news article that they encounter~\cite{readheadlines}, 
a radical modification of a headline translates to two or more groups of people who have consumed the same news from the same sources, yet arrive at potentially different worldviews. 

In this paper, we explore the phenomenon of post-publication modifications of news article 
headlines. We monitor the titles of articles from several top news publishers, 
and capture all post-publication changes for later analysis.
We analyze our compiled dataset to create a taxonomy of title changes using an automated NLP pipeline allowing us to determine
the reasons for article title changes, the entities responsible, and the
tangible benefits the publisher may receive for doing so.
% Using our compiled dataset, we determine
% the common culprits of article title changes, the article topics that are
% more likely to receive a title change, and construct a statistical model
% of title change categories. We also determine the reasons for these modifications, the 
% entities responsible, and the tangible benefits they receive for doing so.
Further, we examine the spread of news over social networks by collecting time-series impression 
statistics (e.g., number of retweets/likes) of tweets containing links to news articles for the Twitter accounts of news publishers.

\noindent The primary contributions of our work are as follows:
\begin{itemize}
    \item \textit{Temporal News Headline Data Set}: We have developed %\footnote{We thank the owner of Twitter bot @nyt\_diff for sharing their data with us.} 
    a temporal news headline corpus consisting of \numArticlesWithTitleChange{} distinct news articles where the headline changed at least once.  Our dataset of \datasetSize{} total altered headlines pairs (some articles change headlines more than once) provides a unique resource for future studies of news trustworthiness as well as user trust, which we will make available upon publication. %\todo{any concern with news copyright? No -- the articles are a problem but not the headlines.}
    \item \textit{Characterizing Changes in News Headlines}: We create an NLP categorization pipeline for assessing headline changes by the proposed nine-class taxonomy, based on journalism domain knowledge and a state-of-the-art language model (BERTScore ~\cite{zhang2019bertscore}), from which we successfully discover that 23.13\% headline edits are perceived as harmful.  This analysis enables us to quantify why headline changes occur in practice, and how policies differ with particular news agencies. 
    \item \textit{Estimating Effective Timing over Twitter}: We present a temporal analysis showing how rapidly news is fully propagated over social networks. We observe that most news tweets are shared/retweeted within the first 10 hours following their initial posting, before gradually fading out due to losing the public's attention, or achieving their maximum audience. This suggests that an effective headline correction must occur quickly to avoid propagating misinformation.
    %before the previous one already reached its maximum population.
\end{itemize}

% This paper is organized as follows. Section \ref{sec:related_work} describes previous work in analysis of news headlines modifications from the perspective of journalism and information security. Section~\ref{sec:data_prepare} describes our data set collection procedure. Section~\ref{sec:results}  reports the results of headline change categorization and the news propagation over Twitter. Section~\ref{sec:discussion} includes discussion of our findings. Finally, we present the conclusion and future work in Section~\ref{sec:conclusion} .

%to articles with title changes on the Twitter accounts of each news publisher.

%Additionally, we examine the spread of misinformation due to article title
%changes by collecting the impression statistics of tweets containing links
%to articles with title changes on the Twitter accounts of each news publisher.

% \input{chapter3-TitleModification/related_work}

\section{Related Work}
\label{sec:related_work}
A genuine news headline summarizes the content and enables readers to draw quick conclusions~\cite{rieis2015breaking, ifantidou2009newspaper, dor2003newspaper}. However, in the digital media era, false information or fake news~\cite{fazio2020pausing,shu2017fake, kumar2018false, zhou2019fake} threatens the public's information consumption by inducing readers with clickbait headlines and fabricated content. \citet{vargas2020detection} developed techniques to distinguish legitimate activity on Twitter from disinformation campaigns using coordination network analysis.
Additionally, a large research line \cite{zhou2019fake, reis2019supervised, karimi2018multi, golbeck2018fake} focus on fake news detection, satire detection\cite{rubin2016fake, de2018attending}, and clickbait detection~\cite{potthast2016clickbait}. \citet{hounsel2020identifying} proposed methods to discover disinformation websites using characteristics of their hosting infrastructure. 
In this work, we investigate news outlets that, in the process of publishing legitimate articles, modify their headlines.
%Some news articles may alter headlines after attracting a large enough audience. There is few study on the reasons behind new headline modifications.

The headline modifications may be harmless (e.g., updating competition score) or malicious (e.g., making the headline clickbaity~\cite{blom2015click}, or starting with an inaccurate headline that maximizes user views and eventually changing it to an accurate one). Hagar and Diakopoulos~\cite{hagar2019optimizing} discussed A/B  testing on news headlines and gather audiences' feedback on several headline writing practices (e.g., starting headlines with "why" or "how" and subjective ideals) and reported that headlines are optimized for specific metric (e.g., click-through rate) from A/B testing. \citet{kuiken2017effective} investigated the effectiveness of headlines by comparing click-through rate of the original title and its of the rewritten one, suggesting that clickbait features led to statistically significant increase in number of clicks.  However, the size of their dataset was limited, with 1,836 pairs of headlines that were rewritten from a single groups of editors.  

%\cite{molyneux2020aggregation} clickbait may lower perceptions of credibility and quality

Another related research line is text edit classification. Previous work \cite{daxenberger2013automatically} analyzed linguistic features for the task of edit category classification, organizing English Wikipedia's edits into 21 categories, from spelling corrections to vandalism. \citet{yang2017identifying} investigated the intentions behind text edits and created a 13-category taxonomy of edit intentions, then developed supervised learning models for automatic identification. Similarly, \citet{yin2018learning, marrese2020variational} employed encoder-decoder deep learning framework to learn the edit representation and predict its categories. Due to data availability, previous research mainly focus on Wikipedia post-edit instead of news headlines. Given our unique, large-scale dataset of 41K pairs of title changes, we aim to understand real-world headline modifications and derive a taxonomy for automated edit categorizations. In addition, we analyze the speed of news propagation over social networks, how they relate to post-publication headline changes, and offer guidance on the timings of responsible post-publication modification.

%Wikipedia - neutral point of view. not news headline. preventing vandalism editing. They collect data from Wikipedia edit history, and extract those edits containing explicit signal words (e.g, \textit{NPOV}) in the edit comments. 

%\noindent{\textbf{Journalism: Headline modification}}

%\noindent{\textbf{Clickbait and Click-through Rate}}

%\noindent{\textbf{Edit changes classification}}

%\input{background}

% \input{chapter3-TitleModification/methodology}
\section{Dataset Preparation}

\begin{figure}[t]
    \centering
    \includegraphics[width=0.8\columnwidth]{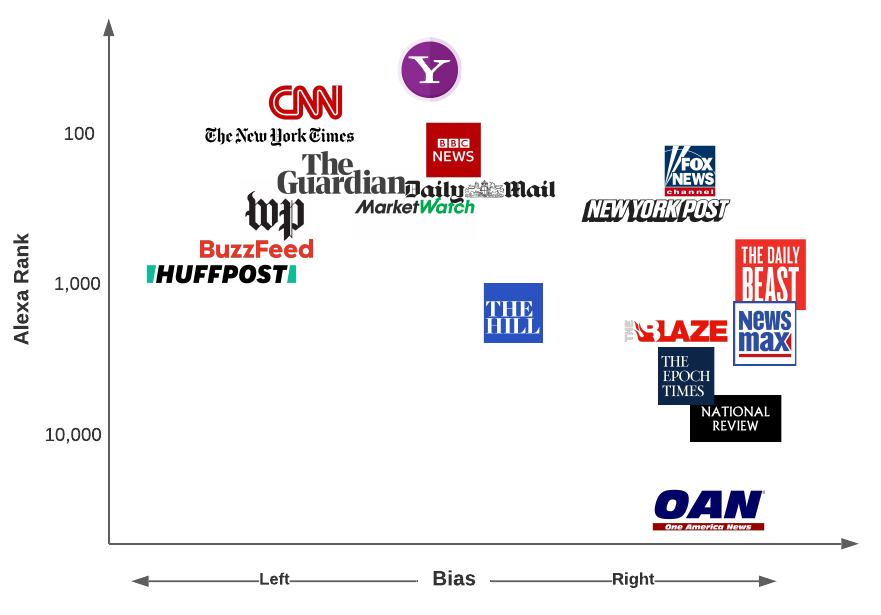}
    \caption{News publishers chosen based on their political bias and
    Alexa ranking.}
    \label{fig:newsPublishers}
    \vspace{-5mm}
\end{figure}

\label{sec:data_prepare} 
In this section, we discuss our process for selecting news publishers
to include in our study. We then describe the infrastructure we designed and implemented
to capture news articles, as well as detect post-publication headline changes and measure the spread of news on social media.

\subsection{News Publisher Identification}

Prior to capturing and studying post-publication article headline changes,
we must first identify a set of news publishers to collect article data
from that are representative of a broad range of audiences, biases, and 
platform sizes.
To this end, we constructed a set of news publishers by first consulting
the Media Bias Chart created by Ad Fontes Media~\cite{mediaBiasChart}.
This chart maps news publishers onto a two-dimensional plane representing
political bias, and reputability. We focus only on the political bias of
each news publisher, selecting an equal number of publishers from each 
region of this axis. 

% Prior to capturing and analyzing the phenomenon of news article title
% changes, we first had to identify a set of news publishers to collect
% article information from, that represent a broad range of audiences,
% biases, and platforms. Doing so would allow us to determine which kinds
% of news agencies are more likely to modify the titles of the articles
% they publish.

\begin{figure*}[ht]
\centering
    \includegraphics[width=0.95\textwidth]{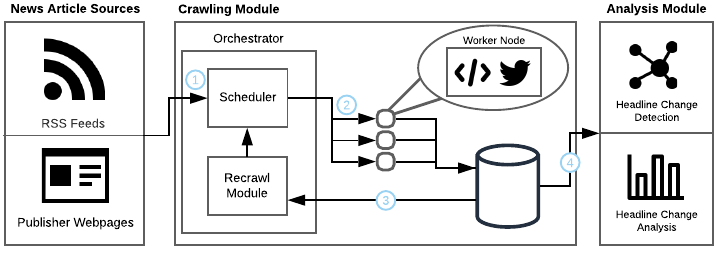}
    \caption{The overview of news crawling infrastructure}\label{fig:infraArchitecture}
\end{figure*}

Rather than utilize the reputability ranking of each
publisher on this chart, we opt instead to measure reputability with the 
Alexa ranking~\cite{alexarankings} of each publisher's website. We reason that the Alexa ranking
of each publisher is an effective and unbiased proxy for a publisher; with
more reputable publishers drawing a larger audience and higher rank. We group
publishers into buckets representing different Alexa rank ranges aiming for 
an equal number of publishers from each bucket. 
Figure \ref{fig:newsPublishers} shows the news publishers we chose, and 
their positions in the two-dimensional plane we described.

\subsection{Data Collection Infrastructure}

% To this end, we choose news sources based on the Media Bias Chart created
% and maintained by Ad Fontes Media~\cite{mediaBiasChart}. This chart maps
% news publishers on a 2D-axis representing political bias (from left to right), 
% and reputability. For our purposes, we focus only on the political bias 
% of each publisher, selecting an equal number of publishers from both left
% and right political biases. Rather than use the reputability ranking of each
% publisher from this chart, we opt instead to measure reputability by the Alexa
% ranking of each publisher website. We reason that the reputability of each
% publisher will be represented by it's Alexa rank, with more reputable websites
% bubbling up the rankings over time. We group news publisher websites into 
% buckets depending on their Alexa ranking. Figure~\ref{fig:newsPublishers}
% shows the news publishers we chose, and their positions in the 2D grid we
% described.

% In addition to selecting news publisher to study, we also choose which article
% categories to collect in order to study the effect of topic on title change
% frequency. Thus, we decide upon three categories we hypothesize will
% contain the greatest number of title changes: Politics, Business, Entertainment,
% and Technology; and a control group that we do not expect to contain many title
% changes, but will be used to compare to the other topic groups: Family.

To measure the extent in which news publishers modify the titles of their articles
post-publication, we created a data collection infrastructure that visits the
web pages of news articles as they are posted and continually monitors them 
to detect changes to their titles and content. Additionally this infrastructure monitors the spread of news on Twitter.
%Additionally this infrastructure monitors the spread of misinformation on Twitter due to the title changes of linked articles.

\noindent\textbf{Post-publication Article Headline Changes} Figure~\ref{fig:infraArchitecture} provides an overview of our infrastructure.
(i) News article URLs are collected using a set of input modules corresponding to the
source in which the URLs are gathered from. For all news publishers we study, URLs
are collected using the RSS feeds provided by each publisher. However, our
infrastructure can be easily extended to support additional URL sources, such as
scraping the home page of a news publisher, by creating a new module for that source.
Our RSS module queries the RSS servers of each news publisher periodically to gather the
URLs of new articles shortly after they are posted.

The URLs for each new article are placed onto a queue in the article crawling module (ii)
where workers consume new URLs by visiting each URL and parsing metadata from the article
web page. We collect and save each HTML web page for future processing, but specifically
parse out the article title using the Python Newspaper library~\cite{python-newspaper}.

In order to detect title changes in news articles, (iii) we recrawl each article web page periodically for the two days following its original publication and parse out the same information as previously described. We arrived at this two-day threshold via a pilot crawling experiment where we established that any headline modifications were typically occurring in the first few hours after an article was published.

\noindent\textbf{Measuring the Spread of News on Twitter} In many cases, modifications to news articles occur \emph{after} these articles have already been consumed and shared. Social media platforms only exacerbate this problem due to the speed in which information propagates among users. 
We chose to study the propagation of news on Twitter because, unlike other social media platforms, Twitter prevents users from editing tweets after they are posted, leading to discrepancies in the information shared in the tweets of news agencies, and their related articles.
When crawling an article found to have a modified title, our infrastructure also searches the Twitter feed of the
relevant news publisher for a tweet containing a link to the article in question. If such a tweet is discovered, it is
recrawled at the same interval as the article itself. For each crawl of a tweet, our infrastructure records the tweet text, as well as the current number of favorites and retweets.
To measure the base level of interaction tweets from each news publisher receive, our data collection infrastructure also crawls a random sample of tweets from each publisher.
%Our infrastructure periodically crawls a random sample of tweets from each news publisher's twitter accounts, the crawler records the tweet text, along with the current number of favorites and retweets. 

%From that point on, each time the article is re-crawled, the tweet associated with that article is re-crawled as well.

% Thus, when the article crawling module detects that an article has produced a title change, 
% the article URL is added to a queue in the Twitter crawling module (iv). This module searches
% the timeline of the Twitter account associated with the article's news publisher and associates
% a tweet to the article if it contains a link to that article. The crawler records the tweet text,
% along with the current number of favorites and retweets. From that point on, each time the article
% is re-crawled, the tweet associated with that article is re-crawled as well.

% In addition to crawling the tweets associated with articles that had a title change, our
% infrastructure also periodically crawls a random sample of tweets from each news publisher's twitter
% accounts that are not associated with title changes. This group of tweets serves as a control group
% in which we can use to compare to the reach of tweets containing misinformation.

% \input{chapter3-TitleModification/title_change}

% tell total tracked number by each agency. (table of total  and sub-total )
% how long is a news title? (histogram)
% how many changed pairs detect? how many unique title changes? (combine with the table above)
% how about edit distance by news agency and overall. 
\section{Post-Publication Headline Changes}\label{sec:results}

Using our data collection infrastructure, we monitored 411,070 articles published by the news agencies listed in Figure~\ref{fig:newsPublishers}
from \dataCollectionStart{} to \dataCollectionEnd{}.
In total, we observed \numArticlesWithTitleChange{} (7.5\%) articles modify their headlines at least once, resulting in \datasetSize{} changed pairs\footnote{We consider one pair as two consecutive headline versions. Additionally, we list news categories of changed pairs per agency in Table \ref{tab:in_changed_news_cate_ratio_agency} in Appendix}. 
Figure~\ref{fig:news_edit_dist_change_timing} shows that over 90\% of changes occur within the first 10 hours after publication.
Readers who encounter an article shortly after publication will likely come away with a different opinion on the subject-matter than a reader
who encounters the same article after the headline has changed. Furthermore, if the headline change is to correct invalid information,
this can lead to rapid spread of misinformation through channels such as social media. In this section, we explore the magnitude of headline
changes in our dataset, assign labels to these changes corresponding to the nature of the modification, and compare the consistency of
headlines published by popular news outlets.
%Additonally, we present the percentage of the tracked news category and modification ratio in Table \ref{tab:in_changed_news_cate_ratio_agency} in appendix.

%In the following section, we discuss the general characteristics post-publication headline modifications, including the

%\begin{figure}[htbp]
%    \centering
%    \includegraphics[width=0.5\textwidth]{Figures/fig_token_length_hist.pdf}
%    \caption{Length of the news titles (in token). Histogram (\textit{left}) shows that most titles are around 10-token long. CDF(\textit{right}) shows that approx. 95\% titles are shorter than %20 tokens.  }
%    \label{fig:news_title_length}
%\end{figure}

\begin{figure}[h]
    \centering
    \includegraphics[width=1.0\columnwidth]{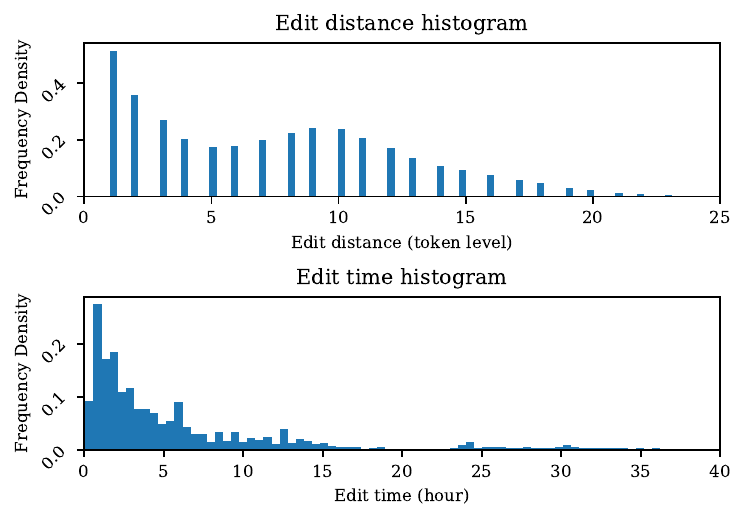}
    \caption{Edit distance histogram (\textit{top}): a large fraction of edits are local edits with distance 1 to 3. Timing of title changes (\textit{bottom}): Most title changes occur within 10 hours after initial publishing. We present plots per publisher in Figure~\ref{fig:edit-time-per-agent} in the appendix.}
    \label{fig:news_edit_dist_change_timing}
\end{figure}

% assign credibility to the news agency by what kind of editing and how often they apply? 
% correlation with some public news agency credibility score? 
\subsection{Headline Change Magnitude}
\label{edit-dis-1}

The inclusion, exclusion, or modification of a single word or phrase in a headline can lead to large differences in reader perception. 
Figure~\ref{fig:news_edit_dist_change_timing} shows the edit distance of headline modifications we detected during our data collection period. 
In the context of headline modifications, we define a token as a single word. We observe a bi-modal distribution in our dataset, with a majority of headline changes resulting in an edit
distance of a single token, and a second local peak at approximately 10 tokens. Since we expect headline modifications in these two groups to be the result
of distinct contributing factors, we discuss them separately.%\todo{add other category}

%d1-only  (6036, 48)
%d1-replace-only: (3767, 48)

\noindent\textbf{Single-token Edit Distance.}
During our data-collection period, we observed 6,036 single-token edit changes. Among them, 3,767 are word substitution. Table~\ref{table:d1_pos_tag_change} shows the top five most common part-of-speech (POS) changes in headlines . 
We find that headlines with single-token substituted typically correspond to either updates of ongoing events, or error corrections. 

\begin{table}[ht]
\caption{The top-5 changes in part of speech}
    \label{table:d1_pos_tag_change}
    \centering
    \scalebox{1.0}
{
\begin{tabular}{l|r}
\toprule
POS Changes &  \# \\
\midrule
(NOUN, NOUN)   &         861 \\
(VERB, VERB)   &         573 \\
(NUM, NUM)     &         396 \\
(PUNCT, PUNCT) &         242 \\
(ADJ, ADJ)     &         216 \\
Others           &       1,479\\
% (PART, PART)   &         183 \\
% (PROPN, PROPN) &         172 \\
% (ADP, ADP)     &         118 \\
% (NOUN, VERB)   &          90 \\
% (ADJ, NOUN)    &          73 \\
% (VERB, NOUN)   &          69 \\
% (AUX, AUX)     &          69 \\
% (NOUN, PROPN)  &          59 \\
% (NOUN, ADJ)    &          51 \\
% (PROPN, PRON)  &          44 \\
\bottomrule
\end{tabular}}
\end{table}

Specifically for word substitution, 44.1\% of noun changes and 54.6\% of verb changes are synonyms, hyponymns, hypernymns, or otherwise share the same lemma.
Tables~\ref{tab:noun_changes} and~\ref{tab:verb_changes} in the appendix list these breakdowns for nouns and verbs, respectively. These modifications typically occur at the end of a newsworthy event; with modified verbs transitioning from present to past tense, indicating the end of the event.
Similarly, changes in headlines correspond to updates in ongoing events (e.g. score changes in sporting events). We observe that
68.39\% of numerical headline changes result in the number increasing by some value (e.g. reflecting more discovered injuries/casualties from an ongoing 
natural disaster). Table~\ref{tab:numerical_changes} in the appendix shows the full breakdown of all numerical changes we observed.

% 396 numbers -> 231 recognizable -> 158 increased

\begin{figure}[ht]
    \centering
    \includegraphics[width=1.0\columnwidth]{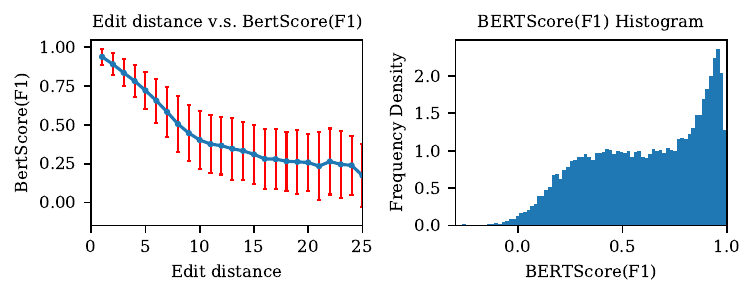}
    \caption{Figure \textit{(left)}: BERTScore decreases as more edits being applied until the  pre/post-edit titles become irrelevant (BERTScore reaches to 0).  Figure \textit{(right)} shows that there is only few complete rewrite in news titles. }
    \label{fig:bertscore_stats}
\end{figure}

\noindent\textbf{Arbitrary Edit Distance.}\label{edit-dis-arbitrary}
To model headline modifications of arbitrary edit distance, we compute the difference of lexicon in changed pairs to distinguish whether it is a minor update or a complete rewrite. 
BERT-based word alignment score --\textit{BERTScore}~\citep{zhang2019bertscore} as defined in Eq.\ref{eq:bertf1} -- is a lexical alignment statistic using contextual word embeddings with attention mechanism. BERTScore can capture the semantic similarity in two sentences of dissimilar structures and lexicons, whereas an edit distance only measures superficial lexical differences.
We extend our analysis to all changed pairs with $BERTScore_{F1}\in [-1,1]$ as the semantic similarity measure. 
The relationship between edit distance and $BERTScore_{F1}$ is shown in Figure~\ref{fig:bertscore_stats}~\textit{(left)}.
%The overall statistics of $BERTScore_{F1}$ is shown in Figure \ref{fig:bertscore_stats} \textit{(left)}.

\begin{equation}
\begin{aligned}
    R = \frac{1}{|x|} \sum_{x_i \in x} {max_{\hat{x_j} \in \hat{x} } x_i \hat{x}_j  } &;\ \ 
    P = \frac{1}{|\hat{x}|} \sum_{\hat{x_j} \in \hat{x}} {max_{x_i \in x } x_i \hat{x}_j  } \\
    BERTScore_{F1} &=  2\frac{P*R}{R+P} \in [-1, 1],  
\end{aligned}
\label{eq:bertf1}
\end{equation}  where $x_i,\hat{x_i}$ denote $i^{th}$ token in two sentences $x,\hat{x}$.

% \begin{figure}[htbp]
%     \centering
%     \subfloat[BERTScore vs. Edit Distance.\label{fig:bertscore_lev_dist}] {\includegraphics[width=0.126\textwidth]{Figures/figure_raw_feature_description_1_2_3_4_5_6_7_8_9_10_11_12_13_14_15_16_17_18_19_edit_bertscore_errorbar.pdf}}\hfill
%     \subfloat[PDF \label{fig:bertscore_hist}] {\includegraphics[width=0.13\textwidth]{Figures/figure_raw_feature_description_1_2_3_4_5_6_7_8_9_10_11_12_13_14_15_16_17_18_19_BERTScore_hist.pdf}}\hfill
%     \subfloat[CDF\label{fig:bertscore_cdf}] {\includegraphics[width=0.13\textwidth]{Figures/figure_raw_feature_description_1_2_3_4_5_6_7_8_9_10_11_12_13_14_15_16_17_18_19_BERTScore_cdf.pdf}}\hfill
%     \caption{Semantic similarity between changed title pairs. The peak in Figure \ref{fig:bertscore_hist} shows that the most common changes are minor updates. Figure \ref{fig:bertscore_cdf} shows that there is only few complete rewrite in news titles. }
%     \label{fig:bertscore_stats}
% \end{figure}

%Figure \ref{fig:bertscore_lev_dist} shows the semantic similarity (BERTScore) against word-level edit distance. As we expected, semantic similarity keeps decreasing as more edits are applied. 
Figure~\ref{fig:bertscore_stats}~\textit{(right)} shows the distribution of BERTScore where the peak around 0.9 indicates that most edits do not change the semantic meaning of the headline, 
%(like the examples in Table \ref{tab:numerical_changes}), 
while the samples centered around 0 represent the post-publication edit completely changing the headline's semantic meaning. We present examples associated with BERTScore in Table~\ref{tab:category-example} of the appendix.
Using a BERTScore threshold of 0.25, we find approximately 10\% of the changed pairs in our dataset are significantly rewritten. 
%More examples of various BERTScore is in Table \ref{tab:category-example} in Appendix.
%\footnote{We present examples of various BERTScore in Table \ref{tab:category-example} in Appendix.}
We consider these headline changes to be the most damaging to readers as it is very likely that the message conveyed by the headline will change drastically after modification.

\begin{table}[ht]
    \centering
    \caption{The statistics of semantic similarity (BERTScore) by agency (Ranked by the mean value). To clarify the similarity rankings from news reputation rankings, the ordered list does not reflect credibility. \textit{Tracked} is the total number of tracked articles, \textit{Mod. Ratio} is the percentage of the articles with at least one altered headline.}
    \label{tab:bertscore_agency}
    \footnotesize
%\hspace*{-25pt}
% copy from: tab_bertscore_by_agency.tex
    \begin{tabular}{lrrrr}
    \toprule
    Agency &   Mean &     Median &    Tracked &  Mod.Ratio \\
    \midrule
    Huffington Post & 0.8292 &  0.8966 &    4,875 &         0.0357 \\
    The Epoch Times & 0.8260 &   0.8908 &    4,765 &         0.1442 \\
    The Hill        & 0.8087 &   0.8644 &    5,676 &         0.0592 \\
    Fox News        & 0.7659 &   0.8645 &   11,003 &         0.0486 \\
    New York Post   & 0.7305 &   0.8174 &   13,881 &         0.0665 \\
    National Review & 0.7268 &   0.8305 &    3,324 &         0.0126 \\
    The Blaze       & 0.7200 &   0.8145 &    4,401 &         0.0400 \\
    CNN             & 0.6888 &   0.7493 &    6,344 &         0.2030 \\
    Washington Post & 0.6840 &   0.7916 &    6,076 &         0.1508 \\
    Newsmax         & 0.6733 &   0.7319 &    1,115 &         0.0233 \\
    Daily Beast     & 0.6584 &   0.7693 &    4,359 &         0.1308 \\
    MarketWatch     & 0.6551 &   0.7011 &    7,503 &         0.3453 \\
    BBC             & 0.6031 &   0.6260 &    5,854 &         0.2277 \\
    Yahoo News      & 0.5985 &   0.6160 &  227,246 &         0.0846 \\
    Daily Mail      & 0.5936 &   0.6348 &   75,118 &         0.0756 \\
    OAN             & 0.5830 &   0.5910 &    8,323 &         0.2616 \\
    The Guardian    & 0.5579 &   0.5826 &    6,916 &         0.1975 \\
    New York Times  & 0.5234 &   0.5595 &    9,746 &         0.1892 \\
    BuzzFeed        & 0.5085 &   0.4872 &    4,545 &         0.4438 \\
    \bottomrule
\end{tabular}
\vspace{-2ex}
\end{table}

\xingzhiswallow{
\begin{table}[t]
\small
    \centering
    \caption{The statistics of semantic similarity (BERTScore) by agency (Ranked by the mean value). To clarify the similarity rankings from news reputation rankings, the ordered list does not reflect credibility.}
    \label{tab:bertscore_agency}
\hspace*{-25pt}
% copy from: tab_bertscore_by_agency.tex
    \begin{tabular}{lrrrrr}
    \toprule
    Agency &   Mean &   Std. &  Median &    Tracked &  Mod.Ratio \\
    \midrule
    Huffington Post & 0.8292 & 0.1730 &  0.8966 &    4,875 &         0.0357 \\
    The Epoch Times & 0.8260 & 0.1740 &  0.8908 &    4,765 &         0.1442 \\
    The Hill        & 0.8087 & 0.1826 &  0.8644 &    5,676 &         0.0592 \\
    Fox News        & 0.7659 & 0.2310 &  0.8645 &   11,003 &         0.0486 \\
    New York Post   & 0.7305 & 0.2424 &  0.8174 &   13,881 &         0.0665 \\
    National Review & 0.7268 & 0.2660 &  0.8305 &    3,324 &         0.0126 \\
    The Blaze       & 0.7200 & 0.2758 &  0.8145 &    4,401 &         0.0400 \\
    CNN             & 0.6888 & 0.2531 &  0.7493 &    6,344 &         0.2030 \\
    Washington Post & 0.6840 & 0.2778 &  0.7916 &    6,076 &         0.1508 \\
    Newsmax         & 0.6733 & 0.2714 &  0.7319 &    1,115 &         0.0233 \\
    Daily Beast     & 0.6584 & 0.2915 &  0.7693 &    4,359 &         0.1308 \\
    MarketWatch     & 0.6551 & 0.2608 &  0.7011 &    7,503 &         0.3453 \\
    BBC             & 0.6031 & 0.2671 &  0.6260 &    5,854 &         0.2277 \\
    Yahoo News      & 0.5985 & 0.2801 &  0.6160 &  227,246 &         0.0846 \\
    Daily Mail      & 0.5936 & 0.2741 &  0.6348 &   75,118 &         0.0756 \\
    OAN             & 0.5830 & 0.2590 &  0.5910 &    8,323 &         0.2616 \\
    The Guardian    & 0.5579 & 0.2896 &  0.5826 &    6,916 &         0.1975 \\
    New York Times  & 0.5234 & 0.3283 &  0.5595 &    9,746 &         0.1892 \\
    BuzzFeed        & 0.5085 & 0.2529 &  0.4872 &    4,545 &         0.4438 \\
    \bottomrule
\end{tabular}
\vspace{-5ex}
\end{table}
}

%We breakdown the title changes by news agency, and measure which agency changes their news titles more significantly. 
Table~\ref{tab:bertscore_agency} shows the ordered list of news agencies by the average similarity score across all changed headlines. Generally, higher similarity indicates minor changes that do not alter the original semantic meaning. 
%In Table \ref{tab:bertscore_agency}, the "notorious" \textit{BuzzFeed} \footnote{\textit{BuzzFeed} is well-known for eliciting users' 
%attention in a clickbaity way.} is ranked last, which is consistent to our consensus. 
We find many well-regarded publishers (e.g., BBC, The Guardian, NYT) frequently modify headline semantics. This unexpected finding suggests that the popularity of a news outlet does not necessarily indicate greater restraint with post-publication headline changes. Contrastingly, we observe publishers such as The Epoch Times with an Alexa rank of close to ten thousand modify headlines at a much lower rate than publishers with a much larger online presence. 

Additionally, by noting the ratio of modified headlines with the average BERTScore for all modified headlines, we can deduce the
expected consistency of a particular news outlet. For instance, we find that BuzzFeed modified the headlines of 44\% of their articles during our data-collection period. Of those articles, we observe the headline changes produce the lowest average BERTScore (i.e. the modified headlines depart significantly from the original ones). Contrastingly, during our data collection period, The Huffington Post only modified the headlines of 3\% of observed articles; achieving the highest
average BERTScore with those changes. We can therefore conclude that, in terms of headline modifications, The Huffington Post is a more consistent news outlet compared to BuzzFeed. We note however, that this consistency of headlines does not necessarily imply accuracy of headlines.

%Table~\ref{tab:lowbert_example} shows the title examples of low similarity from different agencies. We observe that news agencies change 
%writing style or emphasize different content when rewriting the titles, which may give readers different first impressions. 
% \begin{quote}
% Before: "\textit{Hancock defends India travel ban delay as Covid variant cases mount} "\\
% After: "\textit{Hancock: most Bolton Covid patients eligible for jab but haven’t had it}" \\
% Similarity: 0.067
% \end{quote}
%For example, Although the news topic of the pair from \textit{The Guardian} in Table~\ref{tab:lowbert_example} does not change, the original title emphasizes on the action \textit{defend}, while the edited version emphasizes on vaccination rate. The shift of title
%s focus may greatly change based on editor's judgement. % \todo{add some reference here}.
%Table~\ref{tab:change_category_by_agency} shows the percentage of edit types that each agency applied. We found that BBC has significantly higher ratio of dynamic update, reflecting their strong %ability to report and update.

\subsection{Categorization of News Title Changes}

\begin{table*}[ht]
    \centering
    \scriptsize
    \caption{Examples of the altered titles sampled from our dataset by the modification category. The example of the label "Other" involves out-of-vocabulary words (J.\&J) and changes in tone. The example of "Concision" removes the specific person, while the "Elaboration" example expands the original title with an attributive clause. "Forward-Reference" adds a question mark and "Personalization" substitute reader into the title, which both trigger curiosity from the readers. The "Neutralize" example remove \textit{"long"} from the original title, making it less subjectivity. While "Emotional" example adds \textit{"tease, big"}, which makes the title more inflammatory.  }
    \label{tab:category-example-mainbody}
    \scalebox{1.0}{
% \begin{tabular}{>{\raggedright\arraybackslash}p{0.35\textwidth} >{\raggedright\arraybackslash}p{0.35\textwidth} >{\raggedleft\arraybackslash}p{0.1\textwidth}>{\raggedleft\arraybackslash}p{0.1\textwidth}}
\begin{tabular}{p{0.35\textwidth}p{0.35\textwidth}p{0.07\textwidth}p{0.1\textwidth}}
\toprule
                                                                                        Before &                                                                                                             After &  BERT-F1 &       Label \\
\midrule
One Dose of J.\&J. Vaccine Is Ineffective Against Delta... &	J.\&J. Vaccine May Be Less Effective Against Delta ... & 0.6830 & Other \\
% Nike chief executive says firm is 'of China and for China'	&Nike boss defends firm's business in China & 0.4559 & Other \\
                                           Senator Capito says Republicans plan new US infrastructure offer &                                                                          U.S. Senate Republicans prepare new infrastructure offer &         0.5434 &   Concision \\
                                           Raul Castro confirms he's resigning, ending long era in Cuba &                                                                        Raul Castro resigns as Communist chief, ending era in Cuba &         0.7310 &  Neutralize \\
                                        Fourth stimulus check update: Your next payment could be these &                                                                 Fourth stimulus check? These payments are already in the pipeline &         0.4479 & Forward Ref. \\
                                                        The Latest: UN: 38,000 Palestinians displaced in Gaza &                                                                         The Latest: Biden expresses 'support' for Gaza cease-fire &         0.4043 &  Dyn. Update \\

MAGA 2.0&	“A new bargain”: Biden’s 2024 tease bets big on nostalgia & 0.0257 & Emotional \\
                                                                               23 Amazing Jokes From Hot Fuzz &                                                      23 Jokes From "Hot Fuzz" That Humans Will Laugh At For The Next 10,000 Years &         0.4715 & Elaboration \\
                                            %   Suspect arrested following series of Arizona traffic shootings &                                                     Arizona gunman goes on traffic shooting rampage, leaving one dead, 12 injured &         0.2247 &   Emotional \\
                                                 Watch Jeff Bezos’ ‘Blue Origin’ launch into space live today &                                                Everything you need to know as Jeff Bezos’ ‘Blue Origin’ launches into space today &         0.6179 & Personalize \\
                  Afghan guard killed: Firefight leaves at least one dead and others injured at Kabul airport &                  'It would be better to die under Taliban rule than face airport crush', say US embassy's 'betrayed' Afghan staff &         0.0656 &    Citation \\
\bottomrule
\end{tabular}
}
\end{table*}

% \subsection{Taxonomy of Editing Categories}
To discover trends in article headline changes, and the behaviors of the publishers responsible, we create a nine-class taxonomy of editing types based on journalism knowledge (e.g. Emotionalism, etc.), existing taxonomies for Wikipedia edits\cite{daxenberger2013automatically, yang2017identifying}, and the most common edits from our observations (e.g., Paraphrase, Dynamic update, etc.) . 
We design our taxonomy as follows:
%\textbf{Taxonomy Creation Pipeline}: 
\begin{itemize}
    \item \textbf{Paraphrase:} We assign the label \textit{Paraphrase} if the changed pairs have a similarity score (BERTScore-F1) greater than 0.8.
    \item \textbf{Dynamic Update:} If both old and new title contain keywords such as \textit{Monday brief, update, live, stream, etc}. The keywords were manually selected. 
    \item \textbf{Emotionalism\cite{kilgo2016six}:} if only the news title contains any of words in the subjective dictionary. 
    \item \textbf{Neutralization:} if only the old title contains subjective words mentioned above.
    \item \textbf{Forward Reference:} Forward Reference\cite{blom2015click, kilgo2016six} is a common feature in news titles, piquing reader curiosity. We assign this label only if the changed title contains the keywords such as \textit{why, when, which, how, etc.}
    \item \textbf{Personalization\cite{kilgo2016six}:} Personalization is used to retain audiences and make the readers feel involved in the news. We assign this label only if the new title contains keywords such as \textit{you, we, s/he, your, etc.}
    \item \textbf{Citation:} Citation is a common technique in news headlines that make them look more reliable. We assign this label only if the new title contains keywords like \textit{said, says, told, etc.}
    \item \textbf{Concision:} If the new title removes some text, and the remaining text aligns with the old title (P $< 0.6$, R $>0.5$). BERT-Score Precision/Recall is defined in Eq.\ref{eq:bertf1}
    \item \textbf{Elaboration:} If the new title adds text, and is semantically aligned with the old title: P $> 0.6$, R $< 0.5$). 
\end{itemize}

\begin{figure}[ht]
    \centering
    % \subfloat[Histogram of Editing Categories.  \label{fig:category_stats_1}] {\includegraphics[width=0.2\textwidth]{Figures/figure_raw_feature_description_1_2_3_4_5_6_7_8_9_10_11_12_13_14_15_16_17_18_19_edit_category.pdf}}\hfill
    % \subfloat[Coverage Assigned Category  \label{fig:category_stats_2}] {\includegraphics[width=0.2\textwidth]{Figures/figure_raw_feature_description_1_2_3_4_5_6_7_8_9_10_11_12_13_14_15_16_17_18_19_edit_category_coverage.pdf}}
    \includegraphics[width=0.80\columnwidth]{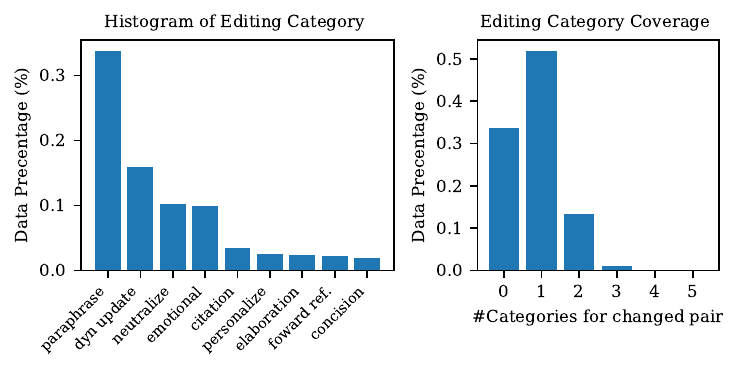}
    \caption{Assigned category statistics. Figure (\textit{left}) shows the majority change types are paraphrase and dynamic update. Figure (\textit{right})  shows that approx. 70\% of samples have at least one category assigned.  50.3\% modifications associate with benign edit categories (paraphrase, updates, elaboration, concision), while 23.13\% modifications are linked with the other less benign edits. }
    \label{fig:category_stats}
\end{figure}

% \hspace*{2cm}
% \clearpage
\begin{table}[htbp]
% \begin{sidewaystable}[ph!]

\centering
    \caption{The breakdown of edit types per news agency. Each headline modification may belong to multiple labels. We highlight the top one in each column.}
    \label{tab:change_category_by_agency}
    \scriptsize
    \scalebox{0.90}{
    \begin{tabular}{lp{0.08\textwidth}p{0.08\textwidth}p{0.08\textwidth}p{0.08\textwidth}p{0.08\textwidth}p{0.08\textwidth}p{0.08\textwidth}p{0.08\textwidth}p{0.06\textwidth}}
    \toprule
    Agency &  Paraphrase &  Dynamic Update &  Elaboration &  Concision &  Emotional &  Neutralize &  Forward Refer. &  Personalize &  Citation  \\
    \midrule
    BBC             &      0.3196 &      0.3848 &       0.0278 &     0.0113 &     0.1043 &      0.0818 &       0.0180 &       0.0188 &    0.0540  \\
    BuzzFeed        &      0.1587 &      0.0203 &       0.0312 &     \textbf{0.1101} &     0.1026 &      \textbf{0.1894} &       \textbf{0.0709} &       \textbf{0.0699} &    0.0238  \\
    CNN             &      0.4262 &      0.0916 &       0.0404 &     0.0148 &     0.0699 &      0.0738 &       0.0171 &       0.0342 &    0.0435 \\
    Daily Beast     &      0.4754 &      0.1035 &       0.0246 &     0.0175 &     0.1123 &      0.0912 &       0.0386 &       0.0351 &    0.0298  \\
    Daily Mail      &      0.2801 &      0.1158 &       0.0357 &     0.0144 &     0.0993 &      0.1218 &       0.0241 &       0.0521 &    0.0422  \\
    Fox News        &      0.6056 &      0.1925 &       0.0187 &     0.0093 &     0.0710 &      0.0617 &       0.0168 &       0.0224 &    0.0262  \\
    Huffington Post &      \textbf{0.7356} &      0.0690 &       0.0000 &     0.0000 &     0.0345 &      0.0632 &       0.0000 &       0.0345 &    0.0230  \\
    MarketWatch     &      0.3867 &      0.0872 &       0.0413 &     0.0178 &     0.0965 &      0.0799 &       0.0363 &       0.0154 &    0.0247  \\
    National Review &      0.5476 &      0.3810 &       0.0476 &     0.0000 &     0.0714 &      0.0476 &       0.0000 &       0.0238 &    0.0000  \\
    New York Post   &      0.5211 &      0.0997 &       0.0336 &     0.0076 &     0.0715 &      0.0618 &       0.0336 &       0.0293 &    0.0141  \\
    New York Times  &      0.2923 &      0.1171 &       0.0108 &     0.0108 &     0.1123 &      0.1432 &       0.0380 &       0.0184 &    0.0331  \\
    Newsmax         &      0.3846 &      0.3077 &       0.0385 &     0.0000 &     0.0000 &      0.0769 &       0.0000 &       0.0000 &    0.0385  \\
    OAN             &      0.2692 &      0.0156 &       0.0280 &     0.0184 &     0.1029 &      0.0988 &       0.0032 &       0.0023 &    0.0271  \\
    The Blaze       &      0.5341 &      0.3239 &       \textbf{0.0625} &     0.0227 &     0.0682 &      0.0398 &       0.0227 &       0.0398 &    0.0455  \\
    The Epoch Times &      0.6856 &      0.1659 &       0.0146 &     0.0204 &     0.0277 &      0.0422 &       0.0044 &       0.0087 &    0.0364  \\
    The Guardian    &      0.2950 &      \textbf{0.4100} &       0.0271 &     0.0124 &     \textbf{0.1215} &      0.1164 &       0.0190 &       0.0190 &    \textbf{0.0688} \\
    The Hill        &      0.6637 &      0.2024 &       0.0268 &     0.0327 &     0.0208 &      0.0446 &       0.0089 &       0.0060 &    0.0179 \\
    Washington Post &      0.4891 &      0.2205 &       0.0186 &     0.0109 &     0.0884 &      0.0797 &       0.0207 &       0.0437 &    0.0371  \\
    Yahoo News      &      0.3256 &      0.1846 &       0.0150 &     0.0126 &     0.1047 &      0.0977 &       0.0156 &       0.0161 &    0.0340  \\
    \bottomrule
    \end{tabular}
    }
% \end{sidewaystable}
\end{table}

% \clearpage

These nine categories allow us to better understand headline modifications by classifying them into groups corresponding to their perceived purpose. 
With the combination of BERTScore, Stanza Pipeline~\cite{qi2020stanza}, word sentiment dictionary~\cite{riloff2003learning}, 
and hand-crafted rules,  we are able to automatically assign labels using our proposed NLP pipeline. We empirically determined the thresholds to use 
by finding those that made the most sense with manual inspection.  
Table~\ref{tab:category-example} of the appendix present more examples of each of these categories.
%Figure~\ref{fig:category_stats} show the histogram and coverage of the taxonomy. Each sample may have multiple categories if the headline meets the criteria for more 
%than one category (e.g., Elaboration and Emotionalism).

Figure~\ref{fig:category_stats}~\textit{(right)} shows the resulting label coverage. Our proposed categorization rules cover approximately 70\% of news title changes (the changed pair has at least one assigned category). 
Among the 30\% of modifications not covered by our pipeline, we find that the missing cases usually have a BERTScore of less than 0.8, involve more complex sentence structure changes (double negation), or out-of-vocabulary words (most of them are COVID-related “COVID-19, J.\&J., AstraZenec”, which debuted after the release of the pre-trained model). 
We listed several examples in Table \ref{tab:category-example-mainbody} and more examples in Table \ref{tab:category-example} in the Appendix.

% In Table , we show two examples of the label 'Other'.
%\todo{Why can't we categorize the other headlines?}.

 We also find that the most common changes belong to the ``Paraphrase'' and ``Dynamic Update'' categories (44.24\%), which do not significantly change the semantic meaning of the headlines. Changes in these categories typically correspond to updates of ongoing events and grammatical/spelling corrections. However, we note that all other 
categories in our taxonomy may contain headline changes that can be perceived as harmful to at least some readers.
%['lb_paraphrase_indicator', 'lb_dyn_update_indicator']
%18540 (41906,) 0.44241874671884696

Over the course of our data collection period, we observed an equal distribution of ``Emotionalism'' and ``Neutralization'' headline changes, 
suggesting that publishers are just as likely to add emotional words to attract more readers after initial publication as they are to remove them.
However, since both of these categories transition the headline from a provocative to non-provocative state, or vice-versa, these two groups can have
the same effect on different groups of readers. In the case of ``Emotionalism'' changes, readers who view a headline after it has been modified, will 
likely have a more emotional response to the subject-matter, while ``Neutralization'' changes will have the same effect on readers who view a headline pre-modification.
Together, these two groups make up 20.13\% of all headline changes in our dataset.

%['lb_emotionalism_indicator', 'lb_neutralization_indicator']
%8437 (41906,) 0.2013315515677946
Table~\ref{tab:change_category_by_agency} shows the percentage of articles by each publisher belonging to each of the nine categories in our taxonomy. 
Since it reflects the explanation of headline changes, the percentage does not reflect the general media bias, which may be embedded in the static news headlines. 
%\todo{What does this sentence mean? I thought these percentages have nothing to do with semantic similarity}
We observe that over 70\% of the articles published by the Huffington Post, The Epoch Times and The Hill belong to either the ``Paraphrase'' or ``Dynamic Update'' categories, demonstrating that their headlines will typically not change 
in ways which will lead to drastically different opinions among readers. We also notice that 
\textit{BuzzFeed} leads in three change categories, including the categories most closely related to click-bait: ``Forward Reference'', and ``Personalization''. 

% ['lb_paraphrase_indicator', 'lb_dyn_update_indicator']
% BBC 753 1333 0.5648912228057015
% BuzzFeed 347 2017 0.17203767972235995
% CNN 597 1288 0.4635093167701863
% Daily Beast 294 570 0.5157894736842106
% Daily Mail 2013 5681 0.354339024819574
% Fox News 358 535 0.6691588785046729
% Huffington Post 133 174 0.764367816091954
% MarketWatch 1112 2591 0.4291779235816287
% National Review 28 42 0.6666666666666666
% New York Post 512 923 0.5547128927410617
% New York Times 720 1844 0.39045553145336226
% Newsmax 15 26 0.5769230769230769
% OAN 601 2177 0.2760679834634819
% The Blaze 115 176 0.6534090909090909
% The Epoch Times 503 687 0.7321688500727802
% The Guardian 856 1366 0.6266471449487555
% The Hill 243 336 0.7232142857142857
% Washington Post 567 916 0.618995633187773
% Yahoo News 8773 19224 0.4563566375364128

By observing the categorizations of each publisher, a model can be developed on their overall headline-modification strategies. For instance, the ``Citation'' category
of our taxonomy allows us to deduce that \textit{The Guardian} is the most likely to include witness or expert testimony in headlines post-publication, with 6.8\% of their articles falling into that category.
Conversely, we find that the publisher National Review is very unlikely to make such a headline change as we did not observe any instances of modification by either publisher during our data collection period. These observations quantify the regularity with which publishers modify their headlines, and can be used as a supplementary feature for future research on digital journalism and public trust in news.

\subsection{News Propagation over Twitter}

In recent years, social media services have emerged as a vital tool for news publishers to quickly share stories with the public. Due to the ability of
users to share and ``repost'' links to articles with others connected to them online, information can spread faster than ever before. Because of this 
user-assisted amplification, false or misleading information in article headlines can propagate to exponentially many readers before headline changes 
occur, magnifying the negative repercussions of unanticipated headline changes. 

We track the reachability of news on Twitter in terms of user engagement. As mentioned in Section~\ref{sec:data_prepare}, our data-collection infrastructure 
records the number of likes and retweets of tweets associated with news articles with modified headlines from the accounts of each publisher. In total,
we detect 5,384 tweets corresponding to articles with modified headlines during our data collection period. We note that these tweets correspond to the
original article headlines, with their immutability preventing publishers from simply editing their contents to reflect headline changes. Rather,
publishers must first delete the original, outdated tweet, and replace it with a new tweet. Alarmingly however, we find that only 0.15\% of all such tweets 
were deleted subsequent to a headline change in the corresponding article. Moreover, we find that these tweets on average garner 39 retweets and 137 favorites. 
This means that the vast majority of tweets associated with outdated and potentially misleading information remain online indefinitely, and receive considerable
attention from users who then spread this information to others.

%In 
%Figure~\ref{fig:news_twitter_prop_time}, red curves individually represent the normalized reached population~\footnote{Scaled by the maximally 
%reached population of each news article on Twitter} versus elapsed time (hour) after the initial release. We plot the fitted function in blue solid line.
%We use a sigmoid-like function to fit the curve as plotted in blue.

Figure~\ref{fig:news_twitter_prop_time} shows the distribution of time in which tweets associated with headline changes were ``retweeted'' and ``favorited''
by users.
We observe a clear trend that news stories gradually lose popularity on social networks as they become less novel.
People usually share (retweet) tweets associated with news stories within the first ten hours and react (favorite) within first five hours. 
Based on the observations in Figure~\ref{fig:news_edit_dist_change_timing}, we found that almost 80\% title changes happened within 
the first five hours after publishing. These findings are alarming as they show that many headline changes occur \emph{after} most
social media activity regarding the particular article ceases. This means that social media users are currently either spreading false or 
misleading information before the publisher can remedy it by modifying the headline, or sharing articles that will portray an entirely
different emotional response when their followers view it at a later point in time. Both of these scenarios can lead to situations 
where different readers of the same article or headline will come to different conclusions depending on when they encountered it.

\xingzhiswallow{

\begin{figure}[t]
    \centering
    % \subfloat[retweet vs time] {\includegraphics[width=0.2\textwidth]{Figures/retweet_vs_time.pdf}}
    % \subfloat[likes vs time] {\includegraphics[width=0.2\textwidth]{Figures/like_vs_time.pdf}}
    \includegraphics[width=0.95\columnwidth]{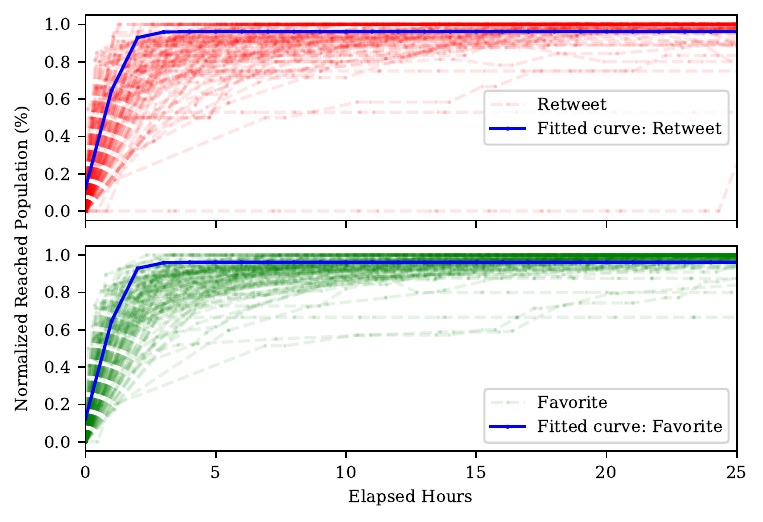}
    \caption{The normalized retweets and likes vs news published time (red curves). The blue curve is the fitted sigmoid-like curve (Eq.\ref{eq:sigfit}), approximating how quick the news achieves its max influence. }
    \label{fig:news_twitter_prop_time}
    %The learnt parameters of retweet-model is [0.9611 2.6353 0.7282].
\end{figure}

}

\begin{figure}[!t]
    \centering
    %\vspace{-5mm}
    % \subfloat[retweet vs time] {\includegraphics[width=0.2\textwidth]{Figures/retweet_vs_time.pdf}}
    % \subfloat[likes vs time] {\includegraphics[width=0.2\textwidth]{Figures/like_vs_time.pdf}}
    \includegraphics[width=1.0\columnwidth]{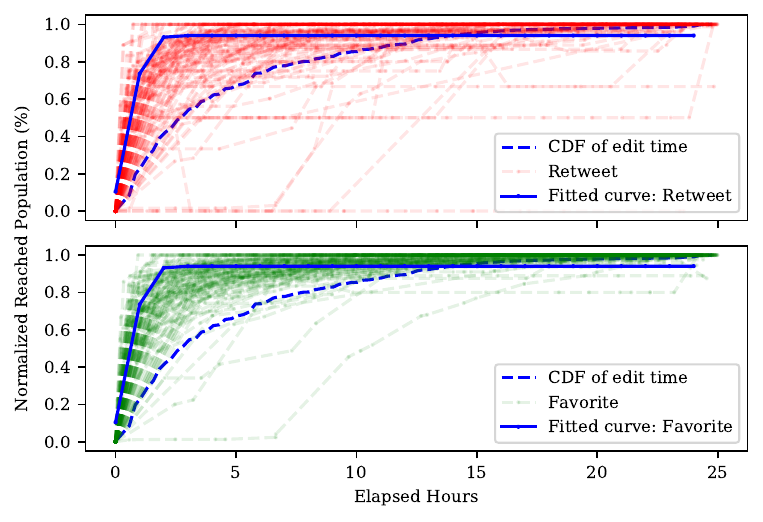}
    \vspace{-4mm}
    \caption{The normalized retweets (top) and favorites (bottom) V.S. news published time. The solid blue curve is the fitted curve 
    $y = a \cdot (1+e^{-b(x-c)})^{-1}$
 , approximating how quick the news achieves its max influence. }
 %We find the tweets generally reach its maximal audience before the most post-edits happen, demonstrating the potential risk of misleading information propagation.  }
    \label{fig:news_twitter_prop_time}
    \vspace{-3mm}
    %The learnt parameters of retweet-model is [0.9611 2.6353 0.7282].
\end{figure}

\begin{figure}[t]
     \centering
     \subfloat{\includegraphics[width=7cm]{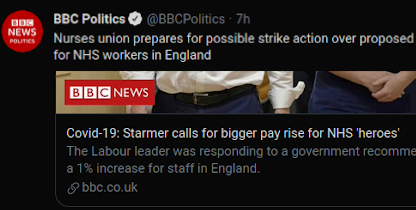}}
    %  \begin{subfigure}{}
    %  {\includegraphics[width=7cm]{Figures/twitterScreenshot-crop-short-2.png} }
    %  \end{subfigure}
    \quad%
    \subfloat{{\includegraphics[width=7cm]{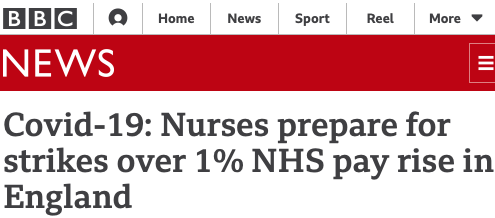}}}
    %  \begin{subfigure}{}
    %  {\includegraphics[width=7cm]{Figures/bbcScreenshot-crop-short.png} }
    %  \end{subfigure}
     \caption{Twitter URL preview of article posted by BBC Politics. Article headline outdated on Twitter due to caching behavior of URL cards.}
     \label{fig:tweetCardScreenshot}
 \end{figure}

\noindent\textbf{URL Preview Cards} A popular feature of Twitter is the rendering of 
cards that preview the content located at the URL linked in a tweet. These cards work
by parsing and rendering the Open Graph~\cite{openGraph} metadata from the HTML source of the linked webpage 
directly below the tweet text. Typically, a webpage thumbnail, title, and description are rendered.
This metadata is cached for each particular URL for approximately one week; only updating when the URL
is modified~\cite{twitterPreviewCard}. This caching behavior can amplify the negative repercussions of
post-publication headline changes as any tweet containing the URL of an article that was published prior
to a headline change will display the outdated headline for at least one week. This is a behavior we
have observed on the Twitter accounts of popular news publishers. Figure~\ref{fig:tweetCardScreenshot}
demonstrates an example of this behavior with a headline change from the BBC, along with a tweet containing
a link to the article posted by the official BBC Twitter account. Twitter users who only read the headline and description
located on the rendered card will consume outdated information on the situation, as opposed to those who click through to the
article webpage.

%  \begin{figure}[t]
%     \vspace{-0mm}
%      \centering
%      \subfloat{{\includegraphics[width=7cm]{Figures/twitterScreenshot-crop-short-2.png} }}
%      \quad
%      \subfloat{{\includegraphics[width=7cm]{Figures/bbcScreenshot-crop-short.png} }}
%      \vspace{-0mm}
%      \caption{Twitter URL preview of article posted by BBC Politics. Article headline outdated on Twitter due to caching behavior of URL cards.}
%      \label{fig:tweetCardScreenshot}
%      \vspace{-5mm}
%  \end{figure}

%\vspace{-2cm}
% \begin{figure}[t]
%     \centering
%     \subfloat{{\includegraphics[width=5cm]{Figures/twitterScreenshot.png} }}
%     \qquad
%     \subfloat{{\includegraphics[width=5cm]{Figures/bbcScreenshot.png} }}
%     \caption{Twitter URL preview of article posted by BBC Politics. Article headline outdated on Twitter due to caching behavior of URL cards.}
%     \label{fig:tweetCardScreenshot}
%     \vspace{-5mm}
% \end{figure}

%Observation I: The information spreads fastest via retweet within the first 10 hours. \\
%Observation II:  The number of likes usually saturates within the first 5 hours.
% \input{chapter3-TitleModification/discussion}

% \input{chapter3-TitleModification/conclusion}

\section{Conclusion}
\label{sec:conclusion} 
In 2022, trust in the media is at an all-time low. In this paper, we explored one dimension that we argue has the potential to further reduce the public's trust in news outlets: post-publication headline changes. By monitoring over 411K articles for seven months across tens of news outlets, we discovered that 7.5\% of titles changed at least once after they were published. This rate of headline modifications is anything but uniform, with certain popular outlets changing almost half of their headlines \emph{after} publication. We used the BERTScore metric and devised a taxonomy to automatically characterize the type of post-publication change. 
We find that 49.7\% of changes go beyond benign corrections and updates, with 23.13\% corresponding to categories such as emotionalism, neutralization, and personalization.
%Even though the majority of changes (50.3\%) correspond to clarifications, dynamic updates of breaking news, and other benign categories, we find that 23.13\% of the modifications belong to categories that are significantly less benign, such as, emotionalism, neutralization, and personalization. 
We also characterized the effects of post-publication headline changes in relation to social media. Among others, we discovered that maximal spread of news on Twitter happens within ten hours after publication and therefore a delayed headline correction will come \emph{after} users have consumed and amplified inaccurate headlines. Finally, we discussed the issue of content caching in social networks and how it can further exacerbate the propagation of stale headlines.

\noindent\textbf{Acknowledgements:} This work was supported by the National Science Foundation (NSF) under grants CMMI-1842020, CNS-1941617, CNS-2126654, IIS-1926781, IIS-1927227, IIS-1546113 and OAC-1919752.

\noindent\textbf{Availability} To assist in the understanding of the spread of misinformation on the web, we are open-sourcing our news headline dataset: \url{https://scripta-volant.github.io/}

%Additionally, we will continue monitoring the news articles and report on future headline changes. This, along with our dataset, will be available post-publication.
%can be found on our website: %\url{https://verbavolant.github.io}\todo{Put real URL here}

\section{Discussion}\label{sec:discussion}

\noindent\textbf{Headline Changes.} The Internet has dramatically
decreased the time between newsworthy events, and the consumption of
information regarding those events by the public. This, along with the
propensity of people to simply skim headlines rather than read full articles~\cite{readheadlines}, 
has fostered an environment where post-publication changes to
article headlines can lead to diverging world-views among readers.
In this work, we studied the behavior of news publishers and
developed a taxonomy of article headline modifications. We find that
popular news publishers regularly make post-publication changes to their
headlines, with some modifying almost half of them. 

Additionally, we categorize these changes
into well-defined groups, allowing us to quantify the behaviors of each
publisher. We find that the majority of headline changes correspond to
updates of ongoing events. Alarmingly however, 20.13\% of headline changes are made to
either increase or decrease the provocativeness of the headline, presenting
a distorted view of the subject-matter to different groups of readers. Using
this taxonomy, we are able to quantify the different headline update strategies
of each news outlet studied, discovering a divergence in the perceived motives
between publishers.

While the news outlets presented in this study do not appear to have outright
malicious intentions, their actions contribute to the overall decrease in integrity
of information consumed by millions of people. 
In order to slow the spread of misinformation in society, news outlets should
limit the rate in which their article headlines are modified. With the speed in 
which news travels, headline changes can have devastating effects to reader 
understanding. We argue that, if a headline must be changed, the new headline should maintain
high lexical similarity to the original headline, only adding new information
to enhance reader understanding. 

%Moreover, news outlets should strive to alert
%readers that a headline has changed after original publication. We note that 
%some publishers provide readers with both a \emph{Published} and \emph{Updated}
%timestamp on each article webpage. However, we argue that this insufficient as many
%readers simply skim article headlines on the homepages of new outlets, rather than 
%click-through to each article's webpage. Rather, news outlets should include this information in all places an 
%article is referenced on their site, allowing a reader to revisit an article upon
%noticing its headline has been updated. 

\noindent\textbf{Information Travel on Social Media. }The rapid propagation
of information through shares and retweets on social media has only exasperated
the negative ramifications of news article headline modifications. By observing
the engagement of tweets associated with news articles with modified headlines,
we determined that the majority of favorites and retweets occur within ten hours 
of a tweet's publication. Due to the caching of article metadata and previews on
Twitter, most article shares will contain a different headline than what is active
on the publisher's website. While it is infeasible for social media platforms to 
constantly monitor each and every webpage linked to from all posts, decreasing the
caching of content can help reduce the spread of misinformation online. Moreover,
it would be worthwhile to explore ideas around the penalization of aggressive
headline changes. That is, if an article received 10K retweets \emph{before} a
significant headline change, is it appropriate for that article to keep all the 
clout that the previous title generated? Penalizing large changes has the potential
to act as a deterrent of unwanted publishing behavior and encourage publishers to
be more considerate about their content choices \emph{before} an article is published.

%\vspace{-1cm}

% \bibliographystyle{ACM-Reference-Format}
% \bibliography{references}

% \clearpage
% \appendix
% \input{chapter3-TitleModification/appendix}
\section{Appendix}
\label{sec:appendix}
%\subsection{Changes in number}

\begin{table}[htbp]
    \centering
    \small
    \caption{Numerical difference between pre/post-edits in changed news titles: Most of them concern news dynamic updates of injuries and competition scores. High \textit{BERTScore} shows that numerical changes does not alter the meaning.}    
    \label{tab:numerical_changes}
    \tiny
   \begin{tabular}{p{0.05\textwidth}p{0.08\textwidth}p{0.35\textwidth}p{0.35\textwidth}p{0.1\textwidth}}
    \toprule
     Diff. &  Number & Before & After & BERTScore\\
     \midrule
      1.0 &     116  & louisiana floods lead to 6 deaths & louisiana floods lead to 7 deaths &        0.99 \\
      2.0 &      45 &yankees lead astros 1-0 in game 1: live score and updates & yankees lead astros 3-0 in game 1: live score and updates &       0.96 \\
      -1.0 &      31 &ransomware attack hits 23 texas towns, authorities say & ransomware attack hits 22 texas towns, authorities say &       0.99 \\
      3.0 &      24 &super bowl 2020 live: chiefs lead the 49ers, 7-3 & super bowl 2020 live: chiefs lead the 49ers, 10-3 &       0.93 \\
      4.0 &      17 &iraq: suicide bombing kills at least 18 in baghdad & iraq: suicide bombing kills at least 22 in baghdad &       0.98 \\
      5.0 &      14 & albania earthquake kills at least 8 & albania earthquake kills at least 13 &       0.97 \\
      100.0 &       6 & iran-iraq earthquake kills more than 300 & iran-iraq earthquake kills more than 400 &       0.98 \\
    \bottomrule
    \end{tabular}
\end{table}

\begin{table*}[htbp]
        \centering
        \caption{Breakdown of changed nouns. One pair may belong to more than one category}
        \label{tab:noun_changes}
        \tiny
    \hspace*{-0.00\textwidth}\begin{tabular}{ >{\raggedright\arraybackslash}p{0.1\textwidth} >{\raggedright\arraybackslash}p{0.05\textwidth} >{\raggedleft\arraybackslash}p{0.05\textwidth} >{\raggedright\arraybackslash}p{0.35\textwidth} >{\raggedright\arraybackslash}p{0.35\textwidth}}
    \toprule
     Category & Total & \%  & \makecell[l]{Examples (Before, After, \#)} & Description \\
    \midrule
     Share lemma & 132 & 15.33\% & \makecell[l]{(report, reports, 4); (riots, riot, 3)\\} & The change noun shares the lemma. \\
    \hline 
     Synonym & 65 & 7.54\% & \makecell[l]{(advisor, adviser, 3); (investigate, probe,2)\\} & The change noun is synonym. \\
    \hline 
     Minor & 109 & 12.65\% & \makecell[l]{(protestors, protesters, 5); \\(reelection, re-election,4)} & Only one character is changed (excluding same lemma pairs ) \\
    \hline 
     Hyponymn  & 33 & 3.83\% & \makecell[l]{(housing, home, 3); (semiconductor, chip, 2), \\(official, officer, 2); (supporter, believer, 1)} & Become more specific\\
    \hline 
     Hypernymn & 41 & 4.76\% & \makecell[l]{(snaps, photos, 5);(budget, plan, 2)} & Become more general  \\
     \hline 
     Others  & 481 & 55.86\% & \makecell[l]{(demo, democrats, 7); (4th, fourth, 4)\\(live, close, 4); (valuation, value, 3) } & All other noun changes \\
     \bottomrule
     All & 861 & 100\% & \makecell[l]{--} & All detected noun substitutions \\
     \bottomrule
    \end{tabular}
\end{table*}
    
\begin{table*}[htbp]
        \caption{Breakdown of changed verbs. One pair may belong to more than one category}
        \label{tab:verb_changes}

        \centering
        \tiny
    \hspace*{-0.00\textwidth}\begin{tabular}{ >{\raggedright\arraybackslash}p{0.1\textwidth} >{\raggedright\arraybackslash}p{0.05\textwidth} >{\raggedleft\arraybackslash}p{0.05\textwidth} >{\raggedright\arraybackslash}p{0.35\textwidth} >{\raggedright\arraybackslash}p{0.35\textwidth}}
    \toprule
     Category & Total & \%  & \makecell[l]{Examples (Before, After, \#)} & Description \\
    \midrule
     Share lemma & 135 & 23.56\% & \makecell[l]{(cancelled, canceled, 7); (shuts, shut, 2)\\} & The change noun shares the lemma. \\
    \hline 
     Synonym & 59 & 10.29\% & \makecell[l]{(rise, climb, 2); (blew, botched, 2)\\} & The change noun is synonym. \\
    \hline 
     Minor & 35 & 6.10\% & \makecell[l]{(eying, eyeing, 2); (targetting, targeting,2)} & Only one character is changed (excluding same lemma pairs ) \\
    \hline 
     Hyponymn  & 38 & 6.63\% & \makecell[l]{(drops, tumbles, 3); (pull, drag, 2)\\(mulls, considers, 2); (rejects, rebuffs, 2)} & Become more specific\\
    \hline 
     Hypernymn & 46 & 8.02\% & \makecell[l]{(passes, advances, 3);(linger, remain, 3)\\(breaking, damaging, 2); (reveals, shows, 1)} & Become more general  \\
     \hline 
     Others  & 260 & 45.37\% & \makecell[l]{(drop, slip, 3); (reveals, says, 3) } & All other noun changes \\
     \bottomrule
     All & 573 & 100\% & \makecell[l]{--} & All detected verb substitutions \\
     \bottomrule
    \end{tabular}
\end{table*}

% \begin{adjustbox}{angle=90}

% \end{adjustbox}

\begin{table}[htbp]
    \caption{\textit{Top}: The tracked news categories per agency. \textit{Bottom}: The headline modification ratio in each news category}
    \label{tab:in_changed_news_cate_ratio_agency}
    \begin{minipage}{1.0\textwidth}
    \centering
    \tiny
    \begin{tabular}{lrrrrrr}
        \toprule
                 Domain &  Politics &  Business &  Entertainment &  Technology &  Top Stories &  Total \\
        \midrule
                    BBC &      1973 &      1562 &           1495 &         824 &            0 &   5854 \\
               BuzzFeed &       330 &        92 &           3928 &         195 &            0 &   4545 \\
                    CNN &      4548 &         1 &           1166 &         629 &            0 &   6344 \\
            Daily Beast &         0 &         0 &              0 &           0 &         4359 &   4359 \\
             Daily Mail &         0 &      3670 &          36312 &        1263 &        33873 &  75118 \\
               Fox News &      6307 &       108 &           4273 &         315 &            0 &  11003 \\
        Huffington Post &      2909 &        57 &           1902 &           7 &            0 &   4875 \\
            MarketWatch &         0 &         0 &              0 &           0 &         7503 &   7503 \\
        National Review &         0 &         0 &              0 &           0 &         3324 &   3324 \\
          New York Post &         0 &      1458 &           2018 &         514 &         9891 &  13881 \\
         New York Times &      2374 &      2983 &           3851 &         538 &            0 &   9746 \\
                Newsmax &       897 &        25 &            155 &          38 &            0 &   1115 \\
                    OAN &      2922 &      3545 &            528 &        1328 &            0 &   8323 \\
              The Blaze &         0 &         0 &              0 &           0 &         4401 &   4401 \\
        The Epoch Times &      3179 &      1077 &            193 &         316 &            0 &   4765 \\
           The Guardian &      3456 &       627 &           1874 &         959 &            0 &   6916 \\
               The Hill &         0 &         0 &              0 &           0 &         5676 &   5676 \\
        Washington Post &      2126 &      2014 &           1458 &         478 &            0 &   6076 \\
             Yahoo News &     11074 &    183653 &              7 &       32512 &            0 & 227246 \\
        \bottomrule
        \end{tabular}
    \end{minipage}
    \begin{minipage}{1.0\textwidth}
        \centering
        \tiny
        \begin{tabular}{lrrrrrr}
        \toprule
        Domain &  Politics &  Business &  Entertainment &  Technology &  Top Stories &  Total \\
        \midrule
        BBC             &    0.3082 &    0.2855 &         0.1371 &      0.0898 &       0.0000 & 0.2277 \\
        BuzzFeed        &    0.2576 &    0.0000 &         0.4812 &      0.2154 &       0.0000 & 0.4438 \\
        CNN             &    0.2419 &    0.0000 &         0.1612 &      0.0000 &       0.0000 & 0.2030 \\
        Daily Beast     &    0.0000 &    0.0000 &         0.0000 &      0.0000 &       0.1308 & 0.1308 \\
        Daily Mail      &    0.0000 &    0.0515 &         0.0592 &      0.0451 &       0.0970 & 0.0756 \\
        Fox News        &    0.0591 &    0.0926 &         0.0314 &      0.0571 &       0.0000 & 0.0486 \\
        Huffington Post &    0.0399 &    0.0877 &         0.0273 &      0.1429 &       0.0000 & 0.0357 \\
        MarketWatch     &    0.0000 &    0.0000 &         0.0000 &      0.0000 &       0.3453 & 0.3453 \\
        National Review &    0.0000 &    0.0000 &         0.0000 &      0.0000 &       0.0126 & 0.0126 \\
        New York Post   &    0.0000 &    0.0638 &         0.0852 &      0.0292 &       0.0650 & 0.0665 \\
        New York Times  &    0.2784 &    0.3144 &         0.0444 &      0.1375 &       0.0000 & 0.1892 \\
        Newsmax         &    0.0268 &    0.0400 &         0.0065 &      0.0000 &       0.0000 & 0.0233 \\
        OAN             &    0.0404 &    0.4886 &         0.0890 &      0.2108 &       0.0000 & 0.2616 \\
        The Blaze       &    0.0000 &    0.0000 &         0.0000 &      0.0000 &       0.0400 & 0.0400 \\
        The Epoch Times &    0.1787 &    0.0706 &         0.0259 &      0.1203 &       0.0000 & 0.1442 \\
        The Guardian    &    0.2922 &    0.3142 &         0.0326 &      0.1022 &       0.0000 & 0.1975 \\
        The Hill        &    0.0000 &    0.0000 &         0.0000 &      0.0000 &       0.0592 & 0.0592 \\
        Washington Post &    0.2413 &    0.1226 &         0.0501 &      0.1736 &       0.0000 & 0.1508 \\
        Yahoo News      &    0.2390 &    0.0713 &         0.0000 &      0.1073 &       0.0000 & 0.0846 \\
        \bottomrule
        \end{tabular}

    \end{minipage}
\end{table}

\begin{table}[htbp]

%\begin{sidewaystable}[htbp]
    \centering
    \scriptsize
    \caption{Examples of altered headlines. Separated by the double line, the first part is the manually selected modifications. The second part is sampled from our dataset by type }
    \label{tab:category-example}
    \scalebox{0.85}{
\begin{tabular}{>{\raggedright\arraybackslash}p{0.5\textwidth} >{\raggedright\arraybackslash}p{0.5\textwidth} >{\raggedleft\arraybackslash}p{0.05\textwidth}>{\raggedleft\arraybackslash}p{0.15\textwidth}}
\toprule
                                                                                        Before &                                                                                                             After &  BERT-F1 &       Type \\
\midrule \midrule
One Dose of J.\&J. Vaccine Is Ineffective Against Delta, Study Suggests &	J.\&J. Vaccine May Be Less Effective Against Delta, Study Suggests & 0.6830 & other \\ \hline
Malawi burns thousands of Covid-19 vaccine doses&	Malawi burns thousands of expired AstraZeneca Covid-19 vaccine doses & 0.7217 & other \\ \hline
Biden to huddle with Senate Democrats on Covid relief ahead of push for passage&	Biden urges Senate Democrats to reject poison pills that could sink relief plan ahead of push for passage & 0.3777 & other \\ \hline
Australia to investigate two deaths for possible links to COVID-19 vaccine&	Australia says two deaths not likely to be linked to COVID-19 vaccine & 0.7236 & citation \\ \hline
Canada adds blood clot warning to AstraZeneca's COVID-19 vaccine &	Canada says AstraZeneca COVID-19 vaccine safe, but adds blood clot warning & 0.7546 & citation \\ \hline
China's Zhurong rover will land on Mars TONIGHT	& China's Zhurong rover will land on Mars 'in the next five days & 0.7234 & other \\ \hline
Nike chief executive says firm is 'of China and for China'	&Nike boss defends firm's business in China & 0.4559 & other \\ \hline
MAGA 2.0&	“A new bargain”: Biden’s 2024 tease bets big on nostalgia & 0.0257 & emotional \\
\midrule \midrule
                                                    At Least Eight People Died In A School Shooting In Russia &                                                                          At Least Nine People Died In A School Shooting In Russia &         0.9866 &  paraphrase \\ \hline
            India sees cases officially drop below 300,000 a day but now country threatened by killer cyclone &                                      India cases officially drop below 300,000 a day but now country threatened by killer cyclone &         0.9034 &  paraphrase \\ \hline
                                                %   90\% of all adults eligible for COVID-19 vaccine by April 19 &                                                                     90\% of all U.S. adults eligible for COVID vaccine by April 19 &         0.8448 &  paraphrase \\ \hline
                                            %   A 19th Century Theory Explains Why Consumers Might Not Splurge &                                                                      A 19th Century Theory Explains Why Consumers May Not Splurge &         0.9857 &  paraphrase \\ \hline
                                                %  The cost of living posts biggest surge since 2008, CPI shows &                                The cost of living posts biggest surge since 2008, CPI shows, as inflation spreads through economy &         0.8202 &  paraphrase \\ \hline
                                                        The Latest: UN: 38,000 Palestinians displaced in Gaza &                                                                         The Latest: Biden expresses 'support' for Gaza cease-fire &         0.4043 &  dyn update \\ \hline
                                    %   Dyson leak row: Don't lock ministers in ivory towers, says senior Tory &                                                                                  Starmer: PM can't distance himself from 'sleaze' &         0.2030 &  dyn update \\ \hline
                               August 5 at 2PM ET: Join Engine Media CEO and Executive Chairman Fireside Chat &                                                 August 5 at 2PM ET: Join Engine Media CEO and Executive Chairman in Fireside Chat &         0.9393 &  dyn update \\ \hline
                                            %  Plant-Based Food Use Rising ‘Unbelievably Fast’: Catalyst Update &                                                                    FTX Chief Sees Enduring Volatility for Crypto: Catalyst Update &         0.1229 &  dyn update \\ \hline
                                                %  Futures Steady Before Payrolls; Treasuries Dip: Markets Wrap &                                                                  Stocks, Treasury Yields Rise on Robust Jobs Report: Markets Wrap &         0.3799 &  dyn update \\ \hline
                                                                               23 Amazing Jokes From Hot Fuzz &                                                      23 Jokes From "Hot Fuzz" That Humans Will Laugh At For The Next 10,000 Years &         0.4715 & elaboration \\ \hline
                                % Dad’s Girlfriend Arrested After Child’s Body Found in Tote Bag at Texas Motel &                                   Dad of Boy Found Dead in Tote Bag at Texas Motel ‘Can Barely Breathe’ as Girlfriend Is Arrested &         0.5259 & elaboration \\ \hline
                                        %   Secrets of Prince William and Kate Middleton's big day 10 years ago &                                  Royal wedding secrets from Prince William and Kate Middleton's big day on their 10th anniversary &         0.5444 & elaboration \\ \hline
                                                 Official: Haiti President Jovenel Moïse assassinated at home &                                               Haiti President Jovenel Moïse assassinated at home; Biden calls it 'very worrisome' &         0.5179 & elaboration \\ \hline
                                %  'More explosions could occur': Thousands flee as volcano erupts in Caribbean &           'More explosions could occur': La Soufriere volcano eruption sends thousands fleeing on Caribbean island of St. Vincent &         0.6160 & elaboration \\ \hline
                                           Senator Capito says Republicans plan new U.S. infrastructure offer &                                                                          U.S. Senate Republicans prepare new infrastructure offer &         0.5434 &   concision \\ \hline
                                            % Caroline Jurie: Mrs World arrested over Sri Lanka pageant bust-up &                                                                                 Sri Lanka Mrs World arrested over pageant bust-up &         0.6387 &   concision \\ \hline
                            %   U.S. stock futures take retail sales, PPI data in stride as investors await Fed &                                                                 U.S. stocks edge lower after economic data as investors await Fed &         0.5774 &   concision \\ \hline
                      U.S. second-quarter economic growth revised slightly higher; weekly jobless claims rise &                                                                        U.S. second-quarter growth raised; corporate profits surge &         0.5540 &   concision \\ \hline
                                            % Right-Wing Media Attacks ‘Weak’ Simone Biles: ‘Selfish Sociopath’ &                                                                         Right-Wing Media Launches Unhinged Attack on Simone Biles &         0.4696 &   concision \\ \hline
                                                                                %   FKA twigs steps out in NYC &                                     FKA twigs steps out in NYC after claiming it was a 'relief' to speak out against Shia LaBeouf &         0.5539 &   emotional \\ \hline
                    %   Phoebe Dynevor Almost Quit Acting Before Her "Bridgerton" Audition With Regé-Jean Page &                                      Phoebe Dynevor Wanted To "Throw In The Towel" On Her Acting Career Right Before "Bridgerton" &         0.4885 &   emotional \\ \hline
                                                 Undercover police officers spied on Peter Hain over 25 years &                                                  Peter Hain accuses undercover police of lying over reports on apartheid campaign &         0.2465 &   emotional \\ \hline
% US surpasses 100M vaccinations; Pfizer vaccine appears effective against asymptomatic cases: COVID-19 updates &                        US surpasses 100M vaccinations; IRS begins sending first round of \$1,400 relief payments: COVID-19 updates &         0.4424 &   emotional \\ \hline
                                               Suspect arrested following series of Arizona traffic shootings &                                                     Arizona gunman goes on traffic shooting rampage, leaving one dead, 12 injured &         0.2247 &   emotional \\ \hline
                                            % 23 Funny "Brooklyn Nine-Nine" Memes To Get You Ready For Season 8 &                                                                                                Brooklyn Nine Nine Memes And Jokes &         0.2271 &  neutralize \\ \hline
                % Hundreds chanting 'Kill the Bill' protest in Bristol over controversial Police and Crime bill &                                    Protesters smash police windows and attack officers' van amid 'Kill the Bill' march in Bristol &         0.4220 &  neutralize \\ \hline
                %  Prince Philip will become 25th Royal in a 200-year-old vault hidden below St George's Chapel &                               Prince Philip to be buried in 200-year-old vault below St George's Chapel but only until Queen dies &         0.6037 &  neutralize \\ \hline
                                                 Raul Castro confirms he's resigning, ending long era in Cuba &                                                                        Raul Castro resigns as Communist chief, ending era in Cuba &         0.7310 &  neutralize \\ \hline
         Nikki Grahame's life will be celebrated in a new Channel 4 documentary four months on from her death &                                                                   Nikki Grahame to be commemorated in a new Channel 4 documentary &         0.6698 &  neutralize \\ \hline
                                        Fourth stimulus check update: Your next payment could be one of these &                                                                 Fourth stimulus check? These payments are already in the pipeline &         0.4479 & foward ref. \\ \hline
                                                                       12 Movie Opinions You Might Agree With &                                                     Here Are 12 More Movie Opinions I Strongly Believe, But Do You Agree With Me? &         0.4246 & foward ref. \\ \hline
                    % A park next to the White House that was closed amid protests last year partially reopens. &                                   Lafayette Park, which was closed amid protests at the White House last year, partially reopens. &         0.7430 & foward ref. \\ \hline
                                        %  Biden to share ‘next steps’ for COVID-19 booster program later today & People in the U.S. who were vaccinated with Moderna and Pfizer’s COVID-19 vaccines will be eligible for a third dose in September &         0.2475 & foward ref. \\ \hline
                                                %   ‘How I Got In’: A Harvard/Stanford/Wharton Admit Tells All &                                                                          How I Got In: A Harvard/Stanford/Wharton Admit Tells All &         0.7685 & foward ref. \\ \hline
                                                 Watch Jeff Bezos’ ‘Blue Origin’ launch into space live today &                                                Everything you need to know as Jeff Bezos’ ‘Blue Origin’ launches into space today &         0.6179 & personalize \\ \hline
                                The "Degrassi: TNG" Cast Is Reuniting In Honor Of The Show's 20th Anniversary &                                                      The "Degrassi" Cast Is Reuniting And I Am So Excited To Relive My Teen Years &         0.4838 & personalize \\ \hline
                        Stimulus Update: States Give Out Thousands of Bonus \$1,000 Checks To School Employees &                                             Stimulus Update: States Give Out Thousands of Bonus \$1,000 Checks – Will You Get One? &         0.7416 & personalize \\ \hline
        %  Chrissy Teigen shares embarrassing encounter with Michael Keaton while playing game on NBC talk show &                                   Chrissy Teigen reveals she once confused Hollywood legend Michael Keaton for a champagne WAITER &         0.3173 & personalize \\ \hline
    %   ‘Strong perception’ that Tory party is insensitive to Muslim communities, report finds – politics live &                       Boris Johnson says he would not repeat ‘offending language’ about Muslims as prime minister – politics live &         0.3395 & personalize \\ \hline
                                    Kendall Jenner Has Some Pretty Strong Thoughts About The Kardashian Curse &                                   Kendall Jenner Talked About The Kardashian Curse And Said "The Men Need To Take Responsibility" &         0.5229 &    citation \\ \hline
                                            %  Howard Beckett: Labour suspends union official after Patel tweet &                                                   Howard Beckett: Patel laws 'disgusting' says union boss after Labour suspension &         0.4986 &    citation \\ \hline
                  Afghan guard killed: Firefight leaves at least one dead and others injured at Kabul airport &                  'It would be better to die under Taliban rule than face airport crush', say US embassy's 'betrayed' Afghan staff &         0.0656 &    citation \\ \hline
                                                    %   Police: 3 officers stable after being shot in Delaware &                                                                       Police say threat over after 3 officers wounded in Delaware &         0.6105 &    citation \\ \hline
                                           Apple beats sales expectations on iPhone, services, China strength &                                                                    Apple says chip shortage reaches iPhone, growth forecast slows &         0.2611 &    citation \\
\bottomrule
\end{tabular}
}
%\end{sidewaystable}
\end{table}

\begin{figure}[htbp]
% \begin{sidewaysfigure}[htbp]
    \centering
    \includegraphics[width=0.8\textwidth]{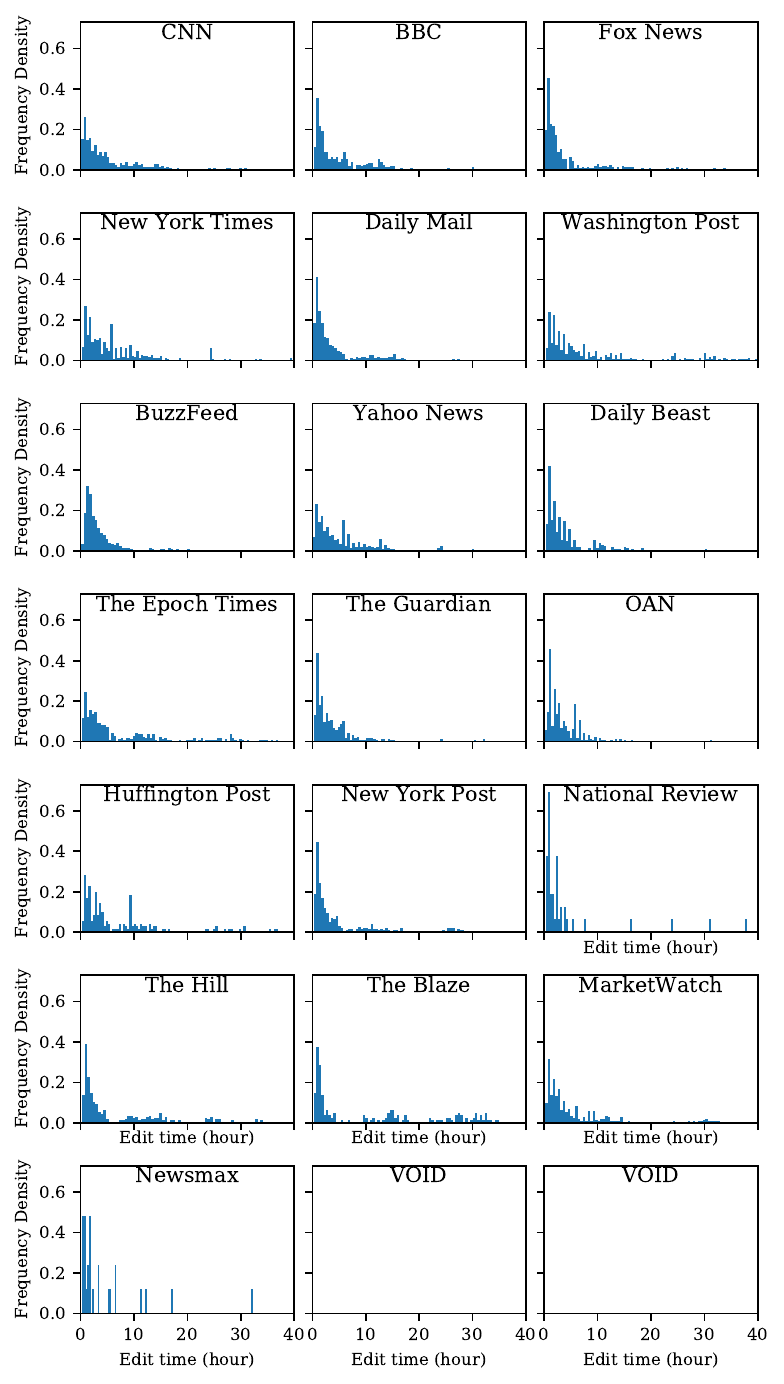}
    \caption{The post-publication edit time per news publisher. }
    \label{fig:edit-time-per-agent}
\end{figure}

\chapter[Information and Evolution of Personal Biographies]{Information and Evolution of Personal Biographies \protect\footnote{Guo, Xingzhi, Dakota Handzlik, Jason J. Jones, and Steven Skiena. Biographies and Their Evolution: Occupational Identities over Twitter. In Submission.}}
\label{chapter:bio-change}
\graphicspath{{chapter4-BiosModification/}}

\xingzhiswallow{
    \begin{abstract}
    Occupational identity concerns the self-image of an individual's affinities and socioeconomic class, and directs how a person should behave in certain ways. Understanding the establishment of occupational identity is important to study work-related behaviors. However,  large-scale quantitative studies of occupational identity are difficult to perform due to its indirect observable nature. 
    But profile biographies on social media contain concise yet rich descriptions about self-identity. Analysis of these self-descriptions provides powerful insights concerning how people see themselves and how they change over time. 
    
    In this paper, we present and analyze a longitudinal corpus recording the self-authored public biographies of 51.18 million Twitter users as they evolve over a six-year period from 2015-2021.
    In particular, we investigate the demographics and
    social approval (e.g., job prestige and salary) effects in how people self-disclose occupational identities, quantifying over-represented occupations as well as the occupational transitions w.r.t. job prestige over time. 
    We show that males are more likely to define themselves by their occupations than females do, and that self-reported jobs and job transitions are biased toward more prestigious occupations.
    We also present an intriguing case study about how self-reported jobs changed amid COVID-19 and the subsequent \textit{"Great Resignation"} trend with the latest full year data in 2022.
    These results demonstrate that social media biographies are a rich source of data for quantitative social science studies, allowing unobtrusive observation of the intersections and transitions obtained in online self-presentation.
    \end{abstract}
    
}

% \todo{The abstract talks about 2015-2021: maybe things should be written for the longer year span of the last part. : claimed to use the latest 2022 data in case study.}

\section{Introduction}
\label{introduction}
Self-identity is a concept of \textit{who we think we are} that directs us how "to behave in a certain way" \cite{vignoles2008identity}. The formation of self-identity is rooted in people's autobiography memory, often cultivated from personal life experiences.
Occupational work is arguably the life domain to which we devote the greatest part of our time awake, thus, it is an essential part of our experience, embedding itself into our motivation, life goals and personal sense of identity.
\textit{"One life-long objective is work per se"}; as \citet{butler1992way} suggested "Every man's work .. is always a portrait of himself.";
However, not every person will develop a primary occupational identity (\textit{e.g., "I think of myself as a mother, rather than a bank clerk."}).
Likewise, not every job ultimately leads to the establishment of occupational identity (\textit{e.g., "I do not define myself as a deliverer, although I do it for living."}).

In this paper, we consider the question of \textit{what factors contribute to the formation and transition of occupational identity?} 
%\versiontwo{and \textit{what are the rising work identities?}}
%
However, research into personal identity is difficult because self-identity is not directly observable. Although an individual's employment status is explicitly stated in tax records and other datasets, an individual's occupational identity is an internal and invisible presentation of \textit{"who I am"}, and generally not publicly available. 

But recent research \cite{rogers2021using} has proposed a new methodology, using Twitter and other social media biographies as the proxy for self-identity. 
The concise (on Twitter, at most 160 character) personal biography is a feature of many social media platforms, and provides an outlet for users to report their current self-identity.  
On Twitter, this biography string is part of every user's public profile, and research has explored the various and changing aspects of identity that users elect to associate themselves with \cite{semertzidis2013people, rogers2021using}. 

Although users differ in what they disclose, many social media biographies express relational identity (\textit{e.g. father/mother of, alumni of}), demographics, occupations and political affiliations.  
\citet{semertzidis2013people} further categorized biography content in terms of average word fraction: Occupation (7.5\%), Interests/Preferences/Hobbies(4.6\%),  Personal Info (2.6\%), etc.
%, Social Nets/Internet (2.3\%), Community (1.5\%) and General. 
For example, the biography:
\begin{quote}
    25. black man. student. Films. "We are guardians of the image \& that's how I see our role as DP's"
    %- Geoff Boyle \#ReleasetheSnyderCut }
\end{quote}
contains information about age (\textit{25-year old}), occupation (\textit{student}), ethnicity (\textit{black}), gender (\textit{man}) and interests (\textit{films}). 

\begin{wraptable}{r}{1.0\textwidth}
    % \begin{table}[htbp]
        \centering
        \small
        \caption{Job transition pairs observed in Twitter biographies. These transitions reflect personal career advancement (e.g., from \textit{assistant professor} to \textit{associate professor}), graduation (e.g., from \textit{student} to \textit{engineer}) or work-status change (e.g, from \textit{designer} to \textit{freelance designer}).  In this paper, we investigate why people self-present with such occupational identities.
        %The dummy node is ignored. 
    \     }
    % \vspace{-5mm}
    \begin{tabular}{p{0.3\textwidth}p{0.3\textwidth}p{0.2\textwidth} }
    \toprule
                     Title A &                 Title B  &  \# (A$\leftrightarrow$B)   \\
    \midrule
        co-founder &        founder &        10149 \\
               ceo &        founder &         8577 \\
            % owner &        founder &         7384  \\
            % owner &            ceo &         5917 \\
        %   director &        founder &         2987 \\
    % vice president &             vp &         2734 \\
    % vice president &      president &         2545  \\
    director &             executive director &         2351 \\
    assistant professor	& associate professor & 2154\\
    graphic designer & freelance graphic designer & 1975\\
    software engineer &	senior software engineer &1043 \\
    student	& nursing student & 770\\
    accountant&	chartered accountant & 661 \\
    student	& mechanical engineer& 575 \\
    \bottomrule
    \end{tabular}
    % \vspace{-4mm}
        \label{table:top_job_transition}
    % \end{table}
    
\end{wraptable}

Longitudinal changes in a subject's self-reported biography allows analysis of changing self-conception over the full life course. 
For example, the aforementioned biography is likely to updated to reflect age and career path (e.g., \textit{25}$\rightarrow$ \textit{26}, \textit{student}$\rightarrow$ \textit{cameraman}).  
This would record an occupational identity change from \textit{student} to \textit{cameraman} upon graduation. 
Our dataset records 435.7 million twitter biography edits, and extracts more than 287,000 job transitions between over 25,000 distinct job titles. We present examples of the frequent job transitions in Table \ref{table:top_job_transition}.

% We explore the dynamic nature of self-identity.  
Our large-scale dataset of public biographies gives researchers a new opportunity to study the self-presented identities and their evolution among Twitter users.
Previous studies involve much smaller datasets, and lack the longitudinal element enabling us to track changes in the self-identity of individuals over time.
\xingzhiswallow{
    In particular, we want to investigate three aspects in biographies:
    \begin{itemize}
    %\small
    \item[1] \textit{Validate the Predictive Power of Biographies:} How much personal identities (e.g. demographics, hobbies) do Twitter users self-disclose in one short biography than in tweets? 
    \item[2] \textit{Understand Bias in Self-reported Occupations on Twitter:} Do the self-presented occupational identities reflect the real-world statistics? If not, what factors make the discrepancy? 
    \item[3] \textit{Interpret Dynamics in Biographies:} As time goes on, what are the changes in the personal biographies? 
    
    % do user want to self-disclose in one short biography when compared with tweets. 
    \end{itemize}
}
\citet{pathak2021method} extracted a set of words from biographies as personal identifiers for future social identities studies.
Other studies \cite{sloan2015tweets} use automatic tools to profile Twitter users, and observe that jobs in the creative sector (e.g., writer, artist) are over-represented in UK Twitter users. 
However, they only report population biases in Twitter users (e.g., over-represented occupational jobs), but do not further investigate the reasons behind them.  
Meanwhile, these studies generally analyze the population at one fixed time point, and ignore the evolution in each individual over time. 

% Inspired by this promising methodology, we aim to investigate the occupational identities in Twitter biographies and quantify the factors contributing to their establishment.
In our study, we collect and analyze a multimillion-subject scale temporal Twitter biographies dataset, and systematically explore self-authored occupational identities.
We present our quantifiable findings to better understand how people consider occupations as a part of self-identity. 
To further clarify our motivation, this paper focuses on the self-disclosed \textit{occupational identity}, instead of the mere employment status over the Twitter population.

The primary contributions of this paper include:
\begin{itemize}
    \item {\em A Temporal Self-Description Data Set} -- We have assembled a longitudinal corpus of self-reported biographies as they evolve annually over the period from 2015-2021, 
    \versiontwo{
    with  %51.18 
    1.35 million distinct users whose biographies are written in English. However, we are legally only able to share the word embeddings of bios due to privacy and terms-of-service concerns. %\footnote{To obtain full raw text, a data usage agreement must be signed case by case. } 
    We will also release an occupational transition graph with 25,033 unique jobs and 287,425 unique edges.} %changed at least once over this period.
    %\todo{NOTE: It is not necessarily true that every user history contains at least one change.  We should check the data if we want to make that claim.  It is true that all of the users tweeted at least once each year, which provides us annual snapshots of their bio in every year. (Jason)}     
    Our dataset provides a unique resource for studying changes in self-identity, and will be made available on publication. 
    % \versiontwo{However,  we are only able to share  the word embeddings of bios (from 1.35M users across six years) rather than raw text due to privacy concerns. }
    %\todo{JJJ: We should not share any dataset with user IDs in it. I don't think we should promise a dataset at all.} \todo{Twitter TOS: can only share 1,500,000 Tweet IDs within 30 days }
    
    %\item {\em The Predictive Power of Biographies} -- We demonstrate that brief (160 character) self-reported biographies have considerable predictive power on the task of gender (78.0\%) and ethnicity (49.3\%) prediction, as well as categorizing users’ personal interests (64.9\%). We estimate that one bio is worth approximately 16 tweets when predicting gender and 8 tweets when predicting ethnicity, implying that people are more willing to self-disclose their demographics in biographies than in tweets. 
    %\todo{We need a better description of the final results.}
    
    \item {\em Identifying the Reasons Behind Occupational Identity} --   With this rich dataset, we present a large-scale case study of occupational identities among Twitter users. \remove{We discover that males are roughly 32\% more likely to define themselves with occupations than females.}  Further, we quantify which occupations are over-represented in self-biographies relative to the number of people holding such positions.
    We show that 74.04\% of over-represented jobs are above the overall median occupational prestige score, versus only 39.48\% for under-represented jobs.  We demonstrate that over-representation is highly correlated with job prestige, with a Spearman correlation of $\rho= 0.4858$. The correlation of over-representation with prestige is stronger than that with income, where $\rho= 0.3961$. 
    
    % \todo{how does prestige correlate compared to money?   My expectation is that high prestige beats high income -- do we have this data?} -- yes, for annual income \rho  - = 0.396129

    \item {\em The Evolution of Identity} --
    We present a temporal analysis explaining how and why self-descriptions change, distinguishing life events (e.g. job promotion, graduation, etc) from lateral moves. In our study of occupation identity transitions, we discover that there is an inherent directionality towards increased occupational prestige with most transitions (e.g., from \textit{assistant professor} to \textit{associate professor}).  \versiontwo{Moreover, we also present an intriguing case study, reflecting changes in observed occupational identity amid COVID-19 and its aftermath}.

\end{itemize}

This paper is organized as follows.  Section \ref{related-work} describes previous work in analysis of self-identity, particularly for social media.
Section \ref{dataset} describes our Twitter biography data set, followed by a discussion of research ethics in Section \ref{ethics}.
Section \ref{sec:occupation-bios} presents our observations concerning occupational identities on Twitter, and its relation to job prestige and income.
We study the changes in user biographies over time and present the case study in Section \ref{biography-evolution}.
Finally, we present our conclusions and future work in Section \ref{future-work}.

\section{Background and Related Work}
\label{related-work}
% what is self-identity
Self-identity is a concept of \textit{who we are} and directs us how to "behave in certain way" \cite{vignoles2008identity}, fostered in our autobiographical memory \cite{bluck2005tale}. Self-identity can be realized as a knowledge representation \cite{kihlstrom1994self} of the past events with personal experience, often characterized as personal life story. 
% hard to observe. 
However, due to the nature of its inobservability in practice, it is notoriously challenging to conduct quantitative research of self-identity on a large scale. 
% jason introduced new method/data
Recently, \cite{rogers2021using, jones2021dataset} proposed a new methodology to discover quantifiable insights of political identities from millions of Twitter biographies, leveraging them as a directly observable source of self-identities.  Such research methodology makes it possible to study self-identity in a large-scale and computational manner. 

In this section, we describe self-identity related theories, methods and insights from the existing research of psychology and computational social science.

\paragraph{Self-identity in theory}:
% describe the proposed theories about self-identity/occupational identity.
% how they do experiment
Generally speaking, there are two kinds of self-identity: 1) \textit{Personal identity} describes how an individual defines him/her-self as a unique human-being; 2) \textit{Social identity} represents how an individual feels belonging to certain groups as a member. Occupational identity lies somewhere in between (e.g., a unique self belonging to a work group).
\citep{christiansen1999defining} first connected occupation to identity, suggesting that occupation contributes to identity shaping (e.g., \textit{I am a sociologist}), and occupation provides a context for creating meaningful life, which promotes well-being and life-satisfaction.
To understand the work-related identity, the current studies  \cite{ashforth2008identification} aim to answer these key questions:
1). \textit{What is work-related identity?};
2). \textit{Why does identity matters?};
3). \textit{How does identity evolve?}
To address such questions, \citet{kielhofner2002model} proposed the term \textit{"Occupational Identity"} and defined it as "a composite sense of who one is and wishes to become as an occupational being". 

\paragraph{Establishment of occupational identity}: Kielhofner \cite{kielhofner2002model} further argued that the occupation which are satisfying and recognized within one's environment are more likely to become meaningful and central to one's life, sustaining to self-identity. According to social identification theory \cite{ellemers2004motivating,van2004my, jackson2002intergroup}, a healthy self and social identity tend to thrive toward positive self-evaluation and link to a positive social group identity. As emphasized in occupational identity research, social approval is also key to the identity establishment.
For example, key occupational identity theorists suggest that \cite{christiansen2004occupation, phelan2009occupational} "identities are fostered when individuals perceive that their chosen occupations win approval from the greater society". 
One quantitative measure of social approval is perceived occupation prestige \cite{treiman2013occupational, hodge1964occupational,hout2015prestige}, derived from job-specific factors like income and education levels. 
Specifically, \citet{hout2015prestige} released occupational prestige scores based on the interviews with 1,001 individuals during 2012. These ratings range from  0 (bottom) to 100 (top), covering 860 occupational titles or 539 job categories in the 2010 Standard Occupational Classification (SOC) \cite{elias2010soc2010}. 
This unique resource enables us to quantitatively measure the correlation between the occupations in identity and its social approval.

\paragraph{Self-identity in Twitter biographies:}
% twitter bio research: different from general user profiling. 
Social media users often present their self-identities in bios. \citet{semertzidis2013people} investigated how people express themselves in Twitter biographies 
%and found users frequently describe interests (e.g., music, food) and profession (e.g., marketing, writer).  
and discovered that the most common content in biographies are occupations.
\citet{pathak2021method} extracted several clusters of personal identifiers from Twitter biographies by matching text pattern, and suggest it as a resource for future social identity studies. 
Similarly, \citet{sloan2015tweets} extracted job titles within the bios of UK Twitter users, and found that the jobs in creative sector (e.g., artist, writer) are over-represented. They explained this observation by the fact that Twitter is used by people who work in the creative industries as a promotional tool, and perhaps as a technical artifact from mis-classifying hobbies and occupations (e.g., leisure or professional writer) . 
Other researchers \cite{mislove2011understanding} studied the demographic distribution of the Twitter population by analyzing self-reported user names and biographies, enabling large-scale social studies involving demographic characteristics.
Note that self-identity researchers focus on how people self-present themselves, rather than working to improve the accuracy of user profiling algorithms \cite{preoctiuc2015analysis,pan2019twitter}. 

\paragraph{Self-identity transitions over time}: Personal biographies are dynamic, reflecting the life or mind changes in a person over time. 
\citet{shima2017users} analyzed the timing and reasons for Twitter bio changes and found that English users most frequently make changes on their birthdays. \citet{rogers2021using} discovered that Americans defined themselves more politically by integrating more political words in their Twitter bios over recent years. 
% \versiontwo{Add related works in raising work (Tiktok, NFT, ...)}
In our temporal analysis, we examine the reasons behind these modifications and discover interesting bio transitions for graduation, anniversaries and job transitions. To our best knowledge, this is also the largest resource available for studying dynamics in occupational identities.

\section{Biography Dataset Collection and Pre-processing}
\label{dataset}
% \begin{figure}[h]
%     \centering
%     \includegraphics[width=0.2\textwidth]{Figures/twitter_bios_raw_chunk/figure_raw_bio_length_cdf.pdf}
%     \caption{Bio length}
%     \label{fig:bio_len}
% \end{figure}

% data crawling techniques 
% raw data

% To measure the predictive power of self-descriptions and conduct temporal analysis, 

% {'num_total_person': 51183859, 'num_detect_gender_person': 12907860, 'num_total_names': 143499043, 'num_detect_gender_names': 15186858, 'num_name_in_dict_name2male': 7529400}
% total_exp_male 8529749.88497942
% total_pop 14340011
% male ratio in population 0.5948217114323985

We conducted a study on longitudinal occupational identity by extracting self-reported job titles from the biographies of Twitter users over six years, and further mapping the job titles to Standard Occupational Classification (SOC) 
\footnote{The U.S. SOC: \url{https://www.bls.gov/soc/}}
for specific job category information (e.g., average salary, job prestige scores, etc). 
We provide a thorough description of our data processing procedures below.

\begin{figure}
\vspace{0mm}
\centering
\includegraphics[width=0.50\textwidth]{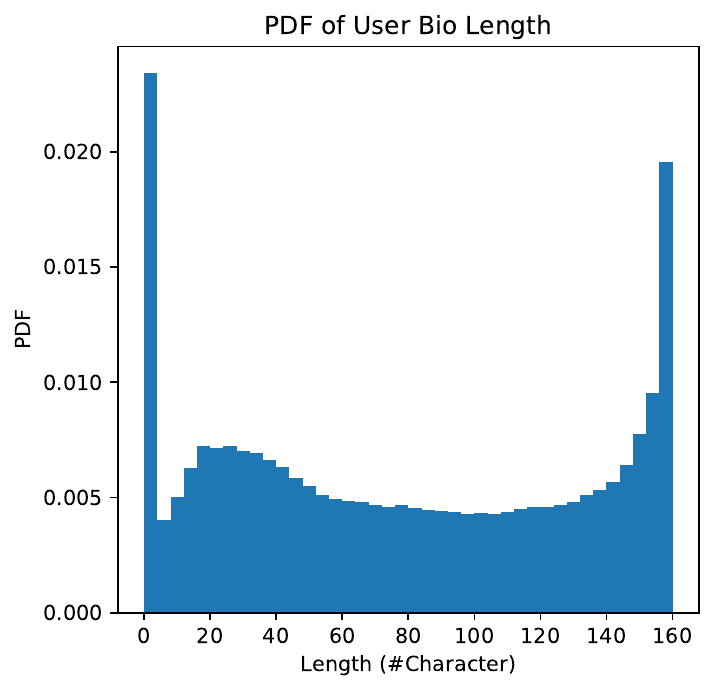}
% \vspace{-8mm}
\caption{The statistics of sampled biography length. The two peaks indicate that people tend to either skip or write complete bios. More than $\sim$80\% users have bios longer than 25 characters.}
\label{fig:bio_len_summary}
% \end{wrapfigure}
\end{figure}

\paragraph{Collecting longitudinal Twitter biographies:}
We collect the Twitter user biography associated with each of  51.18 million English-language  \footnote{We use the FastText language identification tool \cite{joulin2016fasttext, joulin2016bag} to filter out non-English bios.} re/tweets, from Feb. 2015 to July 2021 using Twitter API. 
We have been pulling 1\% of all tweets from the Twitter stream on a daily basis over this period, and been recording personal biographies with each tweet's observed timestamp. 
Our analysis detects 435.70 million biography changes over this corpus. 
Figure \ref{fig:bio_len_summary} shows the bio length distribution, demonstrating that more than 80\% of users take advantage of the biography field to report a substantial self-description.

% \begin{wrapfigure}{R}{0.95\textwidth}
% \setlength\intextsep{0pt}

%, a similar observation (the peaks on two sides and one hump around 20) can be found in \cite{semertzidis2013people}. 

%\todo{JJJ: Did you filter on location? If not, then English-language Twitter users is a better term than American users.} 

% \begin{figure}[htbp]
%     \centering
%     \subfloat[PDF (in character)\label{fig:bio_len_pdf}] {\includegraphics[width=0.24\textwidth, height=0.1\textheight ]{Figures/fig_raw_all/figure_raw_bio_length_pdf.pdf}}\hfill
%     \subfloat[CDF (in character)\label{fig:bio_len}] {\includegraphics[width=0.24\textwidth, height=0.1\textheight ]{Figures/twitter_bios_raw_chunk/figure_raw_bio_length_cdf.pdf}}
%     \caption{The statistics of sampled bio length. Fig.\ref{fig:bio_len_pdf} shows two peaks, indicating people tend to either skip or write complete bios. Fig.\ref{fig:bio_len} shows that more than $\sim$80\% users have bios longer than 25 characters.}
%     \label{fig:bio_len_summary}
% \end{figure}

% \begin{figure}[htbp]
%     \centering
%     \includegraphics[width=0.25\linewidth]{Figures/fig_raw_all/figure_raw_bio_length_pdf.pdf}
%     \caption{The statistics of sampled bio length. The two peaks indicate that people tend to either skip or write complete bios. More than $\sim$80\% users have bios longer than 25 characters.}
%     \label{fig:bio_len_summary}
% \end{figure}

% job extractor: from all 
% prediction: long bio, person, ethnicity, personal interest.
% temporal: from all 

% \paragraph{Demographics Inference from User Names}

\paragraph{Large-scale demographics estimation from user names}: 
    Arguably, a human's name serves as the very first identity. English user names reveal a lot of demographic information. For example, first names (e.g., \textit{Jack, Rose}) may encode the gender, while last names (e.g., \textit{Smith, Yao}) may indicate ethnicity. 
    As a common practice for large-scale social media analysis \cite{mislove2011understanding},  we estimate the ground truth from users' self-reported names.
    % However, not all users provide their genuine names (). 
    To reduce the noise introduced by the non-genuine names (e.g., \textit{Skies-of-Glory, Uber-Man, Morning Brew, ...}),  we exclude most of the non-human-like names by matching the first/last names based on the 2000 U.S. Census Data \footnote{2000 U.S. Census Data  contains 92,600 first name and 151,671 last names. \url{ https://github.com/rossgoodwin\/american-names}}, then manually inspect the first 500 most common first/last names to avoid systematic false positive errors.
    In total, we found 7,529,400 unique genuine names associated with 12,907,860 users (or 25.22\% of the users in the raw data).

    % \todo{Why not all users?  It is probably not important but it seems weird.}
    
    Similar to methods used in \cite{mislove2011understanding,wood2020using,preoctiuc2015analysis}, we employ name embedding-based classifiers \cite{sood2018predicting, ye2017nationality,ye2019secret}  to infer ethnicity and gender \footnote{Gender detector:\url{https://github.com/kensk8er/chicksexer}} purely from names.  To properly handle the estimation uncertainty, instead of using a single label (e.g, male vs female), we represent the demographics attributes in a probabilistic manner. For example, The first name \textit{Kellen} may have a probability of 80\% male, 20\% female.  Ethnicity attributes are also presented probabilistically in four coarse ethnicity groups (White/African-American/Hispanic/Asian). 
    
    %\todo{African is misspelled in your pie chart.}

    \begin{figure}[htbp]
        \centering
        % \subfloat[Gender distributions among users \label{fig:overall_gender}] {\includegraphics[width=0.2\textwidth]{Figures/overall_gender.pdf}}\hfill
        % \subfloat[Ethnicity distributions among users \label{fig:overall_ethnicity}] {\includegraphics[width=0.2\textwidth]{Figures/overall_ethnicity.pdf}}
        \subfloat[Gender ] {\includegraphics[width=0.365\textwidth]{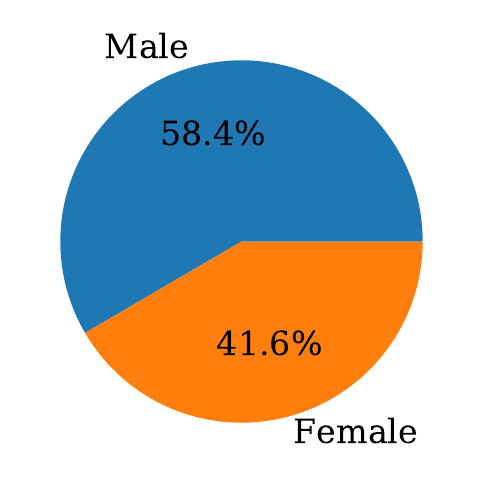}}\quad
        \subfloat[Ethnicity] {\includegraphics[width=0.4\textwidth]{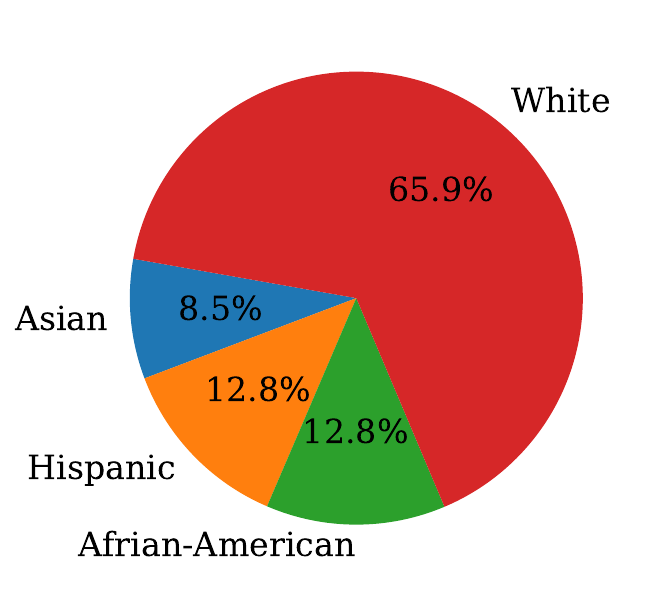}}
        % \vspace{-3mm}
        \caption{The overall estimated demographics distribution among 12.91M users. According to Statista, the gender ratio of Twitter users worldwide is 56.4\% v.s. 43.6\%. Since English Twitter users are dominated by U.S, the ethnicity distribution is similar to its in the U.S..  }
        % \vspace{-3mm}
        \label{fig:overall_demographics}
    \end{figure}

\paragraph{Extracting occupation identities from biographies:} Biographies contain rich occupational information \cite{semertzidis2013people}. To demonstrate the usefulness of our dataset, we extract\footnote{Job title extractor: \url{https://github.com/fluquid/find_job_titles}} and analyze self-reported job titles on a large scale. From biographies sampled over a six year period (2015-2021), we detected 25,033 unique occupational titles from 6,028,581 Twitter users, which covers 11.78\% of whole population in our dataset.  
%In addition, our dataset span a six-year range, from 2015-2021. 
Subjects often report multiple job titles over this period, reflecting promotion or career changes, as will be addressed in Section \ref{biography-evolution}.

\paragraph{Mapping job titles to Standardized Occupational Codes (SOCs) with job prestige scores}:
% The job title matching process is detailed in Section \ref{sec:job-match-appendix} in Appendix. 
% In Table \ref{tab:job-match-prestige-short}
Note that the detected job titles are pure text and unstructured. In order to gain richer insights about each job title (e.g., salary, popularity, ...), we organized them according to the U.S. Standard Occupational Classification (SOC), enriching with supplementary data (e.g., employment and income distribution).
\footnote{\url{https://www.bls.gov/soc/2018/home.htm}} and job prestige data \cite{hout2015prestige} (which estimates the occupation prestige in SOC-indexed work categories).
\footnote{\url{http://gss.norc.org/Documents/other/PRESTG10SEI10_supplement.xls}}.
We first use exact matching to map all Twitter job titles to \textit{SOC(2018) Direct Match Title}, which exhaustively lists 6,593 job title examples in all SOC occupational categories. 
% Then, we use ROUGE score \cite{lin2004rouge} (based on text overlap) to match each job title to the all SOC provided job examples, then reuse the SOC category of the example with the highest match score. 
%
Then, we use the ROUGE score \cite{lin2004rouge}, which is based on text overlap, to \textit{softly} match each job title to the examples provided in the Standard Occupational Classification (SOC). The SOC category of the example with the highest match score is then assigned to the job title.
To ensure the precision of job category mapping, we consider the job title unmatched if the ROUGE-F1 score is lower than 0.8.
%We filter out the job titles whose maximum aligned ROUGE-F1 score is lower than 0.8. 
% Finally, we select the top-aligned occupation category as the result. 
For those filtered unmatched titles, we exclude them in the analysis to avoid noise. Since the job prestige is surveyed based on OCC(2010) format,  we convert SOC(2018) to OCC(2010) via the official mappings, resulting in 1,834 unique job titles associated with high precision occupational prestige scores.  
%\versiontwo{ Note that it covers \todo{fill-in-blank} \% of our population.}
We present several alignment examples with associated prestige scores in Table \ref{tab:job-match-prestige}.

\xingzhiswallow{
    Because the aforementioned data are in different job coding schema, we list the SOC version conversion and the reference document below:
    \begin{itemize}
        \small
        \item  Twitter job titles $\rightarrow$\footnote{\url{https://www.bls.gov/soc/2018/soc_2018_direct_match_title_file.xlsx}} SOC(2018)  $\rightarrow$\footnote{\url{https://www.bls.gov/soc/2018/soc_2010_to_2018_crosswalk.xlsx}}  SOC(2010) $\rightarrow$ \footnote{ \url{https://www.bls.gov/cps/cenocc2010.htm}} OCC-2010
        \item For real-world employee numbers: Twitter job titles $\rightarrow$ SOC(2018)
        \item For job prestige mapping: Twitter job titles $\rightarrow$ OCC(2010)
    \end{itemize}
}

\begin{table*}[htbp]
    \centering
    % \scriptsize
    \caption{Twitter job title mappings and job prestige estimates, ranked by match score and frequency of mention. Note that multiple job titles may map to the same SOC category. For example, both the \textit{ceo, coo} titles fall into the category \textit{Chief executives}, and share the same occupation prestige score. Match score is ROUGE-based with the range $[0.0,1.0]$, where 1.0 indicates an exact match. We assign the occupation prestige score to those titles with the match score greater than 0.8 to avoid noise from job mapping noise.}
    \footnotesize
\begin{tabular}{p{0.25\linewidth}p{0.35\linewidth}p{0.045\linewidth}p{0.065\linewidth}p{0.1\linewidth}}
\toprule
                Job Title (Bios) &                                                  Occupation Category (SOC) &  Match Score & Mention Freq. &  Occupation Prestige \\
\midrule
                ceo &                          Chief executives &         1.00 &      325261 &         71.58 \\
     civil engineer &                           Civil engineers &         1.00 &       46412 &         65.33 \\
                coo &                          Chief executives &         1.00 &       41722 &         71.58 \\
mechanical engineer &                      Mechanical engineers &         1.00 &       38117 &         70.31 \\
         accountant &                  Accountants and auditors &         1.00 &       36614 &         59.72 \\
      web developer &                            Web developers &         1.00 &       32202 &         55.43 \\
      web designer &                            Web developers &         1.00 &       16653 &         55.43 \\
  community manager & Public relations and fundraising managers &         1.00 &       15279 &         51.64 \\
    general manager &           General and operations managers &         1.00 &       14309 &         49.64 \\
   registered nurse &                         Registered nurses &         1.00 &       14116 &         64.43 \\ 
   ... & ... & ... & ... & ...\\
 tamping machine operator &        Paving, surfacing, and tamping equipment operators &         0.86 &           1 &         34.12 \\
        appliance service technician &                                  Home appliance repairers &         0.86 &           1 &         35.00 \\
      sheet metal worker apprentice &                                       Sheet metal workers &         0.86 &           1 &         34.52 \\
  software engineer web application &    Software developers, applications and systems software &         0.86 &           1 &         60.12 \\
    chief juvenile probation officer & Probation officers and correctional treatment specialists &         0.86 &           1 &         46.71 \\
      director, religious education &             Directors, religious activities and education &         0.86 &           1 &         37.36 \\
     business service representative &                Sales representatives, services, all other &         0.86 &           1 &         43.18 \\
     assistant food service director &                                     Food service managers &         0.86 &           1 &         39.43 \\
  medical and health service manager &                      Medical and health services managers &         0.80 &           2 &         64.06 \\
waste water treatment plant operator & Water and wastewater treatment plant and system operators &         0.80 &           1 &         38.87 \\
\bottomrule

\bottomrule
\end{tabular}
    \label{tab:job-match-prestige}
\end{table*}

\xingzhiswallow{
    \begin{table}[htbp]
        \centering
        \footnotesize
        \caption{Twitter job title mappings and job prestige estimation, ranked by Match score and the mentioned frequency.   }
    \begin{tabular}{p{0.23\linewidth}p{0.23\linewidth}p{0.05\linewidth}p{0.10\linewidth}p{0.1\linewidth}}
    \toprule
                    Job Title (Bios) &                                                  Job Category (SOC) &  Match Score & Mention Freq. &  Prestige Score \\
    \midrule
                    ceo &                          Chief executives &         1.00 &      325,261 &         71.58 \\
         civil engineer &                           Civil engineers &         1.00 &       46,412 &         65.33 \\
                    coo &                          Chief executives &         1.00 &       41,722 &         71.58 \\
            %  accountant &                  Accountants and auditors &         1.00 &       36,614 &         59.72 \\
        %   web developer &                            Web developers &         1.00 &       32202 &         55.43 \\
    %   community manager & Public relations and fundraising managers &         1.00 &       15279 &         51.64 \\
        % general manager &           General and operations managers &         1.00 &       14309 &         49.64 \\
      registered nurse &                         Registered nurses &         1.00 &       14,116 &         64.43 \\ 
       \hline
       
     tamping machine operator &        Paving equipment operators &         0.86 &           1 &         34.12 \\
            appliance service technician &                                  Home appliance repairers &         0.86 &           1 &         35.00 \\
           sheet metal worker apprentice &                                       Sheet metal workers &         0.86 &           1 &         34.52 \\
    %   software engineer web application &    Software developers, applications and systems software &         0.86 &           1 &         60.12 \\
        % chief juvenile probation officer & Probation officers and correctional treatment specialists &         0.86 &           1 &         46.71 \\
        %   director, religious education &             Directors, religious activities and education &         0.86 &           1 &         37.36 \\
    %   medical and health service manager &                      Medical and health services managers &         0.80 &           2 &         64.06 \\
    
    \bottomrule
    \end{tabular}
        \label{tab:job-match-prestige-short}
    \end{table}

}
% \red{How to deal with those unmatched jobs?}

\section{Ethical Considerations}
\label{ethics}
% current 
Although Twitter data is public and often falls under public data exemptions in IRB consideration, sensitive information such as user's gender, ethnicity, health, political affiliation or beliefs must be used with caution.  
Inference on the individual level is restricted use case per the Twitter Developer Agreement, however: \footnote{More about restricted uses of the Twitter APIs: \url{https://developer.twitter.com/en/developer-terms/more-on-restricted-use-cases}}
\begin{quote}
    Aggregate analysis of Twitter content that does not store any personal data (for example, user IDs, usernames, and other identifiers) is permitted, ...
\end{quote}
All of our results are presented at an aggregate level, so that no individual identifier or original content can be traced back or reconstructed from our results. 
All potentially identifiable examples given in our results have been artificially modified to enforce the privacy protection.
%We only share data regarding population-level analysis, not any analysis of individual users that may be abused.  
The IRB of our institution has affirmed that the use of public information does not constitute human subjects research, in accordance with regulations from the US Department of Health and Human Services.

\paragraph{Opportunities  and limitations in our dataset}:  To our best knowledge, our dataset is the first and largest resource for both 1). conducting temporal analysis of biographies  2). studying occupational identities on Twitter. 
It also provides a big picture of the Twitter population composition in multiple socioeconomic views.
However, we also note several potential limitations in our current approach: 
1). The difference in self-disclosure may cause the result biased towards  the population who are more willing to reveal themselves.
2). The job title mapping are not perfect. Although we manually validate part of the results, how to accurately annotate users on a large scale is still a challenging problem in both computer and social science studies. 
3). All real-world data (job prestige, occupation distribution, standard occupation category) is based on the U.S. \versiontwo{Among the subset of users with a non-empty Twitter location field it was observed that over 99\% of them are classified as being within the U.S. using a regex-based location parser, making the results not generalizable to other regions.}
4). \versiontwo{
To protect user privacy,  we can only share word embeddings of biography text and the full job transition graph. 
% Raw biography text can only be obtained on a case-by-case basis by signing a data usage agreement.
}
% \textcolor{red}{should add other concerns/ assumptions/ data quality}

% \todo{Do we need to talk about sharing raw text at all?}

\xingzhiswallow{

    \begin{figure*}[htbp]
        \centering
        \subfloat {\includegraphics[width=0.28\textwidth]{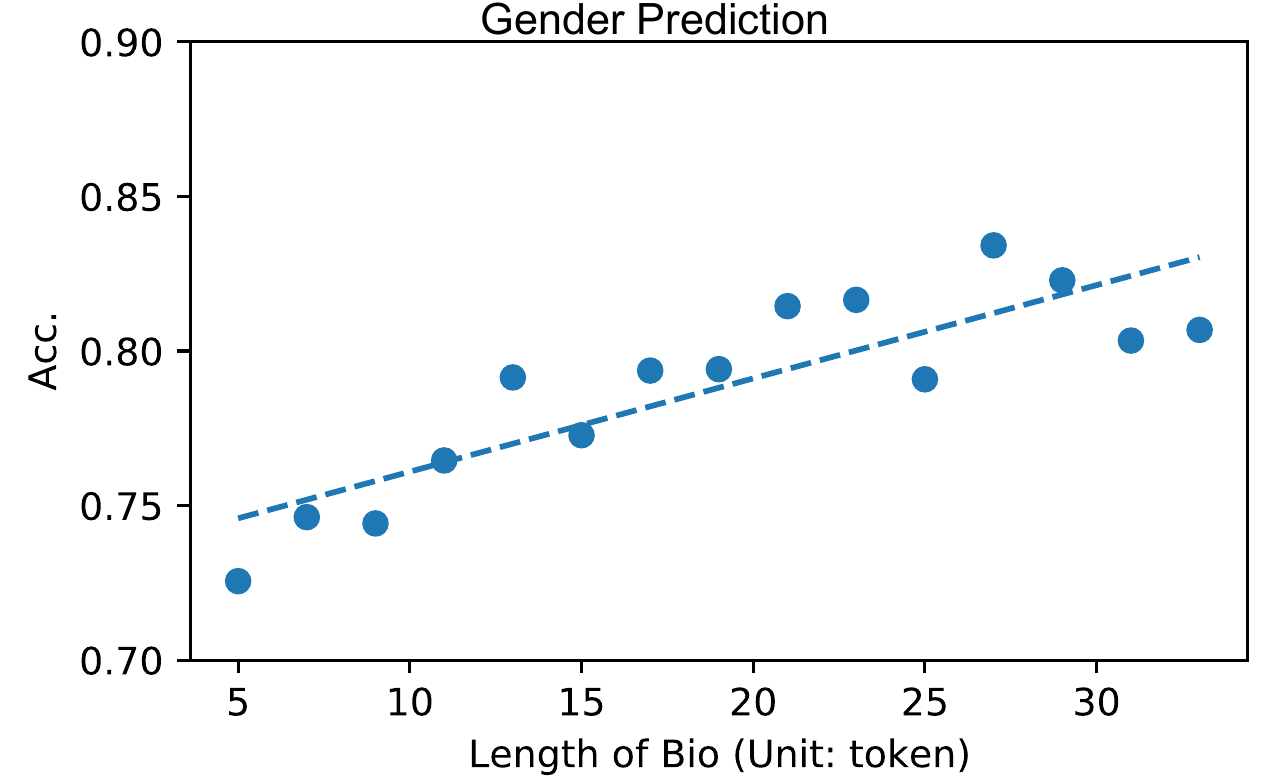}}\hfill
        \subfloat {\includegraphics[width=0.28\textwidth]{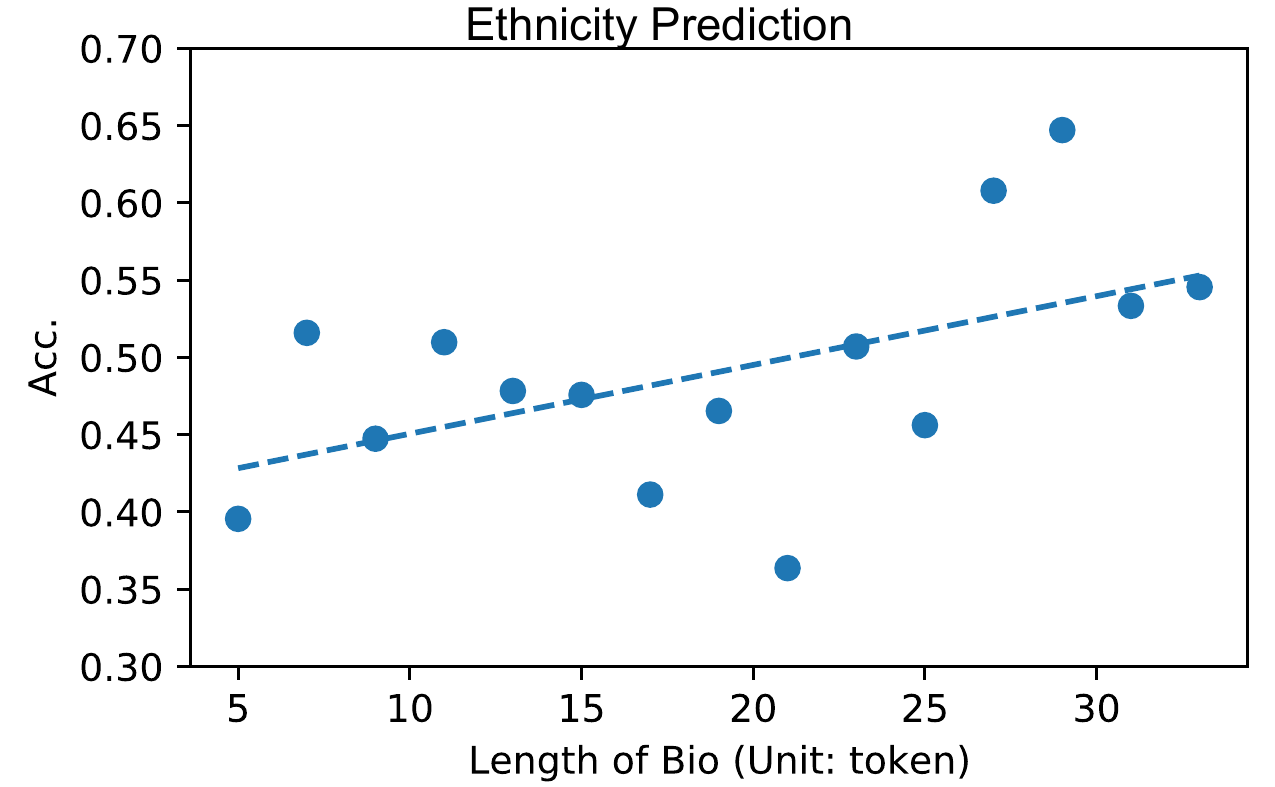}}\hfill
        \subfloat{\includegraphics[width=0.28\textwidth]{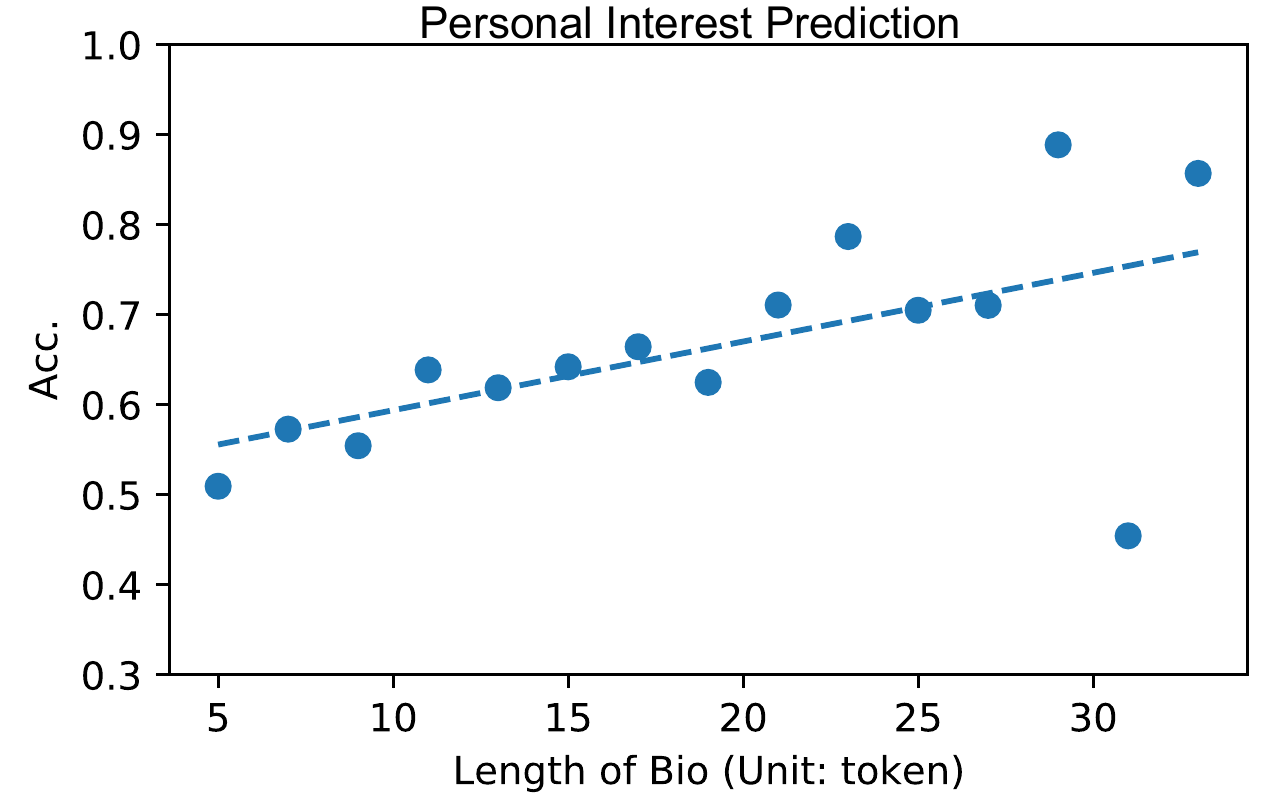}}\hfill
        %\hfill
        % \vspace{-4mm}
        \caption{Bio Length vs. Prediction Performance: The classifier could predict more accurately, as people write more content in their bios. Short bios usually contain a list of informative words (e.g., \textit{mom, father, asian american, MAGA}), which explains the initial surprisingly good performance. }
        \label{fig:prediction_bio_len_res}
    \end{figure*}
    %The breakeven point between tweets and Bios is 16 ($2^4$).
    
    \begin{figure*}[htbp]
        \centering
        \subfloat{\includegraphics[width=0.30\textwidth]{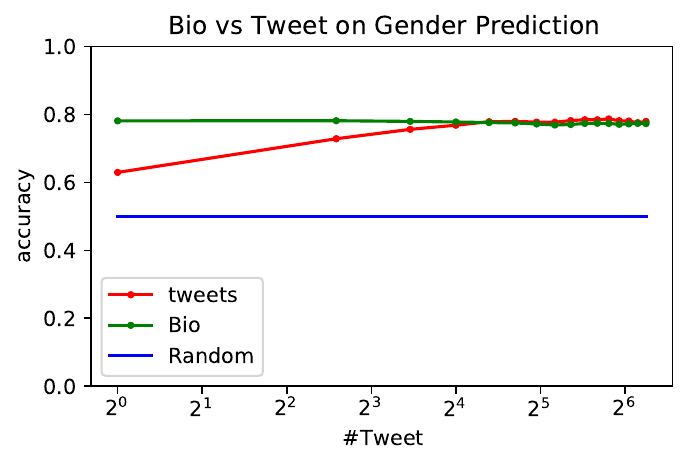}}\hfill
        %\subfloat[Ethnicity.\label{fig:ethnicity_nb}] {\includegraphics[width=0.23\textwidth]{Figures/ethnicity_figures/accuracy_nb_repeat_5.pdf}}\hfill
        \subfloat {\includegraphics[width=0.30\textwidth]{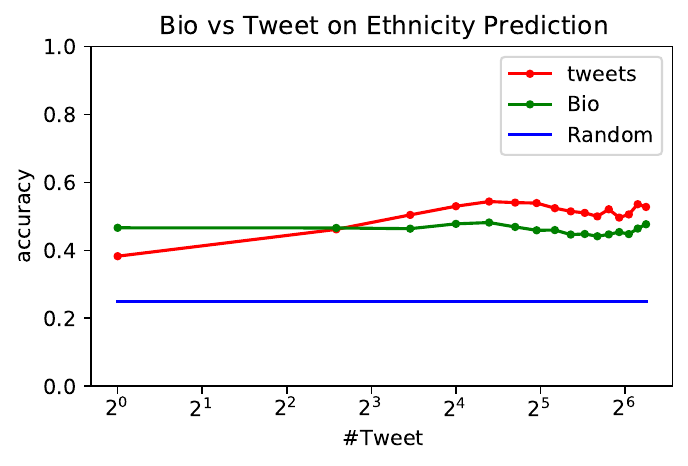}}\hfill
        %\subfloat[Personal Interests.\label{fig:interest_nb}] {\includegraphics[width=0.23\textwidth]{Figures/interests_figures/accuracy_nb_repeat_5.pdf}}\hfill
        \subfloat {\includegraphics[width=0.30\textwidth]{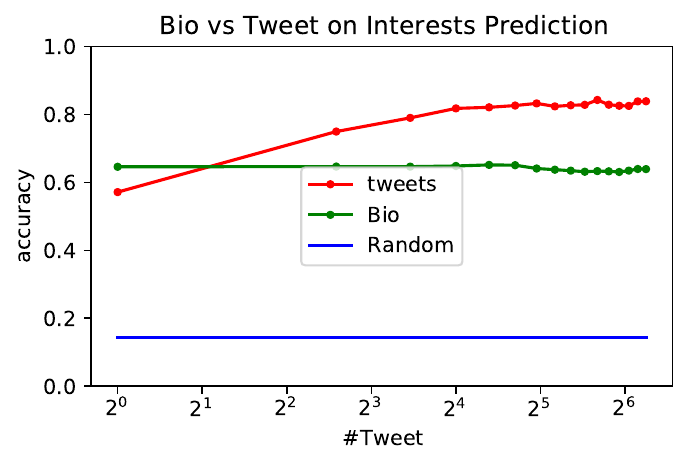}}\hfill
        %\hfill
        % \vspace{-5mm}
    
        \caption{Prediction performance comparison between bio and tweets: it demonstrates the rich self-identity in bio.
        Gender Prediction (\textit{left}): Bios contain rich clues of gender. Approximately 1 bio equals to 16 ($2^4$) tweets. 
        Ethnicity Prediction (\textit{middle}): More tweets reveal more about ethnic background.  The break-even point is about 8 ($2^3$) tweets.
        Personal Interest Prediction (\textit{right}): a few tweets (approx. 2) suffice to reveal user's interest as one bio does.
        }
        \label{fig:prediction_res}
    \end{figure*}

    \section{The Predictive Power of Biography}
    \label{single-bio}
    
    Although biographies have been proven to be useful for user profiling, we want to quantify how revealing a single bio is in the predictive tasks of demographics and personal hobbies. 
    To demonstrate the conciseness and richness of biographies, we compare the prediction performance of biographies with different lengths and to different numbers of tweets, discovering that the biographies reveal more about demographics than tweets, but less about personal interests.

    % We set up the classification tasks and compare the performance 
    % we are interested in the difference of predictive power between bios and tweets toward different attributes, specially for demographics (gender, ethnicity) and personal hobbies.

    \paragraph{Experiment setup:} 
    We randomly sample the users with demographics estimation together with one bio in year 2020, and set the demographic labels as ground-truth if the  label probability is greater than 95\% for gender (either Male/Female) and 50\% for ethnicity (either in White/Black/Hispanic/Asian).  If the probability score does not fall in the designated range, we consider it as an uncertain sample and exclude it from these experiments. 
    Hashtags in the tweets are strong evidence of users' interested topics. Similar to \cite{conover2011predicting}, we select 187 representative hashtags from the most frequent ones and categorize them into 7 groups as shown in Table.\ref{tab:hashtag_category}. For each user, we assign the group of the most mentioned hashtags as the ground-truth label. To compare the predictive power between biographies and tweets, we also randomly sample up to 200 tweets associated to each user.
    Since the original label distribution is highly imbalanced, we further under-sample the majority classes to keep it balanced, resulting in a total of 229,460/12,192/14,483 samples in the prediction task for gender, ethnicity and personal interest, respectively. 
    
    Although the aforementioned label discretization (especially for male/female binarization) does not promise the perfectly correct ground-truth, we want to note that the experiments are for predictive power comparison between biographies and tweets, the findings should be valid unless the labels are purely random.

    We use Naive Bayes and BERTweet\cite{nguyen2020bertweet} as classifiers for interpretability and the best performance comparison.  We brief the model configuration as follows. More detail is included in Table \ref{tab:model_cofig} in Appendix.

    \begin{itemize}
        \item  \textbf{Naive Bayes} : Multinomial Naive Bayes classifier with tf-idf weighted 1 to 5-gram bag-of-words features of max vocabulary size 70,000. We use biographies, and concatenate different numbers of tweets for training and predicting.
        %\item \textbf{AvgEmb.}: Logistic regression classifier with averaged GLoVe pre-trained Twitter word embeddings \cite{pennington2014glove}
        %\item \textbf{BiLSTM}: Two layers of Bi-directional LSTM \cite{hochreiter1997long, schuster1997bidirectional} with two linear layers for classification.
        %\item \textbf{RoBerta-CLS}: Pre-trained RoBerta (\textit{roberta-base}) from HuggingFace transformer\cite{wolf-etal-2020-transformers} library. Use [CLS] embeddings from the last layer for classification. We use averaged tweets embeddings as user-level features for training and predicting. 
        %We use 2-layer linear layers with ReLU as activation function without freezing any pre-trained layer.
        %\item \textbf{RoBerta-Mean}: Similar to RoBerta-CLS, but using average token embeddings from last layer as sentence representation for classification.
        % \item \textbf{BERTweet}\cite{nguyen2020bertweet}: Similar to RoBerta, but we replace the pre-trained model with \textit{BERTweet}, which was pre-trained on 850M English Tweets. 
        \item \textbf{BERTweet}: Similar to RoBerta \cite{wolf-etal-2020-transformers} structure, but it was pre-trained on 850M English Tweets. We use \textit{[CLS]} embeddings of each biography, and  the averaged  \textit{[CLS]} embeddings of tweets as user-level features for training and predicting.
    \end{itemize}

    % We present the results in Table \ref{table:bio_pred_res} and the comparison results in Figure \ref{fig:}
    
    % To train a user-level classifier from varying number of personal tweets, we extract sentence embeddings for all the tweets of a user,  use average pooling to obtain the user-level embeddings, then train the classifier for the downstream tasks. In testing phase, we use the trained classifier and feed pooled embeddings of arbitrary number of tweets from a user. 

    %Another option is to concatenate tweets as a long sentence, but it may be truncated due to 512-token limitation in BERT family models.
    %As the number of tweets increases, there become a challenge of how to obtain sensitive enough document-level embeddings\cite{lynn2020hierarchical} (eg. weighted pooling or message-level attention). It is an interesting topic, but falls out of the scope of this paper. \todo{delete this?}

    \begin{table}[htbp]
    \centering
    %\small
    \caption{The prediction performance using Biography only. The simplest Naive Bayes model achieves decent performance, demonstrating the encoded information in the biographies are straightforward. }
    
    \begin{tabular}{p{0.15\textwidth} p{0.08\textwidth} p{0.08\textwidth} p{0.08\textwidth}}
    \toprule 
    Task: &Gender &Ethnicity & Interests \\
    \midrule
    \#Classes &2 & 4 &7 \\
    \midrule
    \#Samples & 229,460& 12,192 & 14,483 \\
    \midrule 
     Random & 0.500  & 0.250  & 0.142  \\
    
     Naive Bayes & 0.738  & 0.414  & 0.578  \\
    
     %AvgEmb. & 0.707  & 0.403  & 0.573  \\
    
     %BiLSTM & 0.743  & 0.416  & 0.594  \\
    
    % RoBerta - CLS &0.735  & 0.376  & 0.525  \\
    
     %RoBerta - Mean &0.735  &-- & 0.528  \\
     %\hline
     BERTweet  &\textbf{0.780}    &\textbf{0.493}   & \textbf{0.649} \\
    \bottomrule
    \end{tabular}
    \label{table:bio_pred_res}
    \end{table}
    
    %\todo{Accuracy vs. bio length  --- using the best model}
    
    % confusion matrix ??
    % \begin{figure}[H]
    %     \centering
    %     \includegraphics[width=0.3\textwidth]{Figures/gender_figures/accuracy_Naive_Bayes_repeat_5.pdf}
    %     \caption{Naive Bayes classifier (repeat five times).}
    %     \label{fig:gender_nb}
    % \end{figure}

    % \begin{figure}[H]
    %     \centering
    %     \includegraphics[width=0.3\textwidth]{Figures/ethnicity_figures/accuracy_nb_repeat_5.pdf}
    %     \caption{Naive Bayes}
    %     \label{fig:ethnicity_nb}
    % \end{figure}

    % \begin{figure}[H]
    %     \centering
    %     \includegraphics[width=0.3\textwidth]{Figures/interests_figures/accuracy_nb_repeat_5.pdf}
    %     \caption{Naive Bayes}
    %     \label{fig:interest_nb}
    % \end{figure}

    \xingzhiswallow{ 
        \begin{figure}[htbp]
            \centering
            \subfloat[Female.\label{fig:1a}] {\includegraphics[width=0.23\textwidth,height=0.18\textwidth]{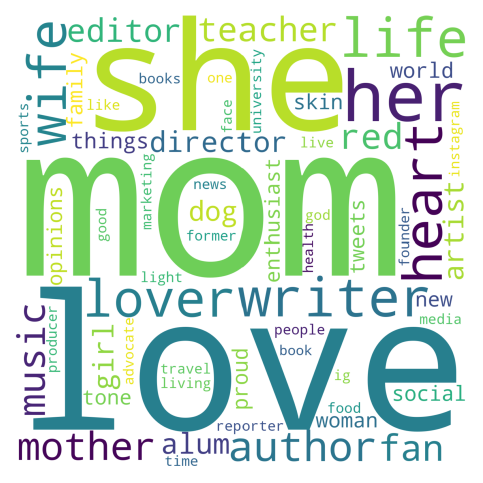}}\hfill
            \subfloat[Male.\label{fig:1}] {\includegraphics[width=0.23\textwidth,height=0.18\textwidth]{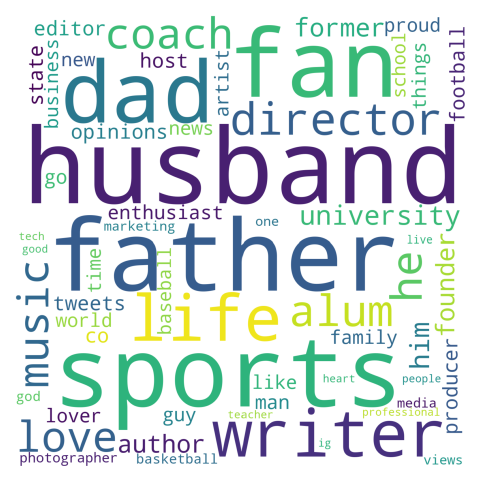}}\hfill
            %\hfill
            \caption{Top predictive words of Male/Female in bios from Naive Bayes Classifier. Stop words are excluded. }
            \label{fig:gender_keywords}
        \end{figure}
    }
    
    \xingzhiswallow{
        \begin{figure}[htbp]
            \centering
            \subfloat[Business.\label{fig:interest_key_bussiness}] {\includegraphics[width=0.23\textwidth,height=0.18\textwidth]{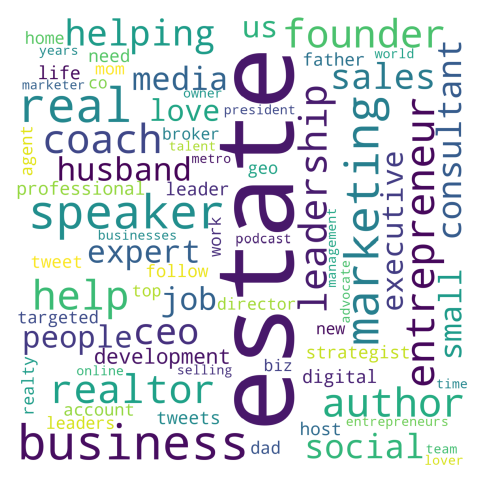}}\hfill
            %\subfloat[Entertainment.\label{fig:interest_key_entertainment}] {\includegraphics[width=0.23\textwidth]{Figures/interests_figures/interests_pos_entertainment_lb_bio_keywords.pdf}}\hfill
            %\subfloat[Literature.\label{fig:interest_key_literature}] {\includegraphics[width=0.23\textwidth]{Figures/interests_figures/interests_pos_literature_lb_bio_keywords.pdf}}\hfill
            %\subfloat[Music.\label{fig:interest_key_music}] {\includegraphics[width=0.23\textwidth]{Figures/interests_figures/interests_pos_music_lb_bio_keywords.pdf}}\hfill
            \subfloat[Politics.\label{fig:interest_key_politics}] {\includegraphics[width=0.23\textwidth,height=0.18\textwidth]{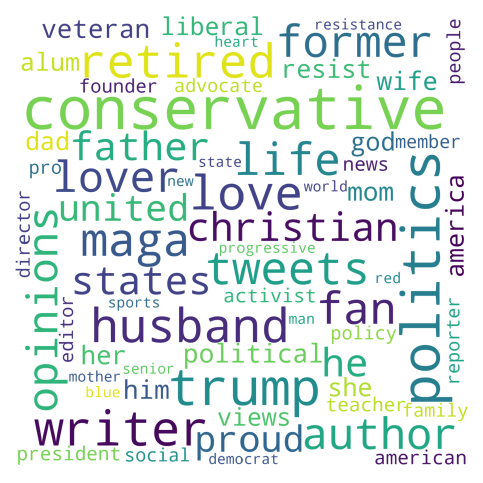}}\hfill
            %\subfloat[Sports.\label{fig:interest_key_sports}] {\includegraphics[width=0.23\textwidth]{Figures/interests_figures/interests_pos_sports_lb_bio_keywords.pdf}}\hfill
            %\subfloat[Tech.\label{fig:interest_key_tech}] {\includegraphics[width=0.23\textwidth]{Figures/interests_figures/interests_pos_tech_lb_bio_keywords.pdf}}\hfill
            %\hfill
            \caption{Top predictive words of Personal Interests in bios from Naive Bayes Classifier }
            \label{fig:interests_keywords}
        \end{figure}
    
    }

    % \subsection{Length of Biography}
    \paragraph{Biography alone is powerful:} The results in Table \ref{table:bio_pred_res} shows that demographics and personal hobbies can be predicted with a decent accuracy from biography alone, achieving accuracy of 78.0\%, 49.3\% and 64.9\% for gender, ethnicity and interest prediction.
    We also show the most predictive words in bios (extracted from Naive Bayes) in Figure \ref{fig:gender_keywords}, consisting of words with very strong gender components (e.g.,\textit{ mom, father, husband}).  For ethnicity prediction, we observed predictive (but usually stereotypical) words, such as \textit{engineer} in Asian, \textit{Texas, Mexico} in Hispanic group and \textit{black} in African-American group. Similar feature analysis results were also observed in \cite{preoctiuc2018user}.
    
    % we found predictive (but usually stereotypical) words: such as \textit{engineer} in Asian, \textit{Texas, Mexico} in Hispanic group and \textit{black} in African-American group. Similar feature analysis results were also observed in \cite{preoctiuc2018user}.
    
    In addition, we breakdown the performance corresponding to the bio length in Figure \ref{fig:prediction_bio_len_res}. In general, the prediction accuracy improves as the bio length increases, reflecting that longer self-description reveals more clues about the person. Interestingly, we found that even the biographies of 5 tokens could reveal demographics and interests of significantly higher chances than random, which further verifies its conciseness and richness.
    
    \paragraph{Biography reveals more demographics than tweets:} Figure \ref{fig:prediction_res} presents the  prediction accuracy (with BERTweet Model) of bios against number of tweets . 
    We divided testing samples into bins according to the number of utilized re/tweets (in x-axis), then report the prediction accuracy with bio and tweets separately. 
    We exclude the users whose total tweets is less than the  designated number (e.g., $2^4$), which makes the bio curve slightly fluctuating. 
    % To make the results comparable, we exclude the users who have the tweets less than the designated number (e.g., $2^4$), and only report the accuracy (bio-alone and tweets) for the rest of them. It makes the bio curve slightly fluctuating as we increase utilized tweets and exclude those don't have sufficient tweets.
    
    As we observed in Figure \ref{fig:prediction_res},  biographies contain much demographics information while tweets cannot compete with.  Approximately the break-even point is about 16 and 8 tweets for gender and ethnicity, respectively. 
    However, a few tweets generally reveal more about personal hobbies because the vocabulary used in (re-)tweets well describes the users' frequently engaged topics, thus, reflecting the personal interests. 
    % \red{people reveal more about demographics than personal interests in bio? the rank/importance of identities? }
    
    Besides the demographics and personal interest, we also observed many occupational identities (e.g., \textit{ director, coach, writer, etc} in Figure \ref{fig:gender_keywords}), which will be analyzed in the following section.
    %we analyze the occupational identities on Twitter in the next section.
    
    %revealing the compositions of twitter population.  

    % It implies that the predictive clue in biography is 
    % For example, when \textit{\#Tweet} equals to one, we use just single randomly selected tweet to predict gender for each user. As we saw in Figure \ref{fig:prediction_res}, a single bio contains richer gender information than a single tweet does: it needs approximately 16 tweets to achieve the similar performance of a bio. 

    % If a user have less than the designated number of tweets (e.g., $2^4$), we exclude such samples when calculating the accuracy  to make the results comparable.
    
    % If a user has more than the designated number of tweets, we down sample the tweets and report the performance.
    
    % \footnote{We keep test samples which has tweets more than the designated number, so the bio performance curve may fluctuate.}, 
    
    % For example, when \textit{\#Tweet} equals to one, we use just single randomly selected tweet to predict gender for each user. As we saw in Figure \ref{fig:prediction_res}, a single bio contains richer gender information than a single tweet does: it needs approximately 16 tweets to achieve the similar performance of a bio. 

    \xingzhiswallow{

        \subsection{Gender Prediction}
        % first, we show the signal words (excluding stop words) from NB's top featured words. P(x|y). We can see clear gender distinctive words for male and female. It shows the bio contains many clues about users' gender in the self-identity.
        The results in Table \ref{table:bio_pred_res} shows that gender can be predicted with a decent accuracy (78.0\%) from biography alone.  
        %\todo{NOTE: There are still many words that appear on both sides (writer, love, music).  Why?  If they are positive predictors on one side and negative on the other, we should say so.  Or better, only include the strongest positive predictors in each. (Jason)}
        We show the most predictive words in bios (extracted from Naive Bayes) in Figure \ref{fig:gender_keywords}, consists of words with very strong gender components (e.g., mom, father, husband).
        %\todo{NOTE: Were these words predictive in the bios or the tweets or both?  We should say explicitly. (Jason)} 
        
        In addition, Figure \ref{fig:prediction_bio_len_res} shows that prediction accuracy improves as the bio length (in token) increases, reflecting that longer self-description reveals more clues about the person.
        
        % \begin{table}[!h]
        %         \centering
        % \begin{tabular}{|p{0.08\textwidth}|p{0.05\textwidth}|p{0.05\textwidth}|p{0.05\textwidth}|p{0.05\textwidth}|p{0.05\textwidth}|p{0.05\textwidth}|}
        % \hline 
        %   & Asian \ vs \ Hispanic & Asian vs Black & Asian vs White & Hispanic \ \ \ vs Black & Hispanic \ \ \ vs White & Black vs White \\
        % \hline 
        %  Bio  & 0.650 & 0.655 & 0.620 & 0.650 & 0.607 & 0.653 \\
        % \hline 
        %  \makecell[l]{\#Tweets} & \textasciitilde 2 & \textasciitilde 1 & \textasciitilde 2 & \textasciitilde 2 & \textasciitilde 2 & \textasciitilde 2 \\
        %  \hline
        % \end{tabular}
        % \end{table}

        We also investigate how much gender-related information is encoded in daily re/tweets.
        In Figure \ref{fig:prediction_res}, we divided users into groups according to the number of utilized re/tweets (in x-axis)\footnote{We keep test samples which has tweets more than the designated number, so the bio performance curve may fluctuate.}, and predict the gender given the tweets. For example, when \textit{\#Tweet} equals to one, we use just single randomly selected tweet to predict gender for each user. As we saw in Figure \ref{fig:prediction_res}, a single bio contains richer gender information than a single tweet does: it needs approximately 16 tweets to achieve the similar performance of a bio.

        %Error analysis: 
        
        % \begin{table}[htbp]
        %     \centering
        %     \begin{tabular}{c|c|c}
        %     \toprule
        %     Bio & Ground truth & Prediction
        %     \midrule
        %     work, hard, pray, harder, romans, mechanical, engineering, student, nc, at & Female & Male  \\ 
        %     \hline
        %     nigeria, united, states, born, co, founder, ceo, co, host, podcast, tuesday, s, pm, wat, on, twitter, youtube, live & Female & Male  \\ 
        %     \hline
        %     \bottomrule
        %     \end{tabular}
        %     \caption{Error Cases}
        %     \label{tab:gender_err_case}
        % \end{table}
        
        \begin{comment}
        \begin{figure}[htbp]
            \centering
            \hspace{-0.034\textwidth} \subfloat {\includegraphics[width=0.23\textwidth]{Figures/ethnicity_figures/fig_pairwise_acc_bert.pdf}}\hfill
            \subfloat {\includegraphics[width=0.23\textwidth]{Figures/interests_figures/fig_pairwise_acc_bert.pdf}}\hfill
            \caption{Pairwise prediction accuracy. Blue color indicates low accuracy. Ethnicity (\textit{left}): It is difficult to distinguish Asian from White. Personal Interests (\textit{right}): It is hard to distinguish sports and entertainment.}
            \label{fig:prediction_res_breakdown}
        \end{figure}
        
        \end{comment}
        
        \subsection{Ethnicity Prediction}
        % Ethnicity is implicitly encoded in bio. We found few people explicitly indicate their ethnicity identity in bio. So most of key words in different ethnicity group looks very similar. But still we found some stereotypical key words  (engineer in Asian, Texas/Mexico in Hispanic)
        
        As shown in Table \ref{table:bio_pred_res} and Figure \ref{fig:prediction_bio_len_res}, the best model can predict 4-class ethnicity with 49.3\% accuracy, and the accuracy increases as users write more in their bios.
        In general, it is more challenging to predict ethnicity from bio, because we don't observe many highly ethnicity biased words. However, some anecdotes can be found, for example, we found predictive (but usually stereotypical) words: such as \textit{engineer} in Asian, \textit{Texas, Mexico} in Hispanic group and \textit{black} in African-American group. Similar feature analysis results were also observed in \cite{preoctiuc2018user}.
        But, as Figure \ref{fig:prediction_res} shows, a single bio has more predictive power when compared to one random tweet. Approximately 8 tweets are required to achieve the same prediction accuracy as a single bio.
        %, and it gains more accuracy as seeing more tweets.

        %We present pair-wise (one-vs-one) ethnicity prediction accuracy in Figure \ref{fig:prediction_res_breakdown}.  
        %We also include pair-wise Bios and Tweets comparison in Figure \ref{fig:prediction_res_ethnicity_breakdown} in Appendix. 
        
        %\todo{add the figure in appendix -- add reference to ethnicity difference analysis?}
        
        %https://www.aclweb.org/anthology/D11-1120.pdf
        % 1. label is not clean 
        
        % \begin{figure}[htbp]
        %     \centering
        %     \subfloat[Asian.\label{fig:eth_key_asian}] {\includegraphics[width=0.23\textwidth]{Figures/ethnicity_figures/ethnicity_pos_asian_bio_keywords.pdf}}\hfill
        %     \subfloat[Black.\label{fig:eth_key_black}] {\includegraphics[width=0.23\textwidth]{Figures/ethnicity_figures/ethnicity_pos_nh_black_bio_keywords.pdf}}\hfill
        %     \subfloat[Hispanic.\label{fig:eth_key_hispanic}] {\includegraphics[width=0.23\textwidth]{Figures/ethnicity_figures/ethnicity_pos_hispanic_bio_keywords.pdf}}\hfill
        %     \subfloat[White\label{fig:eth_key_white}] {\includegraphics[width=0.23\textwidth]{Figures/ethnicity_figures/ethnicity_pos_nh_white_bio_keywords.pdf}}\hfill
        %     %\hfill
        %     \caption{Top predictive words of Ethnicity from Naive Bayes Classifier, excluding stop words. }
        %     \label{fig:ethnicity_keywords}
        % \end{figure}
        
        \subsection{Personal Interests Prediction}
        % We show the example of predictive words of personal interests from bio text. Many of them is directly related with the topics. (maga in politics, artist in music.  young people (university) in sports.)
        
        Biographies may contain personal interests as shown in Figure \ref{fig:interests_keywords}.
        The people of different interest groups use distinctive words in their bios. As Table \ref{table:bio_pred_res} and Figure \ref{fig:prediction_bio_len_res} shows, we can predict personal interest with 64.9\% accuracy, and the accuracy keeps improving as the bio length grows. 
        % In Figure  \ref{fig:prediction_res_breakdown}, we observe that it's easier to distinguish a user interested in literature from one interested in business, while it is more difficult to discern sports interest from entertainment.
        
        In addition, re/tweets also contains rich information as people often engage in the conversations (re/tweets) matching to their interests. 
        From the experimental results in Figure \ref{fig:prediction_res}, roughly 16 random tweets could tell the personal interest with 80\% accuracy. A single bio still outperforms a single tweet.

        % 8 pages,
        % score of bio 
        % routine classification  (2-3 pages)
        % how bio evolves (3 pages) big novelty, promotion demotion ...
        % (ex-..) how often that happened. things they were proud of.
        %  former, ex-
        % break down of different bio

        }

}

\section[Occupations in Twitter Bios]{Occupational Identities Reflected Through Twitter Biographies}
\label{sec:occupation-bios}
% \textcolor{blue}{Occupational information mentioned in biography reflects one's occupational identity.}

Work is an essential event in our daily life, which contributes to our self-identity. 
Sociologists and occupational scientists have proposed several theories about the establishment of occupational identity. One fundamental factor is 
%personal traits (e.g., demographics \cite{sluss2007relational}) and 
the quest for social approval \cite{collin2008constraints, taylor2017doing, kielhofner2002model} (e.g., job prestige, income).  
Although we have a wide variety of public labor statistics (e.g., employment by sector), 
%
%the invisible and ambiguous nature of occupational identity makes the self-identity study difficult, and extremely troublesome for quantitative research. 
%
Twitter biographies provide a unique and directly observable view of how users define \textit{themselves} by occupation, displaying whether the users feel their current work reflects an important part of their \textit{self-identity}.
%By estimating demographics from user names and detecting job titles in Twitter biographies
After filtering out the potential organizational and bot accounts \footnote{
%About 19.25\% of the Twitter users reveal their genuine names. 
%
For this analysis, 
we only include the users who report human-like names which can be matched to US Census data. }, 
we obtain a total of 23,267 distinct self-reported job titles.
%which is slightly smaller than full data sample). 
In addition, we match the job titles to Standard Occupation Code and associated job prestige scores, as illustrated in Table \ref{tab:job-match-prestige}.  
% To our best knowledge, we are first to present such large-scale quantitative analysis by leveraging Twitter bios, and this is the largest resources for occupational identity studies. 
%\ref{tab:top_job_titles}. %12907860 - 10422816 = 2485044

In the following subsections, we first walk through a few extracted occupation examples from Twitter biographies. Then, we propose to use the \textit{Occupation Over-Represented Ratio ($\kappa$)} statistic to describe the relative difference in the frequency of occupations in Twitter biographies compared to actual real-world presence, which captures people's sense that the occupation as a part of self-identity.
Finally, we analyze the job prestige variables with respect to the establishment of occupational identity.

\subsection{Over/Under-represented Occupations in Twitter Biographies}

The distribution of self-mentioned occupations in Twitter biographies is significantly different from that as occurs in the U.S.\ population at large\footnote{We use the employment statistics from U.S. Labor Department}.
The most/least frequently mentioned jobs in Table \ref{tab:top_job_titles} seem contradictory to our sense of the labor market in terms of volume and socioeconomic distribution: for example, we expect more workers than executives.
Prestigious occupations appear much more likely to be included as part of occupational identity, and are thus over-represented in our data. 
We can quantify this observation through data on the relative prestige of jobs/titles.
We present our findings below.
%In Table \ref{tab:top_job_titles}, the most mentioned occupational identity is \textit{student}, reflecting the user's age bias in Twitter's population. 

\begin{table}[hbtp]
    \centering
    \small
    \caption{The most/least popular self-reported jobs in biographies.}
    % \vspace{-3mm}
     \begin{tabular}{@{}l @{}l}
        
        \begin{tabular}{
                |p{0.03\linewidth} % rank 
                p{0.15\linewidth} % job
                p{0.07\linewidth}| % percentage
                }
                \hline
                      \#& Occupation & (\%) \\
                \hline
                      1&    student &  12.91 \\
                      2&    founder &   6.08 \\
                      3&  director &   6.06 \\
                      4&      owner &   5.37 \\
                      5&        ceo &   4.55 \\
                      6&  president &   4.52 \\
                %       co-founder &   2.5716 \\
                %          youtuber &   2.3110 \\
                %               vp &   2.1969 \\
                %           partner &   2.1091 \\
                %           prover &   1.6968 \\
                %               cco &   1.3404 \\
                %  graphic designer &   1.3321 \\
                %         associate &   1.2903 \\
                %          investor &   1.2102 \\
                % software engineer &   1.0185 \\
                %               cto &   0.8914 \\
                %         scientist &   0.7290 \\
                %   vice president &   0.7043 \\
                \hline
            \end{tabular}
        &
        \begin{tabular}{
                |p{0.06\linewidth} % rank 
                p{0.25\linewidth} % job
                p{0.07\linewidth}| % percentage
                }
                \hline
                      \# & Occupation & (\%) \\
                \hline
        % 23247 &                     hair assistant &   0.0000 \\
        23248 &                  restaurant busser &   <0.01 \\
        % 23249 &         electronic system engineer &   0.0000 \\
        % 23250 &  satellite installation technician &   0.0000 \\
        % 23251 &            director cardiovascular &   0.0000 \\
        23252 &                      track sweeper &   <0.01 \\
        % 23253 &            business loan processor &   0.0000 \\
        % 23254 &                    case specialist &   0.0000 \\
        % 23255 &             senior java programmer &   0.0000 \\
        % 23256 &                        rn staffing &   0.0000 \\
        23257 &                     referral clerk &   <0.01 \\
        % 23258 &           director quality systems &   0.0000 \\
        % 23259 &       plant maintenance technician &   0.0000 \\
        % 23260 &         patient service specialist &   0.0000 \\
        % 23261 &       route service representative &   0.0000 \\
        23262 &                        soil expert &   <0.01 \\
        % 23263 &               senior group manager &   0.0000 \\
        23264 &                  grocery deliverer &   <0.01 \\
        % 23265 &          rehabilitation counsellor &   0.0000 \\
        23266 &                    culinary worker &   <0.01 \\
                \hline
            \end{tabular}\\
    \end{tabular}
    \label{tab:top_job_titles}
\end{table}

\begin{figure}[htbp]
    \centering
    \includegraphics[width=0.6\textwidth]{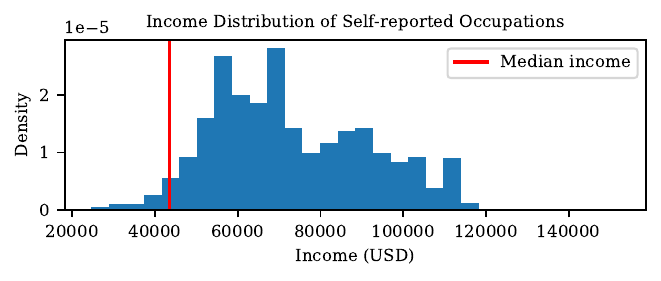}
    % \vspace{-5mm}
    % \small
    \caption{The salaries of self-reported jobs are significantly higher than the overall median (\$43,431 USD) in New York, US. We use New York's job income statistics as standard to avoid salary discrepancy across regions. }
    \label{fig:match_job_salary}
    % \vspace{-5mm}
\end{figure}

\paragraph{The salary of the self-reported occupations is significantly higher}: It is known that the average income of Twitter users is higher than that of the general population \cite{wojcik2019sizing, blank2017digital}. We further find that the difference in self-reported occupations in biographies makes this gap even more significant.
% Those low-income users may not use Twitter or hide their occupations, which makes the them even under-representative. 
By matching the self-reported job titles to their incomes, Figure \ref{fig:match_job_salary} presents the histogram of the salary of the occupations in bios, showing that most users have higher salary than the median income
\footnote{
%We use New York income list as the single source to avoid salary discrepancy across regions. 
We use Rouge-L F1 score to softly align reported occupations to New York median income list. %e.g., Preschool Teacher:\$40,650. 
resulting in 21,946 job titles (out of 25,033) from 3,408,916 users associated with their occupation income.
To avoid the income discrepancy in different regions, we use New York's job income list as the standard. 
Data source: \url{https://catalog.data.gov/dataset/occupational-employment-statistics}}. 
This finding provides a clue that a job's overall salary impacts its role in developing occupational identity.
% whether people would like to include more well-paid jobs as their occupational identity.  
%In the next step, we take into account the real-world job prior and measure whether individual jobs are over-represented in Twitter bios, then investigate the correlation between the over-representation and the job prestige. 

\paragraph{Identifying over-represented occupations in Twitter biographies:}
\versiontwo{Titles can be aspirational: being CEO of Ford Motor Company means something different than CEO of a one-person startup.}
One interesting observation is that more people define themselves as CEOs (\#5) on Twitter than Grocery Deliverers (\#23264), despite the fact that there are raw fewer CEOs in the general population.
This difference can be explained by the fact that: 1) grocery delivers may be less likely have Twitter accounts, and/or 2) grocery delivers do not consider the job as part of their self-identity and exclude it from personal biographies. 
In our analysis, 
%the first factor makes the under-represented occupations (e.g., delivers) even less represented, while 
it is the latter possibility that reveals insights about the job-specific factors toward occupational identity. 

To quantify over-represented occupational families in Twitter biographies, we use the \textit{ver-representation atio} $\kappa$ to measure how each self-mentioned occupation exceeds the expected frequency according to the real-world workforce percentage. Specifically, $\kappa$ is defined by:

% \vspace{-5mm}
\[   
% \footnotesize
\kappa = \frac{P(\text{Occupation|Twitter})}{P(\text{Occupation|Real-World})} 
\left\{
\begin{array}{ll}
       >1.0 &, \text{Over-represented} \\
      =1.0 &, \text{Balanced} \\
      <1.0 &, \text{Under-represented} \\
\end{array} 
\right. \]

% For each occupational family, we aggregate the counts of the observed jobs according to the job title maps. 
% For example, in \textit{Chief Executive} category, we calculate total number of the self-reported jobs like \textit{CEO, COO, President, ..., etc}, then divide it by the real-world portion of \textit{Chief Executive}. 

% \begin{wrapfigure}{R}{0.95\textwidth}
\begin{figure}[hbtp]
    \centering
    \vspace{-0mm}
    % \subfloat[Real-world]{
    % \includegraphics[width=0.48\textwidth]{Figures/fig-job-overrep-prestige-v2.pdf}}
    % \includegraphics[width=0.49\textwidth]{Figures/fig-job-overrep-prestige-v2.pdf}
    \includegraphics[width=0.51\textwidth]{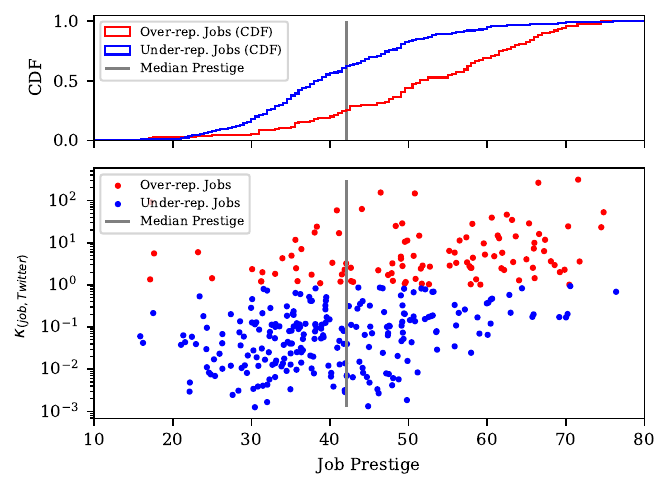}
    % \vspace{-5mm}
    \small
    \caption{ 
    %Job prestige and  $\kappa$ is positively correlated.
    %with Spearman correlation $\rho = 0.4858$
    % The Spearman correlation between $\kappa$ and Job prestige is 0.4858. 
    \textit{Bottom:} Job prestige scores measure the socioeconomic status of an occupation according to its salary and public perception.
    $\kappa$ measures the ratio of how an occupation is over-represented on Twitter compared to its real-world labor distribution.  Prestige and $\kappa$ correlate with Spearman's correlation $\rho = 0.4858$.
    Over-represented jobs (red) generally have higher than the median prestige. 
    \textit{Upper:} The prestige distribution between over/under-represented jobs are statistically different with K-S test $d = 0.4284$ and p-value  $\approx 2.07 \times 10^{-12}$.
    %The 74.04\% of the over-represented jobs (red) are higher than the median, which is significantly greater than the under-represented jobs of only 39.48\%. 
    %The prestige distribution between over/under-represented jobs are significantly different with K-S test 0.42844 with p-value  $\approx 2.07 \times 10^{-12}$
    }
    \label{fig:twitter-job-prestige}
    % \vspace{-9mm}
\end{figure}
% \end{wrapfigure}

In an extreme case, if one occupational family was never mentioned in all biographies, then  $\kappa = 0.0$, indicating a completely under-represented occupational category (e.g., \textit{track sweeper, restaurant busser, grocery deliverer} ). 
In contrast, $\kappa \gg 1.0$ indicates a highly over-represented category on Twitter (e.g., \textit{CEO, COO, ...} \textit{Chief Executive} category ).
Note that the biases in Twitter's user population (e.g., having generally higher education and salary) can only amplify the over/under-representativenes, thus not it is a direct concern regarding to our goal.
We emphasize that our proposal is \textbf{not} to uncover the real-world labor statistics from Twitter but to discover which occupations are more likely to form a part of a user's self-identity (over-represented). Then, we quantitatively investigate its correlation to other hypothesized factors like social approval.

\subsection{Social Approval and Demographics in Occupational Identity Establishment}
% People who mention jobs in their biographies may cultivate occupational identity in themselves. 
The construction of occupational identity is a complex process involving personal variables and social approval.
Based on the observable evidence in Twitter bios, we seek to answer the question: \textit{Are high-prestige jobs more likely to be embedded into occupational identity?}
%
% \begin{itemize}
%     \item \remove{Personal traits: Are men more likely to embed jobs into self-identity? Does ethnicity matter? }
%     \item Social approval: Are high-prestige jobs more likely to be embedded into occupational identity? 
% \end{itemize}

% Several theories argue without solid evidence.  Our statistics justify both of the claims.

    \begin{table}[ht]
        \centering
        
        \caption{
            The breakdown of the percentage of users who disclosed occupational identities by gender and ethnicity. Overall, males are 32.27\% more likely to mention jobs than females do. 
            }
        
        % \scriptsize
        \begin{tabular}{lrrrr|r}
        \toprule
                &  Asian &  Hispanic &  Black &         White &         All Race \\
        \midrule
        Female & 15.01\% &    10.28\% &    16.20\% &    16.89\%  &       15.80\% \\
        Male   & 23.19\% &    16.77\% &    21.07\% &    22.27\%  &      21.69\% \\
        \hline
        All gender   & 20.70\% &    13.64\% &    18.78\% &    20.05\%  &     19.25\%  \\
        \bottomrule
        \end{tabular}
        \label{tab:who-mention}
    \end{table}

\paragraph{Men are more likely to be embedded with occupational identity than women:
 }  
     Men are general more into career and power than women, as reported in 
    \cite{greenhaus2012relations}. 
    % We observed such gender effect in occupational identity, as presented in Table \ref{tab:who-mention}. 
    Table \ref{tab:who-mention} presents the percentage of the Twitter users who disclosed  occupational identity in biographies, breaking down by ethnicity and gender.
    Across all races, we discover a consistent pattern that relatively 32.27\% more males would mention jobs in the bios than females do. This finding is consistent to the hypothesis that men are more job-oriented than women.

    \begin{figure}[ht]
        \centering
        \includegraphics[width=0.65\textwidth]{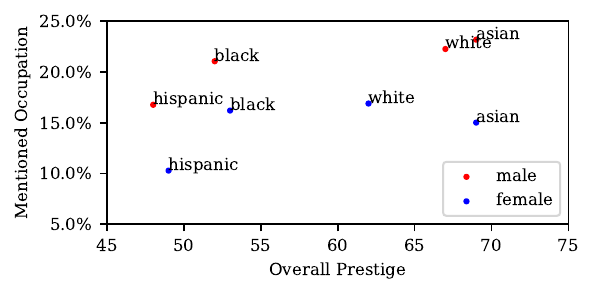}
        % \vspace{-3mm}
        \caption{
            Men are more career-oriented regardless of ethnicity. The group with more prestigious jobs are more likely to mention occupations. 
        }
        \label{fig:occ-mention-gender-race}
    \end{figure}

\paragraph{
Separate ethnicity group shows different occupation disclosure rate, probably, due to disparate occupation prestige:}
    Table \ref{tab:who-mention} reports that significantly lower percentages in Hispanic community mention their occupations in both the fe/male cohorts, compared to other ethnicity. 
    In a stratification lens, \citet{jennings2022occupational} recently revealed that the occupational prestige in Hispanics is lower than other groups.
    %s, indicating American's persistent inequality in racial and gender compositions. 
    %
    This observation coincides with our findings in Figure \ref{fig:occ-mention-gender-race}. Given the disparity of job prestige in Hispanic community, they tend to not to identify with occupation as frequently as other groups do. 

%

% Asian women is far less than man.

% We aim at better understanding the demographics biases in occupations (e.g., more male engineers than female) and the biases in the self-reported occupations itself w.r.t. job prestige (e.g., more presidents than cashiers on Twitter).
%Twitter population and revealing the difference between the self-reported identities of Twitter users and the real-world statistics.  

\paragraph{Self-reported jobs reflect the occupational concentration of demographics }
Table  \ref{tab:top_gender_occ_short} and \ref{tab:ethnicity_over_under_occ_short} shows the most over/under-represented occupations in each gender or ethnicity group based on the biased ratio $\kappa_\text{(job, ethnicity)}$ defined  as following:
\[   
\kappa_{\text{(job, ethnicity)}} = \frac{P(\text{ethnicity | job})}{P(\text{ethnicity})} 
\left\{
\begin{array}{ll}
       >1.0 &, \text{Over-represented} \\
      =1.0 &, \text{Balanced} \\
      <1.0 &, \text{Under-represented} \\
\end{array} 
\right. \]
%p(\text{ethnicity|job}) = p(\text{ethnicity})

% \begin{align}
%     % \kappa_{\text{(job, ethnicity)}} = \frac{\text{\#(job, ethnicity)}}{\text{\#(job)}\times \text{ethnicity prior}} = \frac{p(\text{ethnicity|job})}{p(\text{ethnicity})}
%     \kappa_{\text{(job, ethnicity)}} = \frac{p(\text{ethnicity|job})}{p(\text{ethnicity})}\\
%     p(\text{ethnicity|job}) = p(\text{ethnicity})
% \end{align}

Note that if there is no ethnicity bias in job, $p(\text{ethnicity|job}) = p(\text{ethnicity})$, making  $\kappa_{\text{(job, ethnicity)}} = 1$.  The similar measurement is also applied to job-gender bias.
% Noted that we represent demographics attribution in  probabilistic manner. 

\begin{table}[htbp]
    \centering
    \small
    \caption{The top gender biased occupations in real world and Twitter}
    % \vspace{-3mm}
\begin{tabular}{p{0.3\textwidth}p{0.08\textwidth} | p{0.3\textwidth}p{0.08\textwidth}}
\toprule
                           \textbf{Real world} &  $\kappa_{\text{male}}$ &                  \textbf{Twitter} &  $\kappa_\text{male}$\\
\midrule
\multicolumn{4}{c}{Female-biased Occupations}\\
\midrule

                              Skincare specialists &   0.03  &  doula &          0.27 \\
                             kindergarten teachers &   0.06 & esthetician &          0.29 \\
                            Executive secretaries  &   0.06 & kindergarten teacher &          0.31 \\
                                 Dental hygienists &   0.09 & registered dietitian &          0.32 \\
                                Speech pathologists &   0.09 &  makeup artist &          0.34 \\
                                Medical secretaries  &   0.09 &  early childhood &          0.34 \\
                                 Childcare workers &   0.10 & independent consul. &          0.36 \\
                        Cosmetologists &   0.14 &  yoga teacher &          0.39 \\
\midrule
\multicolumn{4}{c}{Male-biased Occupations}\\
\midrule
Concrete finishers & 1.89 & asst. baseball coach &          1.65 \\
    Power-line installers &  1.87 & asst. football coach &          1.64 \\
                         Crane operators &  1.87 & basketball coach &          1.64 \\
Heavy vehicle technician &   1.86 & offensive coordinator &          1.64 \\
Bus and truck mechanics &   1.86 &  baseball player &          1.63 \\
Heating, AC technician &   1.86 &lead pastor &          1.60 \\
                                      Electricians &   1.85 & software architect &          1.60  \\
           Plumbers &  1.85 & golf professional &          1.57 \\
%                          Tree trimmers and pruners &  97.8 \\
%          Brickmasons, blockmasons, and stonemasons &  97.8 \\
%       Automotive service technicians and mechanics &  97.7 \\
%              Automotive body and related repairers &  97.6 \\
%                                           Roofers &  97.1 \\
%                   Construction equipment operators &  97.0 \\
%                                         Carpenters &  96.9 \\
% Drywall installers, ceiling tile installers, an... &  96.9 \\
%         Aircraft mechanics and service technicians &  96.8 \\
%                                         Machinists &  96.6 \\

\bottomrule
\end{tabular}
    \label{tab:top_gender_occ_short}
\end{table}

Table \ref{tab:top_gender_occ_short} presents the gender bias discovered from the real-world occupation survey and our Twitter data side-by-side. 
We found that females prevail in cosmetics, health and general education domain, which is consistent with the real-world survey results. 
Males are closely related to sports, politics and technology, while the real-world data shows different but still reasonable results, and we will discuss the reasons behind this difference later.

\begin{table*}[htbp]
\centering
    \caption{The examples of over/under-represented occupations in each ethnicity group. We normalize the raw count for each ethnicity and occupation group. The column \textit{Ratio} measures how many time the calculated value exceeds the expected value based on population prior. If ethnicity distribution of an occupation meets the prior distribution, the ratio value is 1.}
    % \vspace{-3mm}
    % \tiny
    \begin{tabular}{p{0.17\textwidth}p{0.04\textwidth}p{0.15\textwidth}p{0.04\textwidth}p{0.17\textwidth}p{0.04\textwidth}p{0.13\textwidth}p{0.04\textwidth}}
    \toprule
                           White &     $\kappa$  &                    Asian &     $\kappa$ &                      Hispanic &     $\kappa$ &                          African-American &     $\kappa$ \\
    \midrule
    \multicolumn{8}{c}{Over-represented}\\
    \midrule
                  golf professional &                   2.12 &                research scholar &                   6.60 &             industrial engineer &                      3.48 &               performing artist &                  1.82 \\
                         golf coach &                   2.10 &            chartered accountant &                   5.76 &               volleyball player &                      2.30 &                    forex trader &                  1.81 \\
                     pitching coach &                   2.09 &               political analyst &                   4.08 &          industrial engineering &                      2.18 &            motivational speaker &                  1.79 \\
           assistant superintendent &                   2.07 &                   social worker &                   3.82 &                   soccer player &                      2.14 &               radio personality &                  1.75 \\
             social studies teacher &                   2.07 &            electronics engineer &                   3.49 &               medical assistant &                      1.74 &                    music artist &                  1.73 \\
    \midrule
    \multicolumn{8}{c}{Under-represented}\\
    \midrule
                   research scholar &                   0.84 &                  softball coach &                   0.17 &               golf professional &                      0.25 &                research scholar &                  0.74 \\
               chartered accountant &                   0.98 &        assistant football coach &                   0.18 &                  support worker &                      0.29 &             industrial engineer &                  0.76 \\
                industrial engineer &                   1.07 &        assistant baseball coach &                   0.18 &      asst. basketball coach &                      0.30 &              software architect &                  0.77 \\
               electronics engineer &                   1.12 &                  baseball coach &                   0.19 &          men's basketball coach &                      0.31 &             assistant professor &                  0.81 \\
                  android developer &                   1.17 &               athletic director &                   0.20 &                      golf coach &                      0.33 &             postdoctoral fellow &                  0.83 \\
    \bottomrule
    \end{tabular}
    \label{tab:ethnicity_over_under_occ_short}
\end{table*}

Similarly, Table \ref{tab:ethnicity_over_under_occ_short} presents that certain ethnicity groups participate more in sports, academic, technology and art industry (Table \ref{tab:ethnicity_over_under_occ_long} in Appendix presents more samples).
These results are demonstrative that our dataset captures complex aspects of society, and could be useful for quantitative social studies.

\xingzhiswallow{

\begin{table}[htbp]
    \centering
    \small
\begin{tabular}{p{0.16\textwidth}p{0.04\textwidth} | p{0.16\textwidth}p{0.04\textwidth}}
\toprule
                           Male &  Male\% &                  Female &  Male\% \\
\midrule
       asst. baseball coach &          0.98 &                   doula &          0.16 \\
       asst. football coach &          0.98 &             esthetician &          0.17 \\
              basketball coach &          0.98 &    kindergarten teacher &          0.18 \\
          offensive coordinator &          0.97 &    registered dietitian &          0.19 \\
                 baseball coach &          0.97 &           makeup artist &          0.20 \\
                baseball player &          0.96 &         early childhood &          0.20 \\
                    lead pastor &          0.95 &  independent consul. &          0.22 \\
                 software architect &          0.95 &            yoga teacher &          0.24 \\
                %  football coach &          0.95 &            yoga teacher &          0.23 \\
            %  software architect &          0.95 &       virtual asst. &          0.24 \\
            %   golf professional &          0.93 &     instructional coach &          0.24 \\
            %      pitching coach &          0.93 &         yoga instructor &          0.25 \\
            %       ios developer &          0.93 &  occupational therapist &          0.25 \\
\bottomrule
\end{tabular}
    \caption{The top gender biased occupations.}
    \label{tab:top_gender_occ_short}
\end{table}
}
%
% \paragraph{Self-reported jobs are highly biased toward the prestigious occupations:}
% The difference between real-world data and Twitter is intriguing because a job is not always being presented in one's self-identity. 
%Although our results capture the demographic concentration in occupations, we also observe the difference between the self-reported occupations and real-world surveys. 
%However, we also observed the differences between self-report jobs and occupation distribution of real-world surveys.
% For example, Table \ref{tab:top_job_titles} shows that most of the popular jobs are dis-proportionally high.
%(e.g., \textit{Founder, CEO} are more than \textit{Grocery deliverers})
% , and Table \ref{tab:top_gender_occ_short} misses many of the popular male-biased occupations (\textit{e.g., Concrete finishers}).
% Prior works \cite{wojcik2019sizing} argued that Twitter population generally have higher education/salaries than real-world average, attributing the differences to the population difference. 
%
\paragraph{Prestigious jobs are more likely to be embedded into self-identity}: 
The anecdotal evidence of Table \ref{tab:top_job_titles}, suggests that \textit{prestigious} jobs (e.g. \textit{CEO,COO, ...}) tend to be mentioned more frequently in personal biographies. 
This coincides with the hypothesis about social approval, that is, the social status of the job (including salary, perceived prestige) is an important factor contributing to the occupational identity establishment. 
In Figure \ref{fig:twitter-job-prestige}, we plot the over/under-represented groups as red/blue dots against the job prestige scores.
% we further divide occupations into over/under-represented groups based on $\kappa$, as visualized by red/blue dots, respectively, in Figure \ref{fig:twitter-job-prestige}.
The result demonstrates that the job prestige score is highly correlated with $\kappa$ of Spearman correlation $\rho = 0.4858$.  
It also shows that 74.04\% of over-represented jobs are above the overall median prestige score, versus 39.48\% for under-represented jobs.  
Specifically, the prestige distributions of over/under-represented jobs are significantly different with K-S test $D = 0.4284$ and p-value  $\approx 2.07 \times 10^{-12}$. 

Interestingly, we find that the correlation of over-representation to annual income ($\rho = 0.3961$) is weaker than it is to prestige, implying that social approval is more important than pure material return. 
For example, \textit{Legislators, Clergy} and \textit{Firefighters} share high prestige score (with percentile > 80\% in prestige), although their income percentiles are only 22.55\%, 55.93\% and 57.27\%, respectively.  But they are all over-represented with $\kappa>1$ based on our observations.
%$\kappa = 2.5582, 9.9287 and 1.0152 $
%
This evidence that more prestigious jobs appear to be included more frequently in Twitter biographies as a part of one's identity.

\xingzhiswallow{
\begin{figure}[hbtp]
    \centering
    % \subfloat[Real-world]{
    % \includegraphics[width=0.48\textwidth]{Figures/fig-job-overrep-prestige-v2.pdf}}
    \includegraphics[width=0.49\textwidth]{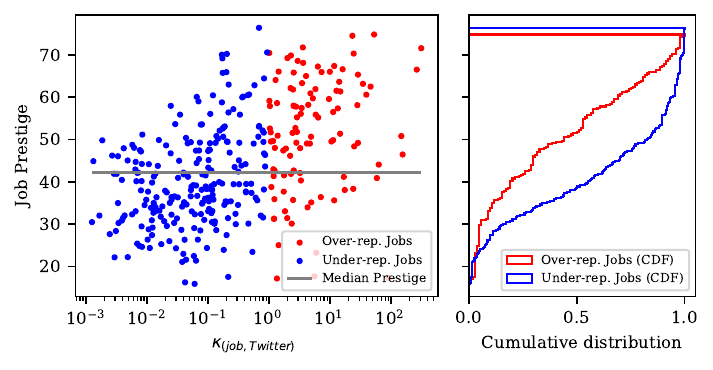}
    \vspace{-9mm}
    \small
    \caption{ Job prestige and  $\kappa$ is positively correlated.
    %with Spearman correlation $\rho = 0.4858$
    % The Spearman correlation between $\kappa$ and Job prestige is 0.4858. 
    Most of the over-represented jobs (red) are higher than the median prestige. The prestige distribution between over/under-represented jobs are statistically different.
    %The 74.04\% of the over-represented jobs (red) are higher than the median, which is significantly greater than the under-represented jobs of only 39.48\%. 
    %The prestige distribution between over/under-represented jobs are significantly different with K-S test 0.42844 with p-value  $\approx 2.07 \times 10^{-12}$
    }
    \label{fig:twitter-job-prestige}
\end{figure}

}

\xingzhiswallow{
\noindent \textbf{Do self-reported jobs reflect actual occupational concentration of demographics?}
Table  \ref{tab:top_gender_occ_short} and \ref{tab:ethnicity_over_under_occ_short} shows the most gender-biased occupations and the over/under-represented occupations in each ethnicity group. We found certain ethnicity groups participate more in sports, academic, technology and art industry (Table \ref{tab:ethnicity_over_under_occ_long} in Appendix presents more samples).
In addition, we found males are closely related to sports, politics and technology, while females prevail in cosmetics, health and general education domain.
These results are demonstrative that bios capture complex aspects of society, and could be useful for quantitative social studies.

}

%In next We use partial data from year 2020 for the prediction tasks, and leverage full data for the temporal exploration. 

\section{Occupational Identity Evolution}
\label{biography-evolution}
If Twitter biographies capture the notions of self-identity, then changes in biographies reflect changes in self-identify. Our dataset captures identity changes for millions of people over the six years from 2015-2021. 
In this section, we first characterize and analyze the general biography edits, then focus quantitatively on issues of occupational identity transitions.

\begin{figure}[htbp]
    \centering
    %\scriptsize
    % \vspace{-3mm}
    \subfloat[Edit distance histogram \label{fig:news_lev_dist_hist}]         {\includegraphics[width=0.25\textwidth]{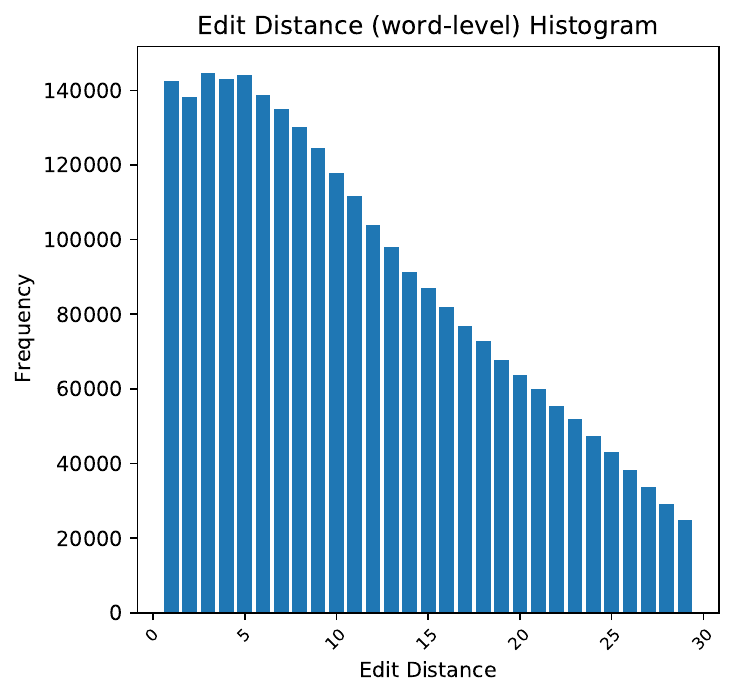}}\qquad
    \subfloat[BERTScore vs. Edit Dis.\label{fig:bertscore_lev_dist}] {\includegraphics[width=0.25\textwidth]{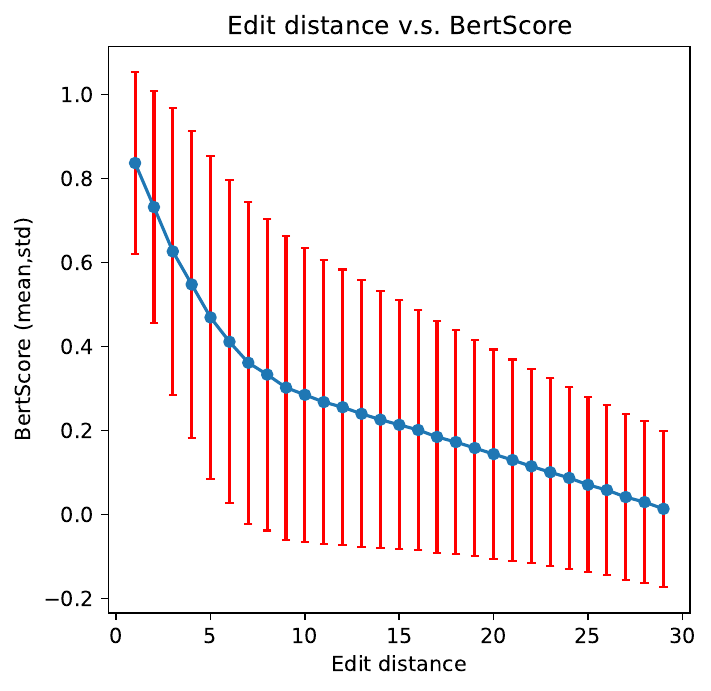}}\qquad
    %\subfloat[BERTScore v.s. Edit Distance: As the text changes more, they share less semantics.\label{fig:bertscore_lev_dist_dot}] {\includegraphics[width=0.2\textwidth]{Figures/twitter_bios_raw_chunk/figure_raw_feature_description_1_2_3_4_5_6_7_8_9_10_11_12_13_14_15_16_17_18_19_20_21_22_23_24_25_26_27_28_29_edit_bertscore_hist}}\hfill
    \subfloat[BERTScore distribution \label{fig:bertscore_hist}] {\includegraphics[width=0.25\textwidth]{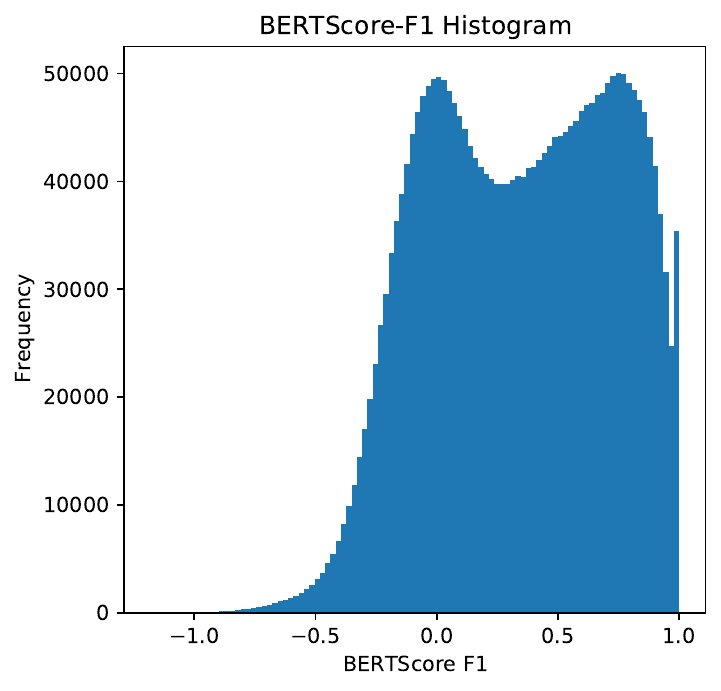}}\hfill
    %\subfloat[The statistics of BERTScore: Roughly 80\% of the pairs is greater than 0; 40\% is greater than 0.5.\label{fig:bertscore_lev_dist}] {\includegraphics[width=0.2\textwidth]{Figures/twitter_bios_raw_chunk/figure_raw_feature_description_1_2_3_4_5_6_7_8_9_10_11_12_13_14_15_16_17_18_19_20_21_22_23_24_25_26_27_28_29_BERTScore_cdf.pdf}}\hfill
    % \vspace{-4mm}
    \caption{\textbf{Left:} Small bio changes are more common than complete rewrite. \textbf{Middle \& Right:} Semantic similarity between changed bio pairs. Two peaks in \ref{fig:bertscore_hist} shows that the most common changes are minor updates and completely rewrite. }
    \label{fig:bertscore_stats}
    % \vspace{-3mm}
\end{figure}

\xingzhiswallow{
    \begin{figure}[htbp]
        \centering
        \includegraphics[width=0.35\textwidth]{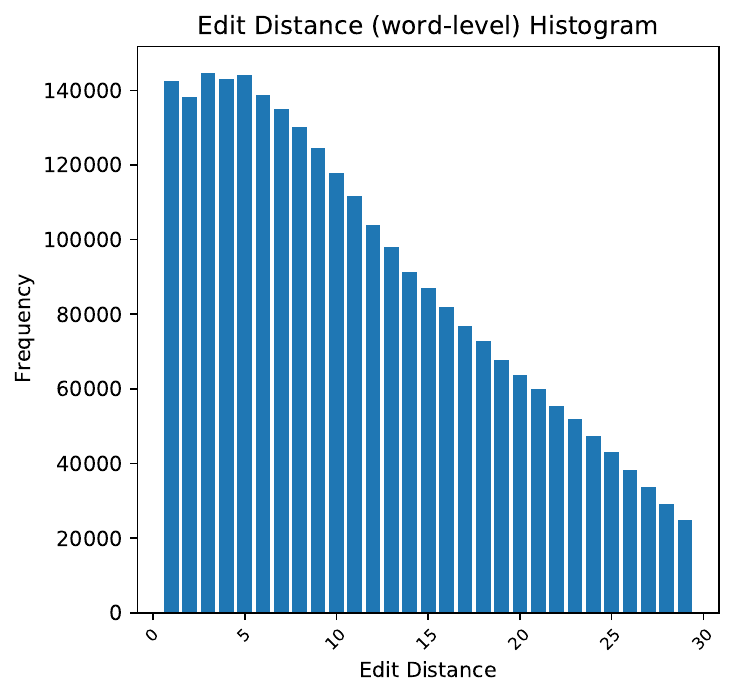}
        % \vspace{-3mm}
        \caption{Edit distance histogram: Small bio changes are more common than complete rewrite.}
        \label{fig:news_lev_dist_hist}
    \end{figure}
    
}

We randomly sample 1,353,325 of our 51.18M individuals and acquire a subset of 2,597,550 consecutive changed pairs whose word edit distance is within 30 characters. 
% Figure \ref{fig:news_lev_dist_hist} shows the histogram of word-level Levenshtein distance. Most people update their bios by 1-5 words at a time.  
Figure \ref{fig:news_lev_dist_hist} shows that most people update their bios by 1-5 words at a time.  
But character differences do not completely reflect semantic differences: an "artist" transformed in a "painter" represents a smaller change than the edit distance suggests.
To better capture semantic similarity, we extend this analysis to all changed pairs with $BERTScore_{F1}\in [-1,1]$ (\cite{zhang2019bertscore}) as the semantic similarity measure.
%The overall statistics of $BERTScore_{F1}$ is shown in Figure \ref{fig:bertscore_stats}.
Figure \ref{fig:bertscore_lev_dist} shows the BERTScore against word-level Levenshtein distance. 
As expected, semantic similarity decreases as more edits are applied.  
Figure \ref{fig:bertscore_hist} shows the distribution of BERTScore, where the hump centered at 0 implies that a significant fraction of people completely rewrite their biographies with each transition, while the hump at 0.8 shows another large group of people making minor edits to their bios (e.g: age or job title updates).

%in Section \ref{sec:special_case_d1}, 
% then extend the analysis to all pairs of arbitrary edit distance with BERTScore \cite{zhang2019bertscore}.
%as the semantic similarity measure.  
% Finally we analyze the old/new pairs with interpretable meaning especially for occupational transitions over time. 
%Finally we partition the old/new pairs into buckets based on the meaning of the changes.

% \begin{wraptable}{r}{0.95\textwidth}
\begin{table}[ht]
    \small
    % \vspace{-5mm}
    \centering
    \caption{The top part-of-speech changes: The most popular change is between numbers, reflecting age updates on birthdays or anniversaries. X is not recognized, e.g.: Emoji or unrecognized abbreviation. Total number of replaced pairs is 89,838. }
    \label{table:d1_pos_tag_change}
    % \vspace{-3mm}
    \begin{tabular}{p{0.4\textwidth} >{\raggedleft\arraybackslash}p{0.1\textwidth}}
    \toprule
    POS change (before, after) & Count\  \\
    \midrule
    (NUM, NUM)     &       23,335 \\ \hline
    (NOUN, NOUN)   &       15,850 \\ \hline
    (X, X)         &       12,889 \\ \hline
    % (PROPN, PROPN) &        9,183 \\ \hline
    % (PUNCT, PUNCT) &        3,933 \\ \hline
    % (ADJ, ADJ)     &        2,443 \\ 
    % (NOUN, PROPN)  &        2,164 \\ \hline
    % (PROPN, NOUN)  &        1,844 \\ \hline
    % (VERB, VERB)   &        1,347 \\ \hline
    % (NOUN, ADJ)    &        1,343 \\ \hline
    % (ADJ, NOUN)    &        1274 \\ \hline
    % (NOUN, X)      &         704 \\ \hline
    % (PROPN, X)     &         612 \\ \hline
    % (X, NUM)       &         565 \\ \hline
    % (X, PROPN)     &         532 \\ \hline
    \bottomrule
    \end{tabular}
    \vspace{2mm}
\end{table}

% \end{wraptable} 

% \subsection{Edits of Single Word Substitution}
\subsection{Minor Edits in Biographies }
\label{sec:special_case_d1}
To understand the nature of the edits distribution, we look into the changed biography pairs where only single words have been substituted (89,838 pairs), and discover many occupational transitions over time.
% For the bio pairs where only one word was substituted, we aim at identifying the reasons behind the changes.
% \begin{enumerate}
%     \item What is the POS tag of the changed word (Are they verbs? nouns? adjectives? or numbers?)
%     \item If changed words are numbers, are the they increased or decreased?
%     \item Are the changed words synonyms / hypernyms / hyponyms?
% \end{enumerate}
%\noindent{\textbf{The POS tags of the changed words}}: 
First, we present the top part-of-speech (POS) changes in Table \ref{table:d1_pos_tag_change}. These changes usually share the same POS tag, indicating the alternation of word choices.

\xingzhiswallow{

\noindent{\textbf{Numerical Changes}}: we show examples of the numerical difference between the post-edited and the original bios in Table \ref{tab:d1_num_change_list}.
%\todo{In the Counts column, the last two rows have a value of 100 exactly.  Is that correct?} 
The number usually increases as the life events (e.g: age, anniversary) being updated.

% \vspace{-2mm}
\begin{table}[htbp]
    \centering
    \small
    \caption{Examples of top number changes}
    \vspace{-3mm}
\begin{tabular}{p{0.03\textwidth} p{0.05\textwidth} p{0.16\textwidth} p{0.15\textwidth}}
    \toprule
     Diff. &  Counts &  Bio(Before) &  Bio(After)\\
    \midrule
       1 &   13,982 & 27 Year old & 28 Year old\\
       %2 &    3179 & 20 years of \#website building & 22 years of \#website building\\
       %3 &     550 & 22, NYC. & 25, NYC.\\
      -1 &     465 & Northwestern '19 & Northwestern '18\\
       5 &     345 & 170 yrs of service & 175 yrs of service\\
       %4 &     263 & I am 16 yr & I am 20 yr\\
      10 &     191 & 40-ish mom & 50-ish mom\\
      %-2 &     105 & VT ‘22 & VT ‘20\\
    1000 &     100 & wrote 3,000 poems & wrote 4,000 poems\\
     100 &     100 & 400 shows a year & 500 shows a year\\
    \bottomrule
    \end{tabular}
    \label{tab:d1_num_change_list}
\end{table}

}

\noindent{\textbf{Breakdown of Nouns:}} 
To better understand the reasons behind word replacement, we organize interpretable relations between changed nouns in Table \ref{tab:noun_changes}. 
We assign five categories of noun changes based on syntactic features (e.g: same lemma).
Besides \textit{Others} group, most of them are minor changes (with single character change), reflecting job promotion  (e.g. \textit{vp} to \textit{svp}) or age updates. In addition, we observed many career transitions (e.g., from \textit{manager} to \textit{director}) in all categories, which will be addressed in Section \ref{occupation-analysis}.

\begin{table}[htbp]
    % \vspace{2mm}
    \centering
    \caption{Breakdown of changed nouns in biography edits. One pair may belong to more than one category. Some of them are related to the change of occupational identity (e.g., \textit{(vp -> svp, coordinate -> manager).}).}
    % \vspace{-2mm}
    \small
\begin{tabular}{ >{\raggedright\arraybackslash}p{0.2\textwidth} >{\raggedright\arraybackslash}p{0.06\textwidth} >{\raggedleft\arraybackslash}p{0.06\textwidth} >{\raggedright\arraybackslash}p{0.55\textwidth}}
\toprule
 Category & Total & \%  & \makecell[l]{Examples (Before, After, \#)}  \\
\midrule
 Share lemma & 693 & 4.54\% & \makecell[l]{(alumni, alumnus, 60); (dog, dogs, 31)\\} \\
\hline 
 Synonym & 669 & 4.38\% & \makecell[l]{(alumni, alum, 46); (manager, director,41)\\}  \\
\hline 
 One Token Change & 2327 & 15.89\% & \makecell[l]{(vp, svp, 20); (22yrs, 23yrs, 14)\\}  \\
\hline 
 Hyponymn  & 335 & 2.19\% & \makecell[l]{(student, graduate, 46); (student, grad, 25)\\(student, alum, 15); (father, dad, 9)} \\
\hline 
 Hypernymn & 266 & 1.17\% & \makecell[l]{(teacher, educator, 17);  (journalist, writer, 8); \\ (actress, actor, 7); }   \\
 \hline 
 Others  & 10898 & 71.36\% & \makecell[l]{(student, candidate, 104); (sophomore, junior, 68); \\(coordinate, manager, 28)}  \\
 \midrule
 All & 15272 & \makecell[l]{100\%} & \makecell[l]{--}  \\
 \bottomrule
\end{tabular}
    \label{tab:noun_changes}
    \vspace{-3mm}
\end{table}

\xingzhiswallow{
    \begin{table}[htbp]
        \centering
        \footnotesize
        \caption{Breakdown of changed nouns. One pair may belong to more than one category}
    \begin{tabular}{ >{\raggedright\arraybackslash}p{0.06\textwidth} >{\raggedright\arraybackslash}p{0.03\textwidth} >{\raggedleft\arraybackslash}p{0.03\textwidth} >{\raggedright\arraybackslash}p{0.15\textwidth} >{\raggedright\arraybackslash}p{0.15\textwidth}}
    \toprule
     Category & Total & \%  & \makecell[l]{Examples (Before, After, \#)} & Description \\
    \midrule
     Share lemma & 693 & 4.54\% & \makecell[l]{(alumni, alumnus, 60); (dog, dogs, 31)\\} & The change noun shares the lemma. \\
    \hline 
     Synonym & 669 & 4.38\% & \makecell[l]{(alumni, alum, 46); (manager, director,41)\\} & The change noun is synonym. \\
    \hline 
     Minor & 2327 & 15.89\% & \makecell[l]{(vp, svp, 20); (22yrs, 23yrs, 14)\\} & Only one character is changed (excluding same lemma pairs ) \\
    \hline 
     Hyponymn  & 335 & 2.19\% & \makecell[l]{(student, graduate, 46); (student, grad, 25)\\(student, alum, 15); (father, dad, 9)} & Become more specific\\
    \hline 
     Hypernymn & 266 & 1.17\% & \makecell[l]{(teacher, educator, 17); (journalist, writer, 8)\\(actress, actor, 7); (dad, father, 7)} & Become more general  \\
     \hline 
     Others  & 10898 & 71.36\% & \makecell[l]{(student, candidate, 104); (sophomore, junior, 68)\\(coordinate, manager, 28)} & All other noun changes \\
     \bottomrule
     All & 15272 & 100\% & \makecell[l]{--} & All detected noun substitutions \\
     \bottomrule
    \end{tabular}
    
        \label{tab:noun_changes}
    \end{table}
}

%For the rest of the questions, we plot tree-like diagrams to answer the questions with a few examples in Figure.\ref{fig:new_d1_noun}-\ref{fig:new_d1_verb} for Nouns, Adjectives and Verbs, respectively.

\xingzhiswallow{
    \subsection{For Arbitrary Edit Distance}
    \label{sec:general_analysis}
    Here we compare semantic difference in a pair of changed bios to distinguish whether it is a minor update or a completely rewrite. 
    %Lexical overlap based metric like ROUGE\cite{lin2004rouge} or BLEU\cite{papineni2002bleu} score calculates the sentence similarity by comparing the surface text. 
    BERT-based word alignment score -\textit{BERTScore} \cite{zhang2019bertscore}- presented a soft lexical alignment method using contextual word embeddings and the attention mechanism. It can capture the semantic similarity in two sentences of different structures and lexicons.
    We extend the analysis to all changed pairs with $BERTScore_{F1}\in [-1,1]$ as the semantic similarity measure. The overall statistics of $BERTScore_{F1}$ is shown in Figure \ref{fig:bertscore_stats}.
    
    %\begin{comment}
    \xingzhiswallow{
    \begin{equation}
    \hspace{-0.00\textwidth}
    \begin{aligned}
        &R = \frac{1}{|x|} \sum_{x_i \in x} {max_{\hat{x_j} \in \hat{x} } x_i \hat{x}_j  } ;
        P = \frac{1}{|\hat{x}|} \sum_{\hat{x_j} \in \hat{x}} {max_{x_i \in x } x_i \hat{x}_j  } \\
        &BERTScore_{F1} =  2\frac{P*R}{R+P},  
    \end{aligned}
    \label{eq:bertf1}
    \end{equation}  where $x_i,\hat{x_i}$ denote $i^{th}$ token in two sentences $x,\hat{x}$.
    }
    
    \begin{figure}[htbp]
        \centering
        \subfloat[BERTScore vs. Edit Distance.\label{fig:bertscore_lev_dist}] {\includegraphics[width=0.23\textwidth]{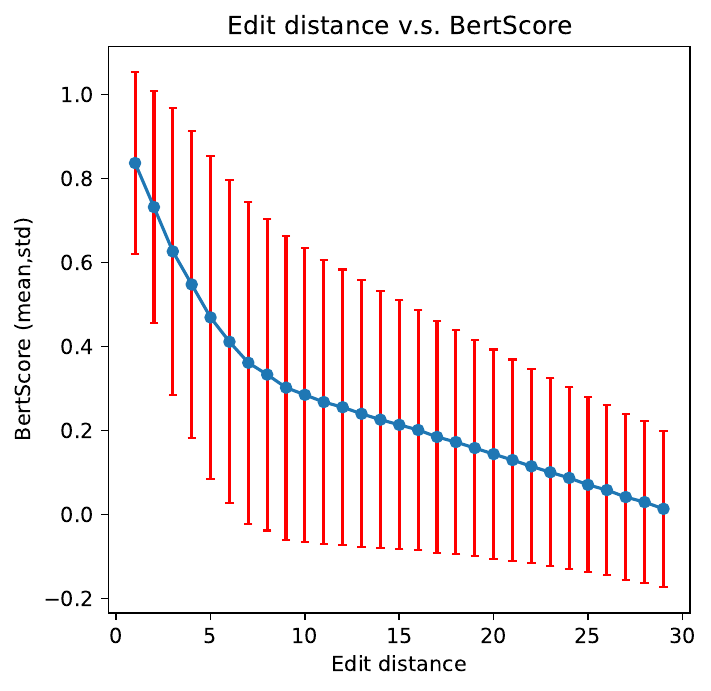}}\hfill
        %\subfloat[BERTScore v.s. Edit Distance: As the text changes more, they share less semantics.\label{fig:bertscore_lev_dist_dot}] {\includegraphics[width=0.2\textwidth]{Figures/twitter_bios_raw_chunk/figure_raw_feature_description_1_2_3_4_5_6_7_8_9_10_11_12_13_14_15_16_17_18_19_20_21_22_23_24_25_26_27_28_29_edit_bertscore_hist}}\hfill
        \subfloat[The distribution of BERTScore\label{fig:bertscore_hist}] {\includegraphics[width=0.23\textwidth]{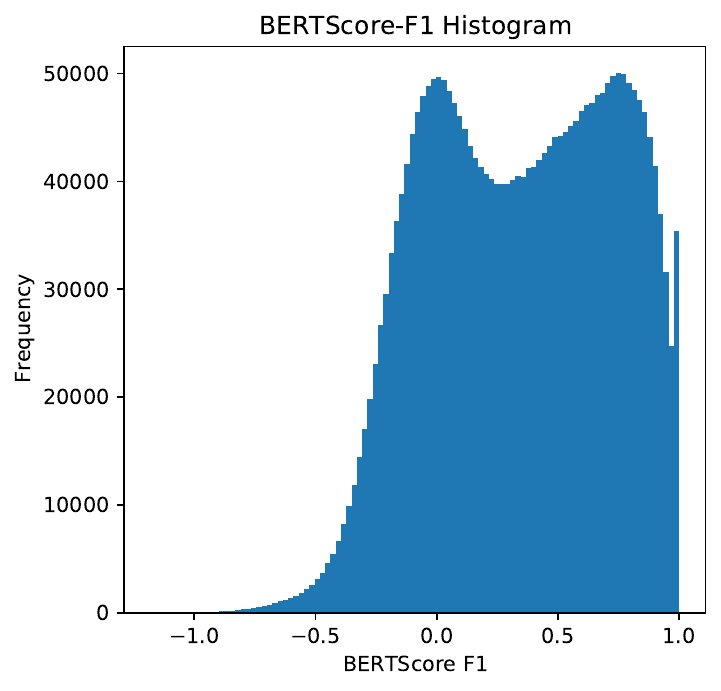}}\hfill
        %\subfloat[The statistics of BERTScore: Roughly 80\% of the pairs is greater than 0; 40\% is greater than 0.5.\label{fig:bertscore_lev_dist}] {\includegraphics[width=0.2\textwidth]{Figures/twitter_bios_raw_chunk/figure_raw_feature_description_1_2_3_4_5_6_7_8_9_10_11_12_13_14_15_16_17_18_19_20_21_22_23_24_25_26_27_28_29_BERTScore_cdf.pdf}}\hfill
        \caption{Semantic similarity between changed bio pairs. Two peaks in \ref{fig:bertscore_hist} shows that the most common changes are minor updates and completely rewrite. }
        \label{fig:bertscore_stats}
    \end{figure}

    Figure \ref{fig:bertscore_lev_dist} shows the BERTScore against word-level Levenshtein distance. 
    Within our expectation, semantic similarity keeps decreasing as more edits are applied.  Figure \ref{fig:bertscore_hist} shows the distribution of BERTScore, the humps centered at 0 tells that a significant fraction of people completely rewrite their bios, while the hump at 0.8 shows another large group of people slightly edit their bios (e.g: age or occupation updates as mentioned before).

    %However, in Figure.\ref{fig:bertscore_lev_dist_dot} we observed that a portion of highly edited sentences (edit distance $>10$) still retain the original semantics with BERTScore greater than 0.6.  We show some examples from A,B,C,D areas in Table.\ref{table:Text_Area_A}-\ref{table:Text_Area_D} in appendix.
    
    %in Table.\ref{table:Text_Area_A},Table.\ref{table:Text_Area_B},Table.\ref{table:Text_Area_C},Table.\ref{table:Text_Area_D} with pre/post-edited sentences, word-level edit distance and BERTScore.
}

\subsection{Transitions Between Occupational Identities}
\label{occupation-analysis}

% \begin{wrapfigure}{r}{0.95\textwidth}
% \setlength\intextsep{0pt}
\begin{figure}[htbp]
    \centering
\includegraphics[width=0.65\textwidth]{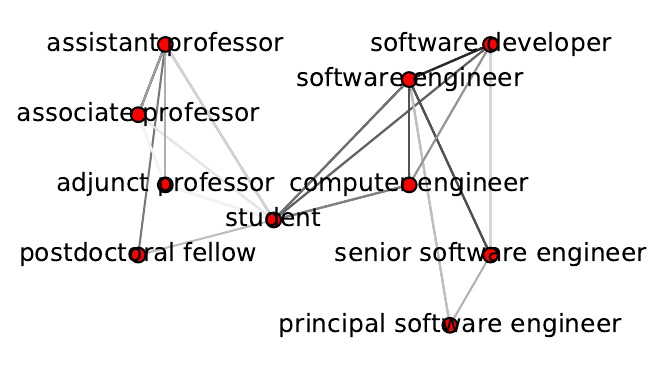}
    % \vspace{-10mm}
    \caption{Examples of job transitions in the graph. Darker edges represent more frequently observed transitions. }
    \label{fig:job_trans_graph}
    % \vspace{-2mm}
\end{figure}
% \end{wrapfigure}
%
To better understand the changes of occupational identities, we extract and analyze those biography pairs in which any occupational title has been added or removed. %, resulting in \versiontwo{fill-in-blank} biographies pairs \todo{add number}. 
%, and detect the occupational transitions as people update their bio. 

\noindent \textbf{Represent occupational transitions as a graph}: 
We formulate occupational transitions as a directed weighted graph $G(V, E)$,  where $V$ contains all unique job titles as vertices, $E$ contains all job title transitions $e_{i,j} = (v_i,v_j, w_{i,j})$, where edge weight $w_{i,j}$ represents the total number of the observed transitions from $i$ to $j$. For example, a biography changed from \textit{"Student, Volunteer"} to \textit{"Software Engineer, Volunteer"} increases the edge weight of \textit{Student} $\rightarrow$ \textit{Software Engineer}. Our final graph contains 25,033 nodes and 287,425 unique edges, consisting of seven weakly-connected components. The largest component contains 25,021 nodes (99.95\% of all). We illustrate a subgraph in Figure \ref{fig:job_trans_graph}.

% Specifically, we insert a dummy node ("none") representing no job was detected, and define a transition from the detected job title set $ s_{t-1}, s_{t} $ at time $t-1$ to $t$ as $S_{\delta(t-1,t1)} \in \{ {e_{i,j}=(v_i, v_j) | v_k \in V, |v_k|=1, i \neq j \} }$,  where $v^{t-1}_i \in s_{t-1} \setminus s_{t}$, and $v^{t}_i \in S_{t} \setminus S_{t-1}$. For example: 

% $S_{t-1} \in $ \{\textit{Student, Volunteer}\} , 

% $S_{t} \in $ \{\textit{Software Engineer, Volunteer}\},

% $S_{\delta(t-1,t1)} \in$ \{ \textit{(Student, Software Engineer)} \}

%job_one_waytransition_freq

\noindent \textbf{Community of Occupational Identity Transitions}: The job transition graph has an appealing property of clusterness, which reflects the proximity between occupational titles. For example, in the rooted clusters from \textit{Student} in Figure \ref{fig:job_trans_graph}, the roles of academia are naturally clustered on the left-hand side (the source), while IT industrial roles are intertwined on the right-hand side (the sink), but well-separated from the academic counterpart. 

Graph partitioning \cite{clauset2004finding} is well studied for community detection. We partition the simple job graph (unweighted, undirected) based on modularity, discovering densely intra-connected communities. %with varying resolution $\gamma \in \{ 1, 2, 4 ,..., 512 \}$.  
%As Figure \ref{fig:group_gamma} shows, a larger $\gamma$ usually leads to discover more and finer communities. 
%Figure \ref{fig:groupsize_gamma} shows that most discovered sub-clusters have less than 100 nodes when\footnote{We show examples when $\gamma=8$ since it's close to the changing point in Figure \ref{fig:group_gamma}. Though there could be finer sub-groups in clusters, we left it for future work. } $\gamma=8$. 
\xingzhiswallow{
    
    \begin{figure}[htbp]
        \centering
        \subfloat[\#Groups vs $\gamma$ \label{fig:group_gamma}] {\includegraphics[width=0.23\textwidth]{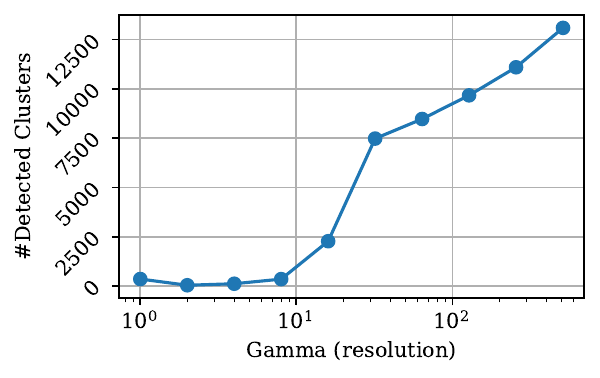}}\hfill
        \subfloat[Group nodes size ($\gamma=8$) \label{fig:groupsize_gamma}] {\includegraphics[width=0.23\textwidth]{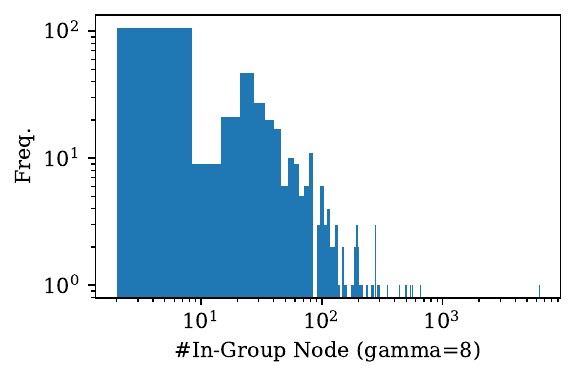}}\hfill
        \caption{\textit{Left}: More fine-grained communities are discovered as $\gamma$ increases. \textit{Right}: The histogram of \#nodes in each job community after graph partition ($\gamma=8$). Most of the community has less than 100 job titles}
        \label{fig:job_trans_graph_community}
    \end{figure}
}
% We also present the job titles in sampled groups
Table \ref{table:job_title_community} shows that the related job titles generally fall into the same cluster.
%\footnote{Cluster size: The number of nodes in the clusters. E.g.: There are 95 clusters of size 2 in all 350 discovered clusters.}
Some of these share common text patterns (e.g., \textit{"hadoop developer"} and \textit{"hadoop architect"}), while others are inherently related but without significant text overlap (e.g., \textit{"highway maintenance worker"} and \textit{"bridge inspector"}). 
These detected transitions and clusters can provide more context to better understand personal social dynamics. 
The full graph can serve as a unique resource for better job title clustering, which we will release upon publication.

\begin{table}[htbp]
    \centering
    % \small
    \vspace{-2mm}
    \caption{Each cluster (row) contains several highly-related job titles derived from our occupational transition graph.}
    \vspace{-2mm}
\begin{tabular}{p{0.95\textwidth}}
\toprule
 \centering\arraybackslash \textbf{Sampled Self-reported Titles in Each Clusters} \\
\midrule
%  \centering\arraybackslash Cluster size : 2 (95/350) \\ 
% \midrule
\textcolor{black}{health information director, health information technician} \\ \hline
% \textcolor{red}{certified registered locksmith, certified master locksmith} \\ \hline
\textcolor{black}{consumer lending manager, secondary market manager} \\ \hline
\textcolor{black}{loader operator, front end loader operator} \\ \hline
% \midrule
%  \centering\arraybackslash Cluster size: 3 - 30 (101/350)\\ 
% \midrule
% \textcolor{red}{regional sales coordinator, district sales coordinator, special projects coordinator },
\textcolor{black}{highway maintenance worker, bridge inspector, toll collector, highway inspector}
\\ \hline
% \textcolor{red}{privacy director, privacy analyst algorithm developer}, 
\textcolor{black}{hadoop developer, hadoop architect, linux architect, windows server administrator, web systems developer, mainframe systems administrator}
\\ \hline
% \textcolor{red}{university administrator, university president, graduate school dean, director of campus recreation},
% \textcolor{blue}{senior strategy manager, program analyst, business system consultant, senior program analyst }
% \\ \hline

\textcolor{black}{occupational safety specialist, safety and health manager, safety trainer},
% \textcolor{blue}{family worker, community board member, community engagement coordinator},
\textcolor{black}{professor of early childhood education, peer educator, speech language pathologist assistant},
% \textcolor{pink}{asp developer, digital product specialist, tool builder, code official, electrician}
\\ \hline
% \midrule
%  \centering\arraybackslash Cluster size: 31 - 60 (71/350)\\ 
% \midrule
\textcolor{black}{clinical assistant, clinic assistant, clinical associate, psychiatric assistant}
\textcolor{black}{communication instructor, language specialist, public speaking teacher}
% \textcolor{purple}{software development leader, information director}
\\ \hline

% \textcolor{red}{building consultant,  building surveyor},
% \textcolor{blue}{topographical surveyor, survey analyst, survey manager, survey director}
% \textcolor{purple}{project management manager, senior project manager, e-business project manager} \\  \hline

% \textcolor{red}{vice president of advertising, channel director, vice president of news, associate publisher}
% \textcolor{blue}{director of reimbursement, group managing director, business director}
% \\ \hline
\textcolor{black}{customer logistics manager, import manager, export sales manager, export specialist}
% \textcolor{blue}{dairy nutritionist, dairy consultant }
\\ 
%\hline
\bottomrule
\end{tabular}
\label{table:job_title_community}
\vspace{-3mm}
\end{table}

Most importantly, this clusterness demonstrates that mobility in occupational identity is usually limited to the same job category, partially because occupational identity directs us to behave as a member of particular group. For example, a Professor in Computer Science rarely becomes a software engineer even though s/he may be able to do so for greater material return.

\paragraph{Directionality in Self-reported Job Transitions:}
We identify the most common title transitions (changing from Job\textit{-i} to Job\textit{-j} in either direction) in Table \ref{table:job_transition_big} (top), and find that the most frequent transitions relate to title revisions (e.g., \textit{vice president} to \textit{vp}). 
Furthermore, we define the transition {\em directionality}, the fraction of directed transitions from $v_i$ to $v_j$, as :
\begin{equation}
    dir_{i,j} =\frac{w_{i,j}}{w_{i,j} + w_{j,i}} \in [0,1],
\end{equation}

% A job transition of high directionality indicates that Twitter users may want to update their job status 

We sort the transitions by  $dir_{i,j}$ in Table \ref{table:job_transition_big} (middle), showing that the most directed transitions are related to job promotion and reflect a significant role change (eg., \textit{"student to nurse"}, \textit{"postdoc to professor"}). 
Meanwhile, among the \textit{less directed transitions}, most of the job titles are interchangeable with subtle difference in specificity (\textit{scientist} to \textit{neuroscientist}). 
% The discovery of the directionarlity could further quantify the career path over Twitter, 
The discovery of this directionality better quantifies the relationships between jobs, and is helpful for understanding job mobility.

\begin{table}[h]
    \centering
    % \small
    % \footnotesize
    % \scriptsize
    
    \caption{The detected occupation transition pairs: generally the transition reflects career changes, graduation (e.g., \textit{student} to \textit{nurse} ) or just rephrasing (e.g., \textit{vice president} to \textit{vp}). The directionality indicates the transition is biased toward one way, which usually indicates a positive movement (e.g., being promoted from \textit{manager} to \textit{director}. )
    %The dummy node is ignored. 
    }
% \vspace{-3mm}
\begin{tabular}{p{0.35\textwidth}p{0.35\textwidth}p{0.08\textwidth} p{0.08\textwidth}}
\toprule
                 Title i &                 Title j  &  \#  & $dir_{i,j}$\\
\midrule
\multicolumn{4}{c}{Top Mentioned Transitions}\\
\midrule
    %   founder &     co-founder &        10149 &         0.46 \\
    co-founder &        founder &        10149 &         0.54 \\
           ceo &        founder &         8577 &         0.57 \\
%       founder &            ceo &         8577 &         0.43 \\
    %   founder &          owner &         7384 &         0.41 \\
        owner &        founder &         7384 &         0.59 \\
        %   ceo &          owner &         5917 &         0.47 \\
        owner &            ceo &         5917 &         0.53 \\
%  founder \& ceo &        founder &         3290 &         0.51 \\
%       founder &  founder \& ceo &         3290 &         0.49 \\
      director &        founder &         2987 &         0.57 \\
%       founder &       director &         2987 &         0.43 \\
vice president &             vp &         2734 &         0.66 \\
%            vp & vice president &         2734 &         0.34 \\
vice president &      president &         2545 &         0.76 \\
\midrule
\multicolumn{4}{c}{Highly Directed Transitions}\\
\midrule
           student nurse &         staff nurse &          109 &         0.99 \\
                %  student &      medical intern &          196 &         0.99 \\
         nursing student &    registered nurse &          217 &         0.98 \\
     assistant professor & associate professor &         2154 &         0.97 \\
     postdoctoral fellow & assistant professor &          136 &         0.96 \\
         account manager &    account director &          163 &         0.96 \\
  Sr. account manager &    account director &          105 &         0.96 \\
     editorial assistant &    assistant editor &          163 &         0.94 \\
     instructional coach & assistant principal &          154 &         0.94 \\
    %       student nurse &    registered nurse &          122 &         0.93 \\
%      marketing assistant & marketing exec. &          106 &         0.93 \\
%                  student &       private pilot &          145 &         0.93 \\
        associate editor &       senior editor &          174 &         0.92 \\
%         assistant editor &    associate editor &          264 &         0.91 \\
Sr. account exec. &     account manager &          102 &         0.91 \\
\midrule
\multicolumn{4}{c}{Less Directed Transitions}\\
\midrule
                    director &            business director &          142 &         0.50 \\
        %   business director &                     director &          142 &         0.50 \\
                   scientist &               neuroscientist &          126 &         0.50 \\
            %   neuroscientist &                    scientist &          126 &         0.50 \\
             assistant coach &   asst. basketball coach &          110 &         0.50 \\
            graphic designer &                 video editor &          110 &         0.50 \\
                % video editor &             graphic designer &          110 &         0.50 \\
%   asst. basketball coach &              assistant coach &          110 &         0.50 \\
                         cto &                        owner &          110 &         0.50 \\
%                       owner &                          cto &          110 &         0.50 \\
          software developer &                ios developer &          102 &         0.50 \\
%               ios developer &           software developer &          102 &         0.50 \\
            digital marketer & digital marketing spec. &          102 &         0.50 \\
% digital marketing spec. &             digital marketer &          102 &         0.50 \\
\bottomrule
\end{tabular}
    \label{table:job_transition_big}
\end{table}

% \begin{wrapfigure}{r}{0.950\textwidth}
% \setlength\intextsep{0pt}
\begin{figure}[ht]
\centering
    % \vspace{-2mm}
\includegraphics[width=0.6\textwidth]{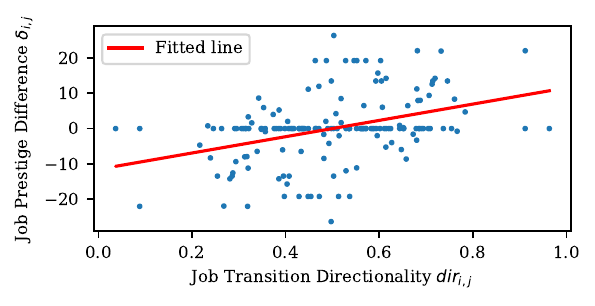}
    % \vspace{-10mm}
    \caption{Highly-directed job transitions generally tend towards more prestigious positions. Some prestige scores do not change because of the limited granularity of the job prestige data (e.g., \textit{assistant prof.} and \textit{associate prof.} have the same job category and prestige).  }
    \label{fig:job-trans-presige}
    % \vspace{-15mm}
\end{figure}

% \end{wrapfigure}

\paragraph{Directionality Correlates with Job Prestige:}
One possible hypothesis is that job updates always reflect increases of job prestige. 
We calculate the prestige difference $\delta_{(i,j)}$ within each job transition pair, and plot against the transition directionality in Figure \ref{fig:job-trans-presige}, where we observed a clear positive correlation with Pearson's $\rho =0.3686$. 
This finding supports that \textit{"we strive for positive work-related allegiances, enhance well-being ... and promote career development"}, as suggested by \citet{gecas1982self} and echoed in \cite{bartel2001ambiguous, cheng2008connecting, ibarra1999provisional}.

% The above analysis demonstrates the occupational transition reflect 

\xingzhiswallow{
    
    \todo[inline]{how transitions are. 
    
    %- moran's I for prestige. 
    
    - desc dataset in detail.
    
    - transition from job to no job. 
    
    - none to student to entry level.
    
    - Exit level job.
    
    What is the 10-most entry/exit level job. 
    
    navy seal --> in lifetime.
    
    -x: twitter age. 
    -y (ratio-job) : two line: nothing-> job, job-> nothing, job -> job.nothing to nothing. per country.
    
    - add a case study: per-country analysis. 
    (add a timeline, tell a good story )
    
    }

}

\newpage
\subsection{Case Study: Transitions in/out Occupational Identities Space amid COVID-19}
\label{sec:case-study}

\begin{table}[ht]
% \begin{wraptable}{r}{0.95\textwidth}
    \centering
    \small
    % \vspace{-5mm}
    \caption{Transition Types. job->none and none->job transitions share nearly equal proportions of total transitions. }
    % \vspace{-3mm}
\begin{tabular}{p{0.15\textwidth} p{0.15\textwidth} p{0.15\textwidth}}
    \toprule
     Type &  Counts &  Proportion \\
    \midrule
       job->job &   847,412 & 0.23\\
      job->none &    1,402,957 & 0.38\\
      none->job &    1,446,278 & 0.39\\
    \bottomrule
    \end{tabular}
    % \vspace{-4mm}
    \label{tab:job-transition-type}
\end{table}
% \end{wraptable}

We performed a case study to qualify how users move in and out of occupational identity space. \versiontwo{In order to cover the complete pandemic period, we augmented our dataset with full year data for 2022}.
%as described in Appendix \ref{sec:app-2022-data}
With regards to the graph structure, the goal was to identify which jobs are most overrepresented in (none->job) and (job->none) transitions.
Table \ref{tab:job-transition-type} shows the breakdown of these transitions by type.
Among 13,980,687 unique U.S. users there were 1,602,731 ($\sim$11.5\%) who listed a recognizable job title in their bio at some point during the observation period. 
%These users performed an average of 2.31 transitions each across the full observation period. This implies that many changes into and out of occupational identity space consist of "flip-flops".

% \begin{wraptable}{r}{0.950\textwidth}
\begin{table}[]ht
    \centering
    % \footnotesize
    % \vspace{-5mm}
    \caption{Top entry/exit jobs by raw counts. All job titles follow the same ranking in both categories. %Combined with an average number of transitions per user of 2.31.
    %this implies that many changes into and out of occupational identity space consist of "flip-flops" (discussed further in section 6.3)
    }
    % \vspace{-4mm}
\begin{tabular}{p{0.1\textwidth} | p{0.25\textwidth} p{0.1\textwidth} | p{0.25\textwidth} p{0.1\textwidth}}
    \toprule
     Rank &  Entry Jobs  & Counts  &  Exit Jobs & Counts \\
    \midrule
      1 &   Student & 84,489 & Student & 95,848     \\
      2 &   Owner & 65,162 & Owner & 60,284         \\
      3 &   President & 58,465 & President & 57,441 \\
      4 &   Founder & 51,650 & Founder & 49,124     \\
      5 &   CEO & 49,004 & CEO & 43,527             \\
      % 6 &   YouTuber & 33,312 & YouTuber & 36,295   \\
      % 7 &   Director & 33,992 & Director & 33,191     \\
      % 8 &   Prover & 22,209 & Prover & 22,047       \\
      % 9 &   VP & 21,877 & VP & 21,800               \\
      % 10 &  Partner & 21,758 & Partner & 18,418     \\
    \bottomrule
    \end{tabular}
    \label{tab:entry_exit_by_count}
    \vspace{3mm}
\end{table}
% \end{wraptable}

% \todo{maybe add this desription to appendix. say something like: To update the results with full 2022 year data, we add additional case study/analysis as an interesting case study. 
% }

 Table \ref{tab:entry_exit_by_count} shows that jobs share the highest entry/exit counts, revealing that the top occupations in each category are the same, and even follow the same ranking order. 
%(see Table \ref{tab:entry_exit_by_count} in Appendix \ref{sec:app-2022-data}) 
Combined with the average number of transitions per user being just over two, this indicates that a large portion of these transitions consist of "flip-flops", where a user changes their bio away from a listed occupation, perhaps for something seasonal or topical, and then returns to their original bio by the next observation. 
For the purposes of this analysis such flip-flops are strictly a source of noise: to compensate for this, raw counts were abandoned in favor of net changes as defined in equation \ref{eq:net-change}. Overrepresented job titles were only considered if they appeared in more than 1/10000 occupational identities.
%as number of entries minus number of exits for a given job over a specified observation period. 
% Table \ref{tab:net_change_jobs} shows the most overrepresented entry/exit jobs over the full observation period.  

% \vspace{-6mm}
\begin{align}
    \delta_{net} (Job) = \#(None \rightarrow Job) -  \#(Job \rightarrow None) 
    \label{eq:net-change}
\end{align}
% \vspace{-3mm}

% \begin{wraptable}{r}{0.95\textwidth}
\begin{table}[htbp]
    \centering
    \footnotesize
    % \vspace{-5mm}
    \caption{Top 10 Overrepresented entry/exit jobs. \versiontwo{ Positive net change $\delta_{net} (Job) >0 $: we observed more people establishing these work identities than giving them up.  }}
    % \vspace{-3mm}
\begin{tabular}{p{0.03\textwidth} | p{0.35\textwidth} p{0.35\textwidth}}
    \toprule
     Rank &  Entry Jobs  &  Exit Jobs \\
    \midrule
       1 &   Esthetician & Soccer Player\\
      2 &    Licensed Esthetician & Multimedia Journalist\\
      3 &    Assistant Baseball Coach & Editor-in-Chief\\
      4 &   First Responder & Independent Consultant\\
      5 &   Doula   & Public Relations \\
      6 & Health Advocate & Snowboarder \\
      7 & Tattoo Artist & Student \\
      8 & Dominatrix & Song Writer \\
      9 & Assistant Professor & YouTuber \\
      10 & Forex Trader & Associate Professor \\
    \bottomrule
    \end{tabular}
    \label{tab:net_change_jobs}
    \vspace{3mm}
\end{table}
% \end{wraptable}

\paragraph{\versiontwo{Personal care and service jobs are rising}}: Table \ref{tab:net_change_jobs} presents the most overrepresented entry/exit jobs over the full observation period. Half of the top 10 entry jobs fall under the category of personal care and service occupations, including esthetician, licensed esthetician, doula, health advocate, and tattoo artist. According to the U.S. Bureau of Labor and Statistics this group of jobs is projected to grow 14\% from 2021 to 2031, nearly double the growth rate of the job market overall \footnote{https://www.bls.gov/ooh/personal-care-and-service/home.htm}. Other titles within this list (dominatrix, forex trader) correspond to more recent growth trends, particularly during the pandemic \footnote{https://www.nytimes.com/2021/04/10/style/findom-kink.html} \footnote{https://financialit.net/news/people-moves/ironfx-hires-200-employees-after-massive-rise-forex-accounts-amid-covid-19}.

\xingzhiswallow{
    \begin{figure}
        \centering
        \subfloat[The number of unique entry job titles is consistently larger than the number of unique exit job titles. 
        However, this relationship begins to falter during the COVID-19 pandemic, indicating that people were removing a wider variety of jobs from their occupational identities during the many employment disruptions surrounding the pandemic and associated lockdowns.]{
        \includegraphics[width=0.40\textwidth]{Figures/fig_entry_exit_jobs.pdf}
        }
        \subfloat[]{
        \includegraphics[width=0.45\textwidth]{Figures/fig_occupational_identity_user_proportion_single_sample.pdf}
        }
        \caption{Caption}
        \label{fig:my_label}
    \end{figure}
    
}

\paragraph{\versiontwo{\textit{Aspirational} identities are easier to be abandoned}}: The top 10 exit jobs largely correspond to media positions (multimedia journalist, editor-in-chief), sports positions (soccer player, snowboarder), and entertainment (song writer, YouTuber). These reflect a category which can be described as "aspirational jobs"; careers with low barriers to entry and title adoption but higher barriers for conversion to a long-term professional vocation. These aspirational jobs are also congruent with the most desired jobs among both children and teens according to several surveys \footnote{https://arstechnica.com/science/2019/07/american-kids-would-much-rather-be-youtubers-than-astronauts/} \footnote{https://www.thesun.co.uk/news/3617062/children-turn-backs-on-traditional-careers-in-favour-of-internet-fame-study-finds/} \footnote{https://today.yougov.com/topics/technology/articles-reports/2021/12/14/influencer-dream-jobs-among-us-teens} 
Aspirational jobs are clearly over-represented on Twitter with respect to self-presented occupational identity.
Other titles on the list, such as student, represent a position traditionally intended to be temporary. In these contexts it matches expecations to see users dropping these titles at a higher rate than more conventional long-term occupations.
%
% \begin{wrapfigure}{R}{0.95\textwidth}
% \setlength\intextsep{0pt}
\begin{figure}[ht]
\centering
    % \vspace{-5mm}
    \includegraphics[width=0.5\textwidth]{Figures/fig_entry_exit_jobs.pdf}
    % \vspace{-5mm}
    \caption{The number of unique entry job titles is consistently larger than the number of unique exit job titles. 
    However, this relationship begins to falter during the COVID-19 pandemic, indicating that people were removing a wider variety of jobs from their occupational identities during the many employment disruptions surrounding the pandemic and associated lockdowns.
    }
    \label{fig:job-entry-exit-pools}
    \vspace{4mm}
    \includegraphics[width=0.55\textwidth]{Figures/fig_occupational_identity_user_proportion_single_sample.pdf}
    % \vspace{-10mm}
    \caption{Among the cohort who have historically included an occupational identity there is a clear local minimum corresponding to the peak U.S. unemployment rate at the beginning of the COVID-19 pandemic.
    %an effect which is not seen in the total user group. 
    This indicates that the reported status of people who already subscribed to an occupational identity does share some correlation with actual employment status.
    }
    \label{fig:occupational_identity_user_single_sample}
    % \vspace{-8mm}
\end{figure}

% Another interesting phenomena was observed regarding the number of entry/exit nodes. 

\paragraph{\versiontwo{Certain types of entry jobs significantly declined amid COVID-19}}: Figure 7 shows the changes in entry/exit job pool sizes over time. When considering unique job titles, the pool of possible entry jobs is consistently larger than the pool of possible exit jobs. Put another way, there are more unique ways to enter into occupational identity space than there are to leave it. This relationship holds true across the full observation period, but during the course of the COVID-19 pandemic the pool sizes draw significantly closer together, with the relationship nearly reversing during peak unemployment \footnote{as reported by the U.S. Bureau of Labor and Statistics, https://www.bls.gov/charts/employment-situation/civilian-unemployment-rate.htm}. 
COVID-19 exacerbated trends which have not yet recovered.

% \begin{wrapfigure}{R}{0.45\textwidth}
% \setlength\intextsep{0pt}
% \centering
%     \vspace{-10mm}
%     \includegraphics[width=0.45\textwidth]{Figures/fig_occupational_identity_user_proportion_single_sample.pdf}
%     \vspace{-10mm}
%     \caption{Among the cohort who have historically included an occupational identity there is a clear local minimum corresponding to the peak U.S. unemployment rate at the beginning of the COVID-19 pandemic.
%     %an effect which is not seen in the total user group. 
%     This indicates that the reported status of people who already subscribed to an occupational identity does share some correlation with actual employement status.
%     }
%     \label{fig:occupational_identity_user_single_sample}
%     % \vspace{-3mm}
% \end{wrapfigure}

%\todo{re-plot Fig 7, make scale right.}

% \begin{figure}[ht]
%     \centering
%     % \vspace{-5mm}
%     \includegraphics[width=0.45\textwidth]{Figures/entry_exit_job_pools.png}
%     % \vspace{-6mm}
%     \caption{The number of unique entry job titles is consistently larger than the number of unique exit job titles. However, this relationship begins to falter during the COVID-19 pandemic, indicating that people were removing a wider variety of jobs from their occupational identities during the many employment disruptions surrounding the pandemic and associated lockdowns.}
% \end{figure}

\paragraph{\versiontwo{Occupational Identity Rates During COVID-19}}:
\xingzhiswallow{
    
    A further question can be raised: how did the COVID-19 pandemic affect people's expression of occupational identity in the aggregate? To address this the full sample was split into two cohorts with important differences. The first cohort consisted of all users, including a large majority (>88\%) who never reported an occupational identity across the entire observation period. The second cohort was a subset of the first, including only users who had at least one transition (minimum two observations) and listed an occupational identity at least once within the observation period. These cohorts are referred to as total users and occupational identity users, respectively. 
    
    Figure \ref{fig:total_user_single_sample} shows the proportions of valid job titles among the total user cohort, and

    \begin{wrapfigure}{R}{0.45\textwidth}
    \setlength\intextsep{0pt}
    \centering
        \vspace{-2mm}
        \includegraphics[width=0.40\textwidth]{Figures/fig_total_user_proportion_single_sample.pdf}
        \vspace{-5mm}
        \caption{Adoption of an occupational identity among the total user cohort shows a monotonic increase over the full sampling period. There is no noticeable response in regards to unemployment values.}
        \label{fig:total_user_single_sample}
        \vspace{-3mm}
    \end{wrapfigure}

}
Figure \ref{fig:occupational_identity_user_single_sample} shows the proportions of valid job titles among the users who exhibit at least one transition (minimum two observations) and listed an occupational identity at least once within the observation period.
% occupational identity cohort.
%These results show that adoption of occupational identities are on the rise in the aggregate.
%, showing monotonic growth across the total user cohort. Among the subset cohort who listed an occupational identity at least once there are more than 60\% of users who include a recognizable job title at any given time. This indicates a level of stability: most users who have historically described themselves via an occupational identity largely tend to continue to do so. 
Users with occupational identities show a greater sensitivity to actual employement rates, an effect which disappears in the total user cohort. 
%The driving factors for this are explored in more detail in Appendix \ref{sec:app-2022-data}. 
Although in this paper our main goal is not to conduct real-world job market reserach using Twitter data, these aforementioned case study results demostrate that our proposed dataset is a reliable and useful resource for the future work-related studies.

\xingzhiswallow{
\begin{table}[htbp]
    \centering
    \small
    \caption{Each cluster (row) contains various highly-related job titles (manually rendered in colors) . }

\begin{tabular}{p{0.50\textwidth}}
\toprule
 \centering\arraybackslash \textbf{Sampled Titles in Job Clusters} (350 discovered clusters, $\gamma=8$) \\ \hline
\midrule
 \centering\arraybackslash Cluster size : 2 (95/350) \\ 
\midrule
\textcolor{red}{health information director, health information technician} \\ \hline
% \textcolor{red}{certified registered locksmith, certified master locksmith} \\ \hline
\textcolor{red}{consumer lending manager, secondary market manager} \\ \hline
\textcolor{red}{loader operator, front end loader operator} \\ \hline
\midrule
 \centering\arraybackslash Cluster size: 3 - 30 (101/350)\\ 
\midrule
\textcolor{red}{regional sales coordinator, district sales coordinator, special projects coordinator },
\textcolor{blue}{highway maintenance worker, bridge inspector, traffic worker, toll collector, highway inspector}
\\ \hline
\textcolor{red}{privacy director, privacy analyst algorithm developer}, 
\textcolor{blue}{hadoop developer, hadoop architect, linux architect, windows server administrator, web systems developer, mainframe systems administrator}
\\ \hline
% \textcolor{red}{university administrator, university president, graduate school dean, director of campus recreation},
% \textcolor{blue}{senior strategy manager, program analyst, business system consultant, senior program analyst }
% \\ \hline

\textcolor{red}{occupational safety specialist, safety and health manager, safety trainer},
\textcolor{blue}{family worker, community board member, community engagement coordinator},
\textcolor{purple}{professor of early childhood education, peer educator, speech language pathologist assistant},
% \textcolor{pink}{asp developer, digital product specialist, tool builder, code official, electrician}
\\ \hline
\midrule
 \centering\arraybackslash Cluster size: 31 - 60 (71/350)\\ 
\midrule
\textcolor{red}{clinical assistant, clinic assistant, clinical associate, psychiatric assistant}
\textcolor{blue}{communication instructor, language specialist, public speaking teacher}
\textcolor{purple}{software development leader, information director}
\\ \hline

\textcolor{red}{building consultant,  building surveyor},
\textcolor{blue}{topographical surveyor, survey analyst, survey manager, survey director}
\textcolor{purple}{project management manager, senior project manager, e-business project manager}
\\ 
\hline

% \textcolor{red}{vice president of advertising, channel director, vice president of news, associate publisher}
% \textcolor{blue}{director of reimbursement, group managing director, business director}
% \\ \hline
\textcolor{red}{customer logistics manager, import manager, export sales manager, export specialist, poultry buyer,}
\textcolor{blue}{dairy nutritionist, dairy consultant }
\\ \hline

\bottomrule
\end{tabular}
    \label{table:job_title_community}
\end{table}
}

\section{Conclusions and Future Work}
\label{future-work}
% We have demonstrated the predictive power of personal self-description (bio) on social media with the task of gender, ethnicity and personal interest classification. 
%We empirically observe that the information encoded in a single bio is equivalent to approximately 16, 8 and 2 tweets for the tasks aforementioned, respectively. 

We have presented a study of occupational identities and their evolution in Twitter biographies from 2015 to 2021, and beyond. 
First, we described a new temporal self-identity dataset, consisting of extracted job titles with associated metadata (e.g., income, prestige score), and will release the occupation transition graph and the embeddings of biographies on publication.
%\footnote{Due to privacy and term-of-service concern, we currently do not release explicit bio text} on publication.   
Secondly, we have quantitatively demonstrated that the high prestige jobs are more likely to be embedded into self-identity, and that occupational transitions are biased towards more prestigious jobs. 
In addition, we presented an interesting case study about how work identities changed during COVID-19, and demonstrated that our dataset captures intriguing real-world phenomenon. 
To our knowledge, this is the most comprehensive study of social media biography dynamics to date. 
%We discover interesting life transition patterns (e.g., graduation, job promotion), and identify the degree of directionality of occupation transitions. 
% Most importantly, we quantify the prestige biases in the over-represented occupations and the directed job transitions.
%
We envision future work building on our results in several directions:
1). Deriving job title embeddings from the occupational graph, to better describe relations among job titles even without any textual similarity, as shown in Table \ref{table:job_title_community}. 
% We could derive node embeddings as more accurate and comprehensive representations for machine leaning-based job category classification.
2). Studying the persistence of self-identities over time. 
We are curious about which identities are ephemeral/eternal and the mechanisms behind self-description revisions.
Occupational identities may change more frequently compared to other identities (e.g., \textit{Father/Mom,...}).

% \todo{Need more detail about future work.}
%predict what occupation transition will happen in the future, 
% \begin{itemize}
%     \item Life event detection, identifying the recent life changes from the bio editing. e.g.: $man\rightarrow father, engaged\rightarrow married, promotion: director\rightarrow chair$, ...

%     \item Life event prediction: forecasting what will change in next 4 years ($student\rightarrow candidate, graduate\rightarrow engaged$, ...)

%     \item life trajectory clustering: looking for the common pattern of the life changes, cluster similar users together.
% \end{itemize}

% \bibliographystyle{ACM-Reference-Format}
% \bibliography{acmart}

% \newpage

\section{Appendix}
\xingzhiswallow{

    \section{Supplementary data set for the year 2022}
    \label{sec:app-2022-data}
    The dataset for the analyses in sections 6.3 and 6.4 was extended to include data from Feb. 2015 to Dec. 2022 with the intuition that interesting phenomena might be observable during the COVID-19 pandemic. Table \ref{tab:entry_exit_by_count} shows the top 10 entry/exit jobs aggregated over this full observation period by raw counts.
    
    % \begin{table}[htbp]
    \begin{wraptable}{r}{0.5\textwidth}
        \centering
        \footnotesize
        \vspace{-1mm}
        \caption{Top entry/exit jobs by raw counts. All job titles follow the same ranking in both categories. Combined with an average number of transitions per user of 2.31, this implies that many changes into and out of occupational identity space consist of "flip-flops" (discussed further in section 6.3)
        }\vspace{-3mm}
    \begin{tabular}{p{0.03\textwidth} | p{0.1\textwidth} p{0.08\textwidth} | p{0.10\textwidth} p{0.08\textwidth}}
        \toprule
         Rank &  Entry Jobs  & Counts  &  Exit Jobs & Counts \\
        \midrule
          1 &   Student & 84,489 & Student & 95,848     \\
          2 &   Owner & 65,162 & Owner & 60,284         \\
          3 &   President & 58,465 & President & 57,441 \\
          4 &   Founder & 51,650 & Founder & 49,124     \\
          5 &   CEO & 49,004 & CEO & 43,527             \\
          % 6 &   YouTuber & 33,312 & YouTuber & 36,295   \\
          % 7 &   Director & 33,992 & Director & 33,191     \\
          % 8 &   Prover & 22,209 & Prover & 22,047       \\
          % 9 &   VP & 21,877 & VP & 21,800               \\
          % 10 &  Partner & 21,758 & Partner & 18,418     \\
        \bottomrule
        \end{tabular}
        \label{tab:entry_exit_by_count}
        % \vspace{-3mm}
    % \end{table}
    \end{wraptable}
    
}

\xingzhiswallow{

    \section{Biannual Sample}
    An addtional sample consisting of more active Twitter users was also considered as a point of comparison for the analysis in Sec \ref{sec:case-study}. The first sample is still the same user set from previous analyses, sampled directly from the 1\% tweet stream with no concern over how frequently each user appears. The second sample is a precompiled longitudinal group where each user appeared in the tweet stream at least once biannually. 
    This latter group represent more active users, indicated by their consistent presence in the randomly sampled tweet stream. These samples are referred to as the full sample and the biannual sample, respectively.
    Each sample was further split into two cohorts following the procedure outlined in section 6.4. Figure \ref{fig:total_user_proportion_two_sample} shows the changes in observed rates of valid job titles over time for the full user cohort for both of these samples, and figure \ref{fig:occupational_identityt_user_proportion_two_sample} shows these changes for the occupational identity cohort for both of these samples.
    
    \begin{figure}[ht]
        \centering
        % \vspace{-5mm}
        \subfloat[\label{fig:total_user_proportion_two_sample}]{
            \includegraphics[width=0.25\textwidth]{Figures/fig_total_user_proportion.pdf}
        }
        \subfloat[\label{fig:occupational_identityt_user_proportion_two_sample}]{
            \includegraphics[width=0.28\textwidth]{Figures/fig_occupational_identity_user_proportion.pdf}
        }
        \caption{
        \textbf{Left:}The total user cohort for both samples shows a steady increase in adoption of occupational identities over time, with the full sample nearly doubling the proportion of users with valid job titles over the full observation period. The full sample shows a monotonic increase over time, and the biannual sample shows a nearly monotonic increase. This cohort does not show any significant response to the peak unemployment of the pandemic in either of the samples. 
        \textbf{right}: The occupational identity user cohort for both samples shows more variation between the samples. Here the proportion of the biannual sample does not show a clear trend over time, likely reflecting the more frequent bio changes from this more active sample group. Both samples have a local minimum during peak unemployment, indicating that the reported status of people who already subscribed to an occupational identity is correlated with actual employement status.}
    
    \end{figure}
}
\xingzhiswallow{
    \begin{figure}[ht]
        \centering
        % \vspace{-5mm}
        \includegraphics[width=0.45\textwidth]{Figures/fig_occupational_identity_user_proportion.pdf}
        % \vspace{-6mm}
        \caption{The occupational identity user cohort for both samples shows more variation between the samples. Here the proportion of the biannual sample does not show a clear trend over time, likely reflecting the more frequent bio changes from this more active sample group. Both samples have a local minimum during peak unemployment, indicating that the reported status of people who already subscribed to an occupational identity is correlated with actual employement status.}
        \label{fig:occupational_identityt_user_proportion_two_sample}
    \end{figure}
    
}

\xingzhiswallow{

\section{Full Unemployment Response}
Figure \ref{fig:occupational_identity_user_proportion_with_unemployment} show the different cohorts of the full sample plotted directly over the relevant U.S. unemployment rates. It appears that the response of the occupational identity cohort is not just in regards to actual unemployment rates; otherwise one would expect to see a more continuous dip as opposed to the immediate return to the increasing trend. Rather, the change in job title rates seems to be more accurately impacted by the immediate changes in unemployment rates as opposed to the raw values.

    \begin{figure}[ht]
        \centering
        % \vspace{-5mm}
        \includegraphics[width=0.45\textwidth]{Figures/fig_occupational_identity_user_with_unemployment_single_sample.pdf}
        % \vspace{-6mm}
        \caption{The dip in the inclusion of valid job titles seems to correspond more closely with a rapid change in unemployment rates as opposed to the rates themselves. After the initial shock of the pandemic the inclusion of an occupational identity quickly returns to its pre-pandemic growth trend}
        \label{fig:occupational_identity_user_proportion_with_unemployment}
    \end{figure}

}

    \begin{table*}[htb]
        \tiny
        \centering
        \begin{tabular}{p{0.18\textwidth}p{0.03\textwidth}p{0.18\textwidth}p{0.03\textwidth}p{0.18\textwidth}p{0.03\textwidth}p{0.18\textwidth}p{0.03\textwidth}}
        \toprule
                               White &     Ratio &                    Asian &     Ratio &                      Hispanic &     Ratio &                          African-American &     Ratio \\
        \midrule
        \multicolumn{8}{c}{Over-represented} \\
        \midrule
                  golf professional &                   2.12 &                research scholar &                   6.60 &             industrial engineer &                      3.48 &               performing artist &                  1.82 \\
                         golf coach &                   2.10 &            chartered accountant &                   5.76 &               volleyball player &                      2.30 &                    forex trader &                  1.81 \\
                     pitching coach &                   2.09 &               political analyst &                   4.08 &          industrial engineering &                      2.18 &            motivational speaker &                  1.79 \\
          assistant superintendent &                   2.07 &                   social worker &                   3.82 &                   soccer player &                      2.14 &               radio personality &                  1.75 \\
             social studies teacher &                   2.07 &            electronics engineer &                   3.49 &               medical assistant &                      1.74 &                    music artist &                  1.73 \\
                  athletic director &                   2.06 &                     it engineer &                   3.43 &            full stack developer &                      1.67 &                recording artist &                  1.71 \\
                         pe teacher &                   2.06 &               android developer &                   3.38 &               computer engineer &                      1.66 &                      rap artist &                  1.71 \\
                     baseball coach &                   2.05 &                 it professional &                   3.21 &                flight attendant &                      1.64 &                     song writer &                  1.70 \\
          assistant baseball coach &                   2.04 &                  prime minister &                   3.18 &                systems engineer &                      1.63 &                      event host &                  1.67 \\
                    sports director &                   2.04 &             electrical engineer &                   2.94 &                  medical doctor &                      1.57 &                 college athlete &                  1.67 \\
                     softball coach &                   2.03 &             mechanical engineer &                   2.92 &                     mma fighter &                      1.54 &                   event planner &                  1.65 \\
                  volleyball coach &                   2.02 &                  content writer &                   2.81 &          environmental engineer &                      1.51 &               basketball player &                  1.61 \\
                      hockey player &                   2.01 &                       secretary &                   2.78 &               community manager &                      1.50 &                  sports analyst &                  1.61 \\
         director of comm. &                   2.00 &                   growth hacker &                   2.78 &                fashion designer &                      1.49 &                 fashion stylist &                  1.60 \\
                  director of sales &                   2.00 &                  civil engineer &                   2.76 &                       networker &                      1.47 &                  hip hop artist &                  1.58 \\
        \midrule
        \multicolumn{8}{c}{Under-represented}\\
        \midrule
                   research scholar &                   0.84 &                  softball coach &                   0.17 &               golf professional &                      0.25 &                research scholar &                  0.74 \\
               chartered accountant &                   0.98 &        assistant football coach &                   0.18 &                  support worker &                      0.29 &             industrial engineer &                  0.76 \\
                industrial engineer &                   1.07 &        asst. baseball coach &                   0.18 &      asst. basketball coach &                      0.30 &              software architect &                  0.77 \\
              electronics engineer &                   1.12 &                  baseball coach &                   0.19 &          men's basketball coach &                      0.31 &             assistant professor &                  0.81 \\
                  android developer &                   1.17 &               athletic director &                   0.20 &                      golf coach &                      0.33 &             postdoctoral fellow &                  0.83 \\
                  political analyst &                   1.17 &                volleyball coach &                   0.21 &             development officer &                      0.33 &                 senior director &                  0.83 \\
                        it engineer &                   1.27 &      asst. basketball coach &                   0.21 &                 primary teacher &                      0.34 &        assistant superintendent &                  0.83 \\
                     medical doctor &                   1.29 &           offensive coordinator &                   0.22 &               athletic director &                      0.36 &                   ios developer &                  0.83 \\
                  computer engineer &                   1.29 &        assistant superintendent &                   0.22 &          recruitment consultant &                      0.37 &                  pitching coach &                  0.84 \\
                      social worker &                   1.30 &                      golf coach &                   0.23 &                 chief executive &                      0.37 &                    sr. director &                  0.84 \\
                         dominatrix &                   1.31 &                  pitching coach &                   0.23 &        assistant football coach &                      0.38 &             associate professor &                  0.85 \\
                     civil engineer &                   1.32 &                     track coach &                   0.23 &          primary school teacher &                      0.38 &        senior software engineer &                  0.85 \\
              full stack developer &                   1.32 &          recruiting coordinator &                   0.23 &                        tar heel &                      0.39 &             engineering manager &                  0.85 \\
                electrical engineer &                   1.34 &               golf professional &                   0.23 &                 radio presenter &                      0.39 &                systems engineer &                  0.86 \\
                  volleyball player &                   1.34 &             assistant principal &                   0.24 &                    estate agent &                      0.40 &           senior vice president &                  0.87 \\
        \bottomrule
        \end{tabular}
            \caption{Top over/under-represented occupations in each ethnicity group (Occupation mentioned time $<$ 60): The column \textit{Ratio} measures how many time the calculated value exceeds the expected balanced value.  We break ties by sorting occupation mentioned times in descending order. }
            \label{tab:ethnicity_over_under_occ_long}
    \end{table*}

% }

\chapter{Conclusion and Future Work}

In this thesis, we focus on the analysis of evolving graphs and texts. Specifically, we proposed a faster dynamic network embedding algorithm (\textit{DynamicPPE}) based on approximate Personalized PageRank, which significantly improves running time by roughly 670 times in our subset node learning setting while maintaining comparable prediction accuracy.

Secondly, we investigate node-level anomaly tracking over dynamic graphs and propose an efficient framework (\textit{DynAnom}) that utilizes PPR changes as anomaly signals and improves the anomaly detection accuracy by roughly 2 times  while being approximately 2.3 times faster.

% Secondly, we investigate node-level anomaly tracking over dynamic graphs and propose an efficient framework (\textit{DynAnom}) that utilizes PPR changes as anomaly signals and improves the anomaly detection accuracy by roughly 2 times while being approximately 2.3 times faster.

Third, we propose \textit{GoPPE}, a framework that combines graph neural networks and PPR-based positional encoding for dynamic graphs. Thanks to recent studies on PPR-equivalent formulation, we can efficiently solve PPR up to $\sim$6 times faster and provide robust node representation under noisy environment, achieving  up to 3 times more accuracy than state-of-the-art (SOTA) baselines.

For the studies on text analysis,  we collect and analyze the post-publication headline changes in major U.S. online news outlets. we discovered that 7.5\% of titles changed at least once after they were published. We devised a taxonomy to automatically characterize the type of post-publication change and found that 49.7\% of changes go beyond benign corrections and updates. Additionally, we find that the maximal spread of news on Twitter happens within 10 hours after publication and therefore a delayed headline correction will come after users have consumed and amplified inaccurate headlines.

Furthermore, we present a study of occupational identities and their evolution in Twitter biographies from 2015 to 2021 and beyond.  We quantitatively demonstrate that high prestige jobs are more likely to be embedded into self-identity, and that occupational transitions are biased towards more prestigious jobs. Additionally, we present an interesting case study about how work identities changed during COVID-19, and demonstrate that our dataset captures intriguing real-world phenomena.

There are three future directions based on our current works as described  below:
\begin{itemize}
    \item  \paragraph{Speeding up PPR approximation under optimization framework}: Local and efficient computation of PPR is fundamental to many downstream tasks and remains an open problem \cite{fountoulakis2022open}. Recent research \cite{zhou2024iterative, chen2023accelerating} has pointed out that the ForwardPush algorithm is essentially a variant of the Gauss–Seidel method. One future direction is to further improve PPR calculation efficiency with its local property under PPR's optimization formulation. The downstream applications include faster knowledge discovery \cite{guo2022hierarchies} and efficient embeddings algorithms for real-time applications, such as Chatbot and online recommendation system \cite{gillespie2020improving, lin2023comet, chiang2014real, guo2019inferring, sultan2022low}.

    \item \paragraph{Categorizing self-presented jobs over social network}    Due to proprietary restrictions and technical difficulties in scraping people's career data from the internet \footnote{e.g., Law case of HiQ Labs v.s. Linkedin about web scraping}, public longitudinal Twitter bios data provides a unique opportunity to study job titles. In particular, one future direction is to build a toolkit to classify arbitrary job titles according to the Standard Occupational Classification (SOC) categories. Many job-related statistics are surveyed based on the SOC category, which consists of about 800 classes. However, many people present their jobs in natural language instead of structured SOC categories.\footnote{ e.g., people would like to describe themselves as \textit{Assistant Professor in CS} instead of \textit{Computer Science Teachers, Postsecondary (25-1021)}} The misalignment between the SOC category and job titles hinders large-scale computational social science studies about jobs. It is valuable to create a pipeline to classify arbitrary job titles into SOC categories based on our collected job title text and the job transition graph \cite{guo2024evolution}.

    \item\ \paragraph{Text edit representation learning}: Building on our previous research on news headline changes, we propose a nine-class taxonomy to better understand modifications to titles. However, this categorization is currently based on a rule-based pipeline. Our goal is to represent edits in a neural way and utilize clustering algorithms to refine the taxonomy and improve classification accuracy. The downstream applications of this research include analyzing modifications to news articles, titles, and social media posts.

\end{itemize}
% 1. To further speed up PPR calculation.
% 2. To build a public job title embeddings for Occupation Classification. 
% 3. To develop edits representation algorithms to better classify news title changes based on our proposed taxonomy. 

%Based on our previous studies on news title changes, we proposed a nine-class taxonomy for understanding the title modifications. However, the categorization is based on a rule-based pipeline. We aim at representing edits in a neural way, and employ cluster algorithms to re-fine the taxonomy and the classification algorithms. The downstream tasks include the modification analysis of news article, title and social network posts.  

\xingzhiswallow{

    \section{Personalized PageRank-based Graph Embeddings}
    The key ingredient in DynamicPPE and its extensions is the localized PPR calculation. Thus, the computational bottleneck depends on the efficiency of the PPR routine and the degree of parallelization.
    
    Recent research\cite{fountoulakis2019variational} has proven that the local push based methods are actually optimizing the objective function with $\ell_1$ regularization:
    
    \begin{align}
        \operatorname{min}_q &\ \epsilon \alpha \| D^{\frac{1}{2}}q \|_{1} + f(q)\\
        f(q) &:= \frac{1}{2} \langle q, Q q \rangle - \alpha \langle s, D^{\frac{1}{2}}q \rangle, \\
        Q &:= D^{-\frac{1}{2}} \left\{ D - \frac{1-\alpha}{2} (D + A) \right\}D^{-\frac{1}{2}}, \\
    \end{align}
    where $D$ is the degree diagonal matrix, $A$ is the , $q$ is the approximate PPV, $\epsilon$ is the approximation hyper-parameter in local push algorithm.
    
    The above findings view the calculation of PPR in the optimization lens, and give the opportunities to apply faster solvers (e.g., FISTA, random coordinate descent) for further speeding up the PPR-based graph embedding algorithms over large graphs. There are potentially three improvements:
    
    \begin{itemize}
        \item Speed up the calculation of dynamic PPR with faster solvers in optimization framework.
        \item Further parallelize the calculation on GPUs: Although the CPU-based implementation can be parallelized by multi-threads, it still suffers from hardware limits ($<$ 100 cores). GPU of massive  cores ($>$4,000 units) would be a good fit to highly parallelize the calculation of the target node set.
        \item Extend the PPR-based algorithms to the feature-enabled dynamic graphs. In particular, introduce PPR as node proximity encoding to tackle heterophily graphs  (e.g.,  \cite{klicpera2018predict}). 
        % \item The benchmark for dynamic graph embeddings.
    \end{itemize}
    
}

% \section{Identity Graph from Social Network}

\xingzhiswallow{

    \section{Categorize Self-presented Jobs over Social Network}
    Since the proprietary restrictions and technical difficulties in scraping people's career data over internet \footnote{e.g., Law case of HiQ Labs v.s. Linkedin about web scraping}, the public longitudinal Twitter bios data gives an unique opportunity to study the job titles. 
    
    In particular,  We want to propose
    \paragraph{Job2SOC}, the toolkit to classify arbitrary job titles according to the SOC occupational category. Many job related statistics are surveyed according to the SOC category (about 800 classes). However, many people present their jobs in natural language instead of structured SOC category.\footnote{ e.g., people would like to describe themselves as \textit{Assistant Professor in CS} instead of \textit{Computer Science Teachers, Postsecondary (25-1021)}} The misalignment between the SOC category and job titles curbs the large-scale computational social science studies about jobs. We aim at building a Job2SOC pipeline to classify arbitrary job titles to SOC categories based on the job title text and the job transition graph.  
}

% Arguably, the stealthy edits are the title/content changes.

% Probe the text summarization models to see which text in main body contributes more to the headlines (called salient content). Then analyze the consistency of the changes between headlines and news content.

% \begin{itemize}
%     \item If headlines got changed, does the salient content get modified to reflect the title change?  
%     \item If headline did not change, but the salient content got modified (arguably stealthy edits), what were the edits (e.g., number changes or rephrase)?
%     \item Do stealthy modifications have more inconsistent ratio? and what are the behavior differences of various news outlets?
% \end{itemize}

% The main idea is to capture the inconsistent changes in headline and content, for example, 
% \section{Dynamic Process (Sergiy)}

% \newpage
% \chapter{Conclusion and Future Directions}

% \bibliographystyle{ACM-Reference-Format}
\bibliographystyle{plainnat}
\bibfont \footnotesize
\bibliography{reference}

\end{document}